\documentclass[11pt]{article}
\usepackage[margin=1in]{geometry}
\usepackage{float}      
\usepackage{placeins} 
\usepackage{booktabs}
\usepackage{amsmath,amsthm,amssymb}
\usepackage{graphicx}
\usepackage{color}
\usepackage{enumerate}
\usepackage{multicol}
\usepackage{tikz}
\usepackage[utf8]{inputenc}
\usepackage{mathrsfs}
\usepackage[
backend=bibtex,
style=numeric,
sorting=ynt
]{biblatex}
\usepackage{titlesec}
\usepackage{enumitem}
\usepackage{longtable}
\titlespacing*{\section}{0pt}{5pt}{5pt}
\titlespacing*{\subsection}{0pt}{2pt}{2pt}

\setlist[enumerate]{topsep=0pt, itemsep=2pt, parsep=0pt, partopsep=0pt}
\setlist[itemize]{itemsep=1pt, topsep=0pt, parsep=0pt}
\newtheorem{axiom}{Axiom}
\numberwithin{axiom}{section}

\newtheorem{incorrectaxiom}[axiom]{Incorrect axiom}
 
\AddToHook{env/incorrectaxiom/after}{\addtocounter{axiom}{-1}}
\newtheorem{theorem}{Theorem}
\newtheorem{definition}{Definition}
\newtheorem{lemma}{Lemma}
\newtheorem{proposition}{Proposition}

\theoremstyle{definition}

\theoremstyle{remark}
\newtheorem*{remark}{Remark}
\DeclareMathOperator{\supp}{supp}

\newcommand{\wt}{w}

\newcommand{\ones}[1]{1_{#1}}

\usepackage[colorlinks=true]{hyperref} 
\usepackage{cleveref}                 
\usepackage{environ}                  
\usepackage{etoolbox}                 

\newif\ifdeferproofs
\deferproofstrue   
\newcounter{appproof}[section]
\renewcommand{\theappproof}{\thesection.\arabic{appproof}}

\newcommand{\AllDeferredProofs}{}

\NewEnviron{deferredproof}[1]{%
  \ifdeferproofs
    \par\noindent\emph{Proof.}%
    \hyperref[prf:#1]{See the proof in Appendix~\ref*{prf:#1}}%
    \par\medskip

    \global\edef\AllDeferredProofs{%
      \unexpanded\expandafter{\AllDeferredProofs}%
      \unexpanded{%
        \par\bigskip
        \refstepcounter{appproof}\label{prf:#1}%
        \subsection*{Proof~\theappproof\; (Proof of \Cref{#1})}%
        \noindent\hyperref[#1]{Back to \Cref*{#1}.}\par\medskip
        \begin{proof}
      }
      \expandafter\unexpanded\expandafter{\BODY}%
      \unexpanded{%
        \end{proof}
      }%
    }%
  \else
    \begin{proof}[Proof of \Cref{#1}]
      \BODY
    \end{proof}
  \fi
}

\author{Wei ShuAng}

\begin{document}

\title{Measure of Morality: A Mathematical Theory of Egalitarian Ethics}

\maketitle
\begin{abstract}
This paper develops a rigorous mathematical framework for egalitarian ethics by integrating formal tools from economics and mathematics. We motivate the formalism by investigating the limitations of conventional informal approaches by constructing examples such as probabilistic variant of the trolley dilemma and comparisons of unequal distributions. Our formal model, based on canonical welfare economics, simultaneously accounts for total utility and the distribution of outcomes. The analysis reveals deficiencies in traditional statistical measures and establishes impossibility theorems for rank-weighted approaches. We derive representation theorems that axiomatize key inequality measures including the Gini coefficient and a generalized Atkinson index, providing a coherent, axiomatic foundation for normative philosophy.
\end{abstract}

How should we weigh the pros and cons of different economic systems like socialism or capitalism? Which rate of marginal tax is ideal under market economy? Is there anything bad with inequality in the distribution of income within and across nation? How should we allocate medical instruments? Should we assign priority to the worst-off patient even though the medication might not be very effective on the patient? should we always save more people?
These topics have haunted and fascinated humans since the inception of thought itself.

Utilitarianism stands out as it supposed to provide a principled answer to all normative questions: we should maximize the total utility \cite{sep-consequentialism}. Utilitarianism tells us that the moral property of act should be solely decided by how it affects the sum of individual’s utility.  Egalitarianism disagrees, the central concern of egalitarianism is how benefits and burdens are distributed across population is also morally important. The core of this paper would be understanding what egalitarianism is, in doing this we would borrow much tool from mathematics and economics.

There had always been strong resistance to employing formal tools in order to preserve the traditional form of philosophy. It might be suggested that we should respect the intellectual division of labor and leave philosophers alone with their thought experiment and conceptual engineering. Concern with formalism should be left with economists or mathematicians. One of the main aim of this paper is to show formalism is not fancy tools for showing off, they are an integral part of the philosophical problem itself. Just as relying solely on natural language would never allow us to grasp the physics underlying many natural phenomena, this paper intend to show in the realm of normative philosophy, there is also large uncharted world to be unlocked by  more rigorous theoretical frameworks.

The focus of this paper is two-fold. First, it aims to provide sufficient background and motivation for the necessity of formalism, transforming traditional ethical discourses in normative philosophy into more concrete and precise frameworks. Second, it examines how egalitarianism can be situated within this formalized framework. The rest would be structured correspondingly: the main part would start with two motivating examples, the first is the familiar trolley problem, we intend to show the conventional wisdom surrounding it would be shaken with minimal formal variable. The second example introduces the problem of unequal inequalities. By introducing utilitarianism as reference Point, We would articulate the challenges facing egalitarianism from this unusual angle. The interim conclusion suggests many ethical theories face problems of incompleteness when it came to articulating their moral reason's implication beyond trivial scenarios that the author arbitrarily developed.

The inconsistency in traditional ethical theory revealed by these examples motivate the introduction of various formal tools and frameworks- Social welfare function in second section. Interpreting SWF as foundations for unifying normative philosophy, Its analytical tools would help illustrate that there exist common defects underlying traditional theories that lead to its problems of incomplete and uninformative that plagues the non-utilitarian theory. We then present some classical results for future reference.
The second part could be understood as test case of formal ethics. The third section would start with minimal framework of egalitarianism, then i would engage with existing approaches to characterizing egalitarianism. The goal is to make sense how the various non-unified framework interconnect with each other and how to best conceptualize the general picture of problems in distributive ethics. My proposal is inequality measurement could be conceptualized as choice between set of functions which extracts the distributional feature of outcomes. 
Following discussion of the challenges within current proposal, the paper culminates in the formulation of two impossibility theorems and two representation theorems. These findings demonstrate that some families of distributive ethics can be logically deduced from non-adhoc axioms, whereas other families exhibit properties that are highly undesirable, thereby disqualifying them as plausible candidates.

\section{Motivating Examples}

\subsection{Probabilistic Trolley Dilemma}
The now world-renowned Trolley problem attempted to illustrate morality is more than maximizing utility through a thought experiment proposed by foot \cite{Foot1967-FOOTPO-2,Thomson1976-THOKLD-2}. An out of control trolley is running along the track towards five people who will be killed if it continues its current path. You are standing next to a lever that can divert the trolley onto a sidetrack, where it will kill one person instead. The question is: should you pull the lever, sacrificing one person to save five? This scenario had been revised to highlight several anti-consequentialist distinctions: Doing vs. Allowing: Is there a moral difference between actively causing harm and passively allowing harm to occur? Intending vs. Foreseeing: Does it matter if the harm caused was a direct intention or merely a foreseen side effect? Negative vs. Positive Rights: is there fundamental difference between the right not to be intervened (negative right) against the obligation to help others (positive right)?  One of another classical revision is proposed by Thomson \cite{thomson1984trolley}: suppose one patient could be killed so his organs could be transplanted to save the lives of another five patient, should you kill him?

Consequentialists had made many excellent defenses, here we purse another strategy by introducing Probabilities as variable. Consider \textbf{Probabilistic Trolley Dilemma:} 
Suppose that if you do nothing, there is a 10\% chance that the five people will survive. If you intervene, there is 50\% chance that five people would survive, but the other one would have 40\% chance of death. To address immediate objections that probability is inadmissible related to guaranteed death in scenarios like organ transplants, we can modify the situation to involve non-lethal interventions, such as blood transfusions or partial kidney removals. These adjustments maintain the moral structure of the problem while mitigating ethical concerns specific to killing.

One possible response to the probabilistic trolley problem is to reaffirm the absolute importance of principles such as negative rights or the doctrine of double effect. Proponents of this view maintain that actively causing harm is always morally impermissible, regardless of the consequences. However, introducing probabilities presents significant challenges to this stance. If moral decisions remain unchanged regardless of varying probabilities, then the position implies that moral choices should be indifferent to benefit and harm, which explains the reason for inaction even when the likelihood of harm changes dramatically. Setting aside ethical considerations, this line of response might not be meaningful once we recall the background fact that risks are unavoidable features of world. Indeed, the common accepted principle is that any logical possibility should be attached to non-zero probability, in the practical realm that concerned ethics the problem is only worse, we would never be certain of any outcome, every epistemically feasible choice facing any individual would involve possible value and negative value. This renders everything incomparable to any other. The second response would along the line of invoking notion like reasonable threshold or potential justification to receiving party of risk \cite{Scanlon1998}, these accounts might pin down to postulate certain cut-off when probability appeared, say when there is more than $20\%$ of violating principle like double effect or negative right, consequentialist reasons would be defeated.

In response we introduced a more generalized framework to reveal the problem, Let \( p \) be the probability that the five will survive if no action is taken, and \( \epsilon_1 \) the increase in their survival probability due to intervention. Similarly, let \( q \) be the probability that the one person will die if no action is taken, and \( \epsilon_2 \) the increase in their probability of dying due to intervention. We can define the expected number of deaths for each option as follows:
\[
\text{Intervention: } 
\mathbb{E} = 
\begin{aligned}&5 \cdot \big[ (p + \epsilon_1)\% \cdot \text{Survival} 
+ \big(1 - (p + \epsilon_1)\big)\% \cdot \text{Death} \big] \\
\quad + &1 \cdot \big[ (q + \epsilon_2)\% \cdot \text{Death} 
+ \big(1 - (q + \epsilon_2)\big)\% \cdot \text{Survival} \big]
\end{aligned}
\]
\[
\text{Do nothing: } 
\mathbb{E} = 
\begin{aligned}
    &5 \cdot \big[ p\% \cdot \text{Survival} + (1 - p)\% \cdot \text{Death} \big] \\
    + &1 \cdot \big[ q\% \cdot \text{Death} + (1 - q)\% \cdot \text{Survival} \big]
\end{aligned}
\]
Assuming consequentialist reasons do have moral weight, the strength of the reasoning for active intervention should depend on \( p \), \( q \), \( \epsilon_1 \), and \( \epsilon_2 \). For example, suppose there exists a certain quantitative threshold 20\% below which consequentialist reasons become admissible, the framework enable us to construct counterexamples to reveal qualitative approach:
\begin{itemize}
    \item Intervention is permissible when \( p = 10\%, p + \epsilon_1 = 20\%, q = 0\%, \text{ and } q + \epsilon_2 = 19\% \).
    \item Intervention is not permissible when \( p = 1\%, p + \epsilon_1 = 99\%, q = 19.9\%, \text{ and } q + \epsilon_2 = 20.01\% \).
\end{itemize}

This doesn't seem plausible. In first case, intervention slightly increases the survival probability of the five from 10\% to 20\% while introducing non-trivial 19\% risk to the one person, resulting in minimal overall benefit but significant risk. Conversely, in the second case, intervention vastly increases the survival probability of the five from 1\% to 99\% while only marginally increasing the risk to the one person by 0.02\%, resulting in great overall benefit with trivial risk. While there is no doubt that non-consequentialist theorist could develop further qualifications to my above challenge case by case, the basic moral should be clear. Background information regarding decisions could be arbitrarily altered to challenge non-consequentialist moral theories. 

Trolley problem is not directly related to the theme of this paper- formalizing egalitarianism, but the inconsistency induced by qualitative thresholds underline key methodological concern of this paper: moral theory must satisfy the minimal requirement of rationality when confronted with formal variables like probabilities. Next we would return to egalitarianism, and show how similar challenges take on even greater importance as we must grapple with \textbf{Unequal Inequalities} and total utility.

\subsection{Not All Inequalities Are Equal}

In this example, we intend to demonstrate that without a well-specified version of egalitarianism, ethical discussions about it might be futile. More specifically, While informal theories can easily state that perfect equality is better than inequality, informal theories struggle when faced with distributions that involve different types or magnitudes of inequality.

To start with minimal controversy, egalitarians might commit to two claims: 1. Utility is good. 2. Decreasing inequality is good. Egalitarians should agree that besides equality, total utility matters, given that they are not indifferent between two outcomes with guaranteed equality, but one is strictly better than the other in terms of total or average utility level. This is a very rough and minimal criterion, but it already gives us different directions from utilitarianism in a prototypical scenario.

Consider two outcomes with utility distributions \( L = (5, 1) \) and \( E = (3, 3) \). Utilitarianism would be indifferent to the inequality in the first choice and perfect equality in the second choice. However, any egalitarians disagree, regardless of the specific reasoning it seems \(E\) is better than \(L\) regarding equality. If one distribution  E is as good as distribution  L with respect to utility, while it is uncontroversially better with respect to equality, then E \text{DOMINATES} L.

Dominance expresses a very simple idea when it comes to evaluation. Suppose we are trying to compare two objects \( A \) and \( B \). \( A \) and \( B \) both consist of proper parts \( A_1, B_1, A_2, B_2 \dots\). For any \( i \), \( A_i \) is at least as good as \( B_i \), Condition \textbf{Dominance} is satisfied.

Dominance-egalitarianism avoids the hard problems: How should we decide when equality and utility point in different directions? For example, most egalitarians would agree that the distribution \( A = (10, 10, 10, \dots, 9, 9, 9) \), where 500 people live a blissfully fulfilling life and another 500 people live a slightly less fulfilling life, is better than the distribution \( (1, 1, 1, \dots) \), where 1000 people are tortured equally for 50 years. Many agree the value of welfare in the first outcome somehow outweighs the value of equality in the second distribution. However, egalitarians might also want to claim that \( (3, 3, 3, \dots) \), where everybody equally lives a mildly happy life, is better than the distribution \( (2\dots 2, 6 \dots 6) \), where half the population lives a miserable life while the other half lives a very happy life, due to the great inequality in the second distribution. The leading egalitarian theories cannot provide clear guidance when facing distributions with absolute equality versus unequal distributions with more utility. This is a salient concern, given that most people who care about distribution also care about understanding the nature of the trade-off.

A more worrisome problem emerges as the notion of equality itself does not give us an obvious scenario when every alternative differs from absolute equality in different ways. Most comparisons regarding inequality don’t involve comparing two objects that contain total equality and inequality. Instead, they involve comparing a set of unequal distributions with different kinds of inequality. An account of egalitarianism that fails to address this would be highly uninformative and uninteresting. For example, which distribution is worse with respect to equality: \( (10, 10, 10, 1) \) or \( (10, 1, 1, 1) \)? There is more deviation from absolute equality in the second distribution since the portion of utility that needs to be redistributed to achieve equality is higher in the second distribution than in the first. Nevertheless, there seems to be something especially bad about the first distribution, as the worst-off people's situation is still very bad while other people lead a very good life as equal fellow human beings. We might feel this reflects a collective attitude of vicious cruelty, and there is something particularly bad if a small group or individuals have to bear all the badness of inequality in contrast with large numbers of equally happy human beings.

When recognizing the complexity of inequality evaluation, it might be tempting to embrace relativism and pluralism. However, much like the predicament faced by anti-Consequentialists previously examined, without further explicating the exact complexity underlying egalitarianism, such pluralism will be vacuous—akin to asserting that a reasonable threshold of risk is the choice that’s reasonable to make. In fact, the toy examples already suggest that there might be far more structure underlying egalitarianism that has been ignored by philosophers due to inadequacy of tools.

Suppose we have to make a choice between three distributions without any explicit aggregation function: \( (5, 9) \), \( (6, 6) \), and \( (6, 7) \). The following naïve reasoning is surely not uncommon:
\begin{enumerate}
    \item We might intuit that distribution \( B = (6, 6) \) is better than \( A = (5, 9) \) since there is absolute equality in \( B \), though the total amount of good is less than in \( A \).
    \item Facing a choice between \( B \) and \( C = (6, 7) \), we might intuitively find that, though there isn’t absolute equality in \( C \), everybody is at least as good as in \( B \), while somebody’s utility has improved from 6 to 7.
    \item When comparing \( A \) to \( C \), we might find \( A \) is better than \( C \) because neither option has equality, but \( A \) contains more total utility than \( C \).
\end{enumerate}
Thus, we arrive at circularity, since intuition tells us \( B \) is better than \( A \), \( C \) is better than \( B \), and \( A \) is better than \( C \). 

In the absence of formal tools, natural language leads us to a place that's difficult to understand. Unlike different attitudes toward consequentialism or negative right, the problems relevant to egalitarianism doesn't seem to involve value divergence rather than simple inconsistency. In next Section, we would introduce  \textbf{Preorder} and \textbf{Social Welfare Function} as they provide analytical tools for understanding the complexities we have encountered so far.

\section{Social Welfare functional}
Here we introduce the framework of social welfare function (henceforth. SWF) and some classical results. My purpose here is to illustrate how the natural language of ethical theory can find a clear representation in the SWF framework. Conversely, the structural emphasis of the SWF framework offers surprisingly strong insights into substantive ethical arguments, bridging the two domains. The first stage objective would be employing SWF frameworks to explain the paradoxes that had been introduced in the last chapters. 

In the study of social welfare functions (SWFs), the primitive tool is using \textbf{preorder relation} \( \preceq \) to  formalize the idea of one alternative being ``at least as good as'' another. This ranking reflects the \textit{betterness relationship} between alternative acts/policies. The ordering structure of SWFs allows us to  derive a ranking of alternatives based on their moral property.

A \textbf{preorder} is a binary relation that satisfies two key properties:
\textbf{Reflexivity}: For any alternative \( A \), \( A \preceq A \) (an alternative is at least as good as itself).
\textbf{Transitivity}: For any alternatives \( A, B, C \), if \( A \preceq B \) and \( B \preceq C \), then \( A \preceq C \).\textbf{Completeness}: For any two alternatives \( A \) and \( B \), either \( A \preceq B \), \( B \preceq A \), or both.

To translate this structure into common language, the preorder \( \preceq \) is interpreted as follows:
\begin{itemize}
    \item If \( A \preceq B \) and \( B \preceq A \), then \( A \) and \( B \) are \textbf{equally good} (\( A \sim B \)).
    \item If \( A \preceq B \) but \( B \npreceq A \), then \( A \) is \textbf{better than} \( B \) (\( A \succ B \)).
    
\end{itemize}

By taking ``at least as good as'' as the primitive notion, the choice criteria between alternatives emerge naturally. This framework provides the foundation for modeling and analyzing moral and social preferences in SWF theory.

\begin{definition}[Welfarist Social Welfare Function]
A \textbf{Social Welfare Function} (SWF) is a function \( F: U^N \rightarrow \mathbb{R} \), where:
\begin{itemize}
    \item \( N = \{1, 2, \dots, n\} \) denotes the set of individuals in society.
    \item \( U^N \) represents the set of all possible utility profiles \( U = (u_1, u_2, \dots, u_n) \), where \( u_i \) is the utility of individual \( i \).
\end{itemize}

The function \( F \) assigns a real number \( F(U) \) to each utility profile \( U \), representing the overall social welfare.

The SWF \( F \) is \textbf{welfarist} if it satisfies the following condition:

\begin{itemize}
    \item \textbf{Dependence Only on Utilities}: For any two utility profiles \( U = (u_1, u_2, \dots, u_n) \) and \( U' = (u'_1, u'_2, \dots, u'_n) \), if \( u_i = u'_i \) for all \( i \in N \), then \( F(U) = F(U') \).
\end{itemize}
\end{definition}

The importance of SWFs manifest themselves immediately— the fundamental defect in our motivating examples could be traced back to the violation of transitivity. The inconsistencies highlighted in the previous section—whether in probabilistic reasoning or comparing different inequalities—underscore the need for a systematic method to rank alternatives. A preorder relation provides the minimal structure necessary to ensure that such rankings are consistent and rational.

Upon first encountering the fundamental structure of social welfare functions, people tend to ask following questions: Why should we care about it? Isn’t SWF already rigged in favor of consequentialism if not directly assumes it correctness?

While within SWF framework the representation of welfarist-consequentialism seems natural, there is no such requirement that the decision rule should be based on utility profiles at all. In our example, the only constraint that we invoked is structural rationality of \(\preceq\): Preorder is a formal concept that originated in set theory, taken by itself, carries the ethical interpretation that whenever you’re deciding anything of moral relevance, you should do it consistently and not contradict yourself. Such notion is requirement of reasoning rather consequentialism.  The substantial restrictive power of transitivity depends on the alternatives we consider. Philosophers' problems can be attributed to the failure to realize they have to make judgments beyond the toy scenarios they happen to consider, The fundamental pivot here is all the controversy surrounding nothing but interpretation, it would be too much to overturn basic set theory to settle ethical dispute.

The pathologies facing traditional ethical theories should be clear through the lens of ranking, as emphasized by SWF. Traditional ethical theories, constrained by the limitations of natural language, lack the systematic tools necessary for coherent moral evaluation. They often rely on arbitrary deliberation over a narrow range of scenarios, leaving a vast space of possibilities unexamined. SWF implicitly impose restrictions upon our ethical theorization, and such restrictions are reasonable but had seldom been noticed within the familiar discourse among philosophers. 

Anti-consequentialist critiques often rely on specific intuitions or hypothetical scenarios to dismiss consequentialism. However, these objections fail to address the broader issue: their proposed surrogates are arbitrary and incomplete, covering only a small fraction of the moral landscape. The SWF framework highlights the importance of formal tools that can evaluate all logically consistent combinations of utility, distribution and risk. By doing so, it goes beyond the constraints of traditional philosophical discourse, which often relies on natural language and intuition. Our examples—probabilistic reasoning in the Trolley Dilemma and the evaluation of inequality—demonstrate that the inconsistencies in traditional methods arise from their inability to formalize these trade-offs systematically. Developing a coherent moral theory requires tools like SWFs, which ensure consistency across a wide range of ethical dilemmas and provide a foundation for resolving the blind spots of traditional approaches.

Given the core status of expected utility theorem (Henceforth, EUT) , Here we would present a maximally rigorous proof of it \cite{fishburn1979utility}. So the readers who are not familiar with formal method could directly get a taste of how it works. The proofs are included solely for technical completeness and to illustrate the foundational framework and techniques used in subsequent sections. In general, the rest of the paper leaves non-original and standard proofs to the appendices, while reserving novel proofs for the main text.

Let \(X=\{x_1,\dots,x_n\}\) be a finite set of prizes. Write \(\Delta(X)\) for the set of all lotteries on \(X\). We identify each prize \(x\in X\) with the degenerate lottery concentrated at \(x\), and we identify compound lotteries with their reduced forms. For a preorder \(\succsim\) on \(\Delta(X)\), write
\[
L\sim M \iff (L\succsim M \text{ and } M\succsim L),
\qquad
L\succ M \iff (L\succsim M \text{ and not } M\succsim L).
\]

\begin{theorem}[Expected Utility Theorem]
\label{thm:eut-finite}
Let \(\succsim\) be a complete preorder on \(\Delta(X)\). Assume:
\begin{enumerate}
\item[(E0)] \textbf{Nontriviality on prizes:} there exist \(x,y\in X\) such that \(x\succ y\).

\item[(E1)] \textbf{Continuity:} whenever \(L\succ M\succ N\), there exists \(\alpha\in(0,1)\) such that
\[
M\sim \alpha L+(1-\alpha)N.
\]

\item[(E2)] \textbf{Independence:} for all \(L,M,N\in\Delta(X)\) and all \(\alpha\in(0,1)\),
\[
L\succsim M
\quad\Longleftrightarrow\quad
\alpha L+(1-\alpha)N \succsim \alpha M+(1-\alpha)N.
\]
\end{enumerate}
Then there exists a function \(u:X\to\mathbb R\) such that for all lotteries
\[
L=\sum_{i=1}^n p_i x_i,
\qquad
M=\sum_{i=1}^n q_i x_i,
\]
one has
\[
L\succsim M
\quad\Longleftrightarrow\quad
\sum_{i=1}^n p_i\,u(x_i)\ge \sum_{i=1}^n q_i\,u(x_i).
\]
Moreover, if \(v:X\to\mathbb R\) is another function with the same representation property, then there exist \(a>0\) and \(b\in\mathbb R\) such that
\[
v(x)=a\,u(x)+b
\qquad(\forall x\in X).
\]
\end{theorem}

\begin{proof} 

\noindent\textbf{Step 1: Best and worst prizes, and the normalization on a two--point scale.}
Because \(X\) is finite and \(\succsim\) is complete and transitive on degenerate lotteries, there exist prizes \(x^+,x^-\in X\) such that
\[
x^+\succsim x\succsim x^-
\qquad(\forall x\in X).
\]
By \textup{(E0)}, choose \(x,y\in X\) such that \(x\succ y\). Since \(x^+\succsim x\) and \(y\succsim x^-\), it follows that \(x^+\succ x^-\): if \(x^-\succsim x^+\), then transitivity would give \(y\succsim x^- \succsim x^+ \succsim x\), contradicting \(x\succ y\).

For \(\alpha\in[0,1]\), define \(B(\alpha):=\alpha x^+ +(1-\alpha)x^-\).

We first derive the strict form of independence implied by \textup{(E2)}: whenever \(L\succ M\), \(N\in\Delta(X)\), and \(\theta\in(0,1)\),
\begin{equation}\label{eq:eut-strict-independence}
\theta L+(1-\theta)N \succ \theta M+(1-\theta)N.
\end{equation}
Indeed, \textup{(E2)} yields \(\theta L+(1-\theta)N \succsim \theta M+(1-\theta)N\). If also \(\theta M+(1-\theta)N \succsim \theta L+(1-\theta)N\), then another application of \textup{(E2)} would give \(M\succsim L\), contradicting \(L\succ M\).

We now prove that
\begin{equation}\label{eq:eut-two-point-monotone}
B(\alpha)\succsim B(\beta)
\quad\Longleftrightarrow\quad
\alpha\ge\beta
\qquad(\alpha,\beta\in[0,1]).
\end{equation}
Suppose first that \(\alpha>\beta\). For every \(\beta<1\), one has \(x^+\succ B(\beta)\): this is immediate when \(\beta=0\), while for \(0<\beta<1\), \eqref{eq:eut-strict-independence} applied to \(x^+\succ x^-\) with common component \(x^+\) and mixing parameter \(1-\beta\) gives
\[
x^+=(1-\beta)x^+ + \beta x^+ \succ (1-\beta)x^- + \beta x^+ = B(\beta).
\]
If \(\alpha=1\), then \(B(\alpha)=x^+\succ B(\beta)\). If \(\beta<\alpha<1\), set \(s:=(\alpha-\beta)/(1-\beta)\in(0,1)\). Then \(s+(1-s)\beta=\alpha\) and \((1-s)(1-\beta)=1-\alpha\), so
\[
B(\alpha)=s x^+ +(1-s)\bigl(\beta x^+ +(1-\beta)x^-\bigr)=s x^+ +(1-s)B(\beta).
\]
Since \(x^+\succ B(\beta)\), another application of \eqref{eq:eut-strict-independence} yields \(B(\alpha)\succ B(\beta)\).

If \(\alpha=\beta\), then \(B(\alpha)=B(\beta)\). If \(\alpha<\beta\), the argument just given with \(\alpha\) and \(\beta\) interchanged shows that \(B(\beta)\succ B(\alpha)\), so it is not the case that \(B(\alpha)\succsim B(\beta)\). Therefore \eqref{eq:eut-two-point-monotone} holds.

Now fix any \(x\in X\). If \(x\sim x^+\), set \(u(x):=1\); if \(x\sim x^-\), set \(u(x):=0\). If \(x^+\succ x\succ x^-\), continuity yields some \(\alpha_x\in(0,1)\) such that \(x\sim \alpha_x x^+ +(1-\alpha_x)x^-\). By \eqref{eq:eut-two-point-monotone}, this \(\alpha_x\) is unique. Let \(u(x):=\alpha_x\). Thus, for every \(x\in X\),
\begin{equation}\label{eq:eut-prize-normalization}
x\sim u(x)x^+ +(1-u(x))x^-,
\qquad
u(x^+)=1,
\qquad
u(x^-)=0.
\end{equation}

\medskip
\noindent\textbf{Step 2: Every lottery is indifferent to a two--point lottery with the same expected utility.}

For a lottery
\(
L=\sum_{i=1}^n p_i x_i\in\Delta(X)
\), define
\(
\lambda(L):=\sum_{i=1}^n p_i\,u(x_i)
\). Since \(u(x_i)\in[0,1]\) for every \(i\), we have \(\lambda(L)\in[0,1]\). We prove
\begin{equation}\label{eq:eut-lottery-reduction}
L\sim \lambda(L)x^+ +(1-\lambda(L))x^-.
\end{equation}

We argue by induction on \(|\operatorname{supp}(L)|\).

If \(|\operatorname{supp}(L)|=1\), then \(L=x\) for some \(x\in X\), and \eqref{eq:eut-lottery-reduction} is exactly \eqref{eq:eut-prize-normalization}.

Assume \eqref{eq:eut-lottery-reduction} has been proved for every lottery whose support has at most \(m-1\) points, and let \(|\operatorname{supp}(L)|=m\ge2\). Choose some \(x_j\in\operatorname{supp}(L)\), and write
\[
p:=p_j\in(0,1),
\qquad
L':=\frac{1}{1-p}\sum_{i\ne j} p_i x_i.
\]
Then
\[
L= p x_j +(1-p)L',
\]
and \(|\operatorname{supp}(L')|=m-1\). By the induction hypothesis,
\[
L'\sim \lambda(L')x^+ +(1-\lambda(L'))x^-.
\]
Also, by \eqref{eq:eut-prize-normalization},
\[
x_j\sim u(x_j)x^+ +(1-u(x_j))x^-.
\]
Applying independence twice, first to substitute \(L'\), and then to substitute \(x_j\), gives
\begin{align*}
L
&=p x_j +(1-p)L'\\
&\sim p x_j +(1-p)\bigl(\lambda(L')x^+ +(1-\lambda(L'))x^-\bigr)\\
&\sim p\bigl(u(x_j)x^+ +(1-u(x_j))x^-\bigr)
+(1-p)\bigl(\lambda(L')x^+ +(1-\lambda(L'))x^-\bigr).
\end{align*}
Reducing the right-hand side yields
\[
L\sim \bigl(pu(x_j)+(1-p)\lambda(L')\bigr)x^+
+\bigl(1-pu(x_j)-(1-p)\lambda(L')\bigr)x^-.
\]
Since
\[
pu(x_j)+(1-p)\lambda(L')
=
pu(x_j)+(1-p)\sum_{i\ne j}\frac{p_i}{1-p}u(x_i)
=
\sum_{i=1}^n p_i u(x_i)
=
\lambda(L),
\]
equation \eqref{eq:eut-lottery-reduction} follows.
 
Let
\(
L=\sum_{i=1}^n p_i x_i,
\qquad
M=\sum_{i=1}^n q_i x_i.
\)
By \eqref{eq:eut-lottery-reduction}, we have
\[
L\sim B(\lambda(L)),
\qquad
M\sim B(\lambda(M)).
\]
Therefore, using transitivity and \eqref{eq:eut-two-point-monotone},
\[
L\succsim M
\Longleftrightarrow
B(\lambda(L))\succsim B(\lambda(M))
\Longleftrightarrow
\lambda(L)\ge \lambda(M).
\]
Since
\[
\lambda(L)=\sum_{i=1}^n p_i u(x_i),
\qquad
\lambda(M)=\sum_{i=1}^n q_i u(x_i),
\]
this is exactly
\[
L\succsim M
\Longleftrightarrow
\sum_{i=1}^n p_i u(x_i)\ge \sum_{i=1}^n q_i u(x_i).
\]

\medskip
\noindent\textbf{Step 3: Uniqueness up to positive affine transformation.}
Let \(v:X\to\mathbb R\) be another function with the same representation property. Because \(x^+\succ x^-\), the representation by \(v\) gives
\[
v(x^+)>v(x^-).
\]
Fix \(x\in X\). By \eqref{eq:eut-prize-normalization},
\[
x\sim u(x)x^+ +(1-u(x))x^-.
\]
Applying the representation by \(v\) to this indifference yields
\[
v(x)=u(x)v(x^+) +(1-u(x))v(x^-).
\]
Therefore
\[
v(x)=\bigl(v(x^+)-v(x^-)\bigr)u(x)+v(x^-).
\]
Setting
\(
a:=v(x^+)-v(x^-)>0,
\quad
b:=v(x^-),
\)
we obtain
\[
v(x)=a\,u(x)+b
\qquad(\forall x\in X).
\]

\end{proof}

\textbf{Conclusion:}

Any two utility functions representing \(\succsim\) are related by a positive affine transformation. Under \textup{(E0)}--\textup{(E2)}, there exists a utility function \(u:X\to\mathbb R\) such that for all \(L,M\in\Delta(X)\),
\[
L\succsim M \quad \text{if and only if} \quad \mathbb E_L[u] \geq \mathbb E_M[u],
\quad
\mathbb E_L[u]:=\sum_{x\in X}L(x)\,u(x).
\]
This utility function is unique up to a positive affine transformation. On the basis of the individual utility representation above, we now introduce Harsanyi's Aggregation Theorem within the same framework, taking the social expected-utility representation as an assumption.

\textbf{Harsanyi's Aggregation Theorem}

For a set \(S\) of social states and a finitely supported lottery \(L\in\Delta(S)\), write
\[
\mathbb E_L[f]:=\sum_{s\in S} L(s)\,f(s)
\]
for the expectation of any function \(f:S\to\mathbb R\) under \(L\). For the social preorder \(\succsim_S\), write
\[
L\sim_S M \iff (L\succsim_S M \text{ and } M\succsim_S L),
\qquad
L\succ_S M \iff (L\succsim_S M \text{ and not } M\succsim_S L).
\]

\begin{theorem}[Harsanyi's Aggregation Theorem]
\label{thm:harsanyi-aggregation}
Let \(S\) be a set of social states, let \(\Delta(S)\) denote the set of finitely supported lotteries on \(S\), and for each \(i=1,\dots,n\) fix a von Neumann--Morgenstern utility \(u_i:S\to\mathbb R\) for individual \(i\). Let \(\succsim_S\) be a social preorder on \(\Delta(S)\). Assume:
\begin{enumerate}
\item[(H1)] \textbf{Social expected utility:} there exists \(U_S:S\to\mathbb R\) such that for all \(L,M\in\Delta(S)\),
\[
L\succsim_S M
\quad\Longleftrightarrow\quad
\mathbb E_L[U_S]\ge \mathbb E_M[U_S].
\]

\item[(H2)] \textbf{Richness of utility profiles:} for every \(r=(r_1,\dots,r_n)\in\mathbb R^n\), fix a state \(s_r\in S\) such that
\[
u_i(s_r)=r_i
\qquad(i=1,\dots,n).
\]

\item[(H3)] \textbf{Pareto indifference:} if
\[
\mathbb E_L[u_i]=\mathbb E_M[u_i]
\qquad(i=1,\dots,n),
\]
then
\[
L\sim_S M.
\]

\item[(H4)] \textbf{Strong Pareto:} if
\[
\mathbb E_L[u_i]\ge \mathbb E_M[u_i]
\qquad(i=1,\dots,n),
\]
and
\[
\mathbb E_L[u_j]>\mathbb E_M[u_j]
\]
for some \(j\), then
\[
L\succ_S M.
\]
\end{enumerate}
Then there exist constants
\[
\lambda_1,\dots,\lambda_n>0,
\qquad
c\in\mathbb R,
\]
such that
\[
U_S(s)=\sum_{i=1}^n \lambda_i u_i(s)+c
\qquad(\forall s\in S).
\]
Consequently, for all lotteries \(L,M\in\Delta(S)\),
\[
L\succsim_S M
\quad\Longleftrightarrow\quad
\sum_{i=1}^n \lambda_i\,\mathbb E_L[u_i]
\ge
\sum_{i=1}^n \lambda_i\,\mathbb E_M[u_i].
\]

If, in addition, the social preorder is \emph{impartial} in the sense that for every permutation \(\sigma\) of \(\{1,\dots,n\}\) and every \(r\in\mathbb R^n\),
\[
s_r\sim_S s_{r_\sigma},
\qquad
r_\sigma:=(r_{\sigma(1)},\dots,r_{\sigma(n)}),
\]
then
\[
\lambda_1=\cdots=\lambda_n.
\]
After positive rescaling, one may therefore write
\[
L\succsim_S M
\quad\Longleftrightarrow\quad
\sum_{i=1}^n \mathbb E_L[u_i]
\ge
\sum_{i=1}^n \mathbb E_M[u_i].
\]
\end{theorem}

\begin{deferredproof}{thm:harsanyi-aggregation}
\noindent\textbf{Step 1: Aggregate utility depends only on individual utility.}
Define
\[
F(r):=U_S(s_r)
\qquad(r\in\mathbb R^n).
\]

We first show that \(F\) depends only on the utility profile. Suppose \(s,t\in S\) satisfy
\[
u_i(s)=u_i(t)
\qquad(i=1,\dots,n).
\]
Viewing \(s\) and \(t\) as degenerate lotteries, we have
\[
\mathbb E_s[u_i]=u_i(s)=u_i(t)=\mathbb E_t[u_i]
\qquad(i=1,\dots,n).
\]
By \textup{(H3)},
\[
s\sim_S t.
\]
Since \(U_S\) represents \(\succsim_S\) by \textup{(H1)}, it follows that
\[
U_S(s)=U_S(t).
\]
Hence, if a different representative had been chosen for the same profile, the value of \(F\) would be unchanged.

Now let \(s\in S\) be arbitrary, and set
\[
r(s):=(u_1(s),\dots,u_n(s)).
\]
Then \(s\) and \(s_{r(s)}\) have exactly the same individual utility profile, so the previous argument gives
\[
U_S(s)=U_S(s_{r(s)})=F(r(s)).
\]
Therefore
\begin{equation}\label{eq:harsanyi-profile-factorization}
U_S(s)=F(u_1(s),\dots,u_n(s))
\qquad(\forall s\in S).
\end{equation}

\medskip
\noindent\textbf{Step 2: The profile map \(F\) is affine.}
Fix \(x,y\in\mathbb R^n\) and \(\alpha\in[0,1]\). Set
\[
z:=\alpha x +(1-\alpha)y.
\]
For each individual \(i\),
\[
u_i(s_z)=z_i=\alpha x_i +(1-\alpha)y_i
=\alpha u_i(s_x)+(1-\alpha)u_i(s_y).
\]
Hence
\[
u_i(s_z)=\mathbb E_{\alpha s_x +(1-\alpha)s_y}[u_i]
\qquad(i=1,\dots,n).
\]
By \textup{(H3)},
\[
s_z\sim_S \alpha s_x +(1-\alpha)s_y.
\]
Applying \textup{(H1)} yields
\[
U_S(s_z)=\alpha U_S(s_x)+(1-\alpha)U_S(s_y).
\]
By definition of \(F\),
\[
F(z)=\alpha F(x)+(1-\alpha)F(y).
\]
Therefore
\begin{equation}\label{eq:harsanyi-affine-F}
F(\alpha x +(1-\alpha)y)=\alpha F(x)+(1-\alpha)F(y)
\qquad(x,y\in\mathbb R^n,\ \alpha\in[0,1]).
\end{equation}

Set
\[
c:=F(0),
\qquad
G(r):=F(r)-c
\qquad(r\in\mathbb R^n).
\]
Then \eqref{eq:harsanyi-affine-F} becomes
\[
G(\alpha x +(1-\alpha)y)=\alpha G(x)+(1-\alpha)G(y)
\qquad(x,y\in\mathbb R^n,\ \alpha\in[0,1]).
\]

Taking \(y=0\), we obtain
\[
G(\alpha x)=\alpha G(x)
\qquad(x\in\mathbb R^n,\ \alpha\in[0,1]).
\]
In particular, with \(\alpha=\frac12\),
\[
G(x)=\frac12 G(2x),
\]
hence
\[
G(2x)=2G(x)
\qquad(x\in\mathbb R^n).
\]

Now for arbitrary \(x,y\in\mathbb R^n\),
\[
G(x+y)=G\!\left(2\cdot \frac{x+y}{2}\right)
=2G\!\left(\frac{x+y}{2}\right)
=2\left(\frac12 G(x)+\frac12 G(y)\right)
=G(x)+G(y).
\]
Thus \(G\) is additive.

We already know \(G(\alpha x)=\alpha G(x)\) for \(\alpha\in[0,1]\). Let \(t>1\). Write
\[
t=m+\theta
\qquad\text{with}\qquad
m\in\mathbb N,\ \theta\in[0,1).
\]
Using additivity and the previous homogeneity,
\[
G(tx)=G(mx+\theta x)=mG(x)+\theta G(x)=tG(x).
\]
If \(t<0\), then additivity gives
\[
0=G(0)=G(tx+(-t)x)=G(tx)+G((-t)x),
\]
so
\[
G(tx)=-G((-t)x)=tG(x).
\]
Hence
\[
G(tx)=tG(x)
\qquad(x\in\mathbb R^n,\ t\in\mathbb R).
\]
Therefore \(G\) is linear on \(\mathbb R^n\).

Let \(e_1,\dots,e_n\) be the standard basis of \(\mathbb R^n\), and define
\[
\lambda_i:=G(e_i)
\qquad(i=1,\dots,n).
\]
For \(r=(r_1,\dots,r_n)\in\mathbb R^n\),
\[
r=\sum_{i=1}^n r_i e_i,
\]
and linearity gives
\[
G(r)=\sum_{i=1}^n r_i\,G(e_i)
=\sum_{i=1}^n \lambda_i r_i.
\]
Thus
\begin{equation}\label{eq:harsanyi-affine-form}
F(r)=\sum_{i=1}^n \lambda_i r_i + c
\qquad(r\in\mathbb R^n).
\end{equation}

Fix \(i\in\{1,\dots,n\}\). The states \(s_{e_i}\) and \(s_0\) satisfy
\[
u_k(s_{e_i})=(e_i)_k\ge 0=u_k(s_0)
\qquad(k=1,\dots,n),
\]
and the inequality is strict when \(k=i\). Applying \textup{(H4)} to the corresponding degenerate lotteries yields
\[
s_{e_i}\succ_S s_0.
\]
Hence, by \textup{(H1)},
\[
F(e_i)=U_S(s_{e_i})>U_S(s_0)=F(0).
\]
Using \eqref{eq:harsanyi-affine-form},
\[
\lambda_i=F(e_i)-F(0)>0.
\]
Therefore
\[
\lambda_i>0
\qquad(i=1,\dots,n).
\]

\medskip
\noindent\textbf{Step 3: Harsanyi's Aggregation.}
Let \(s\in S\). Combining \eqref{eq:harsanyi-profile-factorization} with \eqref{eq:harsanyi-affine-form}, we obtain
\[
U_S(s)=F(u_1(s),\dots,u_n(s))
=\sum_{i=1}^n \lambda_i u_i(s)+c.
\]
Hence, for any lottery \(L\in\Delta(S)\),
\[
\mathbb E_L[U_S]
=
\sum_{s\in S}L(s)\left(\sum_{i=1}^n \lambda_i u_i(s)+c\right)
=
\sum_{i=1}^n \lambda_i\,\mathbb E_L[u_i]+c.
\]
Therefore, for any \(L,M\in\Delta(S)\),
\[
\mathbb E_L[U_S]\ge \mathbb E_M[U_S]
\quad\Longleftrightarrow\quad
\sum_{i=1}^n \lambda_i\,\mathbb E_L[u_i]
\ge
\sum_{i=1}^n \lambda_i\,\mathbb E_M[u_i].
\]
Using \textup{(H1)}, this is exactly
\[
L\succsim_S M
\quad\Longleftrightarrow\quad
\sum_{i=1}^n \lambda_i\,\mathbb E_L[u_i]
\ge
\sum_{i=1}^n \lambda_i\,\mathbb E_M[u_i].
\]

Assume now that the additional impartiality condition holds. Let \(j,k\in\{1,\dots,n\}\) be arbitrary, and let \(\sigma\) be the transposition that swaps \(j\) and \(k\) and fixes all other indices. Then
\[
(e_j)_\sigma=e_k.
\]
Hence impartiality gives
\[
s_{e_j}\sim_S s_{e_k}.
\]
Therefore
\[
F(e_j)=F(e_k).
\]
Using \eqref{eq:harsanyi-affine-form},
\[
\lambda_j + c = \lambda_k + c,
\]
so
\[
\lambda_j=\lambda_k.
\]
Since \(j,k\) were arbitrary, all coefficients are equal:
\[
\lambda_1=\cdots=\lambda_n.
\]
Since multiplying all coefficients by the same positive constant does not change the represented preorder, we may rescale and write
\[
\lambda_1=\cdots=\lambda_n=1.
\]
Thus
\[
L\succsim_S M
\quad\Longleftrightarrow\quad
\sum_{i=1}^n \mathbb E_L[u_i]
\ge
\sum_{i=1}^n \mathbb E_M[u_i].
\]
\end{deferredproof}

These proofs are presented solely for technical completeness and to illustrate the foundational framework and techniques used in the subsequent sections. In general, the rest of the paper would leaves non-original and generic proofs to the appendices and keeps novel proofs in the main text.  We are aware that  Harsanyi's Aggregation Theorem has been the subject of extensive debate—Sen and Weymark\cite{Weymark1991-WEYARO} argue that the result is not utilitarian, while Broome\cite{Broome1991-BROWGE} and McCarthy\cite{McCarthy2016-MCCPIE-2} contend otherwise— we adopt the formalist position: the theorem provides a rigorous characterization of utilitarianism without offering any additional normative defense of it. 

What we care is the technical feature of these classical results, their shared structural form. The Expected Utility Theorem (EUT) begins with rationality axioms on individual preferences over lotteries. From these, it derives a utility function that represents preferences, unique up to positive affine transformations. This technical achievement shows that qualitative judgments over probabilistic outcomes can be captured quantitatively as expected utility. Harsanyi’s Aggregation Theorem extends this structure to the social domain. Starting with individual utility functions derived by EUT, it adds axioms like the Pareto principle and impartiality to demonstrate that social preferences can be represented as a weighted sum of individual utilities. Both results share a common pattern: choosing certain axioms to yield linear utility representations through mathematical tools such as convexity and continuity arguments. Crucially, this framework is entirely normative neutral; while utilitarianism could be formalized within this framework, with few modifications it could naturally represent egalitarian principles or other normative perspectives. 

Representation theorems for egalitarianism use similar methods to formalize inequality-sensitive orderings that ensure specific functional forms, such as those based on the Gini index or Atkinson index. These representations, like those in EUT and Harsanyi’s theorem, transform abstract ordering criteria into well-defined mathematical functions, providing both rigor and generality to egalitarian evaluations.

However, before proceeding to representation theorems, we must first examine the deficiencies of existing approaches to representing egalitarianism. Temkin’s principle of aggregate complaints, while insightful, relies on informal classifications that lack a rigorous foundation and often fail to capture the structural consistency required for a coherent theory. Similarly, statistical formulas (e.g., variance, standard deviation) and rank-based methods suffer from technical flaws. These issues make them inadequate as substitutes for the egalitarian component of an Aggregate Social Welfare Function.

To summarize, egalitarian need to develop theory that's capable of evaluating badness of inequality, then somehow transform it into common scale that's co-measurable with total utility. Such framework should be able to make sense of both equality and different patterns of inequality with different amounts of total utility, so we could rank different distributions that respects the minimal requirement of $\preceq$-rationality. In the following sections, we will critique these existing approaches in detail and motivate the necessity of representation theorems to formalize egalitarian principles rigorously.

\section{Egalitarianism within SWF Framework}
\subsection{Formal set-up}

Equipped with the machinery of Social Welfare Functions (SWFs), we can re-conceptualize the challenges faced by egalitarian theories. The objective is to develop a ranking system for utility profiles that accounts for both the aggregate value of equality and total utility. This is achieved by isolating the relevant aspects of distribution that constitute an egalitarian preorder, thereby enabling coherent rankings of utility profiles. The following framework is inspired by McCarthy \cite{McCarthy2015}.

\begin{definition}[Distribution Pattern]
Let \(U \in \mathcal U^N\) be a utility profile. The \textbf{Distribution Pattern} of \(U\) refers to the arrangement or spread of utilities among individuals, insofar as it captures the equality or inequality present in the distribution.
\end{definition}

To represent egalitarian principles, we evaluate the distribution pattern alongside total utility.

\begin{definition}[Egalitarian Evaluation]
An \textbf{Egalitarian Evaluation} is a function
\[
F_{\text{Egal}}:\mathcal U^N \to \mathbb R
\]
representing an egalitarian preorder \(\preceq_{\text{Egal}}\) on \(\mathcal U^N\). It evaluates the distributional aspect of a utility profile. The relation \(\preceq_{\text{Egal}}\) must satisfy the following properties:
\begin{itemize}
    \item \textbf{Completeness}: For any two utility profiles \(U\) and \(V\), either \(U \preceq_{\text{Egal}} V\), \(V \preceq_{\text{Egal}} U\), or both.
    \item \textbf{Reflexivity}: For any utility profile \(U\), \(U \preceq_{\text{Egal}} U\).
    \item \textbf{Transitivity}: For any utility profiles \(U, V, W\), if \(U \preceq_{\text{Egal}} V\) and \(V \preceq_{\text{Egal}} W\), then \(U \preceq_{\text{Egal}} W\).
\end{itemize}
\end{definition}

The requirement here is that egalitarianism should be able to generate a preorder over utility profiles once other moral concerns are discarded. Now, given that egalitarians care about both total utility and equality, we define an \textbf{Aggregate Social Welfare Function} that represents these two components without yet assuming any specific   relationship between them.

\begin{definition}[Aggregate Social Welfare Function]
An \textbf{Aggregate Social Welfare Function} is a map
\[
F_{\text{agg}}:\mathcal U^N \to \mathbb R^2
\]
defined by
\[
F_{\text{agg}}(U)=\bigl(F_{\text{Util}}(U),F_{\text{Egal}}(U)\bigr).
\]

\end{definition}

This definition deliberately avoids specifying a scalar output for the aggregate evaluation. In other words, it does not yet impose an all-things-considered ordering. Rather, it only suggests that the two dimensions that such ordering should somehow reconcile: total utility and egalitarian evaluation. Indeed, Why not simply write it in more complete functional form like
\[
F_{\text{agg}}(U)=f_{\text{Util}}(U)-\lambda\cdot f_{\text{Egal}}(U),
\qquad \lambda \in \mathbb R?
\]
The explanation will occupy much of what follows.

\subsection{Temkin's Classification}

Temkin's classification of inequality measures offers heuristic arguments that help us approach the problem of inequality-measurement \cite{Temkin1993}. According to his proposal, we can approach the complexity of inequality by identifying \emph{complaints}. The severity of inequality is determined by both the number of people who have complaints about inequality and the magnitude of their complaints.

Three aggregate principles determine who has a complaint, each of which can be combined with three different objects of complaint (such as the best-off level) to determine the size of each term in aggregate complaints.
1.Maximin Principle (Maximin): Focuses exclusively on the worst-off individual. 
2.Additive Principle (AP): Aggregates everyone's complaints equally, treating all complaints as having equal significance in the assessment of inequality.
3. Weighted Additive Principle (WAP): Differentiates the significance of complaints by assigning different weights to individuals' complaints in the final aggregation.

The size of complaints would be determined by the people who have complaints and the objects of those complaints:
1. Relative to the best-off individual: The complaint size is measured based on how far an individual's utility deviates from that of the best-off individual.
2.Relative to the average individual: The complaint is measured based on the difference between an individual's utility and the average utility in the distribution.
3. Relative to all better-off individuals: The complaint is measured in relation to the utilities of all individuals who are better-off than the person in question.

\begin{table}[h!]
\centering
\small 
\setlength{\tabcolsep}{4pt} 
\begin{tabular}{@{}lccc@{}}
\toprule
\textbf{Principles} & \textbf{BO (Best-off)} & \textbf{AVE (Average)} & \textbf{ATBO (All Better-off)} \\ \midrule
Maximin (Maximin) & Maximin+BO & Maximin+AVE & Maximin+ATBO \\
Additive (AP) & AP+BO & AP+AVE & AP+ATBO \\
Weighted Additive (WAP) & WAP+BO & WAP+AVE & WAP+ATBO \\
\textbf{Independent Criteria} & \multicolumn{3}{c}{Deviation, Gratuitousness, Social Responsibility} \\ \bottomrule
\end{tabular}
\caption{ Representation of Temkin's Principles and Complaint Bases}
\end{table}

Although Temkin suggests that the reasoning and calculations underlying these rankings are ``fairly straightforward but tedious''. Concerns arise about the methodology of enumerating inequality-measures by philosophical rationales. First, It is unclear whether 'additive aggregation relative to better-off individuals' can be meaningfully distinguished from 'additive aggregation relative to the average. Second, we are uncertain whether the reasoning covers all interesting distinctions among the space of functions. This issue is particularly salient given that formal concepts such as ordering have already appeared in the initial theorization. To summarize, Temkin’s principle of aggregate complaints offers a valuable heuristic framework, but its reliance on informal classifications lacks the rigor needed for a coherent theory that aligns with our objective.

Despite our concern with Temkin's principles, his informal presentation also hinders further investigation. To avoid further confusion and prevent the state of affairs from becoming intractable, we will instead examine the properties of many statistical formulas, as the correspondence between statistical metrics and aggregation of complaints has been approved by Temkin. Statistical formulas are proposed and studied for their apparent reasonableness and usefulness. Although they may lack rigorous support, their explicit functional forms make their properties and implications more determinate and easier to understand than those expressed in natural language. More importantly,  statistical measures produce values that satisfy the requirement of rational ranking—we can unambiguously determine which numeric values are greater, lesser, or equal even when we don't have any formal theory in the strict sense. However, do they satisfy this seemingly virtue of um-ambiguity in a justifiable way? The existence of many statistical formulas, which necessarily contradict each other, suggests that there are deeper underlying structural interrelationships that are not arbitrary. Understanding how they relate to each other and whether it is possible to select out certain privileged measure non-arbitrarily Would be our goal. The commonly used measures of inequality include:

\begin{longtable}{|p{0.3\textwidth}|p{0.6\textwidth}|}
\hline
\textbf{Measure} & \textbf{Formula} \\ \hline
\endfirsthead
\hline
\textbf{Measure} & \textbf{Formula} \\ \hline
\endhead
\hline
\endfoot
\hline
\endlastfoot

\textbf{Range} & \( R = x_{\max} - x_{\min} \) \\ \hline

\textbf{Variance} & 
\(\sigma^2 = \frac{1}{n} \sum_{i=1}^n (x_i - \bar{x})^2, \quad \bar{x} = \frac{1}{n} \sum_{i=1}^n x_i\) \\ \hline

\textbf{Standard Deviation} & 
\(\sigma = \sqrt{\frac{1}{n} \sum_{i=1}^n (x_i - \bar{x})^2}\) \\ \hline

\textbf{Relative Mean Deviation} & 
\( D = \frac{\frac{1}{n} \sum_{i=1}^n |x_i - \bar{x}|}{\bar{x}} \) \\ \hline

\textbf{Gini Coefficient} & 
\( G = \frac{\sum_{i=1}^n \sum_{j=1}^n |x_i - x_j|}{2n^2 \bar{x}} \) \\ \hline

\textbf{Atkinson Index} & 
\[
A_\varepsilon = 
\begin{cases} 
1 - \frac{\left( \frac{1}{n} \sum_{i=1}^n x_i^{1-\varepsilon} \right)^{\frac{1}{1-\varepsilon}}}{\bar{x}}, & \varepsilon \neq 1 \\[5pt]
1 - \frac{\prod_{i=1}^n x_i^{\frac{1}{n}}}{\bar{x}}, & \varepsilon = 1
\end{cases}
\] \\ \hline

\textbf{Rank-Weighted View} & 
\[
W(X) = \sum_{i=1}^m g(x_i) u_i(x),  X = \{u_1(x), u_2(x), \dots, u_m(x)\}, 
\]
where \( X \) is ordered as \( u_1 < u_2 < \dots < u_m \),  and \( g(x_i) \) is a rank-based coefficient. \\ \hline

\caption{Commonly used Statistical Measures}
\end{longtable}

These measures allow us to re-examine Temkin's principles and explain why the minimal program of egalitarian evaluation introduced makes no commitment to functional form of SWF.

It's easy to observe that Range represents temkin's maximin principle with relative to best-off view. Range focuses solely on the extremes, ignoring the distribution between them. Variance aligns with Temkin's weighted additive principle relative to the average but suffers from being sensitive to scale changes, making it arbitrary without a fixed utility scale. Relative Mean Deviation fail to account for desirable redistributions within the same side of the distribution, violating principles like Pigou-Dalton. This limitation suggests that they only capture part of our pre-theoretic egalitarian intuition.

Given that each mainstream statistical metric can be plausibly criticized or defended, one might be tempted to embrace pluralism: simply choose the measure that best aligns with intuition, as there seems to be no way to adjudicate further between them. However, this conclusion overlooks a critical issue. The next subsection demonstrates that statistical measures such as variance and standard deviation suffer from inherent flaws that render them unsuitable for normative inequality measurement—flaws rooted in their mathematical properties rather than ethical concern. Originally developed for rigorous mathematical applications like proving the Central Limit Theorem, statistical measures like standard deviation are now widely used in descriptive contexts, where they  produce results that seem plausible and intuitive. For instance, when comparing groups of students’ heights,  higher standard deviation often does corresponds to intuitive expectation of greater dispersion, leading people to accept the output without scrutinizing the measure’s properties. However, these overlooked properties become problematic when applied to normative inequality assessment. As the following counterexamples illustrate, such flaws lead to inconsistencies and contradictions, making statistical measures fundamentally incompatible with the demands of characterizing egalitarianism that avoids structural irrationality.

\subsection {Counter-example with Statistical Measures}

In our previous set-up, We acknowledged that it seems natural to go beyond the minimal framework, and assume the egalitarian SWF as:
\[
\text{SWF} = \left(1 - \lambda \cdot f_{\text{Egal}}(u)\right) \times F_{\text{total}}(u), \lambda\in \mathbb{R}
\]

Given what temkin said throughout the book, it seems something resembles this form is what he meant by simple calculation. Using Gini-index as penalty ensures \(f_{\text{Egal}}\) has appropriate behavior, as its bounded nature prevents arbitrary distortions. However, substituting Gini with other statistical measures would immediately lead us to problem.

Here we exhibit the undesirable result following from Using \textbf{variance}, \textbf{standard deviation}, or \textbf{range} as penalty coefficients in the Aggregate Social Welfare Function (SWF)
\[
F_{\text{Aggregate}}(U) = (1 - \lambda \cdot f_{\text{Egal}}(U)) \cdot f_{\text{Util}}(U)
\]
with \(\lambda = 1\) can lead to arbitrary and inconsistent rankings due to their scale-dependent properties.

\textbf{Counter-Example with Variance:}
\textit{Profiles:} 
\[
U = (2, 1, 1, 1, 1, 1), \quad V = (1, 1, 1, 1, 1, 1)
\]
\textit{Calculations:} 
\[
\begin{aligned}
&f_{\text{Util}}(U) = 7, \quad f_{\text{Egal}}(U) = 0.1389, \quad f_{\text{Util}}(V) = 6, \quad f_{\text{Egal}}(V) = 0, \\
&F_{\text{Aggregate}}(U) \approx 6.0278, \quad F_{\text{Aggregate}}(V) = 6, \quad \text{Ordering: } U > V
\end{aligned}
\]
\textit{Scaled Profiles (\(k = 10\)):}
\[
U' = (20, 10, 10, 10, 10, 10), \quad V' = (10, 10, 10, 10, 10, 10)
\]
\[
\begin{aligned}
&f_{\text{Util}}(U') = 70, \quad f_{\text{Egal}}(U') = 13.89, \quad f_{\text{Util}}(V') = 60, \quad f_{\text{Egal}}(V') = 0, \\
&F_{\text{Aggregate}}(U') \approx -902.3, \quad F_{\text{Aggregate}}(V') = 60, \quad \text{Ordering: } V' > U'
\end{aligned}
\]
\textbf{Conclusion:} Initially, \( U > V \). After scaling, \( V' > U' \). \textbf{Reversal occurs} with \textbf{variance} as the egalitarian component.\\

\textbf{Counter-Example with Standard Deviation:}\\
\textit{Profiles:}
\[
U = \left(\frac{1}{10}, \frac{2}{10}, \frac{3}{10}, \frac{4}{10}, \frac{5}{10}\right), \quad V = \left(\frac{1}{10}, \frac{1}{10}, \frac{1}{10}, \frac{1}{10}, \frac{1}{10}\right)
\]
\textit{Calculations:}
\[
\begin{aligned}
&f_{\text{Util}}(U) = 1.5, \quad f_{\text{Egal}}(U) \approx 0.1414, \quad f_{\text{Util}}(V) = 0.5, \quad f_{\text{Egal}}(V) = 0, \\
&F_{\text{Aggregate}}(U) \approx 1.2879, \quad F_{\text{Aggregate}}(V) = 0.5, \quad \text{Ordering: } U > V
\end{aligned}
\]
\textit{Scaled Profiles (\(k = 10\)):}
\[
U' = (1, 2, 3, 4, 5), \quad V' = (1, 1, 1, 1, 1)
\]
\[
\begin{aligned}
&f_{\text{Util}}(U') = 15, \quad f_{\text{Egal}}(U') \approx 1.414, \quad f_{\text{Util}}(V') = 5, \quad f_{\text{Egal}}(V') = 0, \\
&F_{\text{Aggregate}}(U') \approx -6.21, \quad F_{\text{Aggregate}}(V') = 5, \quad \text{Ordering: } V' > U'
\end{aligned}
\]
\textbf{Conclusion:} Initially, \( U > V \). After scaling, \( V' > U' \). \textbf{Reversal occurs} with \textbf{standard deviation} as the egalitarian component.\\

\textbf{Counter-Example with Range:}\\
\textit{Profiles:}
\[
U = \left(\frac{1}{10}, \frac{2}{10}, \frac{3}{10}, \frac{4}{10}, \frac{5}{10}\right), \quad V = \left(\frac{1}{10}, \frac{1}{10}, \frac{1}{10}, \frac{1}{10}, \frac{1}{10}\right)
\]
\textit{Calculations:}
\[
\begin{aligned}
&f_{\text{Util}}(U) = 1.5, \quad f_{\text{Egal}}(U) = 0.4, \quad f_{\text{Util}}(V) = 0.5, \quad f_{\text{Egal}}(V) = 0, \\
&F_{\text{Aggregate}}(U) = 0.9, \quad F_{\text{Aggregate}}(V) = 0.5, \quad \text{Ordering: } U > V
\end{aligned}
\]
\textit{Scaled Profiles (\(k = 10\)):}
\[
U' = (1, 2, 3, 4, 5), \quad V' = (1, 1, 1, 1, 1)
\]
\[
\begin{aligned}
&f_{\text{Util}}(U') = 15, \quad f_{\text{Egal}}(U') = 4, \quad f_{\text{Util}}(V') = 5, \quad f_{\text{Egal}}(V') = 0, \\
&F_{\text{Aggregate}}(U') = -45, \quad F_{\text{Aggregate}}(V') = 5, \quad \text{Ordering: } V' > U'
\end{aligned}
\]
\textbf{Conclusion:} Initially, \( U > V \). After scaling, \( V' > U' \). \textbf{Reversal occurs} with \textbf{range} as the egalitarian component.

Using variance, standard deviation, or range as the egalitarian component \( f_{\text{Egal}}(U) \) in the Aggregate SWF can result in arbitrary and inconsistent rankings upon ratio-scaling utility profiles. For reasons that beyond the scope of this paper, ratio-invariable is generally regarded one of the strongest and reasonable condition for distributive theory. Specifically, these statistical measures are changing in a way that doesn't corresponding to anything of ethical relevance.  This seems to ensure that we would always have the possibility to construct counter-example that exploit their change relative to total utility transformation so they would exhibit highly undesirable properties such as reversal.

Perhaps it's possible to derive certain special form of \(
F_{\text{Aggregate}}(U) )\) or impose further constraint regarding numerical calibration of utility profiles so the final preorder couldn't be changed. The problem is not merely that individual formulas display unsatisfactory properties; rather, we lack a clear understanding of why they behave this way and how they are interrelated. Introducing ad-hoc qualifications or calibrations to resolve these issues seems particularly objectionable—much like applying superficial fixes to the structural problems highlighted by the probabilistic trolley dilemma.
 
In the following subsection we would prove two impossibility theorems for rank-weighed egalitarianism, they had also been proposed as practical substitute for egalitarianism judgment \cite{Adler2011-ADLWAF}.

\subsection{Impossibility Theorems for Rank-Weighted View}

Fix a discount factor \(k\in(0,1)\) and a finite population size \(N\in\mathbb N\).
A utility profile is a vector
\[
x=(x_1,\dots,x_N)\in\mathbb R^N.
\]
Write
\[
x_{(1)}\le \cdots \le x_{(N)}
\]
for the order statistics of \(x\).

\begin{definition}[Individual-Ranking-Based Discounting (IRBD)]
\label{def:IRBD}
For a profile \(x\in\mathbb R^N\), define
\[
W_{\mathrm{IRBD}}(x):=\sum_{i=1}^N k^{\,i-1}x_{(i)}.
\]
\end{definition}

\begin{definition}[Utility-Level-Based Discounting (ULBD)]
\label{def:ULBD}
Let \(x\in\mathbb R^N\), and let
\[
\{u_1<\cdots<u_m\}
\]
be the distinct utility levels attained in \(x\). For each \(j=1,\dots,m\), let
\[
n_j:=\#\{i:\,x_i=u_j\}
\]
be the multiplicity of level \(u_j\). Define
\[
W_{\mathrm{ULBD}}(x):=\sum_{j=1}^m k^{\,j-1}n_j u_j.
\]
\end{definition}

For \(\lambda\in\mathbb N\), define the \(\lambda\)-fold replication of a profile \(x\in\mathbb R^N\) by
\[
x^{(\lambda)}:=\underbrace{(x,\dots,x)}_{\lambda\ \text{copies}}\in\mathbb R^{\lambda N}.
\]
This operation preserves the internal distributional pattern of \(x\): every utility level and every rank is simply repeated \(\lambda\) times.

For profiles \(x,y\in\mathbb R^N\), write \(x\succ_{\mathrm{lex}} y\) if, at the first index \(m\) for which
\[
x_{(m)}\neq y_{(m)},
\]
one has
\[
x_{(m)}>y_{(m)}.
\]
This is the usual leximin order, equivalently a sequential comparison from the worst-off upward.

\begin{theorem}[Weak impossibility for IRBD: pattern-preserving replication can reverse the ranking]
\label{thm:IRBD_replication_reversal}
For any two distinct profiles \(x,y\in\mathbb R^N\), there exists \(\lambda_0\in\mathbb N\) such that for all
\(\lambda\ge \lambda_0\),
\[
W_{\mathrm{IRBD}}(x^{(\lambda)})>W_{\mathrm{IRBD}}(y^{(\lambda)})
\quad\Longleftrightarrow\quad
x\succ_{\mathrm{lex}} y.
\]
In particular, along sufficiently large replications, IRBD collapses to leximin.

Hence, whenever the original IRBD ranking disagrees with leximin, a simple pattern-preserving replication reverses the ranking: if
\[
W_{\mathrm{IRBD}}(x)>W_{\mathrm{IRBD}}(y)
\qquad\text{but}\qquad
y\succ_{\mathrm{lex}} x,
\]
then for all sufficiently large \(\lambda\),
\[
W_{\mathrm{IRBD}}(x^{(\lambda)})<W_{\mathrm{IRBD}}(y^{(\lambda)}).
\]
\end{theorem}

\begin{proof}
Write
\[
a_j:=x_{(j)},
\qquad
b_j:=y_{(j)}
\qquad(j=1,\dots,N),
\]
and let
\[
m:=\min\{j\in\{1,\dots,N\}: a_j\neq b_j\}.
\]

In the ordered list of \(x^{(\lambda)}\), the value \(a_j\) appears exactly at the consecutive ranks
\[
(j-1)\lambda+1,\ (j-1)\lambda+2,\ \dots,\ j\lambda.
\]
Therefore
\begin{align*}
W_{\mathrm{IRBD}}(x^{(\lambda)})
&=\sum_{j=1}^N \sum_{r=0}^{\lambda-1} k^{(j-1)\lambda+r}a_j =\sum_{j=1}^N k^{(j-1)\lambda}a_j\sum_{r=0}^{\lambda-1}k^r \\
&=\frac{1-k^\lambda}{1-k}\sum_{j=1}^N k^{(j-1)\lambda}a_j.\\
W_{\mathrm{IRBD}}(y^{(\lambda)})
&=\frac{1-k^\lambda}{1-k}\sum_{j=1}^N k^{(j-1)\lambda}b_j.
\end{align*}

Subtracting gives
\begin{equation}\label{eq:IRBD-delta-reversal}
\Delta_\lambda
:=W_{\mathrm{IRBD}}(x^{(\lambda)})-W_{\mathrm{IRBD}}(y^{(\lambda)})
=\frac{1-k^\lambda}{1-k}\sum_{j=1}^N k^{(j-1)\lambda}(a_j-b_j).
\end{equation}
Since \(a_j=b_j\) for every \(j<m\),
\[
\sum_{j=1}^N k^{(j-1)\lambda}(a_j-b_j)
=
k^{(m-1)\lambda}(a_m-b_m)
+
\sum_{j=m+1}^N k^{(j-1)\lambda}(a_j-b_j).
\]
Let
\[
C:=\sum_{j=m+1}^N |a_j-b_j|<\infty.
\]
Then the tail is bounded by
\[
\left|
\sum_{j=m+1}^N k^{(j-1)\lambda}(a_j-b_j)
\right|
\le
\sum_{j=m+1}^N k^{(j-1)\lambda}|a_j-b_j|
\le
Ck^{m\lambda}.
\]

Because \(k\in(0,1)\),
\(
\frac{k^{m\lambda}}{k^{(m-1)\lambda}}=k^\lambda\to 0
\qquad(\lambda\to\infty).
\)
Hence there exists \(\lambda_0\in\mathbb N\) such that for all \(\lambda\ge \lambda_0\),
\[
Ck^{m\lambda}\le \frac12 |a_m-b_m|\,k^{(m-1)\lambda}.
\]
For such \(\lambda\), the sign of
\[
k^{(m-1)\lambda}(a_m-b_m)
+
\sum_{j=m+1}^N k^{(j-1)\lambda}(a_j-b_j)
\]
is exactly the sign of \(a_m-b_m\). Since the prefactor
\[
\frac{1-k^\lambda}{1-k}
\]
in \eqref{eq:IRBD-delta-reversal} is positive, \(\Delta_\lambda\) has the same sign as \(a_m-b_m\). Therefore
\[
W_{\mathrm{IRBD}}(x^{(\lambda)})>W_{\mathrm{IRBD}}(y^{(\lambda)})
\quad\Longleftrightarrow\quad
a_m>b_m.
\]
By definition, \(a_m>b_m\) is exactly
\[
x\succ_{\mathrm{lex}} y.
\]
The final reversal claim follows immediately: if the original IRBD ordering disagrees with leximin, then for all sufficiently large replications the ordering must flip.
\end{proof}

\begin{theorem}[Strong impossibility for ULBD: an almost equal high-utility profile can be ranked below a starkly unequal two-level profile]
\label{thm:ULBD_high_utility_almost_equal}
Fix \(k\in(0,1)\). There exists \(N_0\in\mathbb N\) such that for every sufficiently large even \(N\ge N_0\), there exist two profiles \(A,B\in\mathbb R^N\) with the following properties:
\begin{enumerate}
\item \(A\) is highly unequal and has only two utility levels, namely \(1\) and \(10\);
\item \(B\) is almost equal in the intuitive sense that all its utilities lie in the narrow interval \([9,10]\);
\item \(B\) has strictly larger total utility than \(A\);
\item nevertheless
\[
W_{\mathrm{ULBD}}(A)>W_{\mathrm{ULBD}}(B).
\]
\end{enumerate}
\end{theorem}

\begin{proof}
Let \(N\) be even, and define
\[
A=(\underbrace{1,\dots,1}_{N/2\ \text{times}},\underbrace{10,\dots,10}_{N/2\ \text{times}}).
\]
Thus \(A\) has exactly two utility levels:
\[
u_1=1,\qquad u_2=10,
\]
each with multiplicity \(N/2\). By Definition~\ref{def:ULBD},
\[
W_{\mathrm{ULBD}}(A)
=
1\cdot \frac{N}{2}\cdot 1
+
k\cdot \frac{N}{2}\cdot 10
=
\frac{N}{2}+5Nk.
\]
Its total utility is
\[
\sum_{i=1}^N A_i
=
\frac{N}{2}\cdot 1+\frac{N}{2}\cdot 10
=
\frac{11}{2}N.
\]

Now define \(B\in\mathbb R^N\) by
\[
B_i:=9+\frac{i-1}{N-1}
\qquad(i=1,\dots,N).
\]
Then \(B\) has \(N\) distinct utility levels, all lying in the interval \([9,10]\). So \(B\) is intuitively much more equal than \(A\), while still exhibiting many nearby levels.

Since the values of \(B\) form an arithmetic progression from \(9\) to \(10\), its total utility is
\[
\sum_{i=1}^N B_i
=
N\cdot \frac{9+10}{2}
=
\frac{19}{2}N
>
\frac{11}{2}N
=
\sum_{i=1}^N A_i.
\]
Hence \(B\) has strictly larger total utility than \(A\).

It remains to compare the ULBD values. Because \(B\) has \(N\) distinct levels, each with multiplicity \(1\),
\[
W_{\mathrm{ULBD}}(B)
=
\sum_{j=1}^N k^{j-1}\left(9+\frac{j-1}{N-1}\right).
\]
Since
\[
9+\frac{j-1}{N-1}\le 10
\qquad(j=1,\dots,N),
\]
we obtain
\[
W_{\mathrm{ULBD}}(B)
\le
10\sum_{j=1}^N k^{j-1}
\le
10\sum_{j=1}^{\infty} k^{j-1}
=
\frac{10}{1-k}.
\]

Therefore
\[
W_{\mathrm{ULBD}}(A)>W_{\mathrm{ULBD}}(B)
\]
whenever
\[
\frac{N}{2}+5Nk>\frac{10}{1-k}.
\]
Equivalently,
\[
N\left(\frac12+5k\right)>\frac{10}{1-k}.
\]
Since the left-hand side grows linearly in \(N\), there exists \(N_0\in\mathbb N\) such that for every even
\(N\ge N_0\), the inequality holds. For every such \(N\), the pair \(A,B\) constructed above satisfies all four claims.
\end{proof}

\subsection{Summary}
The two impossibility results are intended to highlight two distinct pathologies arising from the seemingly plausible characterization of the rank-weighted view. Note that under IRBD, the problem is not merely that the functional eventually resembles leximin in the abstract, since one might endorse leximin on independent grounds. Rather, the problem is that the ranking between two objects can reverse across scenarios for seemingly irrelevant reasons.  
Indeed, across these scenarios, both the internal distributional patterns and their total utility comparison remain unchanged as this reversal can always be generated by a simple pattern-preserving population replication. Under ULBD, the problem is worse: the functional penalizes a profile simply for having many distinct utility levels, even when those levels are tightly concentrated in a narrow interval such as \([9,10]\). As a result, an almost equal profile with much greater total utility can be ranked below a very unequal two-level profile containing only the values \(1\) and \(10\). Together with the previous counterexample that illustrates the similar arbitrariness of mainstream statistical measures, it seems that the practical approach of using ready-to-hand tools to express egalitarian principles isn't of much use. This further necessitates a rigorous and axiomatic approach to characterizing inequality measures.

Although statistical metrics at hand did not yield a coherent aggregative function due to the lack of consistency check, they provided crucial insights that necessitate the subsequent representation theorem. The mean function appears as a central component in most inequality measures beside range. Additionally,  while mean function plays a central role in measures like variance and standard deviation, its role in the Gini coefficient differs: it functions as a normalizing constant external to the main computational structure. Upon reflection, this is unsurprising, as it's a natural requirement that any generalized mean-function should rank equal distributions equally. Consequently, it is natural that broad inequality measures incorporate the mean function in certain form, and we conjecture that many measures may be parameterized functions of it. Gini coefficient offers a particularly direct geometric interpretation via the Lorenz curve. This observation leads us to conjecture that there exists a fundamental distinction between the Gini index and other measures of inequality.

In the subsequent section, we will present two representation theorems that formalize egalitarianism using Gini index and generalized Atkinson index. These theorems demonstrate how inequality measures can be derived from axioms, providing rigorous foundation for characterizing egalitarianism and overcome the difficulty we identified in this section.

\section{Axiomatic Characterization of Gini-Coefficient}

\subsection{Background}
When facing the problem of measuring inequality, the most natural approach is to establish some kind of metric based on "deviations from equality," so qualitative patterns of inequality become comparable numerical values. As we saw before, standard deviation understanding as AP+AVE isn't suitable for this role for its behavior in various aspects that had been illustrated by last section. However, the problem isn't with the underlying philosophical motivation but rather with the structural property of standard deviation that made it simply irrelevant to the pre-formal ideas. Indeed, the final verdict is Lorenz types ordering are  the unique representations of deviation from equality under assumption of ratio-invariability. 

The basic idea behind the following proof is to model egalitarian evaluation as choice among normalized Lorenz curve rather than outcome/lotteries in the ordinary EUT framework. This implicitly achieve two goals, by assuming Lorenz curve as fundamental object, it built the principle of ratio-invariance into assumption. Moreover, Given Lorenz curve would transform distribution of any total utility into the interval $[0,1]$, it expressed the very idea of extracting \textit{Distribution pattern} that we anticipated, then we could use it as \( f_{\text{Egal}}(U) \) that served as the penal coefficient of Utilitarian SWF .

Similar axioms and representation theorems that aligned in spirit with the above characterization had been attempted by Aaberge in \cite{Aaberge}. Unfortunately, both the theorems and the proof strategies presented in the original paper are incorrect. The main issues arise already at the  first several theorems that concerned with primal and dual one-sided representation, and therefore affect the route by which the present main theorem would be obtained from them. Furthermore, these issues are not fixable by minor adjustments. Indeed, as it will be explain that any representation theorem of Egalitarianism in the same spirit necessarily requires some more advanced techniques. We proceed by first presenting counterexamples, then analyzing the source of the error, and finally describing the proof strategy that would be developed in this paper. 

The obstruction is best understood as the existence of an ill-behaved representing kernel of an integral representation that was intended to be obtained, which could not be excluded from the standard decision-theoretic axioms assumed in the literature. On the domain of Lorenz curves, those axioms determine an affine continuous order, but they do not prevent the representing kernel function from developing a boundary singularity near the equality curve. Indeed, it would be shown a divergent kernel representations remains compatible with mixture independence, dominance, and closed-graph continuity on the domain itself. As a result, such singular phenomena would wreck the condition for finite Stieltjes Integral representation that the theorem asserts. After presentation of the counterexamples to the original characterization. This paper would introduce an novel regularity condition that excludes precisely this kind of blow-up and finally describing the novel proof strategies that would be developed in this paper.

\subsection{Counter-examples}

The axiomatic system that will be used throughout the paper is first introduced. It will serve both to construct counterexamples demonstrating the shortcomings of the Aaberge approach and to establish a series of representation and classification theorems about Lorenz ordering.

In what follows, we will work with the class of functions having these defining properties  of Lorenz curves.

Let $X:=C([0,1])$ endowed with the supremum norm $\|\cdot\|_\infty$, and define the Lorenz domain
\[
\mathcal L:=\Bigl\{L\in X\ \Big|\ L(0)=0,\ L(1)=1,\ L\ \text{is nondecreasing and convex}\Bigr\}.
\]
When invertibility is needed, we denote the strictly increasing subdomain
\[
\mathcal L^\uparrow:=\Bigl\{L\in\mathcal L\ \Big|\ L\ \text{is strictly increasing on }[0,1]\Bigr\},
\]
so that each $L\in\mathcal L^\uparrow$ has a (continuous) inverse $L^{-1}:[0,1]\to[0,1]$. In what follows, we denote the  relevant Lorenz domain by:
\[
\mathcal K\in\{\mathcal L,\mathcal L^\uparrow\}.
\]

A preference relation $\succsim$ is a binary relation on either Lorenz domain $\mathcal L$ or $\mathcal L^\uparrow$, as specified. Its strict part is defined as
\[
L_1\succ L_2
\quad:\Longleftrightarrow\quad
\bigl(L_1\succsim L_2\bigr)\ \text{and}\ \neg\bigl(L_2\succsim L_1\bigr).
\]

For $\lambda\in[0,1]$ and $L_1,L_2\in\mathcal L$, the (primal) mixture is the pointwise convex combination
$(1-\lambda)L_1+\lambda L_2\in\mathcal L$.
On $\mathcal L^\uparrow$ we also use the dual mixture
\[
L_1\oplus_\lambda L_2
:=\Bigl(\lambda L_1^{-1}+(1-\lambda)L_2^{-1}\Bigr)^{-1}\in\mathcal L^\uparrow,
\qquad \lambda\in[0,1].
\]

\begin{axiom}[Complete preorder]\label{ax:HM-order}
$\succsim$ is complete and transitive on the relevant Lorenz domain.
\end{axiom}

\begin{axiom}[Weak dominance]\label{ax:HM-dominance}
For all \(L_1,L_2\in\mathcal K\),
\[
L_1(u)\ge L_2(u)\ \ \forall u\in[0,1]
\quad\Longrightarrow\quad
L_1\succsim L_2.
\]
\end{axiom}

\begin{axiom}[Closed graph Continuity]\label{ax:HM-continuity}
The graph
\[
\{(L_1,L_2)\in\mathcal K\times\mathcal K:\ L_1\succsim L_2\}
\]
is closed in \(\mathcal K\times\mathcal K\), where \(\mathcal K\) carries the topology induced by
\((C([0,1]),\|\cdot\|_1)\).
\end{axiom}

\begin{axiom}[Nontriviality]\label{ax:HM-nontrivial}
There exist \(L^+,L^-\in\mathcal K\) such that \(L^+\succ L^-\).
\end{axiom}

\begin{axiom}[Primal Mixture Independence]\label{ax:HM-independence}
For all $L_1,L_2,L_3\in\mathcal L$ and all $\lambda\in(0,1)$,
\[
L_1\succsim L_2
\iff
(1-\lambda)L_1+\lambda L_3 \succsim (1-\lambda)L_2+\lambda L_3 .
\]
Equivalently, for all $L_1,L_2,L_3\in\mathcal L$ and all $\lambda\in(0,1)$,
\[
L_1\succ L_2
\iff
(1-\lambda)L_1+\lambda L_3 \succ (1-\lambda)L_2+\lambda L_3 .
\]
\end{axiom}

\begin{axiom}[Dual independence]\label{ax:HM-dual-independence}
For all $L_1,L_2,L_3\in\mathcal L^\uparrow$ and all $\lambda\in(0,1)$,
\[
L_1\succsim L_2
\iff
L_1\oplus_\lambda L_3\succsim L_2\oplus_\lambda L_3.
\]
Equivalently, for all $L_1,L_2,L_3\in\mathcal L^\uparrow$ and all $\lambda\in(0,1)$,
\[
L_1\succ L_2
\iff
L_1\oplus_\lambda L_3\succ L_2\oplus_\lambda L_3.
\]
\end{axiom}

Aaberge's route begins with two one-sided representation claims, by adopting axioms 1-5 it will be established that there exists an integral representation of primal ordering. By adopting axioms 1-4 and 6, it will be established that there exists an integral representation of dual ordering. More specifically, on the primal side, the asserted conclusion is that there exist \(c_1\in\mathbb R\) and a continuous nonincreasing kernel \(p:[0,1]\to\mathbb R\) such that
\[
V_p(L):=c_1+\int_0^1 p(u)\,dL(u)
\qquad(L\in\mathcal L)
\]
represents the preorder on \(\mathcal L\).

On the dual side, Aaberge's original formulation is based on the generalized inverse
\[
L^{\leftarrow}(t):=\inf\{u\in[0,1]:L(u)\ge t\},
\qquad t\in[0,1].
\]
The asserted conclusion is that there exist \(c_2\in\mathbb R\) and a continuous nondecreasing kernel \(q:[0,1]\to\mathbb R\) such that
\[
V_q^*(L):=c_2+\int_0^1 q(t)\,dL^{\leftarrow}(t)
\qquad(L\in\mathcal L)
\]
represents the preorder in dual form.

We now show that neither the one-sided primal nor the one-sided dual representation could hold as stated. In fact, we establish a stronger negative result: even after weakening the one-sided representing conclusions from continuous monotone kernels to monotone kernels of bounded variation, both representation claims still fail. The obstructions are endpoint singular behavior of the kernel functions, and the singular kernels produced by the following counterexamples do not belong to \(BV([0,1])\).

Set \(
E(u):=u,
\quad
B(u):=u^2.
\)
For every \(L\in\mathcal L\), convexity and the endpoint conditions imply
\[
0\le L(u)\le (1-u)L(0)+uL(1)=u
\qquad(u\in[0,1]).
\]
Define
\[
\Phi(L):=\int_0^1 \frac{L(u)}{u^{3/2}}\,du
\qquad(L\in\mathcal L).
\]
This is well defined because \(0\le L(u)\le u\) gives
\[
0\le \frac{L(u)}{u^{3/2}}\le \frac{1}{\sqrt u},
\qquad
\int_0^1 \frac{du}{\sqrt u}=2.
\]

Now define a preorder on \(\mathcal L\) by
\[
L_1\succsim L_2
\quad:\Longleftrightarrow\quad
\Phi(L_1)\ge \Phi(L_2).
\tag{$\star$}
\]
Axioms~\ref{ax:HM-order}, \ref{ax:HM-dominance}, and \ref{ax:HM-independence}
are immediate.

To verify Axiom~\ref{ax:HM-continuity}, let \(L,M\in\mathcal L\) and fix \(\delta\in(0,1]\). Then
\[
|\Phi(L)-\Phi(M)|
\le
\int_0^\delta \frac{|L(u)-M(u)|}{u^{3/2}}\,du
+
\int_\delta^1 \frac{|L(u)-M(u)|}{u^{3/2}}\,du.
\]
Since \(0\le L(u),M(u)\le u\), one has \(|L(u)-M(u)|\le u\), hence
\[
|\Phi(L)-\Phi(M)|
\le
\int_0^\delta u^{-1/2}\,du
+
\delta^{-3/2}\|L-M\|_1
=
2\sqrt{\delta}+\delta^{-3/2}\|L-M\|_1.
\]
Also \(\|L-M\|_1\le \int_0^1 u\,du=\tfrac12\). Choosing \(\delta=\|L-M\|_1^{1/2}\) gives
\[
|\Phi(L)-\Phi(M)|
\le
3\,\|L-M\|_1^{1/4}.
\]
Thus \(\Phi\) is continuous on \((\mathcal L,\|\cdot\|_1)\). Since \(\succsim\) is represented by \(\Phi\), its graph is closed in the \(L^1\)-topology.

Nontriviality holds because
\[
\Phi(E)=\int_0^1 u^{-1/2}\,du=2,
\qquad
\Phi(B)=\int_0^1 u^{1/2}\,du=\frac23,
\]
so \(E\succ B\).

Now define, for \(\varepsilon\in(0,1)\),
\[
L_\varepsilon(u):=
\begin{cases}
0,&0\le u\le \varepsilon,\\[4pt]
\dfrac{u-\varepsilon}{1-\varepsilon},&\varepsilon<u\le 1.
\end{cases}
\]
Then \(L_\varepsilon\in\mathcal L\), and
\[
\|E-L_\varepsilon\|_1
=
\int_0^\varepsilon u\,du
+
\int_\varepsilon^1 \frac{\varepsilon(1-u)}{1-\varepsilon}\,du
=
\frac{\varepsilon}{2}.
\tag{1}
\]
A direct calculation gives
\[
\Phi(L_\varepsilon)
=
\int_\varepsilon^1 \frac{(u-\varepsilon)/(1-\varepsilon)}{u^{3/2}}\,du
=
\frac{2(1-\sqrt\varepsilon)^2}{1-\varepsilon},
\]
hence
\[
\Phi(E)-\Phi(L_\varepsilon)
=
\frac{4(\sqrt\varepsilon-\varepsilon)}{1-\varepsilon}.
\tag{2}
\]
For \(\varepsilon\in(0,1/4)\), one has
\[
\Phi(B)<\Phi(L_\varepsilon)<\Phi(E).
\]
Since \(\Phi\) is affine, there is a unique \(t_\varepsilon\in(0,1)\) such that
\[
L_\varepsilon\sim (1-t_\varepsilon)E+t_\varepsilon B,
\]
and
\[
t_\varepsilon
=
\frac{\Phi(E)-\Phi(L_\varepsilon)}{\Phi(E)-\Phi(B)}
=
\frac{3(\sqrt\varepsilon-\varepsilon)}{1-\varepsilon}.
\]
Combining this with \((1)\) gives
\[
\frac{t_\varepsilon}{\|E-L_\varepsilon\|_1}
=
\frac{6(\varepsilon^{-1/2}-1)}{1-\varepsilon}
\xrightarrow[\varepsilon\downarrow0]{}+\infty.
\tag{3}
\]

We now show that \(\succsim\) admits no representation by any nonincreasing kernel of bounded variation.
Assume, towards a contradiction, that there exists a nonincreasing function
\[
p\in BV([0,1])
\]
such that
\[
L_1\succsim L_2
\iff
\int_0^1 p\,dL_1\ge \int_0^1 p\,dL_2
\qquad(L_1,L_2\in\mathcal L).
\]
Since \(p\in BV([0,1])\) and every \(L\in\mathcal L\) is continuous, the Stieltjes integral is well defined. Write
\[
V_p(L):=\int_0^1 p\,dL.
\]
Because this representation induces the same indifference relation as \(\Phi\),
the same coefficient \(t_\varepsilon\) must satisfy
\[
t_\varepsilon
=
\frac{V_p(E)-V_p(L_\varepsilon)}{V_p(E)-V_p(B)}.
\tag{4}
\]
The denominator is strictly positive because \(E\succ B\).

Now \(E-L_\varepsilon\) is continuous and vanishes at both endpoints, so integration by parts gives
\[
V_p(E)-V_p(L_\varepsilon)
=
-\int_0^1 (E-L_\varepsilon)\,dp.
\]
Moreover,
\[
0\le E(u)-L_\varepsilon(u)\le \varepsilon
\qquad(u\in[0,1]),
\]
and from
\[
\|E-L_\varepsilon\|_1=\frac{\varepsilon}{2}
\]
we obtain
\[
0\le E(u)-L_\varepsilon(u)\le 2\|E-L_\varepsilon\|_1.
\]
Therefore
\[
|V_p(E)-V_p(L_\varepsilon)|
\le
2\,\operatorname{Var}(p)\,\|E-L_\varepsilon\|_1.
\]
Substituting this into \((4)\), we get
\[
0<
\frac{t_\varepsilon}{\|E-L_\varepsilon\|_1}
\le
\frac{2\,\operatorname{Var}(p)}{V_p(E)-V_p(B)}.
\]
But earlier we proved
\[
\frac{t_\varepsilon}{\|E-L_\varepsilon\|_1}
=
\frac{6(\varepsilon^{-1/2}-1)}{1-\varepsilon}
\xrightarrow[\varepsilon\downarrow0]{}+\infty,
\]
a contradiction. Hence no finite nonincreasing kernel in \(BV([0,1])\) can represent the preorder defined by \(\Phi\).

Now consider the dual one-sided theorem. Recall that Aaberge formulates the dual representation on \(\mathcal L\) using the generalized inverse
\[
L^{\leftarrow}(t):=\inf\{u\in[0,1]:L(u)\ge t\}.
\]
For the present counterexample it is enough to work on the strictly increasing subset $L^\uparrow$ since in following proof all the involved Lorenz curves
\(
E,\ B,\ M_\varepsilon\in\mathcal L^\uparrow,
\)
On \(\mathcal L^\uparrow\) one has \(L^{\leftarrow}=L^{-1}\). Thus a contradiction on \(\mathcal L^\uparrow\) already refutes the larger-domain statement.

Consider the nondecreasing kernel
\[
q_*(t):=\frac{1}{\sqrt{1-t}},
\qquad 0\le t<1.
\] 
Define for \(L\in\mathcal L^\uparrow\),
\[
\Psi(L):=\int_0^1 q_*(t)\,d(L^{-1})(t)
=\int_0^1 \frac{du}{\sqrt{1-L(u)}}.
\]
 
This integral is finite because every Lorenz curve satisfies \(L(u)\le u\), hence
\[
0\le \frac{1}{\sqrt{1-L(u)}}\le \frac{1}{\sqrt{1-u}},
\qquad
\int_0^1 \frac{du}{\sqrt{1-u}}=2.
\]

Now define a preorder on \(\mathcal L^\uparrow\) by
\[
L_1\succsim L_2
\quad:\Longleftrightarrow\quad
\Psi(L_1)\ge \Psi(L_2).
\]
This is a complete preorder. Weak dominance holds because
\[
L_1(u)\ge L_2(u)\ \forall u
\quad\Longrightarrow\quad
\frac{1}{\sqrt{1-L_1(u)}}\ge \frac{1}{\sqrt{1-L_2(u)}}
\]
and therefore \(\Psi(L_1)\ge\Psi(L_2)\). Dual independence holds because
\[
(L_1\oplus_\lambda L_3)^{-1}
=
\lambda L_1^{-1}+(1-\lambda)L_3^{-1},
\]
so
\[
\Psi(L_1\oplus_\lambda L_3)
=
\lambda\Psi(L_1)+(1-\lambda)\Psi(L_3).
\]

To verify \(L^1\)-continuity, let \(L,M\in\mathcal L^\uparrow\) and fix \(\delta\in(0,1]\). Then
\[
|\Psi(L)-\Psi(M)|
\le
\int_{1-\delta}^1
\left(
\frac{1}{\sqrt{1-L(u)}}+\frac{1}{\sqrt{1-M(u)}}
\right)du
+
\int_0^{1-\delta}
\left|
\frac{1}{\sqrt{1-L(u)}}-\frac{1}{\sqrt{1-M(u)}}
\right|du.
\]
Because \(L(u),M(u)\le u\), the first term is bounded by
\[
2\int_{1-\delta}^1 \frac{du}{\sqrt{1-u}}=4\sqrt{\delta}.
\]
On \([0,1-\delta]\), the derivative of \(x\mapsto(1-x)^{-1/2}\) is bounded by \(1/(2\delta^{3/2})\), so the second term is at most
\[
\frac12\,\delta^{-3/2}\|L-M\|_1.
\]
Choosing \(\delta=\|L-M\|_1^{1/2}\) gives
\[
|\Psi(L)-\Psi(M)|\le \frac92\,\|L-M\|_1^{1/4}.
\]
Thus \(\Psi\) is continuous on \((\mathcal L^\uparrow,\|\cdot\|_1)\), and the graph of the induced preorder is \(L^1\)-closed.

Recall the definitions:
\(
E(u):=u,
\quad
B(u):=u^2
\).
Then
\[
\Psi(E)=\int_0^1 \frac{du}{\sqrt{1-u}}=2,
\qquad
\Psi(B)=\int_0^1 \frac{du}{\sqrt{1-u^2}}=\frac{\pi}{2},
\]
so \(E\succ B\).

For \(0<\varepsilon<1\), define
\[
\begin{aligned}
J_\varepsilon(t)&:=
\begin{cases}
(1+\varepsilon)t,&0\le t\le 1-\varepsilon,\\[4pt]
1-\varepsilon(1-t),&1-\varepsilon\le t\le 1,
\end{cases}
\qquad
M_\varepsilon(u)&=
\begin{cases}
\dfrac{u}{1+\varepsilon},&0\le u\le 1-\varepsilon^2,\\[6pt]
1-\dfrac{1-u}{\varepsilon},&1-\varepsilon^2\le u\le 1.
\end{cases}
\end{aligned}
\]
The left-side function is increasing and concave, with \(
J_\varepsilon(0)=0,\quad J_\varepsilon(1)=1
\).
Simple calculation shows it is the inverse of a unique curve \(M_\varepsilon\in\mathcal L^\uparrow\).  
A direct calculation gives:
\[
\|E-M_\varepsilon\|_1
=
\int_0^{1-\varepsilon^2}\left(u-\frac{u}{1+\varepsilon}\right)du
+
\int_{1-\varepsilon^2}^1\left(u-1+\frac{1-u}{\varepsilon}\right)du
=
\frac{\varepsilon(1-\varepsilon)}{2}.
\tag{D1}
\]

Next,
\[
\Psi(M_\varepsilon)
=
\int_0^1 \frac{1}{\sqrt{1-t}}\,dJ_\varepsilon(t)
=
(1+\varepsilon)\int_0^{1-\varepsilon}\frac{dt}{\sqrt{1-t}}
+
\varepsilon\int_{1-\varepsilon}^1\frac{dt}{\sqrt{1-t}},
\]
so
\[
\Psi(M_\varepsilon)
=
2(1+\varepsilon)(1-\sqrt{\varepsilon})+2\varepsilon^{3/2}
=
2+2\varepsilon-2\sqrt{\varepsilon}.
\]
Therefore
\[
\Psi(E)-\Psi(M_\varepsilon)=2(\sqrt{\varepsilon}-\varepsilon).
\tag{D2}
\]

Since \(\Psi(M_\varepsilon)\to2\) and \(\Psi(B)=\pi/2<2\), for all sufficiently small \(\varepsilon\) one has
\[
\Psi(B)<\Psi(M_\varepsilon)<\Psi(E).
\]
Because \(\Psi\) is affine under the dual mixture, there is a unique \(\tau_\varepsilon\in(0,1)\) such that
\[
M_\varepsilon\sim E\oplus_{1-\tau_\varepsilon}B,
\]
and
\[
\tau_\varepsilon
=
\frac{\Psi(E)-\Psi(M_\varepsilon)}{\Psi(E)-\Psi(B)}
=
\frac{4(\sqrt{\varepsilon}-\varepsilon)}{4-\pi}.
\]
Combining this with \((D1)\), we obtain
\[
\frac{\tau_\varepsilon}{\|E-M_\varepsilon\|_1}
=
\frac{8(\varepsilon^{-1/2}-1)}{(4-\pi)(1-\varepsilon)}
\xrightarrow[\varepsilon\downarrow0]{}+\infty.
\tag{D3}
\]

Assume towards contradiction, that there exists a nondecreasing function \(
q\in BV([0,1])
\)
such that
\[
L_1\succsim L_2
\iff
\int_0^1 q(t)\,d(L_1^{-1})(t)\ge \int_0^1 q(t)\,d(L_2^{-1})(t)
\qquad(L_1,L_2\in\mathcal L^\uparrow).
\]
Since \(q\in BV([0,1])\) and \(L^{-1}\) is continuous on \(\mathcal L^\uparrow\), the Stieltjes integral is well defined. Write
\[
V_q^*(L):=\int_0^1 q(t)\,d(L^{-1})(t).
\]
Because the same indifference relation must hold,
\[
\tau_\varepsilon
=
\frac{V_q^*(E)-V_q^*(M_\varepsilon)}{V_q^*(E)-V_q^*(B)}.
\tag{D4}
\]
The denominator is strictly positive because \(E\succ B\).

Now \(E^{-1}-M_\varepsilon^{-1}\) is continuous and vanishes at both endpoints, so
\[
V_q^*(E)-V_q^*(M_\varepsilon)
=
-\int_0^1 \bigl(E^{-1}(t)-M_\varepsilon^{-1}(t)\bigr)\,dq(t).
\]
Moreover,
\[
E^{-1}(t)=t,
\qquad
M_\varepsilon^{-1}(t)=J_\varepsilon(t),
\]
so
\[
|E^{-1}(t)-M_\varepsilon^{-1}(t)|
=
|t-J_\varepsilon(t)|
\le \varepsilon(1-\varepsilon)
\qquad(t\in[0,1]).
\]
Since
\[
\|E-M_\varepsilon\|_1=\frac{\varepsilon(1-\varepsilon)}{2},
\]
this becomes
\[
|E^{-1}(t)-M_\varepsilon^{-1}(t)|
\le 2\|E-M_\varepsilon\|_1.
\]
Hence
\[
|V_q^*(E)-V_q^*(M_\varepsilon)|
\le
2\,\operatorname{Var}(q)\,\|E-M_\varepsilon\|_1.
\]
Substituting into \((D4)\), we get
\[
0<
\frac{\tau_\varepsilon}{\|E-M_\varepsilon\|_1}
\le
\frac{2\,\operatorname{Var}(q)}{V_q^*(E)-V_q^*(B)}.
\]
But earlier we proved in \((D3)\)
\[
\frac{\tau_\varepsilon}{\|E-M_\varepsilon\|_1}
=
\frac{8(\varepsilon^{-1/2}-1)}{(4-\pi)(1-\varepsilon)}
\xrightarrow[\varepsilon\downarrow0]{}+\infty,
\]
Contradiction again. Hence no finite nondecreasing kernel in \(BV([0,1])\) can represent the dual preorder.

The dual one-sided theorem fails for the same underlying reason as the primal one. The current package of axioms does not exclude the existence of singular endpoint kernel for dual integral representation. On the primal side the singularity occurs at \(u=0\); on the dual side it reappears at \(t=1\). Therefore the standard set of decision-theoretic axioms is not sufficient for establishing non-singular integral representation, and Aaberge's claimed derivation of both primal and dual theorems cannot be corrected without further strengthened assumptions that rule out this kind of blow-up at equality.

Interestingly, while counterexamples to the one-sided primal representation theorem and the dual theorem individually are easy to find, the actual status of the main theorem under the combined axiomatic assumptions remains unclear even without additional regularity strengthening. As there is no obvious counterexample or impossibility theorem. Perhaps the broader project of establishing the affine  kernel integral representation under the standard decision-theoretic package could still survive. 

By this, we mean that the above counterexample only illustrated the invalidity of the conclusions of both the one-sided primal and one-sided dual theorems, thereby undermining the validity of the current proof strategy of the main theorem, as it is based on them. However, it is not actually a counterexample to the main theorem itself. Nevertheless, given the length and intricacy of the following proofs even under additional strengthened conditions, it seems such a proof would be far more difficult than originally conceived.

\subsection{Analytical Preliminaries}
This subsection presents standard techniques and results from real and functional analysis. They are included only to keep the paper self-contained.
We begin this review by presenting the derivations and properties of bump and mollifier functions, as they will be invoked repeatedly. We then proceed to Hahn–Banach–type and Riesz–Markov–type machinery. Interested readers may consult the appendix.
Lemma~\ref{lem:test-functions-determine-measures} is a standard uniqueness fact for locally finite Borel measures tested against smooth compactly supported functions; cf. \cite{FollandRealAnalysis1999,Bogachev2007,Hormander2003}.
Lemmas~\ref{lem:helly-sublinear}--\ref{lem:HB-separation-convex} are standard Hahn--Banach extension/separation results; cf. \cite{ConwayFunctionalAnalysis1990,AliprantisBorder2006}.
Lemma~\ref{lem:R2-positive-Riesz} is a specialization of the Riesz representation theorem on $C([0,1])$; cf. \cite{ RudinRealComplex1987,FollandRealAnalysis1999}.

\begin{lemma}[Smooth bump, standard mollifier, and interval cutoffs]
\label{lem:mollifier-cutoffs}
Define \(\rho:\mathbb R\to\mathbb R\) by
\[
\rho(s):=
\begin{cases}
e^{-1/s},& s>0,\\
0,& s\le 0.
\end{cases}
\]
Then \(\rho\in C^\infty(\mathbb R)\) and \(\rho^{(k)}(0)=0\) for every \(k\in\mathbb N_0\).

Set \(\varphi(t):=\rho(1-t^2)\), \(A:=\int_{\mathbb R}\varphi(t)\,dt\), and \(\psi:=\varphi/A\). Then \(\psi\in C_c^\infty(\mathbb R)\), \(\psi\ge0\), \(\supp(\psi)=[-1,1]\), and \(\int_{\mathbb R}\psi(t)\,dt=1\).

For \(\varepsilon>0\), define \(\psi_\varepsilon(t):=\varepsilon^{-1}\psi(t/\varepsilon)\) for \(t\in\mathbb R\). Then \(\psi_\varepsilon\in C_c^\infty(\mathbb R)\), \(\psi_\varepsilon\ge0\), \(\supp(\psi_\varepsilon)\subset[-\varepsilon,\varepsilon]\), and \(\int_{\mathbb R}\psi_\varepsilon(t)\,dt=1\).

If \(0<a<b<1\) and \(0<\varepsilon<\min\{a,1-b\}\), define
\begin{equation}\label{eq:interval-cutoff}
\eta_{a,b,\varepsilon}(u):=\int_a^b \psi_\varepsilon(u-s)\,ds
\qquad(u\in\mathbb R).
\end{equation}
Then \(\eta_{a,b,\varepsilon}\in C_c^\infty(0,1)\), and
\begin{equation}\label{eq:interval-cutoff-bounds}
0\le \eta_{a,b,\varepsilon}(u)\le 1
\qquad\forall\,u\in\mathbb R,
\end{equation}
\begin{equation}\label{eq:interval-cutoff-support}
\supp(\eta_{a,b,\varepsilon})\subset[a-\varepsilon,b+\varepsilon]\subset(0,1),
\end{equation}
\begin{equation}\label{eq:cutoff-plateau}
\eta_{a,b,\varepsilon}\equiv 1
\qquad\text{on }[a+\varepsilon,b-\varepsilon].
\end{equation}
\end{lemma}

\begin{deferredproof}{lem:mollifier-cutoffs}
We first prove smoothness of \(\rho\). For \(s>0\) and \(k\in\mathbb N_0\), we claim that there exists a polynomial \(P_k\) such that
\begin{equation}\label{eq:rho-deriv-form}
\rho^{(k)}(s)=P_k(1/s)\,e^{-1/s}.
\end{equation}
For \(k=0\), take \(P_0\equiv1\). If \eqref{eq:rho-deriv-form} holds at rank \(k\), then
\[
\rho^{(k+1)}(s)
=
\frac{d}{ds}\bigl(P_k(1/s)e^{-1/s}\bigr)
=
\frac{1}{s^2}\bigl(P_k(1/s)-P_k'(1/s)\bigr)e^{-1/s},
\]
so \eqref{eq:rho-deriv-form} holds at rank \(k+1\) with
\[
P_{k+1}(x):=x^2\bigl(P_k(x)-P_k'(x)\bigr).
\]
This proves the claim.

For every \(m\in\mathbb N\),
\begin{equation}\label{eq:exp-dominates-poly}
\lim_{s\downarrow0}s^{-m}e^{-1/s}=0.
\end{equation}
Indeed, with \(t=1/s\), one has \(s^{-m}e^{-1/s}=t^m e^{-t}=t^m/e^t\), and repeated l'H\^opital gives \(\lim_{t\to\infty}t^m/e^t=0\).

For \(k\in\mathbb N_0\), define
\[
g_k(s):=
\begin{cases}
P_k(1/s)e^{-1/s},& s>0,\\
0,& s\le0.
\end{cases}
\]
Then \(g_k=\rho^{(k)}\) on \((0,\infty)\) by \eqref{eq:rho-deriv-form}, while \(g_k\equiv0\) on \((-\infty,0]\).

Write \(P_k(x)=\sum_{j=0}^d a_jx^j\). For \(s>0\),
\[
g_k(s)=\sum_{j=0}^d a_j\,s^{-j}e^{-1/s},
\]
and each term tends to \(0\) as \(s\downarrow0\): for \(j=0\) this is immediate, and for \(j\ge1\) it follows from \eqref{eq:exp-dominates-poly}. Hence \(g_k(s)\to0=g_k(0)\), so \(g_k\) is continuous at \(0\).

Likewise,
\[
\frac{g_k(h)-g_k(0)}{h}
=
\begin{cases}
0,& h<0,\\[2mm]
\sum_{j=0}^d a_j\,h^{-(j+1)}e^{-1/h},& h>0,
\end{cases}
\]
and every term on the positive side tends to \(0\) by \eqref{eq:exp-dominates-poly}. Thus \(g_k\) is differentiable at \(0\) with \(g_k'(0)=0\).

Away from \(0\), we have \(g_k'=g_{k+1}\) on \((0,\infty)\) by construction, while both functions vanish identically on \((-\infty,0)\). Hence \(g_k\in C^1(\mathbb R)\) and \(g_k'=g_{k+1}\) on \(\mathbb R\). By induction on \(k\), it follows that \(g_0=\rho\in C^\infty(\mathbb R)\) and
\[
\rho^{(k)}(0)=g_k(0)=0
\qquad\forall\,k\in\mathbb N_0.
\]

Now define \(\varphi(t):=\rho(1-t^2)\). Since \(\rho\in C^\infty(\mathbb R)\) and \(t\mapsto1-t^2\) is \(C^\infty\), we have \(\varphi\in C^\infty(\mathbb R)\). Moreover,
\[
\varphi(t)>0\quad\text{for }|t|<1,\qquad
\varphi(t)=0\quad\text{for }|t|\ge1,
\]
because \(\rho(s)>0\) for \(s>0\) and \(\rho(s)=0\) for \(s\le0\). Therefore \(\supp(\varphi)=[-1,1]\).

Since \(\varphi\) is continuous and compactly supported, the integral \(A:=\int_{\mathbb R}\varphi(t)\,dt\) is finite. Also \(A>0\): since \(\varphi(0)=e^{-1}>0\), continuity gives \(c>0\) and \(\delta_0>0\) such that \(\varphi(t)\ge c\) for \(|t|\le\delta_0\), and thus
\[
A\ge \int_{-\delta_0}^{\delta_0}c\,dt=2c\delta_0>0.
\]

Define \(\psi:=\varphi/A\). Then \(\psi\in C_c^\infty(\mathbb R)\), \(\psi\ge0\), \(\supp(\psi)=[-1,1]\), and \(\int_{\mathbb R}\psi(t)\,dt=1\).

Fix \(\varepsilon>0\). The function \(\psi_\varepsilon(t)=\varepsilon^{-1}\psi(t/\varepsilon)\) belongs to \(C_c^\infty(\mathbb R)\) and is nonnegative. If \(t\notin[-\varepsilon,\varepsilon]\), then \(t/\varepsilon\notin[-1,1]=\supp(\psi)\), hence \(\psi_\varepsilon(t)=0\); therefore \(\supp(\psi_\varepsilon)\subset[-\varepsilon,\varepsilon]\). Moreover, with the change of variables \(r=t/\varepsilon\),
\[
\int_{\mathbb R}\psi_\varepsilon(t)\,dt
=
\int_{\mathbb R}\frac1\varepsilon\,\psi(t/\varepsilon)\,dt
=
\int_{\mathbb R}\psi(r)\,dr
=
1.
\]

Now fix \(0<a<b<1\) and \(0<\varepsilon<\min\{a,1-b\}\), and define \(\eta_{a,b,\varepsilon}\) by \eqref{eq:interval-cutoff}. For every \(k\in\mathbb N_0\), the map \((u,s)\mapsto \psi_\varepsilon^{(k)}(u-s)\) is continuous on \(\mathbb R\times[a,b]\), and \([a,b]\) is compact. Differentiation under the integral sign therefore gives
\[
\eta_{a,b,\varepsilon}^{(k)}(u)
=
\int_a^b \psi_\varepsilon^{(k)}(u-s)\,ds
\qquad(u\in\mathbb R),
\]
so \(\eta_{a,b,\varepsilon}\in C^\infty(\mathbb R)\).

Since \(\psi_\varepsilon\ge0\), we have \(\eta_{a,b,\varepsilon}\ge0\). Also
\[
\eta_{a,b,\varepsilon}(u)
=
\int_a^b\psi_\varepsilon(u-s)\,ds
\le
\int_{\mathbb R}\psi_\varepsilon(u-s)\,ds
=
\int_{\mathbb R}\psi_\varepsilon(r)\,dr
=
1,
\]
which proves \eqref{eq:interval-cutoff-bounds}.

If \(u\notin[a-\varepsilon,b+\varepsilon]\), then \(|u-s|>\varepsilon\) for every \(s\in[a,b]\). Since \(\supp(\psi_\varepsilon)\subset[-\varepsilon,\varepsilon]\), this implies \(\psi_\varepsilon(u-s)=0\) for all \(s\in[a,b]\), and therefore
\[
\supp(\eta_{a,b,\varepsilon})\subset[a-\varepsilon,b+\varepsilon].
\]
Because \(0<\varepsilon<\min\{a,1-b\}\), we also have \([a-\varepsilon,b+\varepsilon]\subset(0,1)\). This proves \eqref{eq:interval-cutoff-support}, and after restriction to \((0,1)\) we get \(\eta_{a,b,\varepsilon}\in C_c^\infty(0,1)\).

Finally, let \(u\in[a+\varepsilon,b-\varepsilon]\). Then \([u-\varepsilon,u+\varepsilon]\subset[a,b]\). Since \(\psi_\varepsilon(u-s)=0\) whenever \(|u-s|>\varepsilon\), we have
\[
\eta_{a,b,\varepsilon}(u)
=
\int_{u-\varepsilon}^{u+\varepsilon}\psi_\varepsilon(u-s)\,ds.
\]
With the change of variables \(r=u-s\), this becomes
\[
\eta_{a,b,\varepsilon}(u)
=
\int_{-\varepsilon}^{\varepsilon}\psi_\varepsilon(r)\,dr
=
\int_{\mathbb R}\psi_\varepsilon(r)\,dr
=
1.
\]
Thus \eqref{eq:cutoff-plateau} holds.
\end{deferredproof}

\begin{lemma}[Approximation of interval indicators by interval cutoffs]
\label{lem:indicator-approx-cutoffs}
Fix \(0<a<b<1\), and set
\[
\delta:=\min\{a,\ 1-b,\ (b-a)/8\}>0,\qquad
\varepsilon_n:=\delta/n,\qquad
\chi_n:=\eta_{a+2\varepsilon_n,\ b-2\varepsilon_n,\ \varepsilon_n}
\qquad(n\in\mathbb N).
\]
Then \(\chi_n\in C_c^\infty(0,1)\) for every \(n\), and
\begin{equation}\label{eq:interval-indicator-domination}
0\le \chi_n\le \mathbf 1_{(a,b)}
\qquad\text{on }(0,1)
\qquad(n\in\mathbb N),
\end{equation}
\begin{equation}\label{eq:interval-indicator-pointwise}
\chi_n(u)\longrightarrow \mathbf 1_{(a,b)}(u)
\qquad\forall\,u\in(0,1).
\end{equation}
\end{lemma}

\begin{deferredproof}{lem:indicator-approx-cutoffs}
Fix \(n\in\mathbb N\). Since \(4\varepsilon_n\le 4\delta\le (b-a)/2<b-a\), we have \(a+2\varepsilon_n<b-2\varepsilon_n\). Also \(\varepsilon_n\le\delta\le a\) and \(\varepsilon_n\le\delta\le 1-b\), so
\[
0<\varepsilon_n<a+2\varepsilon_n,\qquad
0<\varepsilon_n<1-(b-2\varepsilon_n).
\]
Hence \(\chi_n\) is well-defined by \eqref{eq:interval-cutoff}, and Lemma~\ref{lem:mollifier-cutoffs} yields \(\chi_n\in C_c^\infty(0,1)\).

Applying \eqref{eq:interval-cutoff-support} with
\[
a\rightsquigarrow a+2\varepsilon_n,\qquad
b\rightsquigarrow b-2\varepsilon_n,\qquad
\varepsilon\rightsquigarrow \varepsilon_n,
\]
we obtain
\[
\supp(\chi_n)\subset[a+\varepsilon_n,b-\varepsilon_n]\subset(a,b).
\]
Together with \eqref{eq:interval-cutoff-bounds}, this implies
\[
0\le \chi_n\le \mathbf 1_{(a,b)}
\qquad\text{on }(0,1),
\]
which proves \eqref{eq:interval-indicator-domination}. Likewise, \eqref{eq:cutoff-plateau} gives
\[
\chi_n\equiv1
\qquad\text{on }[a+3\varepsilon_n,b-3\varepsilon_n].
\]

To prove \eqref{eq:interval-indicator-pointwise}, fix \(u\in(0,1)\). If \(u\notin(a,b)\), then \(u\notin\supp(\chi_n)\) for every \(n\), because \(\supp(\chi_n)\subset(a,b)\). Hence \(\chi_n(u)=0=\mathbf 1_{(a,b)}(u)\) for all \(n\).

Now assume \(u\in(a,b)\), and set \(h:=\min\{u-a,\ b-u\}>0\). Since \(\varepsilon_n\downarrow0\), there exists \(N\in\mathbb N\) such that \(3\varepsilon_n<h\) for all \(n\ge N\). For such \(n\), we have \(a+3\varepsilon_n<u<b-3\varepsilon_n\), so \(u\in[a+3\varepsilon_n,b-3\varepsilon_n]\), and therefore \(\chi_n(u)=1=\mathbf 1_{(a,b)}(u)\) for every \(n\ge N\). This proves \eqref{eq:interval-indicator-pointwise}.
\end{deferredproof}

\begin{lemma}[Determination of locally finite Borel measures by $C_c^\infty(0,1)$]
\label{lem:test-functions-determine-measures}
Let \(\alpha\) and \(\beta\) be nonnegative Borel measures on \((0,1)\). Assume that
\[
\alpha(K)<\infty,\qquad \beta(K)<\infty
\qquad\text{for every compact set } K\Subset(0,1).
\]
Let \(\lambda\in[0,\infty)\). If
\[
\int_{(0,1)} \eta(u)\,\alpha(du)
=
\lambda\int_{(0,1)} \eta(u)\,\beta(du)
\qquad
\forall\,\eta\in C_c^\infty(0,1),
\]
then \(\alpha=\lambda\beta\) as Borel measures on \((0,1)\).
\end{lemma}

\begin{deferredproof}{lem:test-functions-determine-measures}
Define \(\widetilde\beta:=\lambda\beta\). Then \(\widetilde\beta\) is a nonnegative Borel measure on \((0,1)\), and for every compact set \(K\Subset(0,1)\),
\[
\widetilde\beta(K)=\lambda\,\beta(K)<\infty.
\]
Thus the hypothesis becomes
\[
\int_{(0,1)} \eta(u)\,\alpha(du)
=
\int_{(0,1)} \eta(u)\,\widetilde\beta(du)
\qquad
\forall\,\eta\in C_c^\infty(0,1).
\]
It therefore suffices to prove that \(\alpha=\widetilde\beta\) on \(\mathcal B((0,1))\).

\emph{Step 1: equality on open intervals.}
Fix \(0<a<b<1\). By Lemma~\ref{lem:indicator-approx-cutoffs}, there exists a sequence \((\chi_n)_{n\ge1}\subset C_c^\infty(0,1)\) satisfying \eqref{eq:interval-indicator-domination} and \eqref{eq:interval-indicator-pointwise}. Because \([a,b]\Subset(0,1)\), local finiteness gives
\[
\alpha([a,b])<\infty,\qquad \widetilde\beta([a,b])<\infty.
\]
Hence \(\mathbf 1_{(a,b)}\in L^1(\alpha)\cap L^1(\widetilde\beta)\), and dominated convergence yields
\[
\alpha((a,b))
=
\lim_{n\to\infty}\int_{(0,1)}\chi_n(u)\,\alpha(du),\qquad
\widetilde\beta((a,b))
=
\lim_{n\to\infty}\int_{(0,1)}\chi_n(u)\,\widetilde\beta(du).
\]
Since the two integrals agree for every \(n\), passing to the limit gives
\[
\alpha((a,b))=\widetilde\beta((a,b))
\qquad\forall\,0<a<b<1.
\]

\emph{Step 2: equality on every Borel subset of each interior interval \(I_m\).}
Fix \(m\ge3\), and set
\[
I_m:=\left(\frac1m,\,1-\frac1m\right).
\]
Let \(\mathcal B(I_m)\) denote the Borel \(\sigma\)-algebra of the subspace \(I_m\). Every \(B\in\mathcal B(I_m)\) may be viewed canonically as a Borel subset of \((0,1)\) contained in \(I_m\), and we define the restricted measures
\[
\alpha_m(B):=\alpha(B),\qquad
\widetilde\beta_m(B):=\widetilde\beta(B)
\qquad(B\in\mathcal B(I_m)).
\]
Because \(\overline{I_m}=[1/m,1-1/m]\Subset(0,1)\), local finiteness implies
\[
\alpha_m(I_m)=\alpha(I_m)\le \alpha(\overline{I_m})<\infty,\qquad
\widetilde\beta_m(I_m)=\widetilde\beta(I_m)\le \widetilde\beta(\overline{I_m})<\infty.
\]
Hence \(\alpha_m\) and \(\widetilde\beta_m\) are finite measures on \((I_m,\mathcal B(I_m))\).

Define
\[
\mathcal Q_m
:=
\{\varnothing\}
\cup
\left\{
(c,d):
c,d\in\mathbb Q,\ \frac1m<c<d<1-\frac1m
\right\}.
\]
We first show that \(\mathcal Q_m\) is a \(\pi\)-system. If one of \(J_1,J_2\in\mathcal Q_m\) is empty, then \(J_1\cap J_2=\varnothing\in\mathcal Q_m\). Otherwise,
\[
J_1=(c_1,d_1),\qquad J_2=(c_2,d_2)
\]
with \(c_i,d_i\in\mathbb Q\) and \(1/m<c_i<d_i<1-1/m\). Then
\[
J_1\cap J_2=
\begin{cases}
(\max\{c_1,c_2\},\,\min\{d_1,d_2\}),
& \text{if }\max\{c_1,c_2\}<\min\{d_1,d_2\},\\[2mm]
\varnothing,
& \text{if }\max\{c_1,c_2\}\ge \min\{d_1,d_2\},
\end{cases}
\]
and the endpoints are again rational. Thus \(J_1\cap J_2\in\mathcal Q_m\), so \(\mathcal Q_m\) is a \(\pi\)-system.

We next claim that \(\sigma(\mathcal Q_m)=\mathcal B(I_m)\). Since every member of \(\mathcal Q_m\) is open in the relative topology of \(I_m\), one has \(\sigma(\mathcal Q_m)\subset\mathcal B(I_m)\). Conversely, let \(O\subset I_m\) be open in the relative topology, and fix \(x\in O\). There exists \(\varepsilon_0>0\) such that
\[
(x-\varepsilon_0,x+\varepsilon_0)\cap I_m\subset O.
\]
Set
\[
\varepsilon
:=
\min\left\{
\varepsilon_0,\,
\frac12\left(x-\frac1m\right),\,
\frac12\left(1-\frac1m-x\right)
\right\}>0.
\]
Then
\[
(x-\varepsilon,x+\varepsilon)\subset I_m\cap O.
\]
By density of \(\mathbb Q\), choose \(c,d\in\mathbb Q\) such that
\[
x-\varepsilon<c<x<d<x+\varepsilon.
\]
Hence
\[
\frac1m<c<d<1-\frac1m,\qquad
x\in(c,d)\subset(x-\varepsilon,x+\varepsilon)\subset O.
\]
So there exists \(J_x\in\mathcal Q_m\) with \(x\in J_x\subset O\). Since \(x\in O\) was arbitrary,
\[
O=\bigcup_{x\in O}J_x
=
\bigcup\{J\in\mathcal Q_m:J\subset O\}.
\]
Because \(\mathcal Q_m\) is countable, the last union is countable, so \(O\in\sigma(\mathcal Q_m)\). Thus every open subset of \(I_m\) belongs to \(\sigma(\mathcal Q_m)\), and therefore \(\mathcal B(I_m)\subset\sigma(\mathcal Q_m)\). Hence \(\sigma(\mathcal Q_m)=\mathcal B(I_m)\).

Now Step~1 shows that
\[
\alpha_m(J)=\widetilde\beta_m(J)
\qquad\forall\,J\in\mathcal Q_m:
\]
this is trivial for \(J=\varnothing\), while for \(J=(c,d)\) with \(1/m<c<d<1-1/m\), Step~1 gives
\[
\alpha_m(J)=\alpha((c,d))=\widetilde\beta((c,d))=\widetilde\beta_m(J).
\]

Define
\[
\Lambda_m
:=
\{B\in\mathcal B(I_m):\alpha_m(B)=\widetilde\beta_m(B)\}.
\]
We verify that \(\Lambda_m\) is a \(\lambda\)-system on \(I_m\). First, \(I_m\in\Lambda_m\), because \(I_m\) itself is an open interval and Step~1 gives
\[
\alpha_m(I_m)=\alpha(I_m)=\widetilde\beta(I_m)=\widetilde\beta_m(I_m).
\]
Second, if \(B\in\Lambda_m\), then finiteness of \(\alpha_m\) and \(\widetilde\beta_m\) on \(I_m\) gives
\[
\alpha_m(I_m\setminus B)=\alpha_m(I_m)-\alpha_m(B)
=
\widetilde\beta_m(I_m)-\widetilde\beta_m(B)
=
\widetilde\beta_m(I_m\setminus B),
\]
so \(I_m\setminus B\in\Lambda_m\). Third, if \((B_k)_{k\ge1}\subset\Lambda_m\) are pairwise disjoint, then
\[
\alpha_m\left(\bigcup_{k=1}^\infty B_k\right)
=
\sum_{k=1}^\infty \alpha_m(B_k)
=
\sum_{k=1}^\infty \widetilde\beta_m(B_k)
=
\widetilde\beta_m\left(\bigcup_{k=1}^\infty B_k\right),
\]
so \(\bigcup_{k=1}^\infty B_k\in\Lambda_m\). Thus \(\Lambda_m\) is a \(\lambda\)-system on \(I_m\).

Since \(\mathcal Q_m\) is a \(\pi\)-system, \(\sigma(\mathcal Q_m)=\mathcal B(I_m)\), and \(\mathcal Q_m\subset\Lambda_m\), the \(\pi\)-\(\lambda\) theorem implies
\[
\Lambda_m=\mathcal B(I_m).
\]
Therefore
\[
\alpha(B)=\widetilde\beta(B)
\qquad
\forall\,B\in\mathcal B(I_m),\ \forall\,m\ge3.
\]

\emph{Step 3: equality on all Borel subsets of \((0,1)\).}
Let \(B\in\mathcal B((0,1))\), and for \(m\ge3\) define
\[
B_m:=B\cap I_m.
\]
Then \(B_m\in\mathcal B(I_m)\), so Step~2 gives
\[
\alpha(B_m)=\widetilde\beta(B_m)
\qquad\forall\,m\ge3.
\]

We now show that \(B_m\uparrow B\). Since \(I_m\subset I_{m+1}\), one has
\[
B_m=B\cap I_m\subset B\cap I_{m+1}=B_{m+1},
\]
so \((B_m)_{m\ge3}\) is increasing. Conversely, if \(x\in B\), then \(\delta:=\min\{x,1-x\}>0\); choosing \(m\) so large that \(1/m<\delta\) gives \(x\in I_m\), hence \(x\in B_m\). Thus
\[
B=\bigcup_{m=3}^\infty B_m.
\]
By continuity from below,
\[
\alpha(B)
=
\lim_{m\to\infty}\alpha(B_m)
=
\lim_{m\to\infty}\widetilde\beta(B_m)
=
\widetilde\beta(B).
\]
Since \(B\in\mathcal B((0,1))\) was arbitrary, \(\alpha=\widetilde\beta\) on \(\mathcal B((0,1))\). Recalling that \(\widetilde\beta=\lambda\beta\), we obtain
\[
\alpha=\lambda\beta
\qquad\text{on }\mathcal B((0,1)).
\]
\end{deferredproof}

\begin{lemma}[One-dimensional dominated extension for a sublinear functional]\label{lem:helly-sublinear}
Let \(X\) be a real vector space and let \(p:X\to\mathbb R\) be sublinear:
\[
p(\lambda x)=\lambda p(x)\quad(\lambda\ge 0,\ x\in X), \qquad
p(x+y)\le p(x)+p(y)\quad(x,y\in X).
\]
Let \(Y\subset X\) be a linear subspace and let \(F_0:Y\to\mathbb R\) be linear with
\[
F_0(y)\le p(y)\quad(\forall y\in Y).
\]
Let \(x_0\in X\setminus Y\) and set \(Y_1:=Y\oplus\mathbb R x_0\). Define
\[
\alpha := \sup_{y\in Y}\bigl(F_0(y)-p(y-x_0)\bigr), \qquad
\beta := \inf_{y\in Y}\bigl(p(y+x_0)-F_0(y)\bigr).
\]
Then \(\alpha\le\beta\), and for each \(t\in[\alpha,\beta]\) the map \(F_t:Y_1\to\mathbb R\) given by
\[
F_t(y+s x_0) := F_0(y)+s t \qquad (y\in Y,\ s\in\mathbb R)
\]
is a well-defined linear functional such that \(F_t|_Y=F_0\) and
\[
F_t(z)\le p(z)\quad\text{for all }z\in Y_1.
\]
\end{lemma}

\begin{deferredproof}{lem:helly-sublinear}
For \(y\in Y\), subadditivity gives
\[
p(y)=p\bigl((y-x_0)+x_0\bigr)\le p(y-x_0)+p(x_0),\qquad
p(y)=p\bigl((y+x_0)+(-x_0)\bigr)\le p(y+x_0)+p(-x_0).
\]
Since \(F_0(y)\le p(y)\), it follows that
\[
F_0(y)-p(y-x_0)\le F_0(y)-p(y)+p(x_0)\le p(x_0),
\]
\[
p(y+x_0)-F_0(y)\ge p(y)-F_0(y)-p(-x_0)\ge -p(-x_0).
\]
Hence \(\alpha\le p(x_0)<\infty\) and \(\beta\ge -p(-x_0)>-\infty\).

For \(y,z\in Y\),
\[
F_0(y)+F_0(z)
=
F_0(y+z)
\le p(y+z)
=
p\bigl((y-x_0)+(z+x_0)\bigr)
\le p(y-x_0)+p(z+x_0),
\]
so
\[
F_0(y)-p(y-x_0)\le p(z+x_0)-F_0(z)\qquad(y,z\in Y).
\]
Taking the supremum in \(y\) and then the infimum in \(z\) yields \(\alpha\le\beta\).

Fix \(t\in[\alpha,\beta]\). Since \(Y_1=Y\oplus\mathbb R x_0\), each \(z\in Y_1\) has a unique representation \(z=y+s x_0\) with \(y\in Y\) and \(s\in\mathbb R\), so \(F_t\) is well-defined. For \(y_j\in Y\), \(s_j\in\mathbb R\), and \(\lambda\in\mathbb R\),
\begin{align*}
F_t\bigl((y_1+s_1x_0)+(y_2+s_2x_0)\bigr)
&=F_t\bigl((y_1+y_2)+(s_1+s_2)x_0\bigr)\\
&=F_0(y_1+y_2)+(s_1+s_2)t\\
&=F_0(y_1)+F_0(y_2)+s_1t+s_2t\\
&=F_t(y_1+s_1x_0)+F_t(y_2+s_2x_0),
\end{align*}
and similarly \(F_t(\lambda(y+s x_0))=\lambda F_t(y+s x_0)\). Thus \(F_t\) is linear and \(F_t|_Y=F_0\).

It remains to show that \(F_t(z)\le p(z)\) for all \(z\in Y_1\). Let \(z=y+s x_0\) with \(y\in Y\) and \(s\in\mathbb R\). If \(s=0\), then \(z=y\) and \(F_t(z)=F_0(y)\le p(y)=p(z)\).

If \(s>0\), set \(u:=y/s\in Y\). Then \(z=s(u+x_0)\), and
\[
F_t(z)=F_0(y)+s t=s\bigl(F_0(u)+t\bigr).
\]
Since \(t\le\beta\le p(u+x_0)-F_0(u)\), we have \(F_0(u)+t\le p(u+x_0)\), and positive homogeneity gives
\[
F_t(z)\le s\,p(u+x_0)=p\bigl(s(u+x_0)\bigr)=p(z).
\]

If \(s<0\), write \(s=-k\) with \(k>0\) and set \(u:=y/k\in Y\). Then \(z=k(u-x_0)\), and
\[
F_t(z)=F_0(y)+s t=k\bigl(F_0(u)-t\bigr).
\]
Because \(\alpha\le t\), we have \(F_0(u)-p(u-x_0)\le \alpha\le t\), so \(F_0(u)-t\le p(u-x_0)\). Again by positive homogeneity,
\[
F_t(z)\le k\,p(u-x_0)=p\bigl(k(u-x_0)\bigr)=p(z).
\]
Thus \(F_t(z)\le p(z)\) for all \(z\in Y_1\).
\end{deferredproof}

\begin{lemma}[Hahn--Banach extension dominated by sublinear functional]\label{lem:HB-sublinear}
Let \(X\) be a real vector space and let \(p:X\to\mathbb R\) be sublinear:
\[
p(\lambda x)=\lambda p(x)\quad(\lambda\ge 0,\ x\in X),\qquad
p(x+y)\le p(x)+p(y)\quad(x,y\in X).
\]
Let \(Y\subset X\) be a linear subspace and let \(F_0:Y\to\mathbb R\) be linear with
\[
F_0(y)\le p(y)\quad(\forall y\in Y).
\]
Then there exists a linear functional \(F:X\to\mathbb R\) such that
\[
F|_Y = F_0,\qquad F(x)\le p(x)\quad(\forall x\in X).
\]
\end{lemma}

\begin{deferredproof}{lem:HB-sublinear}
Define
\[
\mathcal F
:=
\bigl\{
(E,G):
Y\subset E\subset X \text{ is a linear subspace},\
G:E\to\mathbb R \text{ is linear},\
G|_Y=F_0,\
G(x)\le p(x)\ \forall x\in E
\bigr\}.
\]
Partially order \(\mathcal F\) by
\[
(E_1,G_1)\preceq (E_2,G_2)
\iff
E_1\subset E_2 \text{ and } G_2|_{E_1}=G_1.
\]
Clearly \((Y,F_0)\in\mathcal F\), so \(\mathcal F\neq\varnothing\).

Let \(\mathcal C\subset\mathcal F\) be a chain, and put
\[
E_{\mathcal C}:=\bigcup_{(E,G)\in\mathcal C}E.
\]
Then \(E_{\mathcal C}\) is a linear subspace of \(X\): if \(x,y\in E_{\mathcal C}\) and \(\lambda,\mu\in\mathbb R\), choose \((E_1,G_1),(E_2,G_2)\in\mathcal C\) with \(x\in E_1\) and \(y\in E_2\); since \(\mathcal C\) is totally ordered, say \(E_1\subset E_2\), so \(x,y\in E_2\) and therefore \(\lambda x+\mu y\in E_2\subset E_{\mathcal C}\). Also \(Y\subset E_{\mathcal C}\).

For \(x\in E_{\mathcal C}\), choose \((E,G)\in\mathcal C\) with \(x\in E\) and set
\[
G_{\mathcal C}(x):=G(x).
\]
This is well-defined: if \(x\in E_1\cap E_2\) for \((E_1,G_1),(E_2,G_2)\in\mathcal C\), then, say, \(E_1\subset E_2\), hence \(G_2|_{E_1}=G_1\) and therefore \(G_1(x)=G_2(x)\).

Linearity of \(G_{\mathcal C}\) follows similarly: for \(x,y\in E_{\mathcal C}\) and \(\lambda,\mu\in\mathbb R\), choose \((E,G)\in\mathcal C\) with \(x,y\in E\). Then
\[
G_{\mathcal C}(\lambda x+\mu y)
=
G(\lambda x+\mu y)
=
\lambda G(x)+\mu G(y)
=
\lambda G_{\mathcal C}(x)+\mu G_{\mathcal C}(y).
\]
If \(y\in Y\), then \(G_{\mathcal C}(y)=F_0(y)\) because every member of \(\mathcal C\) extends \(F_0\). Moreover, if \(x\in E_{\mathcal C}\), choosing \((E,G)\in\mathcal C\) with \(x\in E\) gives
\[
G_{\mathcal C}(x)=G(x)\le p(x).
\]
Hence \((E_{\mathcal C},G_{\mathcal C})\in\mathcal F\), and it is an upper bound of \(\mathcal C\).

By Zorn's lemma, \(\mathcal F\) has a maximal element \((E^\ast,G^\ast)\). If \(E^\ast\neq X\), choose \(x_0\in X\setminus E^\ast\) and set
\[
E_1:=E^\ast\oplus\mathbb R x_0.
\]
Applying Lemma~\ref{lem:helly-sublinear} with \(Y:=E^\ast\) and \(F_0:=G^\ast\), we obtain a linear functional \(G_1:E_1\to\mathbb R\) such that
\[
G_1|_{E^\ast}=G^\ast,\qquad G_1(z)\le p(z)\quad(\forall z\in E_1).
\]
Hence \((E_1,G_1)\in\mathcal F\) and \((E^\ast,G^\ast)\prec(E_1,G_1)\), contradicting maximality. Therefore \(E^\ast=X\).

Define \(F:=G^\ast\). Then \(F:X\to\mathbb R\) is linear, \(F|_Y=F_0\), and \(F(x)\le p(x)\) for all \(x\in X\).
\end{deferredproof}

\begin{lemma}[Hahn--Banach separation for convex sets]\label{lem:HB-separation-convex}
Let $(X,|\cdot|)$ be a real normed linear space. For arbitrary nonempty convex sets $A,B\subset X$ with $A\cap B=\varnothing$ and $\operatorname{int}A\neq\varnothing$, there exist a nonzero continuous linear functional $F:X\to\mathbb R$ and a real number $\gamma$ such that
\[
F(a)\ge\gamma\ge F(b)\quad\text{for all }a\in A,\ b\in B.
\]
\end{lemma}

\begin{deferredproof}{lem:HB-separation-convex}
Let $C\subset X$ be convex with $\operatorname{int}C\neq\varnothing$, and let $x_0\notin C$. Choose $c\in\operatorname{int}C$ and set
\[
C':=C-c,\qquad x_0':=x_0-c.
\]
Then $C'$ is convex, $0\in\operatorname{int}C'$, and $x_0'\notin C'$. We first construct a nonzero continuous linear functional $G':X\to\mathbb R$ such that
\begin{equation}\label{eq:sep-Cprime}
G'(x)\le G'(x_0')\quad\text{for all }x\in C'.
\end{equation}

Since $0\in\operatorname{int}C'$, there exists $r>0$ with $B(0,r)\subset C'$. Define $p:X\to[0,\infty)$ by
\[
p(x):=\inf\{t>0 : x\in tC'\}.
\]
For any $x\in X$ and $t>|x|/r$ we have $|x/t|<r$, hence $x/t\in B(0,r)\subset C'$ and thus $x\in tC'$. Therefore $p(x)\le t<\infty$ for all $x$, so $p$ is finite everywhere.

We prove positive homogeneity for $\alpha>0$. Let $\alpha>0$ and $x\in X$. If $x\in tC'$ for some $t>0$, then $\alpha x\in\alpha tC'$, so
\[
p(\alpha x)\le \alpha t\quad\text{for all }t>0\text{ with }x\in tC'.
\]
Taking the infimum over such $t$ gives $p(\alpha x)\le\alpha p(x)$. Conversely, if $\alpha x\in tC'$ for some $t>0$, then $x\in (t/\alpha)C'$, hence $p(x)\le t/\alpha$ and $\alpha p(x)\le t$. Taking the infimum over $t>0$ with $\alpha x\in tC'$ yields $\alpha p(x)\le p(\alpha x)$. Thus
\[
p(\alpha x)=\alpha p(x)\quad(\alpha>0,\ x\in X).
\]

We now prove subadditivity. Let $x,y\in X$ and $\varepsilon>0$. Choose $s>p(x)$ and $t>p(y)$ such that
\[
s<p(x)+\frac{\varepsilon}{2},\qquad t<p(y)+\frac{\varepsilon}{2}.
\]
Then $x\in sC'$ and $y\in tC'$, so there exist $c_1,c_2\in C'$ with $x=sc_1$, $y=tc_2$. Since $C'$ is convex,
\[
\frac{s}{s+t}c_1+\frac{t}{s+t}c_2\in C'.
\]
But
\[
\frac{s}{s+t}c_1+\frac{t}{s+t}c_2=\frac{x+y}{s+t},
\]
so $(x+y)\in (s+t)C'$ and hence $p(x+y)\le s+t<p(x)+p(y)+\varepsilon$. As $\varepsilon>0$ is arbitrary,
\[
p(x+y)\le p(x)+p(y)\quad(x,y\in X).
\]
Thus $p$ is sublinear.

Next, for $|x|\le 1$ we have $rx\in B(0,r)\subset C'$, hence $x\in (1/r)C'$ and $p(x)\le 1/r$. For $x\neq 0$,
\[
p(x)=|x|\,p\!\left(\frac{x}{|x|}\right)\le |x|\cdot\frac{1}{r}=\frac{|x|}{r},
\]
and $p(0)=0$. Therefore
\begin{equation}\label{eq:p-bound}
p(x)\le\frac{|x|}{r}\quad(x\in X).
\end{equation}

Consider the one-dimensional subspace $Y:=\mathbb R x_0'$ and define $f_0:Y\to\mathbb R$ by
\[
f_0(\lambda x_0'):=\lambda\quad(\lambda\in\mathbb R).
\]
We show $f_0(y)\le p(y)$ for all $y\in Y$. For $\lambda>0$, suppose $p(\lambda x_0')<\lambda$. Then there exists $t$ with $0<t<\lambda$ and $\lambda x_0'\in tC'$, so $\lambda x_0'=t c$ for some $c\in C'$. Hence $x_0'=(t/\lambda)c$, where $0<t/\lambda<1$. Since $0,c\in C'$ and $C'$ is convex, $(t/\lambda)c\in C'$, so $x_0'\in C'$, contradicting $x_0'\notin C'$. Thus $p(\lambda x_0')\ge\lambda=f_0(\lambda x_0')$ for all $\lambda>0$. For $\lambda\le 0$, we have $p(\lambda x_0')\ge 0\ge\lambda=f_0(\lambda x_0')$. Hence
\[
f_0(y)\le p(y)\quad\text{for all }y\in Y.
\]

By the Hahn--Banach Sublinear Dominated extension (Lemma~\ref{lem:HB-sublinear}) applied to the sublinear functional $p$ and the linear functional $f_0$, there exists a linear functional $G':X\to\mathbb R$ such that
\[
G'|_Y=f_0,\qquad G'(x)\le p(x)\quad\text{for all }x\in X.
\]
In particular $G'(x_0')=f_0(x_0')=1$, so $G'$ is nonzero. For $x\in C'$ we have $x\in 1\cdot C'$, hence $p(x)\le 1$ and thus
\[
G'(x)\le p(x)\le 1=G'(x_0')\quad(x\in C').
\]
From \eqref{eq:p-bound} we obtain
\[
G'(x)\le p(x)\le\frac{|x|}{r},\qquad G'(-x)\le p(-x)\le\frac{|x|}{r},
\]
so $-G'(x)=G'(-x)\le |x|/r$ and therefore
\[
|G'(x)|\le\frac{|x|}{r}\quad(x\in X).
\]
Thus $G'$ is bounded and hence continuous. This proves \eqref{eq:sep-Cprime}.

Returning to $C$ and $x_0$, for $x\in C$ we have $x-c\in C'$, so by \eqref{eq:sep-Cprime},
\[
G'(x-c)\le G'(x_0')=G'(x_0-c).
\]
Using linearity,
\[
G'(x)-G'(c)\le G'(x_0)-G'(c)\quad\Longrightarrow\quad G'(x)\le G'(x_0)\quad(x\in C).
\]
Hence For every convex $C\subset X$ with $\operatorname{int}C\neq\varnothing$ and every $x_0\notin C$, there exists a nonzero continuous linear functional $G:X\to\mathbb R$ such that $G(x)\le G(x_0)$ for all $x\in C$ (take $G=G'$ above).

Now let $A,B\subset X$ be nonempty convex sets with $A\cap B=\varnothing$ and $\operatorname{int}A\neq\varnothing$. Define
\[
Z:=B-A:=\{b-a : b\in B,\ a\in A\}.
\]
Then $Z$ is convex, being the image of $B\times A$ under the linear map $(b,a)\mapsto b-a$. If $0\in Z$, then $0=b-a$ for some $a\in A$, $b\in B$, so $a=b\in A\cap B$, a contradiction; hence $0\notin Z$.

We show $\operatorname{int}Z\neq\varnothing$. Choose $a_0\in\operatorname{int}A$ and $b_0\in B$. There exists $r>0$ with $B(a_0,r)\subset A$. For any $x\in X$ with $|x|<r$, one has $a_0+x\in A$, so
\[
(b_0-a_0)-x=b_0-(a_0+x)\in Z.
\]
Thus $B(b_0-a_0,r)\subset Z$, hence $b_0-a_0\in\operatorname{int}Z$ and $\operatorname{int}Z\neq\varnothing$.

Apply (*) to $C:=Z$ and $x_0:=0$. There exists a nonzero continuous linear functional $G:X\to\mathbb R$ such that
\[
G(z)\le G(0)=0\quad\text{for all }z\in Z.
\]
For $a\in A$ and $b\in B$ we have $z:=b-a\in Z$, hence
\[
G(b)-G(a)=G(b-a)\le 0\quad\Longrightarrow\quad G(a)\ge G(b)\quad(a\in A,\ b\in B).
\]

Define
\[
\alpha:=\inf_{a\in A}G(a),\qquad \beta:=\sup_{b\in B}G(b).
\]
Since $A,B$ are nonempty, choose $a_0\in A$, $b_0\in B$. From $G(b)\le G(a_0)$ for all $b\in B$ we obtain $\beta\le G(a_0)<\infty$, and from $G(a)\ge G(b_0)$ for all $a\in A$ we obtain $\alpha\ge G(b_0)>-\infty$. Moreover $G(a)\ge G(b)$ for all $a\in A$, $b\in B$ implies $\alpha\ge\beta$. Thus $[\beta,\alpha]$ is a nonempty interval; choose any $\gamma\in[\beta,\alpha]$. Then, for all $a\in A$ and $b\in B$,
\[
G(a)\ge\alpha\ge\gamma\ge\beta\ge G(b),
\]
so, with $F:=G$,
\[
F(a)\ge\gamma\ge F(b)\quad(\forall a\in A,\ \forall b\in B),
\]
and $F$ is a nonzero continuous linear functional on $X$. This is the separation that would be useful to us for 
\end{deferredproof}

\begin{lemma}[Positive Linear Functional Representation on $C(\lbrack0,1\rbrack)$ (right--continuous version)]\label{lem:R2-positive-Riesz}

Let $X:=C(\lbrack0,1\rbrack)$ with the supremum norm $\|f\|_\infty:=\sup_{x\in[0,1]}|f(x)|$.
Let $L:X\to\mathbb R$ be a \emph{continuous} linear functional such that
\[
f\ge 0\ \text{on }[0,1]\quad\Longrightarrow\quad L(f)\ge 0.
\]
Define for $t\in[0,1)$:
\[
\mathcal L_t:=\bigl\{\phi\in X:\ 0\le\phi\le 1,\ \phi(x)=0\ \text{for all }x\in[t,1]\bigr\},
\qquad
\alpha(t):=\sup\{\,L(\phi):\ \phi\in\mathcal L_t\,\}.
\]
Define at $t=1$:
\[
\mathcal L_1:=\{\phi\in X:\ 0\le\phi\le 1\},
\qquad
\alpha(1):=\sup\{\,L(\phi):\ \phi\in\mathcal L_1\,\}.
\]
Then $\alpha:[0,1]\to\mathbb R$ is increasing, the Riemann--Stieltjes integral $\int_0^1 f\,d\alpha$ exists for every $f\in X$, and
\[
L(f)=\int_0^1 f(x)\,d\alpha(x)\qquad\forall f\in X.
\]

Moreover, define the \emph{right--continuous modification}
\begin{equation}\label{eq:alpha-rc}
\alpha^{\mathrm{rc}}(0):=\alpha(0),\qquad
\alpha^{\mathrm{rc}}(t):=\alpha(t+)\ (t\in(0,1)),\qquad
\alpha^{\mathrm{rc}}(1):=\alpha(1),
\end{equation}
where $\alpha(t+):=\lim_{h\downarrow 0}\alpha(t+h)$.
Then $\alpha^{\mathrm{rc}}$ is increasing and right--continuous on $(0,1)$, the integral $\int_0^1 f\,d\alpha^{\mathrm{rc}}$ exists for every
$f\in X$, and
\[
L(f)=\int_0^1 f\,d\alpha^{\mathrm{rc}}\qquad\forall f\in X.
\]
(In particular, after replacing $\alpha$ by $\alpha^{\mathrm{rc}}$ one may assume $\alpha$ is right--continuous on $(0,1)$.)
\end{lemma}

\begin{deferredproof}{lem:R2-positive-Riesz}
For $0\le s\le t<1$ we have $\mathcal L_s\subset \mathcal L_t$ (if $\phi=0$ on $[s,1]$ then $\phi=0$ on $[t,1]$), hence
\[
\alpha(s)\le \alpha(t)\qquad(0\le s\le t<1),
\]
so $\alpha$ is increasing on $[0,1)$.

If $t=0$ then $\mathcal L_0=\{0\}$, hence $\alpha(0)=L(0)=0$.
If $t=1$ then $\mathcal L_1=\{\phi\in X:0\le\phi\le 1\}$.
For every $\phi\in\mathcal L_1$ we have $0\le \phi\le \mathbf 1$, hence $0\le L(\phi)\le L(\mathbf 1)$ by positivity.
Also $\mathbf 1\in\mathcal L_1$, hence
\[
\alpha(1)=L(\mathbf 1).
\]
Thus $\alpha$ is increasing on all of $[0,1]$ and we have the boundary values:
\[
\alpha(0)=0,\qquad \alpha(1)=L(\mathbf 1),\qquad 0\le \alpha(t)\le L(\mathbf 1)\ \ (t\in[0,1]).
\]

For $a\in[0,1)$ define the right limit
\[
\alpha(a+):=\lim_{h\downarrow 0}\alpha(a+h)=\inf_{\delta>0}\alpha(a+\delta).
\]

If $L(\mathbf 1)=0$, then for any $f\in X$ with $f\ge 0$,
\[
0\le f\le \|f\|_\infty\,\mathbf 1\quad\Longrightarrow\quad 0\le L(f)\le \|f\|_\infty\,L(\mathbf 1)=0,
\]
so $L(f)=0$ for all $f\ge 0$, hence $L\equiv 0$ on $X$ by linearity.
Also $\alpha\equiv 0$. Then $L(f)=0=\int_0^1 f\,d\alpha$ for all $f$.
Henceforth assume $L(\mathbf 1)>0$.

We are going to use moving window class of bridging increasing functions to establish the global properties of the unit interval.
For $0\le a<b\le 1$ define
\[
\mathcal B_{(a,b]}:=\bigl\{\psi\in X:\ 0\le\psi\le 1,\ \psi(x)=0\ \text{for all }x\in[0,a]\cup(b,1]\bigr\}.
\]
(If $b<1$, continuity implies $\psi(b)=0$ because $\psi\equiv 0$ on $(b,1]$.)

Now we establish the identity that for every $0\le a<b\le 1$:
\begin{equation}\label{eq:window-claim-correct}
\sup_{\psi\in\mathcal B_{(a,b]}}L(\psi)\ =\ \alpha(b)-\alpha(a+).
\end{equation}

\textbf{Proof of \eqref{eq:window-claim-correct}, ``$\le$'' direction.}
Fix $0\le a<b\le 1$ and $\psi\in\mathcal B_{(a,b]}$. Then $\psi(a)=0$ and $\psi$ is continuous.
Choose arbitrary $\varepsilon>0$, there exists $\delta\in(0,b-a)$ such that
\[
0\le \psi(x)\le \frac{\varepsilon}{2L(\mathbf 1)}\qquad\forall x\in[a,a+\delta].
\]

Moreover, there exists $\phi_0\in\mathcal L_{a+\delta}$ such that
\[
L(\phi_0)\ \ge\ \alpha(a+\delta)-\frac{\varepsilon}{2}.
\]
Define
\[
\phi(x):=\min\{\phi_0(x),\,1-\psi(x)\}\qquad(x\in[0,1]).
\]
Then $\phi\in X$, $0\le \phi\le 1$.
Also $\phi\le \phi_0$ and $\phi_0=0$ on $[a+\delta,1]$, hence $\phi=0$ on $[a+\delta,1]$.
Therefore $\phi\in\mathcal L_{a+\delta}$.

We now establish $\phi+\psi\in\mathcal L_b$. Since $\phi\le 1-\psi$, we have $0\le \phi+\psi\le 1$.
If $b<1$, then $\phi=0$ on $[a+\delta,1]\supset [b,1]$ and $\psi=0$ on $(b,1]$; by continuity $\psi(b)=0$,
so $\psi=0$ on $[b,1]$ and therefore $\phi+\psi=0$ on $[b,1]$, i.e.\ $\phi+\psi\in\mathcal L_b$.
If $b=1$, then $\phi+\psi\in\mathcal L_1$ because $0\le \phi+\psi\le 1$.
In both cases,
\[
L(\phi+\psi)\le \alpha(b).
\]
By linearity,
\[
L(\psi)\le \alpha(b)-L(\phi). \tag{$\star$}
\]

Next we establish the lower bound on $L(\phi)$. Let $d:=\phi_0-\phi\ge 0$. For each $x$:
if $\phi_0(x)\le 1-\psi(x)$ then $\phi(x)=\phi_0(x)$ so $d(x)=0$;
if $\phi_0(x)>1-\psi(x)$ then $\phi(x)=1-\psi(x)$ so
\[
d(x)=\phi_0(x)-(1-\psi(x))\le 1-(1-\psi(x))=\psi(x),
\]
using $\phi_0(x)\le 1$. Hence
\begin{equation}\label{eq:d-le-psi}
0\le d\le \psi \quad\text{pointwise on }[0,1].
\end{equation}
Moreover, $\phi_0=0$ on $[a+\delta,1]$ so $d=0$ there; and $\psi=0$ on $[0,a]$ so $d=0$ there by \eqref{eq:d-le-psi}.
Thus $\supp(d)\subset[a,a+\delta]$, and on that interval $\psi\le \varepsilon/(2L(\mathbf 1))$, so by \eqref{eq:d-le-psi},
\[
0\le d\le \frac{\varepsilon}{2L(\mathbf 1)}\,\mathbf 1.
\]
By positivity,
\[
0\le L(d)\le \frac{\varepsilon}{2L(\mathbf 1)}\,L(\mathbf 1)=\frac{\varepsilon}{2}.
\]
Therefore
\[
L(\phi)=L(\phi_0)-L(d)\ \ge\ \Bigl(\alpha(a+\delta)-\frac{\varepsilon}{2}\Bigr)-\frac{\varepsilon}{2}
=\alpha(a+\delta)-\varepsilon.
\]
Insert into $(\star)$:
\[
L(\psi)\le \alpha(b)-\alpha(a+\delta)+\varepsilon \le \alpha(b)-\alpha(a+)+\varepsilon,
\]
since $\alpha(a+)=\inf_{\delta>0}\alpha(a+\delta)\le \alpha(a+\delta)$.
Let $\varepsilon\downarrow 0$ to obtain
\[
L(\psi)\le \alpha(b)-\alpha(a+).
\]
Taking the supremum over $\psi\in\mathcal B_{(a,b]}$ yields
\[
\sup_{\psi\in\mathcal B_{(a,b]}}L(\psi)\ \le\ \alpha(b)-\alpha(a+).
\]

\textbf{Proof of \eqref{eq:window-claim-correct}, ``$\ge$'' direction.}
Fix $0\le a<b\le 1$ and $\varepsilon>0$.
Choose $\phi_b\in\mathcal L_b$ such that
\[
L(\phi_b)>\alpha(b)-\frac{\varepsilon}{2}.
\]
Choose $\delta\in(0,b-a)$ such that
\[
\alpha(a+\delta)\le \alpha(a+)+\frac{\varepsilon}{2}.
\]
Define the cutoff
\[
\chi_{a,\delta}(x):=
\begin{cases}
1,& 0\le x\le a,\\[2pt]
1-\dfrac{x-a}{\delta},& a<x<a+\delta,\\[6pt]
0,& x\ge a+\delta.
\end{cases}
\]
Then $0\le \chi_{a,\delta}\le 1$ and $\chi_{a,\delta}=0$ on $[a+\delta,1]$, hence $\chi_{a,\delta}\in\mathcal L_{a+\delta}$.
Define
\[
\psi:=\phi_b(1-\chi_{a,\delta}).
\]
Then $0\le \psi\le 1$, and $\psi=0$ on $[0,a]$ because $1-\chi_{a,\delta}=0$ there.
If $b<1$, then $\phi_b=0$ on $[b,1]$, hence $\psi=0$ on $(b,1]$; if $b=1$, there is no constraint on $(b,1]$.
Thus $\psi\in\mathcal B_{(a,b]}$.

Also $\phi_b\chi_{a,\delta}\in\mathcal L_{a+\delta}$, so
\[
L(\phi_b\chi_{a,\delta})\le \alpha(a+\delta).
\]
By linearity,
\[
L(\psi)=L(\phi_b)-L(\phi_b\chi_{a,\delta})
\ge \Bigl(\alpha(b)-\frac{\varepsilon}{2}\Bigr)-\alpha(a+\delta)
\ge \alpha(b)-\alpha(a+)-\varepsilon.
\]
Taking the supremum over $\psi\in\mathcal B_{(a,b]}$ and letting $\varepsilon\downarrow 0$ yields
\[
\sup_{\psi\in\mathcal B_{(a,b]}}L(\psi)\ \ge\ \alpha(b)-\alpha(a+).
\]
Together with Step~3 this proves \eqref{eq:window-claim-correct}.

We are back in the arena of classical Riemann--Stieltjes integral.
Fix $f\in X$. For a partition $P:0=t_0<\dots<t_m=1$ define
\[
m_i:=\inf_{[t_{i-1},t_i]}f,\qquad M_i:=\sup_{[t_{i-1},t_i]}f,
\]
and the Darboux--Stieltjes sums
\[
L(P;f,\alpha):=\sum_{i=1}^m m_i\bigl(\alpha(t_i)-\alpha(t_{i-1})\bigr),\qquad
U(P;f,\alpha):=\sum_{i=1}^m M_i\bigl(\alpha(t_i)-\alpha(t_{i-1})\bigr).
\]
Let $\omega_f(\delta):=\sup\{|f(x)-f(y)|:\ |x-y|\le \delta\}$ be the modulus of continuity of $f$.
For mesh $\|P\|:=\max_i(t_i-t_{i-1})$ we have $M_i-m_i\le \omega_f(\|P\|)$, hence
\[
U(P;f,\alpha)-L(P;f,\alpha)
=\sum_{i=1}^m (M_i-m_i)\bigl(\alpha(t_i)-\alpha(t_{i-1})\bigr)
\le \omega_f(\|P\|)\sum_{i=1}^m \bigl(\alpha(t_i)-\alpha(t_{i-1})\bigr)
=\omega_f(\|P\|)\bigl(\alpha(1)-\alpha(0)\bigr).
\]
Since $f$ is uniformly continuous, $\omega_f(\delta)\to 0$ as $\delta\downarrow 0$.
Thus for every $\varepsilon>0$ there exists a partition $P$ with $U(P;f,\alpha)-L(P;f,\alpha)<\varepsilon$.
Consequently
\[
\sup_P L(P;f,\alpha)=\inf_P U(P;f,\alpha)=:\int_0^1 f\,d\alpha,
\]
so the Riemann--Stieltjes integral exists for every $f\in X$.

We are going to pin down the Representation for $f\ge 0$. Let's Fix $f\in X$ with $f\ge 0$. Set $B:=\|f\|_\infty$.
Define the right-jump function
\[
J(x):=\alpha(x^+)-\alpha(x)\ge 0,\qquad x\in[0,1),
\]
and the jump set $D:=\{x\in[0,1):J(x)>0\}$.

\emph{Claim 1: $D$ is at most countable and $\sum_{x\in D}J(x)<\infty$.}
For $n\in\mathbb N$ set $D_n:=\{x\in[0,1):J(x)\ge 1/n\}$.
If $x_1<\dots<x_k$ are distinct points in $D_n$, choose $h>0$ so small that $x_k+h\le 1$ and the intervals $[x_j,x_j+h]$ are disjoint.
Since $\alpha$ is increasing,
\[
J(x_j)=\alpha(x_j^+)-\alpha(x_j)\le \alpha(x_j+h)-\alpha(x_j).
\]
Define the increasing sequence
\[
r_0:=0,\ r_1:=x_1,\ r_2:=x_1+h,\ r_3:=x_2,\ r_4:=x_2+h,\ \dots,\ r_{2k-1}:=x_k,\ r_{2k}:=x_k+h,\ r_{2k+1}:=1.
\]
Then
\[
\alpha(1)-\alpha(0)=\sum_{\ell=1}^{2k+1}\bigl(\alpha(r_\ell)-\alpha(r_{\ell-1})\bigr)
\ge \sum_{j=1}^k\bigl(\alpha(x_j+h)-\alpha(x_j)\bigr)
\ge \sum_{j=1}^k J(x_j).
\]
But each $J(x_j)\ge 1/n$, hence $k/n\le \alpha(1)-\alpha(0)=L(\mathbf 1)$, so $k<\infty$.
Thus each $D_n$ is finite, hence $D=\bigcup_{n\ge 1}D_n$ is countable.
Enumerate $D=\{d_1,d_2,\dots\}$ (finite or countably infinite) and set $J_k:=J(d_k)$.
By the same finite-set estimate above, the partial sums $\sum_{k=1}^N J_k$ are bounded by $L(\mathbf 1)$, hence
$\sum_{k\ge 1}J_k<\infty$.

Fix $\varepsilon>0$.
Choose $N$ such that
\[
\sum_{k>N}J_k\le \frac{\varepsilon}{B+1},
\]
and set $F:=\{d_1,\dots,d_N\}$ (possibly empty).

\emph{Claim 2: there exists a partition $P:0=t_0<\dots<t_m=1$ such that}
\[
\|P\|\ \text{is arbitrarily small, and}\quad \{t_1,\dots,t_{m-1}\}\cap F=\emptyset.
\]
(Construction: choose $t_1\in(0,\delta]\setminus F$, then $t_2\in(t_1,t_1+\delta]\setminus F$, etc., until reaching $1$;
this is possible since $F$ is finite.)

Choose such a partition $P$ with $\|P\|$ so small that
\begin{equation}\label{eq:mesh-small}
U(P;f,\alpha)-L(P;f,\alpha)<\varepsilon.
\end{equation}

\emph{Lower bound on $L(f)$.}
Fix $\eta>0$. Choose $\phi_1\in\mathcal L_{t_1}$ such that
\[
L(\phi_1)\ge \alpha(t_1)-\eta.
\]
For each $i=2,\dots,m$, use \eqref{eq:window-claim-correct} to choose $\psi_i\in\mathcal B_{(t_{i-1},t_i]}$ such that
\[
L(\psi_i)\ge \bigl(\alpha(t_i)-\alpha(t_{i-1}+)\bigr)-\eta.
\]
Define
\[
\Psi:= m_1\,\phi_1+\sum_{i=2}^m m_i\,\psi_i\in X.
\]
Then $0\le \Psi\le f$ pointwise, hence by positivity,
\[
L(f)\ge L(\Psi).
\]
By linearity,
\[
L(\Psi)
\ge m_1\bigl(\alpha(t_1)-\eta\bigr)+\sum_{i=2}^m m_i\Bigl(\alpha(t_i)-\alpha(t_{i-1}+)-\eta\Bigr).
\]
Thus
\[
L(f)\ge \sum_{i=1}^m m_i\bigl(\alpha(t_i)-\alpha(t_{i-1})\bigr)
-\sum_{i=2}^m m_i\bigl(\alpha(t_{i-1}+)-\alpha(t_{i-1})\bigr)
-\eta\sum_{i=1}^m m_i.
\]
Since $0\le m_i\le B$ and $\{t_1,\dots,t_{m-1}\}\cap F=\emptyset$,
\[
\sum_{i=2}^m m_i\bigl(\alpha(t_{i-1}+)-\alpha(t_{i-1})\bigr)
=\sum_{i=2}^m m_i\,J(t_{i-1})
\le B\sum_{\substack{i=2,\dots,m\\ t_{i-1}\in D\setminus F}} J(t_{i-1})
\le B\sum_{k>N}J_k
\le B\cdot\frac{\varepsilon}{B+1}<\varepsilon.
\]
Choose $\eta>0$ so small that $mB\eta<\varepsilon$. Then $\eta\sum_{i=1}^m m_i\le \eta\,mB<\varepsilon$, hence
\begin{equation}\label{eq:Lf-lower}
L(f)\ge L(P;f,\alpha)-2\varepsilon.
\end{equation}

\emph{Upper bound on $L(f)$.}
Let $g:=B\mathbf 1-f\ge 0$. Apply \eqref{eq:Lf-lower} to $g$ (with the same partition $P$):
\[
L(g)\ge L(P;g,\alpha)-2\varepsilon.
\]
But $L(g)=B\,L(\mathbf 1)-L(f)=B\alpha(1)-L(f)$ and
\[
L(P;g,\alpha)=\sum_{i=1}^m \inf_{[t_{i-1},t_i]}(B-f)\,\bigl(\alpha(t_i)-\alpha(t_{i-1})\bigr)
=\sum_{i=1}^m (B-M_i)\bigl(\alpha(t_i)-\alpha(t_{i-1})\bigr)
=B\alpha(1)-U(P;f,\alpha).
\]
Hence
\[
B\alpha(1)-L(f)\ge B\alpha(1)-U(P;f,\alpha)-2\varepsilon,
\]
so
\begin{equation}\label{eq:Lf-upper}
L(f)\le U(P;f,\alpha)+2\varepsilon.
\end{equation}

Let $I:=\int_0^1 f\,d\alpha$ (which exists by Step~5). Then for every partition,
\[
L(P;f,\alpha)\le I\le U(P;f,\alpha).
\]
With \eqref{eq:mesh-small}, we have $U(P;f,\alpha)-L(P;f,\alpha)<\varepsilon$, hence
\[
I-\varepsilon< L(P;f,\alpha)\le I\le U(P;f,\alpha)<I+\varepsilon.
\]
Combine with \eqref{eq:Lf-lower}--\eqref{eq:Lf-upper}:
\[
I-3\varepsilon< L(f) < I+3\varepsilon.
\]
Let $\varepsilon\downarrow 0$ to conclude
\[
L(f)=\int_0^1 f\,d\alpha\qquad\text{for all }f\in X,\ f\ge 0.
\]

Now let's deal with the case of arbitrary $f\in X$, denote $f=f^+-f^-$ where
\[
f^+(x):=\max\{f(x),0\},\qquad f^-(x):=\max\{-f(x),0\}.
\]
Then $f^\pm\in X$, $f^\pm\ge 0$, and by Step~6,
\[
L(f^\pm)=\int_0^1 f^\pm\,d\alpha.
\]
By linearity of $L$ and of the Riemann--Stieltjes integral on $X$,
\[
L(f)=L(f^+)-L(f^-)=\int_0^1 f^+\,d\alpha-\int_0^1 f^-\,d\alpha=\int_0^1 (f^+-f^-)\,d\alpha=\int_0^1 f\,d\alpha.
\]

The Right--continuous modification is conducted because of the special need of the Main theorems of Lorenz Curve representation. Recall the definition of $\alpha^{\mathrm{rc}}$  in \eqref{eq:alpha-rc}.
Then $\alpha^{\mathrm{rc}}$ is increasing. Moreover, for $t\in(0,1)$,
\[
\lim_{h\downarrow 0}\alpha^{\mathrm{rc}}(t+h)=\lim_{h\downarrow 0}\alpha((t+h)+)=\alpha(t+)=\alpha^{\mathrm{rc}}(t),
\]
so $\alpha^{\mathrm{rc}}$ is right--continuous on $(0,1)$.

Fix $f\in X$. Since $\alpha^{\mathrm{rc}}$ is increasing, Step~5 implies $\int_0^1 f\,d\alpha^{\mathrm{rc}}$ exists.
We show
\begin{equation}\label{eq:int-alpha-rc-equals}
\int_0^1 f\,d\alpha^{\mathrm{rc}}=\int_0^1 f\,d\alpha.
\end{equation}

For a tagged partition $(P,\xi)$ with $P:0=t_0<\cdots<t_m=1$ and $\xi_i\in[t_{i-1},t_i]$, define
\[
S(P,\xi;f,\beta):=\sum_{i=1}^m f(\xi_i)\bigl(\beta(t_i)-\beta(t_{i-1})\bigr),\qquad \beta\in\{\alpha,\alpha^{\mathrm{rc}}\}.
\]
Set $\delta_i:=\alpha^{\mathrm{rc}}(t_i)-\alpha(t_i)$ for $i=0,\dots,m$. Then $\delta_0=0$ and $\delta_m=0$, and
\[
\alpha^{\mathrm{rc}}(t_i)-\alpha^{\mathrm{rc}}(t_{i-1})
=\bigl(\alpha(t_i)-\alpha(t_{i-1})\bigr)+(\delta_i-\delta_{i-1}).
\]
Therefore
\[
S(P,\xi;f,\alpha^{\mathrm{rc}})-S(P,\xi;f,\alpha)=\sum_{i=1}^m f(\xi_i)(\delta_i-\delta_{i-1})
=\sum_{i=1}^{m-1}\bigl(f(\xi_i)-f(\xi_{i+1})\bigr)\delta_i,
\]
since $\delta_0=\delta_m=0$.

Let $\|P\|:=\max_{1\le i\le m}(t_i-t_{i-1})$ and let $\omega_f$ be the modulus of continuity of $f$ as in Step~5.
Because $\xi_i\in[t_{i-1},t_i]$ and $\xi_{i+1}\in[t_i,t_{i+1}]$, we have $|\xi_i-\xi_{i+1}|\le t_{i+1}-t_{i-1}\le 2\|P\|$, hence
\[
|f(\xi_i)-f(\xi_{i+1})|\le \omega_f(2\|P\|).
\]
Thus
\[
\bigl|S(P,\xi;f,\alpha^{\mathrm{rc}})-S(P,\xi;f,\alpha)\bigr|
\le \omega_f(2\|P\|)\sum_{i=1}^{m-1}\delta_i.
\]
Now $\delta_i=\alpha(t_i+)-\alpha(t_i)\ge 0$ for $i=1,\dots,m-1$. For each such $i$, choose $\varepsilon_i>0$ so small that the
intervals $[t_i,t_i+\varepsilon_i]$ are disjoint and contained in $[0,1]$. Then
\[
\delta_i=\alpha(t_i+)-\alpha(t_i)\le \alpha(t_i+\varepsilon_i)-\alpha(t_i).
\]
Summing over $i=1,\dots,m-1$ and using disjointness as in Claim~1 yields
\[
\sum_{i=1}^{m-1}\delta_i\le \alpha(1)-\alpha(0).
\]
Consequently,
\[
\bigl|S(P,\xi;f,\alpha^{\mathrm{rc}})-S(P,\xi;f,\alpha)\bigr|
\le \omega_f(2\|P\|)\bigl(\alpha(1)-\alpha(0)\bigr).
\]
Since $f$ is uniformly continuous, $\omega_f(2\|P\|)\to 0$ as $\|P\|\to 0$.
Therefore the difference between the Riemann--Stieltjes sums for $\alpha$ and $\alpha^{\mathrm{rc}}$ tends to $0$ uniformly in the tags
as the mesh tends to $0$, which proves \eqref{eq:int-alpha-rc-equals}.

Finally, since $L(f)=\int_0^1 f\,d\alpha$ and \eqref{eq:int-alpha-rc-equals} holds, we obtain
\[
L(f)=\int_0^1 f\,d\alpha^{\mathrm{rc}}\qquad(\forall f\in X).
\]

\end{deferredproof}

\begin{proposition}[Affine continuous representation on the Lorenz domain (Herstein--Milnor construction)]\label{prop:HM-affine-Lorenz}
Let $X:=C([0,1])$ with $\|\cdot\|_\infty$, and let
\[
\mathcal L:=\{L\in X:\ L(0)=0,\ L(1)=1,\ L\ \text{is nondecreasing and convex}\}.
\]
Let $\succsim$ be a binary relation on $\mathcal L$ satisfying
Axioms~\ref{ax:HM-order}, \ref{ax:HM-independence}, and \ref{ax:HM-nontrivial}. Moreover, it satisfies the \(\|\cdot\|_\infty\) version of closed-graph continuity:
 
The graph
\[
\{(L_1,L_2)\in\mathcal L\times\mathcal L:\ L_1\succsim L_2\}
\]
is closed in \( \mathcal L  \times \mathcal L \), where $\mathcal L$  carries the topology induced by
\((C([0,1]),\|\cdot\|_\infty)\).
Then there exists a nonconstant function $U:\mathcal L\to\mathbb R$ such that:

\begin{enumerate}
\item[\textup{(i)}] For all $L_1,L_2\in\mathcal L$,
\[
L_1\succsim L_2\quad\Longleftrightarrow\quad U(L_1)\ge U(L_2).
\]
\item[\textup{(ii)}] For all $L_1,L_2\in\mathcal L$ and all $\alpha\in[0,1]$,
\[
U\bigl((1-\alpha)L_1+\alpha L_2\bigr)=(1-\alpha)U(L_1)+\alpha U(L_2).
\]
\item[\textup{(iii)}] $U$ is continuous on $\mathcal L$ with respect to $\|\cdot\|_\infty$.
\item[\textup{(iv)}] If $V:\mathcal L\to\mathbb R$ also satisfies \textup{(i)}--\textup{(ii)}, then there exist $a>0$ and $b\in\mathbb R$ such that
\[
V=aU+b\quad\text{on }\mathcal L.
\]
Moreover, for any fixed $L^1\succ L^0$ one may normalize $U(L^1)=1$ and $U(L^0)=0$.
\end{enumerate}
\end{proposition}

\begin{deferredproof}{prop:HM-affine-Lorenz}
Write $L\succ M$ for $L\succsim M$ and $\neg(M\succsim L)$, and $L\sim M$ for $L\succsim M$ and $M\succsim L$.
Convexity of $\mathcal L$ is immediate: if $L_1,L_2\in\mathcal L$ and $\alpha\in[0,1]$, then $(1-\alpha)L_1+\alpha L_2\in\mathcal L$.

We record the strict form of independence (used repeatedly): for all $L_1,L_2,L_3\in\mathcal L$ and $\alpha\in(0,1)$,
\begin{equation}\label{eq:SI}
L_1\succ L_2\iff (1-\alpha)L_1+\alpha L_3\succ (1-\alpha)L_2+\alpha L_3.
\end{equation}
Proof: if $L_1\succ L_2$, then $L_1\succsim L_2$ and $\neg(L_2\succsim L_1)$; by Strong Independence,
$(1-\alpha)L_1+\alpha L_3\succsim (1-\alpha)L_2+\alpha L_3$, and if also
$(1-\alpha)L_2+\alpha L_3\succsim (1-\alpha)L_1+\alpha L_3$, then Strong Independence (reverse direction) gives
$L_2\succsim L_1$, contradiction; hence the mixture is strict. Conversely, if the mixture is strict, then it is weak, hence
$L_1\succsim L_2$ by the reverse direction; if $L_2\succsim L_1$ then the reverse mixture weak preference holds by Strong
Independence, contradicting strictness.

Fix $L^1,L^0\in\mathcal L$ with $L^1\succ L^0$ (Axiom~\ref{ax:HM-nontrivial}).

Fix $X,Y\in\mathcal L$ with $X\succ Y$ and write $m(\alpha):=\alpha X+(1-\alpha)Y$ for $\alpha\in[0,1]$.
If $0\le\beta<\alpha\le 1$, set $\lambda:=\frac{\alpha-\beta}{1-\beta}\in(0,1]$ and compute
\[
\lambda X+(1-\lambda)m(\beta)
=\lambda X+(1-\lambda)\bigl(\beta X+(1-\beta)Y\bigr)
=\bigl(\beta+\lambda(1-\beta)\bigr)X+\bigl(1-\beta-\lambda(1-\beta)\bigr)Y
=\alpha X+(1-\alpha)Y=m(\alpha).
\]
Since $X\succ Y$, applying \eqref{eq:SI} with $(L_1,L_2,L_3)=(X,Y,X)$ and weight $\beta\in(0,1)$ gives
$X=(1-\beta)X+\beta X\succ (1-\beta)Y+\beta X=m(\beta)$, and then applying \eqref{eq:SI} with
$(L_1,L_2,L_3)=(X,m(\beta),m(\beta))$ and weight $\lambda$ yields $m(\alpha)=\lambda X+(1-\lambda)m(\beta)\succ m(\beta)$.
Thus
\begin{equation}\label{eq:segment-mon}
0\le \beta<\alpha\le 1\quad\Longrightarrow\quad m(\alpha)\succ m(\beta).
\end{equation}

Define the interval $[Y,X]:=\{L\in\mathcal L:\ X\succsim L\succsim Y\}$.
If $L_1,L_2\in[Y,X]$ and $\theta\in[0,1]$, then $X\succsim L_i$ and $L_i\succsim Y$ for $i=1,2$, and Strong Independence gives
\[
X=(1-\theta)X+\theta X\succsim (1-\theta)L_1+\theta X,\qquad
(1-\theta)L_1+\theta X\succsim (1-\theta)L_1+\theta L_2,
\]
hence $X\succsim (1-\theta)L_1+\theta L_2$ by transitivity; similarly,
\[
(1-\theta)L_1+\theta L_2\succsim (1-\theta)Y+\theta L_2,\qquad
(1-\theta)Y+\theta L_2\succsim (1-\theta)Y+\theta Y=Y,
\]
hence $(1-\theta)L_1+\theta L_2\succsim Y$. Therefore $[Y,X]$ is convex.

Now fix $L\in[Y,X]$ and set
\[
A_L:=\{\alpha\in[0,1]:\ m(\alpha)\succsim L\},\qquad
B_L:=\{\alpha\in[0,1]:\ L\succsim m(\alpha)\}.
\]
By completeness, $A_L\cup B_L=[0,1]$, and $1\in A_L$ (since $X\succsim L$) and $0\in B_L$ (since $L\succsim Y$),
so both sets are nonempty. If $\alpha_n\in A_L$ and $\alpha_n\to\alpha$, then $m(\alpha_n)\succsim L$ for all $n$; since
$\alpha\mapsto m(\alpha)$ is continuous in $\|\cdot\|_\infty$ and $\operatorname{graph}(\succsim)$ is closed
(Axiom~\ref{ax:HM-continuity}), we get $m(\alpha)\succsim L$, hence $\alpha\in A_L$; thus $A_L$ is closed, and similarly $B_L$ is closed.
As $[0,1]$ is connected and $A_L,B_L$ are nonempty closed with union $[0,1]$, we have $A_L\cap B_L\neq\varnothing$; pick
$\alpha_L\in A_L\cap B_L$, so $m(\alpha_L)\succsim L\succsim m(\alpha_L)$, i.e.
\begin{equation}\label{eq:indiff-alpha}
L\sim m(\alpha_L).
\end{equation}
If also $L\sim m(\beta)$, then either $\beta=\alpha_L$ or, say, $\beta<\alpha_L$, and \eqref{eq:segment-mon} gives
$m(\alpha_L)\succ m(\beta)$, contradicting transitivity with $L\sim m(\alpha_L)$ and $L\sim m(\beta)$; hence $\alpha_L$ is unique.
Define $u_{X,Y}:[Y,X]\to[0,1]$ by $u_{X,Y}(L):=\alpha_L$.

If $L,M\in[Y,X]$ and $u_{X,Y}(L)=\alpha$, $u_{X,Y}(M)=\beta$, then $L\sim m(\alpha)$ and $M\sim m(\beta)$ by \eqref{eq:indiff-alpha}.
By \eqref{eq:segment-mon} and completeness, $m(\alpha)\succsim m(\beta)$ iff $\alpha\ge\beta$; therefore, by transitivity,
\begin{equation}\label{eq:local-rep}
L\succsim M\iff u_{X,Y}(L)\ge u_{X,Y}(M)\qquad(L,M\in[Y,X]).
\end{equation}
Moreover, if $L,M\in[Y,X]$ and $\theta\in[0,1]$, then convexity of $[Y,X]$ gives $N:=(1-\theta)L+\theta M\in[Y,X]$.
Let $\alpha:=u_{X,Y}(L)$ and $\beta:=u_{X,Y}(M)$, so $L\sim m(\alpha)$ and $M\sim m(\beta)$. Strong Independence implies
\[
(1-\theta)L+\theta M \sim (1-\theta)m(\alpha)+\theta M \sim (1-\theta)m(\alpha)+\theta m(\beta)
= \bigl((1-\theta)\alpha+\theta\beta\bigr)X+\Bigl(1-\bigl((1-\theta)\alpha+\theta\beta\bigr)\Bigr)Y
= m\bigl((1-\theta)\alpha+\theta\beta\bigr),
\]
and by uniqueness in \eqref{eq:indiff-alpha},
\begin{equation}\label{eq:local-aff}
u_{X,Y}\bigl((1-\theta)L+\theta M\bigr)=(1-\theta)u_{X,Y}(L)+\theta u_{X,Y}(M).
\end{equation}

Now assume $X,Y\in\mathcal L$ satisfy $X\succsim L^1\succ L^0\succsim Y$, so $L^1,L^0\in[Y,X]$ and $u_{X,Y}$ is defined there.
Define on $[Y,X]$ the normalized affine map
\[
U_{X,Y}(L):=\frac{u_{X,Y}(L)-u_{X,Y}(L^0)}{u_{X,Y}(L^1)-u_{X,Y}(L^0)}.
\]
Since $u_{X,Y}$ represents $\succsim$ on $[Y,X]$ by \eqref{eq:local-rep} and $L^1\succ L^0$, we have
$u_{X,Y}(L^1)>u_{X,Y}(L^0)$, so $U_{X,Y}$ is well-defined, represents $\succsim$ on $[Y,X]$, is affine on $[Y,X]$
(by \eqref{eq:local-aff}), and satisfies $U_{X,Y}(L^1)=1$, $U_{X,Y}(L^0)=0$.

For each $L\in\mathcal L$, choose some $X,Y\in\mathcal L$ with $X\succsim L^1\succ L^0\succsim Y$ and $L\in[Y,X]$
(which is possible by completeness: if $L\succsim L^1$ take $(X,Y)=(L,L^0)$; if $L^1\succsim L\succsim L^0$ take
$(X,Y)=(L^1,L^0)$; if $L^0\succsim L$ take $(X,Y)=(L^1,L)$), and define $U(L):=U_{X,Y}(L)$.
To see this does not depend on the choice, fix any admissible $(X,Y)$ with $L^1,L^0,L\in[Y,X]$.
If $L^1\succsim L\succsim L^0$, then by the construction applied on $[L^0,L^1]$ there is a unique $\alpha\in[0,1]$ such that
$L\sim \alpha L^1+(1-\alpha)L^0$, hence, by affinity and normalization of $U_{X,Y}$,
\[
U_{X,Y}(L)=U_{X,Y}(\alpha L^1+(1-\alpha)L^0)=\alpha.
\]
If $L\succsim L^1$, then $L\succ L^0$, and applying the same construction on $[L^0,L]$ yields a unique $\alpha\in(0,1]$ such that
$L^1\sim \alpha L+(1-\alpha)L^0$, hence
\[
1=U_{X,Y}(L^1)=U_{X,Y}(\alpha L+(1-\alpha)L^0)=\alpha U_{X,Y}(L)+(1-\alpha)U_{X,Y}(L^0)=\alpha U_{X,Y}(L),
\]
so $U_{X,Y}(L)=1/\alpha$. If $L^0\succsim L$, then $L^1\succ L$, and applying the construction on $[L,L^1]$ yields a unique
$\alpha\in(0,1)$ such that $L^0\sim \alpha L^1+(1-\alpha)L$, hence
\[
0=U_{X,Y}(L^0)=U_{X,Y}(\alpha L^1+(1-\alpha)L)=\alpha U_{X,Y}(L^1)+(1-\alpha)U_{X,Y}(L)=\alpha+(1-\alpha)U_{X,Y}(L),
\]
so $U_{X,Y}(L)=-\alpha/(1-\alpha)$. In each case, the computed value depends only on $L$ and the fixed pair $L^1\succ L^0$,
so $U$ is well-defined on $\mathcal L$ and satisfies $U(L^1)=1$, $U(L^0)=0$, hence is nonconstant.

For (i), given $L_1,L_2\in\mathcal L$, let $X$ be a $\succsim$-maximal element of $\{L^1,L_1,L_2\}$ and
$Y$ a $\succsim$-minimal element of $\{L^0,L_1,L_2\}$ (exist by completeness). Then $X\succsim L^1\succ L^0\succsim Y$
and $L_1,L_2\in[Y,X]$, so $U$ coincides with $U_{X,Y}$ on $[Y,X]$ and $U_{X,Y}$ represents $\succsim$ there; thus
$L_1\succsim L_2\iff U(L_1)\ge U(L_2)$.

For (ii), fix $L_1,L_2\in\mathcal L$ and $\theta\in[0,1]$ and use the same $X,Y$ as above; since $[Y,X]$ is convex,
$(1-\theta)L_1+\theta L_2\in[Y,X]$, and $U=U_{X,Y}$ is affine on $[Y,X]$, so
$U((1-\theta)L_1+\theta L_2)=(1-\theta)U(L_1)+\theta U(L_2)$.

For (iii), let $L_n\to L$ in $\|\cdot\|_\infty$ and suppose $U(L_n)\not\to U(L)$. Then there exist $\varepsilon>0$ and a subsequence
(not relabeled) such that either $U(L_n)\ge U(L)+\varepsilon$ for all $n$ or $U(L_n)\le U(L)-\varepsilon$ for all $n$; assume the first.
Choose $n_0$ and set $H:=L_{n_0}$, so $U(H)\ge U(L)+\varepsilon$ and $U(H)>U(L)$. Let
\[
t:=\frac{\varepsilon/2}{U(H)-U(L)}\in(0,1),\qquad L^\sharp:=(1-t)L+tH\in\mathcal L,
\]
so by affinity $U(L^\sharp)=(1-t)U(L)+tU(H)=U(L)+\varepsilon/2$. Then for all $n$,
$U(L_n)\ge U(L)+\varepsilon>U(L^\sharp)$, hence $L_n\succ L^\sharp$, so $(L_n,L^\sharp)\in\operatorname{graph}(\succsim)$.
By closedness of $\operatorname{graph}(\succsim)$ and $L_n\to L$, we get $(L,L^\sharp)\in\operatorname{graph}(\succsim)$, i.e.\ $L\succsim L^\sharp$,
hence $U(L)\ge U(L^\sharp)$ by (i), contradicting $U(L^\sharp)=U(L)+\varepsilon/2$. Thus $U(L_n)\to U(L)$.

For (iv), let $V:\mathcal L\to\mathbb R$ also satisfy (i)--(ii). Set $a:=V(L^1)-V(L^0)$ and $b:=V(L^0)$; since $L^1\succ L^0$ and $V$ represents
$\succsim$, we have $a>0$. Fix $L\in\mathcal L$. If $L^1\succsim L\succsim L^0$, let $\alpha\in[0,1]$ be the unique number with
$L\sim \alpha L^1+(1-\alpha)L^0$; then by affinity,
$V(L)=\alpha V(L^1)+(1-\alpha)V(L^0)=a\alpha+b$, while by definition $U(L)=\alpha$, so $V(L)=aU(L)+b$.
If $L\succsim L^1$, let $\alpha\in(0,1]$ be the unique number with $L^1\sim \alpha L+(1-\alpha)L^0$; then
$V(L^1)=\alpha V(L)+(1-\alpha)V(L^0)$ gives $V(L)=a(1/\alpha)+b$, while $U(L)=1/\alpha$, so again $V(L)=aU(L)+b$.
If $L^0\succsim L$, let $\alpha\in(0,1)$ be the unique number with $L^0\sim \alpha L^1+(1-\alpha)L$; then
$V(L^0)=\alpha V(L^1)+(1-\alpha)V(L)$ gives $V(L)=b-a\,\alpha/(1-\alpha)$, while $U(L)=-\alpha/(1-\alpha)$, so $V(L)=aU(L)+b$.
Hence $V=aU+b$ on $\mathcal L$, and the normalization $U(L^1)=1$, $U(L^0)=0$ is already in force.
\end{deferredproof}

\subsection{Representation Theorems for Lorenz Curves: Dyadic Discretization and One-sided Representation}

Before formally delivering the main results of this paper, we would like to provide some metaphorical explanation for the ideas behind our new proof structures for the first and second representation theorems for Lorenz ordering. Lorenz curves are inhabitants of an infinite-dimensional function space. However, many arguments that are standard in finite-dimensional settings fail in the infinite-dimensional case. For example, in finite dimensions it is possible to establish the existence of a linear functional representation by means of open-set separation. In the full infinite-dimensional space, the analogous translation set would have empty interior under the $\|\cdot\|_\infty$ norm, and the same interior based separation argument would therefore fail directly.

As we have seen, the absence of such an implicit assumption leads to counterexamples, which imply that the primal and dual theorems in their original form cannot be true. Strictly speaking, this is not directly fatal, given that the final goal is to derive the metric representation of Egalitarianism from the qualitative ordering of curves. However, these two theorems serve as key footholds that we need to leverage in the original project. As we attempt to climb the otherwise smooth mountain of the representation theorem, in the absence of strengthened boundary conditions, we have no auxiliary intermediate representations to rely on. Therefore, if a similar representation result were still to hold, it seems that it would necessarily require more delicate techniques and would have to differ in a fundamental way from relatively common constructions in analysis or measure theory, since the existence of singularities may preclude all the mainstream comparable structures.

On the other hand, it is possible to circumvent these obstructions by directly imposing regularity conditions on the kernel functions themselves. However, such conditions are not intrinsic to the general framework of Lorenz curves. In light of the observation regarding the difficulty of deriving a representation theorem from the current axiomatic system, and drawing on earlier studies on equivalent formulations of continuity, here we try to hit a balance between assuming too much and too little by introducing the following new axiom which is constructed using the intrinsic objects from Lorenz domain:

\begin{axiom}[Archimedean on the Lorenz domain]\label{ax:archimedean}
Let \(E:[0,1]\to[0,1]\) be the equality curve \(E(u):=u\). Let
\(\mathcal K\in\{\mathcal L,\mathcal L^\uparrow\}\) be the relevant Lorenz domain.
There exist an anchor \(B\in\mathcal K\), a constant \(C>0\), and a radius \(\delta>0\) such that
\[
E\succ B,
\]
Then for every \(L\in\mathcal K\) with
\(
0<\|E-L\|_\infty<\delta
\), there exists \(t_L\in[0,1]\) satisfying
\[
t_L\le C\|E-L\|_\infty
\qquad\text{and}\qquad
L\succsim (1-t_L)E+t_L B.
\]
\end{axiom}

It is helpful to describe the architecture of the proofs and their places among current literature. The arguments from  theorem~\ref{thm:dyadic-discrete-Lorenz-Stieltjes} to the main theorem~\ref{thm:main} appear to be new in this setting. Of course, their novelty is not in introducing new general analytical machinery, 
but in combining standard tools from Functional/Convex analysis, measure theory in a way tailored to the Lorenz domain.

More precisely, the paper builds the characterization result through three methodological layers.
The next Theorem~\ref{thm:dyadic-discrete-Lorenz-Stieltjes} is a rather conventional finite-dimensional approximation argument. The continuous Lorenz curve is embedded into a discretized dyadic polygonal subspace in which interior geometry is restored so that classical convex separation becomes available again. The resulting finite-dimensional linear functional is then rewritten as a step-kernel Stieltjes representation by Abel summation.

Then theorems~\ref{thm:lorenz-stieltjes-Lip} and \ref{thm:dual-A5} would be built on that discrete representation as the continuum limit of the previous discretized representation. Under Axiom~\ref{ax:archimedean}, the dyadic kernels satisfy uniform mass bounds, which allows one to pass to the limit by diagonal extraction and the Riesz representation theorem. In the dual case, one first conjugates the preorder by the involution operation \(\mathcal S\), which transforms dual mixtures into ordinary mixtures, and then applies the same discretization-and-limit machinery to the transported problem. Finally, the main Theorem~\ref{thm:main} is different in character. Indeed, the main proof is organized around a parametrized functional construction, with various analytical and measure-theoretic arguments interwoven throughout to bridge the gaps and connect the different steps. The purpose of this construction is to overcome pointwise discontinuities and to impose a global structural rigidity constraint governing the primal and dual measures. Ultimately, these analytical structures that we ourselves constructed will impose a rigidity that forces the kernels  \(p\) and \(q\) to be affine.

The discrete layer is mainly borrowed from convex analysis and piecewise-affine approximation theory, see \cite{Rockafellar1970,RudinFunctionalAnalysis1991,ConwayFunctionalAnalysis1990,KreyszigFunctionalAnalysis1978,AliprantisBorder2006,SchaeferWolff1999,BrennerScott2008,deBoor2001}. The limiting arguments related to standard positive-functional representation and measure-theoretic compactness arguments, see, e.g, \cite{RudinRealComplex1987,RudinPMA1976,Apostol1974,FollandRealAnalysis1999,Bogachev2007,dudley2018real,Billingsley1995,RoydenFitzpatrick2010,SteinShakarchi2005,AliprantisBorder2006}. For the dual side, also see the existing economics literature \cite{Yaari1987,Schmeidler1989,Acerbi2002,Kusuoka2001}. For the general background about variational and \(BV\) methods arguments that appeared in the main rigidity theorem, check \cite{RockafellarWets1998,AmbrosioFuscoPallara2000,brezis2010functional,Hormander2003,FollandRealAnalysis1999,Bogachev2007,EvansPDE2010}.

\begin{theorem}[Polygonal Discretization of Lorenz--Stieltjes Curve Representation]
\label{thm:dyadic-discrete-Lorenz-Stieltjes}
Let
\[
\mathcal K\in\{\mathcal L,\mathcal L^\uparrow\}\subset C([0,1]).
\]
Let $\succsim$ be a complete preorder on $\mathcal K$, and let
\[
U:\mathcal K\to\mathbb R
\]
be an affine $\|\cdot\|_\infty$--continuous, nonconstant representation of $\succsim$.
Assume \emph{weak dominance}:
\[
G_1(u)\ge G_2(u)\ \forall u\in[0,1]\quad\Longrightarrow\quad G_1\succsim G_2 .
\]

For each $m\in\mathbb N$ set
\[
n_m:=2^m,\qquad t_{m,k}:=\frac{k}{n_m}\qquad(k=0,\dots,n_m),
\]
and define the polygonal discretization subspace
\[
X_m:=\Bigl\{g\in C([0,1]) : g \text{ is affine on each }[t_{m,k-1},t_{m,k}]
\ (k=1,\dots,n_m)\Bigr\},
\qquad
\mathcal K_m:=\mathcal K\cap X_m.
\]

Then there exists $m_0\in\mathbb N$ such that for every $m\ge m_0$ there exist
\begin{itemize}
\item a right--continuous step function $p_m:[0,1]\to\mathbb R$ that is constant on each $[t_{m,k-1},t_{m,k})$,
is nonincreasing, and satisfies $p_m(1)=0$;
\item a finite nonnegative Borel measure $\nu_m$ supported on $\{t_{m,1},\dots,t_{m,n_m-1}\}$,
\end{itemize}
such that for every $G,H\in\mathcal K_m$,
\begin{equation}\label{eq:disc-rep-RS}
U(G)-U(H)=\int_0^1 p_m(u)\,d\bigl(G-H\bigr)(u),
\end{equation}
where, for every $g\in X_m$,
\[
\int_0^1 p_m(u)\,dg(u)
:=\sum_{k=1}^{n_m}p(m,k)\bigl(g(t_{m,k})-g(t_{m,k-1})\bigr),
\qquad
p_m|_{[t_{m,k-1},t_{m,k})}\equiv p(m,k),\ \ p(m,n_m)=0.
\]

Moreover, defining the grid \emph{drops}
\[
\alpha(m,k):=p(m,k)-p(m,k+1)\qquad(k=1,\dots,n_m-1),
\]
one may take
\[
\nu_m=\sum_{k=1}^{n_m-1}\alpha(m,k)\,\delta_{t_{m,k}},
\]
and the discrete summation-by-parts identity holds:
\begin{equation}\label{eq:disc-rep-parts}
\int_0^1 p_m\,d(G-H)=\int_{[0,1]}\bigl(G(u)-H(u)\bigr)\,\nu_m(du)
=\sum_{k=1}^{n_m-1}\bigl(G(t_{m,k})-H(t_{m,k})\bigr)\,\nu_m(\{t_{m,k}\})
\end{equation}
for all $G,H\in\mathcal K_m$.
In particular,
\[
\nu_m\ge0,
\qquad
\nu_m([0,1])=p(m,1)\ge0.
\]
\end{theorem}

\begin{proof}
By proposition \ref{prop:HM-affine-Lorenz}, we have an affine continuous representation $U$ that represents the complete preorder on the Lorenz Domain, our goal is to constructively prove the existence of its integral representation form when the domain of $U$ is restricted to  finite-dimensional embedding space of discretized Lorenz curves. The proof would be invoked twice in the future as it applies to both Lorenz representation theorems under independence and Dual independence assumptions, respectively. 

\noindent\textbf{Step 1: Dyadic grids, Polygonal subspaces,  and Cone Interior.}
Let $X:=C([0,1])$ with $\|\cdot\|_\infty$.

For each $m\in\mathbb N$ set
\[
n_m:=2^m,\qquad t_{m,k}:=\frac{k}{n_m}\qquad(k=0,1,\dots,n_m).
\]
Define:
\[
X_m:=\Bigl\{g\in C([0,1]) : g \text{ is affine on each }[t_{m,k-1},t_{m,k}]
\ (k=1,\dots,n_m)\Bigr\},
\]
\[
\mathcal L_m:=\mathcal L\cap X_m,\qquad
\mathcal L_m^{\uparrow}:=\mathcal L^{\uparrow}\cap X_m. \qquad
\mathcal K_m:=\mathcal K\cap X_m.
\]
Define the polygonal interpolation operator $P_m:X\to X_m$ by prescribing nodal values
\[
(P_m f)(t_{m,k})=f(t_{m,k})\qquad(k=0,1,\dots,n_m),
\]
and letting $P_m f$ be affine on each $[t_{m,k-1},t_{m,k}]$.
Then $P_m f\to f$ uniformly for every $f\in X$.

Now fix $L\in\mathcal K$. On each $[t_{m,k-1},t_{m,k}]$ the slope of $P_mL$ equals the secant slope
\[
s_{m,k}:=\frac{L(t_{m,k})-L(t_{m,k-1})}{t_{m,k}-t_{m,k-1}}
=n_m\bigl(L(t_{m,k})-L(t_{m,k-1})\bigr).
\]
Convexity of $L$ implies that $(s_{m,k})_{k=1}^{n_m}$ is nondecreasing.
Because $L$ is increasing, we have $s_{m,k}\ge 0$ for all $k$; if $\mathcal K=\mathcal L^\uparrow$, then in fact
$s_{m,k}>0$ for all $k$.
Also
\[
P_mL(0)=L(0)=0,\qquad P_mL(1)=L(1)=1.
\]
Hence
\[
P_mL\in\mathcal K_m.
\]

\paragraph{Anchoring pair and induced finite-dimensional utility.}
Choose $L^+,L^-\in\mathcal K$ with $U(L^+)>U(L^-)$ and set
\[
\eta:=\frac{U(L^+)-U(L^-)}{4}>0.
\]
By continuity of $U$ there exists $\varepsilon>0$ such that
\[
\|L-L^+\|_\infty<\varepsilon \Longrightarrow |U(L)-U(L^+)|<\eta,
\qquad
\|L-L^-\|_\infty<\varepsilon \Longrightarrow |U(L)-U(L^-)|<\eta.
\]
Choose $m_0$ such that for all $m\ge m_0$,
\[
\|P_mL^+-L^+\|_\infty<\varepsilon,\qquad \|P_mL^--L^-\|_\infty<\varepsilon.
\]
Then for all $m\ge m_0$,
\[
U(P_mL^+)\ge U(L^+)-\eta>U(L^-)+\eta\ge U(P_mL^-).
\]

Fix $m\ge m_0$ and denote \(
n:=n_m,\quad t_k:=t_{m,k}\quad(k=0,1,\dots,n)
\), set
\[
L_m^+:=P_mL^+,\qquad L_m^-:=P_mL^-,
\qquad L_m^0:=\frac{L_m^++L_m^-}{2}\in\mathcal K_m,
\]
and define the centered affine utility on $\mathcal K_m$ by
\[
U_m(L):=U(L)-U(L_m^0)\qquad(L\in\mathcal K_m).
\]
Then $U_m$ is affine on $\mathcal K_m$, represents $\succsim$ on $\mathcal K_m$, and satisfies
\[
U_m(L_m^0)=0,\qquad U_m(L_m^+)>0,\qquad U_m(L_m^-)<0.
\]

Let
\[
X_m^0:=\{h\in X_m:\ h(0)=h(1)=0\},
\]
with $\|\cdot\|_\infty$. Each $h\in X_m^0$ is determined by $(h(t_1),\dots,h(t_{n-1}))\in\mathbb R^{n-1}$.

For $j=1,\dots,n-1$ define $H_{m,j}\in X_m^0$ by
\[
H_{m,j}(t_j)=1,\qquad H_{m,j}(t_k)=0\ (k\neq j),
\qquad H_{m,j}\ \text{affine on each }[t_{k-1},t_k].
\]
Then, for every $h\in X_m^0$,
\begin{equation}\label{eq:FD-hat-expansion}
h=\sum_{j=1}^{n-1} h(t_j)\,H_{m,j}.
\end{equation}
Indeed both sides lie in $X_m^0$ and coincide at all nodes $t_k$, hence coincide on each $[t_{k-1},t_k]$.

Note that for each $m$ and each $g\in X_m$,
\[
\|g\|_\infty=\max_{0\le k\le n_m}|g(t_{m,k})|.
\]
Indeed, on each interval $[t_{m,k-1},t_{m,k}]$ the function $g$ is affine, hence its maximum and minimum are attained at endpoints. In particular, for $h\in X_m^0$,
\begin{equation}\label{eq:FD-sup-at-interior-nodes}
\|h\|_\infty=\max_{1\le k\le n-1}|h(t_k)|
\end{equation}
because $h(t_0)=h(t_n)=0$. Hence $\|H_{m,j}\|_\infty=1$ for each $j$.

Define the cone
\[
C_m:=\{h\in X_m^0:\ h(u)\ge 0\ \forall u\in[0,1]\},
\qquad
C_m^\circ:=\{h\in X_m^0:\ h(t_k)>0\ \forall k=1,\dots,n-1\}.
\]
Then
\begin{equation}\label{eq:FD-cone-interior}
C_m^\circ=\operatorname{int}_{X_m^0}(C_m).
\end{equation}
$(\subset)$ Let $h\in C_m^\circ$ and set $m(h):=\min_{1\le k\le n-1} h(t_k)>0$.
If $\|g-h\|_\infty<m(h)$ then by \eqref{eq:FD-sup-at-interior-nodes},
\[
g(t_k)\ge h(t_k)-\|g-h\|_\infty>0\qquad(k=1,\dots,n-1).
\]
By affinity of $g$ on each $[t_{k-1},t_k]$, positivity at endpoints implies $g\ge 0$ pointwise, so $g\in C_m$.
Thus $B(h,m(h))\subset C_m$ and $h\in \operatorname{int}_{X_m^0}(C_m)$.

$(\supset)$ Let $h\in C_m$ with $h(t_j)=0$ for some $j\in\{1,\dots,n-1\}$.
For any $\delta>0$ set $g:=h-\delta H_{m,j}$. Then
\[
\|g-h\|_\infty=\delta\|H_{m,j}\|_\infty=\delta,\qquad g(t_j)=-\delta<0,
\]
so $g\notin C_m$. Hence no open ball around $h$ is contained in $C_m$, i.e. $h\notin \operatorname{int}_{X_m^0}(C_m)$.
Therefore $\operatorname{int}_{X_m^0}(C_m)\subset C_m^\circ$.

\medskip
\noindent\textbf{Step 2: Translation of Disjoint Open Convex Sets.}
Let
\[
\mathcal D_m:=\{d(L):=L-L_m^0:\ L\in\mathcal K_m\}\subset X_m^0,
\qquad
U_m^0(d(L)):=U_m(L).
\]
Then $\mathcal D_m$ is convex and $U_m^0$ is affine on $\mathcal D_m$. Define convex subsets of $\mathcal D_m$:
\[
\mathcal B_m:=\{d\in\mathcal D_m:\ U_m^0(d)>0\}, \qquad
\mathcal W_m:=\{d\in\mathcal D_m:\ U_m^0(d)<0\}.
\]
These are nonempty because $d(L_m^+)\in\mathcal B_m$ and $d(L_m^-)\in\mathcal W_m$. Now define subsets of $X_m^0$:
\[
\mathcal U_m:=\mathcal B_m+C_m^\circ, \qquad \mathcal V_m:=\mathcal W_m-C_m^\circ.
\]
Because $C_m^\circ$ is open in $X_m^0$ and a cone, $\mathcal U_m$ and $\mathcal V_m$ are open, convex, and nonempty.

We claim that they are disjoint sets. Suppose for contradiction that there exists $z\in\mathcal U_m\cap\mathcal V_m$.
Then there exist $d_b\in\mathcal B_m$, $d_w\in\mathcal W_m$ and $c,c'\in C_m^\circ$ such that
\[
z=d_b+c=d_w-c'.
\]
Let $L_b:=L_m^0+d_b\in\mathcal K_m$ and $L_w:=L_m^0+d_w\in\mathcal K_m$. Then
\[
L_w-L_b=d_w-d_b=c+c'\in C_m,
\]
so $L_w(u)\ge L_b(u)$ for all $u\in[0,1]$. By weak dominance, $L_w\succsim L_b$, hence
$U_m(L_w)\ge U_m(L_b)$ because $U_m$ represents $\succsim$ on $\mathcal K_m$.
But $U_m(L_w)=U_m^0(d_w)<0$ and $U_m(L_b)=U_m^0(d_b)>0$, contradiction. Thus
\begin{equation}\label{eq:FD-disjoint}
\mathcal U_m\cap\mathcal V_m=\varnothing.
\end{equation}

\medskip
\noindent\textbf{Step 3: Hahn--Banach separation on $X_m^0$; positivity on the cone.}
Apply Hahn--Banach Separation Lemma~\ref{lem:HB-separation-convex} in the normed space $(X_m^0,\|\cdot\|_\infty)$
to the disjoint nonempty convex sets $\mathcal U_m$ and $\mathcal V_m$.
Since $\mathcal U_m$ is open, $\operatorname{int}\mathcal U_m=\mathcal U_m\neq\varnothing$. Hence there exist a nonzero
continuous linear functional $\widetilde F_m:X_m^0\to\mathbb R$ and $\gamma_m\in\mathbb R$ such that
\begin{equation}\label{eq:FD-sep-UmVm}
\widetilde F_m(u)\ge\gamma_m\ge \widetilde F_m(v)\qquad(\forall u\in\mathcal U_m,\ \forall v\in\mathcal V_m).
\end{equation}

\emph{(3.1) $\widetilde F_m\ge0$ on $C_m^\circ$.}
Fix $c\in C_m^\circ$ and any $d_b\in\mathcal B_m$.
For every $\lambda>0$ we have $d_b+\lambda c\in\mathcal U_m$ (since $C_m^\circ$ is a cone),
so by \eqref{eq:FD-sep-UmVm},
\[
\widetilde F_m(d_b)+\lambda \widetilde F_m(c)=\widetilde F_m(d_b+\lambda c)\ge\gamma_m\qquad(\forall\lambda>0).
\]
If $\widetilde F_m(c)<0$, the left side tends to $-\infty$ as $\lambda\to\infty$, impossible. Thus
\begin{equation}\label{eq:FD-F-positive-on-cone}
\widetilde F_m(c)\ge0\qquad(\forall c\in C_m^\circ).
\end{equation}

\emph{(3.2) Sign inequalities on $\mathcal B_m$ and $\mathcal W_m$.}
Fix some $c_0\in C_m^\circ$. Let $d_b\in\mathcal B_m$. For every $\lambda>0$,
$d_b+\lambda c_0\in\mathcal U_m$, so \eqref{eq:FD-sep-UmVm} yields
\[
\widetilde F_m(d_b+\lambda c_0)=\widetilde F_m(d_b)+\lambda \widetilde F_m(c_0)\ge\gamma_m\qquad(\forall\lambda>0),
\]
equivalently
\[
\widetilde F_m(d_b)\ge\gamma_m-\lambda \widetilde F_m(c_0)\qquad(\forall\lambda>0).
\]
By \eqref{eq:FD-F-positive-on-cone}, $\widetilde F_m(c_0)\ge0$; letting $\lambda\to0^+$ gives
\[
\widetilde F_m(d_b)\ge\gamma_m.
\]
Similarly, for $d_w\in\mathcal W_m$ and every $\lambda>0$ we have $d_w-\lambda c_0\in\mathcal V_m$, hence
\[
\widetilde F_m(d_w-\lambda c_0)=\widetilde F_m(d_w)-\lambda \widetilde F_m(c_0)\le\gamma_m\qquad(\forall\lambda>0),
\]
so letting $\lambda\to0^+$ again yields $\widetilde F_m(d_w)\le\gamma_m$. Therefore
\begin{equation}\label{eq:FD-F-sign-on-BW}
\widetilde F_m(d)\ge\gamma_m\ (d\in\mathcal B_m), \qquad \widetilde F_m(d)\le\gamma_m\ (d\in\mathcal W_m).
\end{equation}

\medskip
\noindent\textbf{Step 4: Constancy of $\widetilde F_m$ on $\{U_m^0=0\}$.}
Let $d\in\mathcal D_m$ with $U_m^0(d)=0$. Fix any $d^+\in\mathcal B_m$ and $d^-\in\mathcal W_m$.
For each $\lambda\in(0,1)$,
\[
U_m^0\bigl((1-\lambda)d+\lambda d^+\bigr)=(1-\lambda)\cdot 0+\lambda U_m^0(d^+)>0,
\]
so $(1-\lambda)d+\lambda d^+\in\mathcal B_m$; similarly $(1-\lambda)d+\lambda d^-\in\mathcal W_m$.
By \eqref{eq:FD-F-sign-on-BW},
\[
\widetilde F_m\bigl((1-\lambda)d+\lambda d^+\bigr)\ge\gamma_m,\qquad
\widetilde F_m\bigl((1-\lambda)d+\lambda d^-\bigr)\le\gamma_m.
\]
Expanding by linearity and letting $\lambda\downarrow 0$ gives
\[
\widetilde F_m(d)\ge\gamma_m,\qquad \widetilde F_m(d)\le\gamma_m,
\]
hence
\begin{equation}\label{eq:FD-F-on-zero}
U_m^0(d)=0\ \Longrightarrow\ \widetilde F_m(d)=\gamma_m.
\end{equation}
In particular $0\in\mathcal D_m$ (take $L=L_m^0$), and $U_m^0(0)=0$, so \eqref{eq:FD-F-on-zero} yields
\[
0=\widetilde F_m(0)=\gamma_m.
\]
Thus, rewriting \eqref{eq:FD-F-sign-on-BW} with $\gamma_m=0$,
\begin{equation}\label{eq:FD-sep-normalized}
\widetilde F_m(d)\ge0\ (d\in\mathcal B_m), \qquad \widetilde F_m(d)\le0\ (d\in\mathcal W_m),\qquad
\widetilde F_m(c)\ge0\ (c\in C_m^\circ),
\end{equation}
and $\widetilde F_m(d)=0$ whenever $U_m^0(d)=0$.

\medskip
\noindent\textbf{Step 5: Proportionality $\widetilde F_m=a(m)\,U_m^0$ on $\mathcal D_m$ and Interior ball.}

\smallskip
\noindent\emph{(5.1) Proportionality on $\mathcal D_m$.}
Let $I_m:=U_m^0(\mathcal D_m)\subset\mathbb R$. Since $\mathcal D_m$ is convex and $U_m^0$ is affine, $I_m$ is an interval.
Define $\varphi_m:I_m\to\mathbb R$ by
\[
\varphi_m(y):=\widetilde F_m(d)\quad\text{for any }d\in\mathcal D_m\text{ with }U_m^0(d)=y.
\]
We show $\varphi_m$ is well-defined. Let $d_1,d_2\in\mathcal D_m$ satisfy $U_m^0(d_1)=U_m^0(d_2)=y$.

If $y=0$, then $\widetilde F_m(d_1)=\widetilde F_m(d_2)=0$ by \eqref{eq:FD-F-on-zero} (with $\gamma_m=0$).

Assume $y>0$. Choose any $d^-\in\mathcal W_m$ and set
\[
\lambda(y;d^-):=\frac{y}{y-U_m^0(d^-)}\in(0,1),
\]
so that
\[
U_m^0\bigl((1-\lambda(y;d^-))d_i+\lambda(y;d^-)d^-\bigr)=0\qquad(i=1,2).
\]
By \eqref{eq:FD-F-on-zero} and linearity,
\[
0=\widetilde F_m\bigl((1-\lambda)d_i+\lambda d^-\bigr)
=(1-\lambda)\widetilde F_m(d_i)+\lambda \widetilde F_m(d^-)\qquad(i=1,2),
\]
hence $\widetilde F_m(d_1)=\widetilde F_m(d_2)$. The case $y<0$ is identical using a fixed $d^+\in\mathcal B_m$.

Therefore $\varphi_m$ is well-defined. It is affine on $I_m$ because for $y_i\in I_m$ and $d_i\in\mathcal D_m$
with $U_m^0(d_i)=y_i$, we have for $\theta\in[0,1]$:
\[
U_m^0\bigl((1-\theta)d_1+\theta d_2\bigr)=(1-\theta)y_1+\theta y_2,\qquad
\widetilde F_m\bigl((1-\theta)d_1+\theta d_2\bigr)=(1-\theta)\widetilde F_m(d_1)+\theta \widetilde F_m(d_2).
\]
Also $\varphi_m(0)=0$. Hence there exists $a(m)\in\mathbb R$ such that
\begin{equation}\label{eq:FD-F-proportional}
\widetilde F_m(d)=a(m)\,U_m^0(d)\qquad(\forall d\in\mathcal D_m).
\end{equation}
Moreover $a(m)\ge0$ because for any $d^+\in\mathcal B_m$,
\[
0\le \widetilde F_m(d^+)=a(m)\,U_m^0(d^+),\qquad U_m^0(d^+)>0.
\]
\smallskip
\noindent\emph{(5.2) Interior point and radius in $\mathcal D_m$.}
For $j=0,\dots,n-1$ define the corner Lorenz curves
\[
K_{m,j}(u):=\max\!\left\{0,\ \frac{u-t_j}{1-t_j}\right\}\in\mathcal L_m.
\]
Pick any strictly positive weight vector
\[
\lambda=(\lambda_0,\dots,\lambda_{n-1})\in \Delta_{n-1}^\circ
:=\Bigl\{\lambda_j>0,\ \sum_{j=0}^{n-1}\lambda_j=1\Bigr\},
\]
and define
\[
S_m(u;\lambda):=\sum_{j=0}^{n-1}\lambda_j\,K_{m,j}(u)\in X_m.
\]
Define discrete differences for $f\in X_m$:
\[
\Delta_k f:=f(t_k)-f(t_{k-1})\quad(k=1,\dots,n),\qquad
\Delta^{2}_k f:=\Delta_{k+1}f-\Delta_k f
=f(t_{k+1})-2f(t_k)+f(t_{k-1})\quad(k=1,\dots,n-1).
\]
Then for $L\in X_m$ with $L(0)=0$, $L(1)=1$,
\[
L\in\mathcal L_m \ \Longleftrightarrow\ \Delta_k L\ge 0\ (k=1,\dots,n)\ \text{and}\ \Delta^{2}_k L\ge 0\ (k=1,\dots,n-1),
\]
since on $[t_{k-1},t_k]$ the slope equals $n\,\Delta_k L$ and convexity means these slopes are nondecreasing.

A direct calculation gives for $K_{m,j}$:
\[
\Delta_k K_{m,j}=
\begin{cases}
0, & k\le j,\\[3pt]
\dfrac{1}{n-j}, & k\ge j+1,
\end{cases}
\qquad(k=1,\dots,n),
\]
hence
\[
\Delta_k S_m(\cdot;\lambda)=\sum_{j=0}^{k-1}\frac{\lambda_j}{n-j}, \qquad
\Delta^{2}_k S_m(\cdot;\lambda)=\frac{\lambda_k}{n-k}.
\]
Let
\[
\lambda_*(\lambda):=\min_{0\le j\le n-1}\lambda_j>0,\qquad
s(n;\lambda):=\frac{\lambda_*(\lambda)}{n}.
\]
Then for all $k$,
\[
\Delta_k S_m(\cdot;\lambda)\ge \Delta_1 S_m(\cdot;\lambda)=\frac{\lambda_0}{n}\ge s(n;\lambda),\qquad
\Delta^{2}_k S_m(\cdot;\lambda)=\frac{\lambda_k}{n-k}\ge s(n;\lambda).
\]

Now let $h\in X_m^0$ with $\|h\|_\infty<r$. Then for all $k$,
\[
|\Delta_k h| \le |h(t_k)|+|h(t_{k-1})| \le 2\|h\|_\infty<2r,
\qquad
|\Delta^{2}_k h|
\le |h(t_{k+1})|+2|h(t_k)|+|h(t_{k-1})|
\le 4\|h\|_\infty<4r.
\]
Choose the interior radius
\[
r(n;\lambda):=\frac{s(n;\lambda)}{8}=\frac{\lambda_*(\lambda)}{8n}.
\]
For any $\|h\|_\infty<r(n;\lambda)$ we then have, for all $k$,
\[
\Delta_k\bigl(S_m(\cdot;\lambda)+h\bigr)\ge s(n;\lambda)-2r(n;\lambda)=\frac34\,s(n;\lambda)>0,
\]
and for $k=1,\dots,n-1$,
\[
\Delta^{2}_k\bigl(S_m(\cdot;\lambda)+h\bigr)\ge s(n;\lambda)-4r(n;\lambda)=\frac12\,s(n;\lambda)>0.
\]
Since also
\[
\bigl(S_m(\cdot;\lambda)+h\bigr)(0)=0,\qquad \bigl(S_m(\cdot;\lambda)+h\bigr)(1)=1,
\]
it follows that $S_m(\cdot;\lambda)+h$ is strictly increasing and convex.
Hence
\[
S_m(\cdot;\lambda)+h\in\mathcal K_m.
\]
In particular, taking $h=0$ gives $S_m(\cdot;\lambda)\in\mathcal K_m$. Therefore
\begin{equation}\label{eq:FD-interior-ball}
B_{X_m^0}\bigl(S_m(\cdot;\lambda)-L_m^0,\ r(n;\lambda)\bigr)\subset \mathcal D_m,
\qquad \operatorname{int}_{X_m^0}(\mathcal D_m)\neq\varnothing.
\end{equation}

\smallskip
\noindent\emph{(5.3) $a(m)>0$.}
Assume for contradiction that $a(m)=0$. Then \eqref{eq:FD-F-proportional} implies $\widetilde F_m(d)=0$ for all $d\in\mathcal D_m$,
hence $\widetilde F_m$ vanishes on the open ball in \eqref{eq:FD-interior-ball}. Let
\[
d_0:=S_m(\cdot;\lambda)-L_m^0\in\mathcal D_m,
\]
so $\widetilde F_m(d_0)=0$. Fix arbitrary $x\in X_m^0$. If $x\neq 0$, choose $q>0$ such that $q\|x\|_\infty<r(n;\lambda)$.
Then $d_0+q x\in B(d_0,r(n;\lambda))\subset\mathcal D_m$, so
\[
0=\widetilde F_m(d_0+qx)=\widetilde F_m(d_0)+q\widetilde F_m(x)=q\widetilde F_m(x),
\]
hence $\widetilde F_m(x)=0$. Since $x\in X_m^0$ was arbitrary, $\widetilde F_m\equiv 0$ on $X_m^0$, contradicting separation
(which gave $\widetilde F_m\neq 0$). Hence $a(m)\neq0$, and with $a(m)\ge0$ we obtain $a(m)>0$.

Define the normalized functional
\[
F_m:=\frac{1}{a(m)}\,\widetilde F_m.
\]
Then $F_m$ is continuous linear on $X_m^0$ and by \eqref{eq:FD-F-proportional} we have
\begin{equation}\label{eq:FD-F-normalized-on-Dm}
F_m(d)=U_m^0(d)\qquad(\forall d\in\mathcal D_m).
\end{equation}
Consequently, for $L\in\mathcal K_m$,
\[
F_m(L-L_m^0)=U_m^0(L-L_m^0)=U_m(L)=U(L)-U(L_m^0),
\]
and $L\mapsto F_m(L-L_m^0)$ represents $\succsim$ on $\mathcal K_m$.

\medskip
\noindent\textbf{Step 6: Step--kernel representation on the fixed grid.}
Every \(h\in X_m^0\) has the hat expansion \eqref{eq:FD-hat-expansion}. Define
\[
\alpha(m,j):=F_m(H_{m,j})\qquad(j=1,\dots,n-1).
\]
Then by linearity,
\begin{equation}\label{eq:FD-F-nodal}
F_m(h)=\sum_{j=1}^{n-1}\alpha(m,j)\,h(t_j)
\qquad(\forall h\in X_m^0).
\end{equation}

\smallskip
\noindent\emph{(6.1) Abel summation and a right--continuous step kernel.}
Define tail sums
\[
p(m,k):=\sum_{j=k}^{n-1}\alpha(m,j)\quad(k=1,\dots,n-1),
\qquad
p(m,n):=0,
\]
and define the right--continuous step function \(p_m:[0,1]\to\mathbb R\) by
\[
p_m(u):=p(m,k)
\quad\text{for }u\in[t_{k-1},t_k)\ (k=1,\dots,n),
\qquad
p_m(1):=p(m,n)=0.
\]
For \(g\in X_m\) define the grid Riemann--Stieltjes sum
\begin{equation}\label{eq:FD-grid-RS}
\int_0^1 p_m(u)\,dg(u)
:=
\sum_{k=1}^{n}p(m,k)\bigl(g(t_k)-g(t_{k-1})\bigr).
\end{equation}

Let \(h\in X_m^0\), and write \(h_k:=h(t_k)\) for \(k=0,\dots,n\). Then
\(h_0=h_n=0\). Since
\[
p(m,k)-p(m,k+1)=\alpha(m,k)\qquad(k=1,\dots,n-1),
\]
we compute
\begin{align*}
\int_0^1 p_m\,dh
&=\sum_{k=1}^{n}p(m,k)(h_k-h_{k-1})\\
&=\sum_{k=1}^{n}p(m,k)h_k-\sum_{k=1}^{n}p(m,k)h_{k-1}\\
&=\sum_{k=1}^{n-1}p(m,k)h_k-\sum_{k=1}^{n-1}p(m,k+1)h_k
\qquad(\text{because }h_0=h_n=0)\\
&=\sum_{k=1}^{n-1}\bigl(p(m,k)-p(m,k+1)\bigr)h_k\\
&=\sum_{k=1}^{n-1}\alpha(m,k)\,h(t_k).
\end{align*}
Combining this with \eqref{eq:FD-F-nodal}, we obtain
\begin{equation}\label{eq:FD-F-Stieltjes}
F_m(h)=\int_0^1 p_m\,dh
\qquad(\forall h\in X_m^0).
\end{equation}

\smallskip
\noindent\emph{(6.2) Difference form on \(\mathcal K_m\).}
Let \(G,H\in\mathcal K_m\). Then \(G-H\in X_m^0\), so by
\eqref{eq:FD-F-Stieltjes},
\[
F_m(G-H)=\int_0^1 p_m\,d(G-H).
\]
Also,
\[
G-H=(G-L_m^0)-(H-L_m^0),
\]
and \(G-L_m^0,\ H-L_m^0\in\mathcal D_m\). Therefore, by linearity of \(F_m\)
and \eqref{eq:FD-F-normalized-on-Dm},
\begin{align*}
F_m(G-H)
&=F_m(G-L_m^0)-F_m(H-L_m^0)\\
&=U_m^0(G-L_m^0)-U_m^0(H-L_m^0)\\
&=U(G)-U(H).
\end{align*}
Hence
\begin{equation}\label{eq:FD-difference-rep}
U(G)-U(H)=\int_0^1 p_m\,d(G-H)
\qquad(\forall G,H\in\mathcal K_m).
\end{equation}

In particular, the map
\[
L\longmapsto F_m(L-L_m^0)
\]
represents \(\succsim\) on \(\mathcal K_m\), because for \(A,B\in\mathcal K_m\),
\[
A\succsim B
\Longleftrightarrow
U(A)\ge U(B)
\Longleftrightarrow
F_m(A-L_m^0)\ge F_m(B-L_m^0).
\]

\smallskip
\noindent\emph{(6.3) Weak dominance \(\Rightarrow\) monotonicity of \(p_m\).}
Fix \(\lambda\in\Delta_{n-1}^\circ\), and let
\[
S_m:=S_m(\cdot;\lambda)\in\mathcal K_m
\]
and \(r(n;\lambda)>0\) be as in Step~5.2, so that
\[
\|h\|_\infty<r(n;\lambda)\quad\Longrightarrow\quad S_m+h\in\mathcal K_m
\qquad(h\in X_m^0).
\]
Fix \(k\in\{1,\dots,n-1\}\), and set
\[
\varepsilon:=\frac{r(n;\lambda)}{2}>0.
\]
Since \(\|H_{m,k}\|_\infty=1\), we have
\[
\|\varepsilon H_{m,k}\|_\infty=\varepsilon<r(n;\lambda),
\]
hence
\[
S_m+\varepsilon H_{m,k}\in\mathcal K_m.
\]
Moreover \(H_{m,k}\ge0\) pointwise, so
\[
S_m+\varepsilon H_{m,k}\ge S_m
\qquad\text{pointwise on }[0,1].
\]
By weak dominance,
\[
S_m+\varepsilon H_{m,k}\succsim S_m.
\]
Since \(L\mapsto F_m(L-L_m^0)\) represents \(\succsim\) on \(\mathcal K_m\),
\begin{align*}
0
&\le
F_m\bigl((S_m+\varepsilon H_{m,k})-L_m^0\bigr)-F_m(S_m-L_m^0)\\
&=\varepsilon F_m(H_{m,k})
=\varepsilon\alpha(m,k).
\end{align*}
Therefore
\begin{equation}\label{eq:FD-alpha-nonneg}
\alpha(m,k)\ge0
\qquad(k=1,\dots,n-1).
\end{equation}
Since
\[
p(m,k)-p(m,k+1)=\alpha(m,k)\qquad(k=1,\dots,n-1),
\qquad
p(m,n)=0,
\]
it follows that
\[
p(m,1)\ge p(m,2)\ge\cdots\ge p(m,n)=0.
\]
Hence \(p_m\) is nonincreasing on \([0,1]\).

\medskip
\noindent\textbf{Step 7: Finite kernel $\nu_m$ and discrete integration by parts.}
Define discrete drops
\[
\alpha(m,k):=p(m,k)-p(m,k+1)\qquad(k=1,\dots,n-1).
\]
(Equivalently $\alpha(m,k)=F_m(H_{m,k})$ and $\alpha(m,k)\ge0$ by \eqref{eq:FD-alpha-nonneg}.)
Define a finite nonnegative Borel measure $\nu_m$ on $[0,1]$ by
\begin{equation}\label{eq:FD-num-def}
\nu_m(\{t_k\}):=\alpha(m,k)\quad(k=1,\dots,n-1),\qquad \nu_m(\{0\})=0,\qquad \nu_m(\{1\})=0.
\end{equation}
Its total mass is
\begin{equation}\label{eq:FD-mass}
\nu_m([0,1])=\sum_{k=1}^{n-1}\alpha(m,k)=p(m,1)-p(m,n)=p(m,1)\ge0.
\end{equation}

For $G,H\in\mathcal K_m$, set
\[
g:=G-H\in X_m^0.
\]
Then $g(t_0)=g(t_n)=0$, and discrete integration by parts gives
\begin{align}
\int_0^1 p_m\,d(G-H)
&=\sum_{k=1}^{n}p(m,k)\bigl(g(t_k)-g(t_{k-1})\bigr)\nonumber =\sum_{k=1}^{n}p(m,k)g(t_k)-\sum_{k=1}^{n}p(m,k)g(t_{k-1})\nonumber\\
&=\sum_{k=1}^{n-1}p(m,k)g(t_k)-\sum_{k=0}^{n-1}p(m,k+1)g(t_k)
\qquad(\text{since }g(t_0)=g(t_n)=0,\ p(m,n)=0)\nonumber\\
&=\sum_{k=1}^{n-1}g(t_k)\bigl(p(m,k)-p(m,k+1)\bigr)\nonumber =\sum_{k=1}^{n-1}\bigl(G(t_k)-H(t_k)\bigr)\alpha(m,k)\nonumber\\
&=\int_{[0,1]}\bigl(G(u)-H(u)\bigr)\,\nu_m(du).\label{eq:FD-difference-parts}
\end{align}
Combining \eqref{eq:FD-difference-rep} with \eqref{eq:FD-difference-parts}, we obtain
\[
U(G)-U(H)
=
\int_0^1 p_m\,d(G-H)
=
\int_{[0,1]}\bigl(G(u)-H(u)\bigr)\,\nu_m(du)
\qquad(\forall G,H\in\mathcal K_m).
\]
This completes the proof.

\end{proof}

\begin{theorem}[Lorenz--Stieltjes representation under weak dominance and Archimedean]
\label{thm:lorenz-stieltjes-Lip}
Let $\succsim$ be a complete preorder on $\mathcal L$ satisfying:
\begin{enumerate}
\item[(A1)] \textbf{Mixture independence:} Axiom~\ref{ax:HM-independence}.

\item[(A2)] \textbf{Closed graph continuity:}  
The graph
\[
\{(L_1,L_2)\in\mathcal L \times\mathcal L :\ L_1\succsim L_2\}
\]
is closed in \( \mathcal L   \times \mathcal L \), where $\mathcal L $  carries the topology induced by
\((C([0,1]),\|\cdot\|_\infty)\).. 

\item[(A3)] \textbf{Weak dominance:} if $L_1(u)\ge L_2(u)$ for all $u\in[0,1]$, then $L_1\succsim L_2$.

\item[(A4)] \textbf{Nontriviality:} Axiom~\ref{ax:HM-nontrivial} on \(\mathcal L\).

\item[(A5)] \textbf{Archimedean on Lorenz Domain:} Axiom~\ref{ax:archimedean} on \(\mathcal L\).
\end{enumerate}
Then there exist a finite nonnegative Borel measure \(\nu\) on \([0,1]\) and a nonincreasing
bounded--variation kernel \(p:[0,1]\to\mathbb R\), right--continuous on \([0,1)\), such that
\[
p(u)=\nu((u,1])\qquad(0\le u<1),
\]
and fixing the endpoint value by
\[
p(1)=p(1^-),\qquad p(1^-):=\lim_{t\uparrow1}p(t),
\]
one has
\[
V(L):=\int_{[0,1]}L(u)\,\nu(du)=\int_0^1 p(u)\,dL(u)\qquad(L\in\mathcal L).
\]
Moreover, if \(\widetilde p:[0,1]\to\mathbb R\) is another nonincreasing bounded--variation
kernel, right--continuous on \([0,1)\), also satisfying:
\[
\widetilde p(1)=\widetilde p(1^-),\qquad
\widetilde p(1^-):=\lim_{t\uparrow1}\widetilde p(t),
\]
and such that
\[
L_1\succsim L_2
\Longleftrightarrow
\int_0^1 \widetilde p\,dL_1 \ge \int_0^1 \widetilde p\,dL_2
\qquad(\forall L_1,L_2\in\mathcal L),
\]
then there exist \(a>0\) and \(b\in\mathbb R\) such that
\[
\widetilde p(u)=a\,p(u)+b\qquad(u\in[0,1]).
\]
\end{theorem}

\begin{proof}
By Proposition~\ref{prop:HM-affine-Lorenz} (using \textup{(A1)}, \textup{(A2)}, \textup{(A4)}), fix an affine
$\|\cdot\|_\infty$--continuous representation $U:\mathcal L\to\mathbb R$ of $\succsim$.
By \textup{(A4)} (nontriviality), $U$ is nonconstant.

\paragraph{Dyadic setup.}
For each $m\in\mathbb N$ let
\[
n_m:=2^m,\qquad t_{m,k}:=\frac{k}{n_m}\qquad(k=0,1,\dots,n_m),
\]
and define
\[
X_m:=\Bigl\{g\in C([0,1]) : g \text{ is affine on each }[t_{m,k-1},t_{m,k}]
\ (k=1,\dots,n_m)\Bigr\},
\qquad
\mathcal L_m:=\mathcal L\cap X_m.
\]
Let $P_m:C([0,1])\to X_m$ be polygonal interpolation on $\{t_{m,k}\}$; then $P_m f\to f$ uniformly for every
$f\in C([0,1])$, and $P_mL\in\mathcal L_m$ for every $L\in\mathcal L$.

\paragraph{Finite-grid representation.}
Apply Theorem~\ref{thm:dyadic-discrete-Lorenz-Stieltjes} with $\mathcal K=\mathcal L$ to $U$
(weak dominance is \textup{(A3)}).
Then there exists $m_0\in\mathbb N$ such that for every $m\ge m_0$ there exist
a nonincreasing right--continuous dyadic step kernel $p_m$
and a finite nonnegative Borel measure $\nu_m$ supported on the interior grid
$\{t_{m,1},\dots,t_{m,n_m-1}\}$ such that for all $L,H\in\mathcal L_m$,
\begin{equation}\label{eq:disc-on-Lm-difference}
U(L)-U(H)=\int_{[0,1]}(L-H)\,d\nu_m=\int_0^1 p_m\,d(L-H).
\end{equation}

For the remainder of Step~1 fix $m\ge m_0$ and abbreviate
\[
n:=n_m=2^m,\qquad t_k:=t_{m,k}=\frac{k}{n}\qquad(k=0,1,\dots,n).
\]
Write the step kernel values as
\[
p_m(u)=p(m,k)\quad \text{for }u\in[t_{k-1},t_k)\qquad(k=1,\dots,n).
\]
Since $\nu_m$ is supported on $\{t_1,\dots,t_{n-1}\}$, if $u\in[t_{n-1},1)$ then
\[
(u,1]\cap \operatorname{supp}\nu_m=\varnothing,
\]
hence
\[
p_m(u)=\nu_m((u,1])=0.
\]
Therefore
\[
p(m,n)=0.
\]

If $u\in[0,t_1)$, then
\[
(u,1]\cap \operatorname{supp}\nu_m=\{t_1,\dots,t_{n-1}\},
\]
so
\[
p(m,1)=p_m(u)=\nu_m((u,1])=\nu_m([0,1]).
\]
Hence
\begin{equation}\label{eq:mass-as-p11}
\nu_m([0,1])=p(m,1).
\end{equation}

\medskip
\noindent\textbf{Step 1: Uniform Mass Bound for $\nu_m$.}
Assume \textup{(A5)}. Let $E(u):=u$ (so $E\in\mathcal L_m$ for every $m$).  For $j=0,\dots,n-1$ define the dyadic
corner Lorenz curves
\[
K_{m,j}(u):=\max\!\left\{0,\frac{u-t_j}{1-t_j}\right\}\in\mathcal L_m.
\]
Then $E\ge K_{m,1}\ge K_{m,n-1}$ pointwise, hence by weak dominance \textup{(A3)},
\begin{equation}\label{eq:dominance-chain}
U(E)\ge U(K_{m,1})\ge U(K_{m,n-1}).
\end{equation}

\smallskip
\noindent\emph{(1.1) Reference curve $K_{m,n-1}$.}
For $K_{m,n-1}$ we have $\Delta_k K_{m,n-1}=0$ for $k=1,\dots,n-1$ and $\Delta_n K_{m,n-1}=1$, hence
\[
\int_0^1 p_m\,dK_{m,n-1}
=\sum_{k=1}^n p(m,k)\Delta_k K_{m,n-1}
=p(m,n)\cdot 1
=0.
\]
Also, since $\nu_m$ is supported on $\{t_1,\dots,t_{n-1}\}$ and
\[
K_{m,n-1}(t_k)=0\qquad(k=1,\dots,n-1),
\]
we have
\[
\int_{[0,1]}K_{m,n-1}(u)\,\nu_m(du)=0.
\]

\begin{equation}\label{eq:cm-is-U-Klast}
c_m:=U(K_{m,n-1}).
\end{equation}

Applying \eqref{eq:disc-on-Lm-difference} with $H=K_{m,n-1}$ gives, for every $L\in\mathcal L_m$,
\begin{align}
U(L)-c_m
&=U(L)-U(K_{m,n-1})\nonumber\\
&=\int_{[0,1]}\bigl(L(u)-K_{m,n-1}(u)\bigr)\,\nu_m(du)
=\int_0^1 p_m\,d\bigl(L-K_{m,n-1}\bigr)\nonumber\\
&=\int_{[0,1]}L(u)\,\nu_m(du)
=\int_0^1 p_m\,dL(u).\label{eq:disc-on-Lm}
\end{align}

\smallskip
\noindent\emph{(1.2) Express $p(m,1)=\nu_m([0,1])$ in terms of $U(E)-U(K_{m,1})$ and $U(E)-U(K_{m,n-1})$.}
Using \eqref{eq:disc-on-Lm} with $L=E$ and $\Delta_k E=1/n$ for all $k$,
\[
U(E)-c_m=\int_0^1 p_m\,dE=\sum_{k=1}^n p(m,k)\frac1n=\frac1n\sum_{k=1}^n p(m,k),
\]
so
\begin{equation}\label{eq:sum-pk-all}
\sum_{k=1}^n p(m,k)=n\bigl(U(E)-c_m\bigr).
\end{equation}
For $K_{m,1}$ we have $K_{m,1}(t_0)=K_{m,1}(t_1)=0$ and
$K_{m,1}(t_k)=(t_k-t_1)/(1-t_1)=(k-1)/(n-1)$ for $k\ge2$, hence
\[
\Delta_1K_{m,1}=0,\qquad \Delta_kK_{m,1}=\frac1{n-1}\quad(k=2,\dots,n),
\]
and therefore
\[
U(K_{m,1})-c_m=\int_0^1 p_m\,dK_{m,1}=\sum_{k=2}^n p(m,k)\frac1{n-1}.
\]
Equivalently,
\begin{equation}\label{eq:sum-pk-tail}
\sum_{k=2}^n p(m,k)=(n-1)\bigl(U(K_{m,1})-c_m\bigr).
\end{equation}
Subtracting \eqref{eq:sum-pk-tail} from \eqref{eq:sum-pk-all} yields
\begin{align}
p(m,1)
&=\sum_{k=1}^n p(m,k)-\sum_{k=2}^n p(m,k)\nonumber\\
&=\bigl(U(E)-c_m\bigr)+(n-1)\bigl(U(E)-U(K_{m,1})\bigr).\label{eq:p11}
\end{align}
Using \eqref{eq:cm-is-U-Klast} to eliminate $c_m$, we may rewrite \eqref{eq:p11} as
\begin{equation}\label{eq:p11-expanded}
p(m,1)=\bigl(U(E)-U(K_{m,n-1})\bigr)+(n-1)\bigl(U(E)-U(K_{m,1})\bigr).
\end{equation}
By \eqref{eq:mass-as-p11}, this is exactly the total mass $\nu_m([0,1])$.

\smallskip
\noindent\emph{(1.3) Global comparison bound from Axiom~\ref{ax:archimedean}.}
Let \(B\in\mathcal L\) and \(C,\delta>0\) be given by Axiom~\ref{ax:archimedean}, and set
\[
\Delta_B:=U(E)-U(B)>0.
\]
We claim that
\begin{equation}\label{eq:U-sup-global}
U(E)-U(L)\le C\Delta_B\,\|E-L\|_\infty
\qquad(\forall L\in\mathcal L).
\end{equation}
Indeed, if \(0<\|E-L\|_\infty<\delta\), then by Axiom~\ref{ax:archimedean} there exists
\(t_L\in[0,1]\) such that
\[
t_L\le C\|E-L\|_\infty
\qquad\text{and}\qquad
L\succsim (1-t_L)E+t_LB.
\]
Applying the affine representation \(U\) gives
\[
U(L)\ge (1-t_L)U(E)+t_LU(B)=U(E)-t_L\Delta_B,
\]
hence
\[
U(E)-U(L)\le t_L\Delta_B\le C\Delta_B\,\|E-L\|_\infty.
\]
Now let \(L\in\mathcal L\) be arbitrary. If \(L=E\), \eqref{eq:U-sup-global} is trivial. Otherwise define
\[
\varepsilon:=\min\Bigl\{1,\frac{\delta}{2\|E-L\|_\infty}\Bigr\}\in(0,1],
\qquad
L^\varepsilon:=(1-\varepsilon)E+\varepsilon L.
\]
Then
\[
\|E-L^\varepsilon\|_\infty=\varepsilon\|E-L\|_\infty<\delta,
\]
so the previous estimate applied to \(L^\varepsilon\) yields
\[
U(E)-U(L^\varepsilon)\le C\Delta_B\,\|E-L^\varepsilon\|_\infty
= C\Delta_B\,\varepsilon\|E-L\|_\infty.
\]
Since
\[
U(L^\varepsilon)=(1-\varepsilon)U(E)+\varepsilon U(L),
\]
we have
\[
U(E)-U(L^\varepsilon)=\varepsilon\bigl(U(E)-U(L)\bigr).
\]
Cancelling \(\varepsilon>0\) proves \eqref{eq:U-sup-global}.

\smallskip
\noindent\emph{(1.4) Apply \eqref{eq:U-sup-global} to \(K_{m,1}\) and \(K_{m,n-1}\) and conclude.}
Since
\[
\|E-K_{m,1}\|_\infty=\frac1n,
\qquad
\|E-K_{m,n-1}\|_\infty=\frac{n-1}{n}\le 1,
\]
equation \eqref{eq:U-sup-global} gives
\[
U(E)-U(K_{m,1})\le \frac{C\Delta_B}{n},
\qquad
U(E)-U(K_{m,n-1})\le C\Delta_B.
\]
Substituting these bounds into \eqref{eq:p11-expanded}, we obtain
\[
\nu_m([0,1])=p(m,1)
\le C\Delta_B +(n-1)\frac{C\Delta_B}{n}
\le 2C\Delta_B.
\]
Thus
\begin{equation}\label{eq:uniform-mass-bound}
0\le \nu_m([0,1])\le 2C\Delta_B
\qquad(\forall m\ge m_0).
\end{equation}
Also, since \(c_m=U(K_{m,n-1})\), the same estimate yields
\begin{equation}\label{eq:cm-bounded}
U(E)-C\Delta_B\le c_m\le U(E)
\qquad(\forall m\ge m_0).
\end{equation}

\medskip
\noindent\textbf{Step 2: Countable dense dyadic--rational class in $C([0,1])$.}
From this point onward, $m$ becomes a free variable again. For each $m\in\mathbb N$ set
\[
n_m:=2^m,\qquad t_{m,k}:=\frac{k}{n_m}\qquad(k=0,1,\dots,n_m),
\]
and let
\[
X_m:=\Bigl\{g\in C([0,1]) : g \text{ is affine on each }[t_{m,k-1},t_{m,k}]\ (k=1,\dots,n_m)\Bigr\}.
\]
For $g\in X_m$ define the nodal vector
\[
\mathbf g^{(m)}:=(g(t_{m,0}),g(t_{m,1}),\dots,g(t_{m,n_m}))\in\mathbb R^{n_m+1}.
\]
Then $g\mapsto \mathbf g^{(m)}$ is a bijection between $X_m$ and $\mathbb R^{n_m+1}$, since a piecewise affine function
is uniquely determined by its nodal values. Moreover, for each $g\in X_m$,
\begin{equation}\label{eq:sup-at-dyadic-nodes}
\|g\|_\infty=\max_{0\le k\le n_m}|g(t_{m,k})|.
\end{equation}
Indeed, on each interval $[t_{m,k-1},t_{m,k}]$ the function $g$ is affine, hence its maximum and minimum are attained at endpoints;
taking the maximum over all subintervals yields \eqref{eq:sup-at-dyadic-nodes}.

Define the dyadic--rational subclass
\[
\mathscr D_m:=\Bigl\{g\in X_m:\ g(t_{m,k})\in\mathbb Q\ \text{for all }k=0,1,\dots,n_m\Bigr\},
\qquad
\mathscr D:=\bigcup_{m=1}^\infty \mathscr D_m\subset C([0,1]).
\]

\smallskip
\noindent\emph{Claim 2.1: $\mathscr D$ is countable.}
Fix $m$. The nodal map $g\mapsto \mathbf g^{(m)}$ identifies $\mathscr D_m$ with $\mathbb Q^{n_m+1}$.
Since $\mathbb Q$ is countable and $n_m+1<\infty$, the product $\mathbb Q^{n_m+1}$ is countable, hence $\mathscr D_m$ is countable.
Therefore $\mathscr D=\bigcup_{m\ge 1}\mathscr D_m$ is a countable union of countable sets which is countable according to basic set theory.

\smallskip
\noindent\emph{Claim 2.2: $\mathscr D$ is dense in $C([0,1])$ under $\|\cdot\|_\infty$.}
Let $f\in C([0,1])$ and $\varepsilon>0$ be given. For each $m$ define the polygonal interpolation operator
$P_m:C([0,1])\to X_m$ by
\[
(P_m f)(t_{m,k})=f(t_{m,k})\qquad(k=0,1,\dots,n_m),
\quad\text{and}\quad
P_m f \text{ affine on each }[t_{m,k-1},t_{m,k}].
\]
Since $f$ is uniformly continuous, $\|P_m f-f\|_\infty\to 0$ as $m\to\infty$.
Choose $m$ so large that
\[
\|P_m f-f\|_\infty<\frac{\varepsilon}{2}.
\]
For each node $t_{m,k}$ choose a rational $q_k\in\mathbb Q$ such that
\[
|q_k-(P_m f)(t_{m,k})|<\frac{\varepsilon}{2}\qquad(k=0,1,\dots,n_m),
\]
which is possible since $\mathbb Q$ is dense in $\mathbb R$.
Let $g\in X_m$ be the unique polygonal function satisfying $g(t_{m,k})=q_k$ for all $k$.
Then $g\in\mathscr D_m\subset\mathscr D$. Moreover, $g-P_m f\in X_m$, hence by \eqref{eq:sup-at-dyadic-nodes},
\[
\|g-P_m f\|_\infty
=\max_{0\le k\le n_m}\bigl|g(t_{m,k})-(P_m f)(t_{m,k})\bigr|
=\max_{0\le k\le n_m}|q_k-(P_m f)(t_{m,k})|
<\frac{\varepsilon}{2}.
\]
Therefore
\[
\|g-f\|_\infty\le \|g-P_m f\|_\infty+\|P_m f-f\|_\infty
<\frac{\varepsilon}{2}+\frac{\varepsilon}{2}=\varepsilon,
\]
which proves density.

Since $\mathscr D$ is countable, fix an enumeration
\begin{equation}\label{eq:D-enumeration}
\mathscr D=\{d_1,d_2,d_3,\dots\}.
\end{equation}

\medskip
\noindent\textbf{Step 3: Uniformly Bounded Positive Functionals and a diagonal subsequence.}
For each $m\ge m_0$ let $\nu_m$ be as above and let $c_m$ be the corresponding constant from Step~1, so that
\[
U(L)-c_m=\int_{[0,1]} L\,d\nu_m\qquad(\forall L\in\mathcal L_m),
\]
and the total masses are uniformly bounded:
\[
0\le \nu_m([0,1])\le M_\nu\qquad(\forall m\ge m_0),
\]
with $M_\nu<\infty$ fixed.

Define the positive continuous linear functionals $T_m:C([0,1])\to\mathbb R$ by
\begin{equation}\label{eq:Tm-def}
T_m(f):=\int_{[0,1]} f\,d\nu_m\qquad(f\in C([0,1])).
\end{equation}
Then for every $f\in C([0,1])$,
\[
|T_m(f)|\le \|f\|_\infty\,\nu_m([0,1]),
\]
and since each $T_m$ is positive,
\[
\|T_m\|=\sup_{\|f\|_\infty\le 1}|T_m(f)|=T_m(\mathbf 1)=\nu_m([0,1]).
\]
Hence we have the uniform operator norm bound
\begin{equation}\label{eq:Tm-uniform}
\|T_m\|\le M_\nu\qquad(\forall m\ge m_0).
\end{equation}

Let $\mathscr D=\{d_1,d_2,\dots\}$ be the fixed countable dense dyadic--rational class from Step~2,
with enumeration \eqref{eq:D-enumeration}.
We now build a diagonal subsequence along which $T_m(d_r)$ converges for every $r$.

Initialize $m^{(0)}_j:=m_0+j-1$ for $j\ge 1$. Inductively, given an increasing index sequence
$(m^{(r-1)}_j)_{j\ge 1}$, the scalar sequence $\bigl(T_{m^{(r-1)}_j}(d_r)\bigr)_{j\ge 1}$ is bounded by \eqref{eq:Tm-uniform},
so it admits a convergent subsequence; choose an increasing subsequence $(m^{(r)}_j)_{j\ge 1}$ of $(m^{(r-1)}_j)_{j\ge 1}$
such that $\bigl(T_{m^{(r)}_j}(d_r)\bigr)$ converges.

Define the diagonal indices
\begin{equation}\label{eq:diag-indices}
m_j:=m^{(j)}_j\qquad(j\ge 1).
\end{equation}
Then $(m_j)$ is strictly increasing, and for each fixed $r\in\mathbb N$, the tail $(m_j)_{j\ge r}$ is a subsequence of
$(m^{(r)}_j)$, hence
\begin{equation}\label{eq:diag-conv-on-D}
T_{m_j}(d_r)\ \text{converges as }j\to\infty\qquad(\forall r\in\mathbb N).
\end{equation}

Finally, Step~1 shows that $(c_m)_{m\ge m_0}$ is bounded. Passing to a further subsequence of $(m_j)$ (not relabeled),
we may assume that
\begin{equation}\label{eq:c-limit-diag}
c_{m_j}\to c\in\mathbb R.
\end{equation}
The convergence property \eqref{eq:diag-conv-on-D} is preserved under passing to subsequences.

\medskip
\noindent\textbf{Step 4: Limit functional $T$ on $C([0,1])$.}
We claim that for every $f\in C([0,1])$ the real sequence $\bigl(T_{m_j}(f)\bigr)_{j\ge 1}$ is Cauchy, hence convergent.

If $M_\nu=0$, then $\|T_{m_j}\|=0$ for all $j$, so $T_{m_j}\equiv 0$ and convergence is trivial. Hence assume $M_\nu>0$.
Fix $f\in C([0,1])$ and $\varepsilon>0$. By density of $\mathscr D$, choose $r\in\mathbb N$ such that
\[
\|f-d_r\|_\infty<\frac{\varepsilon}{3M_\nu}.
\]
Since $\bigl(T_{m_j}(d_r)\bigr)$ converges by \eqref{eq:diag-conv-on-D}, there exists $J$ such that for all $j,j'\ge J$,
\[
|T_{m_j}(d_r)-T_{m_{j'}}(d_r)|<\frac{\varepsilon}{3}.
\]
Then for all $j,j'\ge J$, using linearity and the operator norm bound \eqref{eq:Tm-uniform},
\begin{align*}
|T_{m_j}(f)-T_{m_{j'}}(f)|
&\le |T_{m_j}(f-d_r)| + |T_{m_j}(d_r)-T_{m_{j'}}(d_r)| + |T_{m_{j'}}(d_r-f)|\\
&\le \|T_{m_j}\|\,\|f-d_r\|_\infty + \frac{\varepsilon}{3} + \|T_{m_{j'}}\|\,\|f-d_r\|_\infty\\
&\le M_\nu\cdot\frac{\varepsilon}{3M_\nu}+\frac{\varepsilon}{3}+M_\nu\cdot\frac{\varepsilon}{3M_\nu}
=\varepsilon.
\end{align*}
Thus $\bigl(T_{m_j}(f)\bigr)$ is Cauchy and hence converges in $\mathbb R$. Define
\begin{equation}\label{eq:T-def}
T(f):=\lim_{j\to\infty}T_{m_j}(f)\qquad(f\in C([0,1])).
\end{equation}

In particular, for each $r$, the limit $T(d_r)$ agrees with the diagonal limit in \eqref{eq:diag-conv-on-D}; i.e. writing
\begin{equation}\label{eq:T-on-D}
\ell_r:=\lim_{j\to\infty}T_{m_j}(d_r)\quad\text{we have}\quad T(d_r)=\ell_r\qquad(r\in\mathbb N).
\end{equation}

\smallskip
\noindent\emph{Linearity, boundedness, positivity.}
For $f,g\in C([0,1])$ and $\alpha,\beta\in\mathbb R$,
\[
T(\alpha f+\beta g)=\lim_{j\to\infty}T_{m_j}(\alpha f+\beta g)
=\alpha\lim_{j\to\infty}T_{m_j}(f)+\beta\lim_{j\to\infty}T_{m_j}(g)
=\alpha T(f)+\beta T(g),
\]
so $T$ is linear. Moreover, by \eqref{eq:Tm-uniform},
\[
|T(f)|=\lim_{j\to\infty}|T_{m_j}(f)|
\le \limsup_{j\to\infty}\|T_{m_j}\|\,\|f\|_\infty
\le M_\nu\,\|f\|_\infty,
\]
so $T$ is bounded and hence continuous on $C([0,1])$. Finally, if $f\ge 0$ then $T_{m_j}(f)\ge 0$ for all $j$
(since each $\nu_{m_j}$ is nonnegative), hence $T(f)\ge 0$.

\medskip
\noindent\textbf{Step 5: Riesz--Stieltjes integral representation for $T$).}
Apply Lemma~\ref{lem:R2-positive-Riesz} to the positive continuous linear functional $T:C([0,1])\to\mathbb R$.
Then there exists an increasing function $\alpha:[0,1]\to\mathbb R$ (taken as right--continuous on $(0,1)$
by the right--continuous modification) such that
\begin{equation}\label{eq:T-Stieltjes}
T(f)=\int_0^1 f\,d\alpha\qquad(\forall f\in C([0,1])).
\end{equation}
Let $\nu$ be the finite nonnegative Borel measure induced by $\alpha$, so that
\[
\nu((a,b])=\alpha(b)-\alpha(a)\qquad(0\le a<b\le 1),
\]
and hence
\begin{equation}\label{eq:T-measure-rep}
T(f)=\int_{[0,1]} f\,d\nu\qquad(\forall f\in C([0,1])).
\end{equation}

\medskip
\noindent\textbf{Step 6: Identify $U$ with the limit integral on $\mathcal L$.}
Fix $L\in\mathcal L$ and set $L^{(j)}:=P_{m_j}L\in\mathcal L_{m_j}$.
Then $\|L^{(j)}-L\|_\infty\to 0$ as $j\to\infty$.
For each $j$, applying \eqref{eq:disc-on-Lm} to $m=m_j$ and $L^{(j)}\in\mathcal L_{m_j}$ yields
\begin{equation}\label{eq:Ucm-equality}
U\bigl(L^{(j)}\bigr)-c_{m_j}=\int_{[0,1]} L^{(j)}\,d\nu_{m_j}=T_{m_j}\!\bigl(L^{(j)}\bigr).
\end{equation}

Since $U$ is continuous on $(\mathcal L,\|\cdot\|_\infty)$ by Proposition~\ref{prop:HM-affine-Lorenz}\textup{(iii)},
\[
U\bigl(L^{(j)}\bigr)\to U(L).
\]
Also $c_{m_j}\to c$ by \eqref{eq:c-limit-diag}. It remains to show
$T_{m_j}(L^{(j)})\to T(L)$.

Write
\[
T_{m_j}\!\bigl(L^{(j)}\bigr)-T(L)
=\Bigl(T_{m_j}\!\bigl(L^{(j)}\bigr)-T_{m_j}(L)\Bigr)+\Bigl(T_{m_j}(L)-T(L)\Bigr)
= T_{m_j}\!\bigl(L^{(j)}-L\bigr)+\bigl(T_{m_j}(L)-T(L)\bigr).
\]
The second term tends to $0$ by \eqref{eq:T-def}. For the first term, \eqref{eq:Tm-uniform} gives
\[
\bigl|T_{m_j}(L^{(j)}-L)\bigr|
\le \|T_{m_j}\|\,\|L^{(j)}-L\|_\infty
\le M_\nu\,\|L^{(j)}-L\|_\infty
\to 0.
\]
Hence $T_{m_j}(L^{(j)})\to T(L)$. Passing to the limit in \eqref{eq:Ucm-equality} yields
\begin{equation}\label{eq:U-minus-c-measure}
U(L)-c=T(L)=\int_{[0,1]} L\,d\nu\qquad(\forall L\in\mathcal L),
\end{equation}
where the last equality uses \eqref{eq:T-measure-rep}.
Define
\[
V(L):=\int_{[0,1]} L\,d\nu\qquad(L\in\mathcal L).
\]
Then $V=U-c$ on $\mathcal L$, and therefore $V$ represents $\succsim$ on $\mathcal L$.

\medskip
\noindent\textbf{Step 7: Tail kernel and Stieltjes form.}
Define \(p:[0,1]\to\mathbb R\) by
\[
p(u):=\nu((u,1])\qquad(0\le u<1),
\qquad
p(1):=p(1^-),
\]
where the left limit exists because the map \(u\mapsto \nu((u,1])\) is nonincreasing on \([0,1)\).

By continuity from above of the finite measure \(\nu\),
\[
p(1)=\lim_{t\uparrow1}\nu((t,1])=\nu\!\left(\bigcap_{t<1}(t,1]\right)=\nu(\{1\}).
\]
For \(0\le u<1\),
\[
p(u)=\nu((u,1])=\nu((u,1))+\nu(\{1\})\ge \nu(\{1\})=p(1).
\]
Hence \(p\) is nonincreasing on \([0,1]\).

Let \(u\in[0,1)\) and let \(u_n\downarrow u\). Then \((u_n,1]\downarrow (u,1]\), so continuity from above gives
\[
p(u_n)=\nu((u_n,1])\longrightarrow \nu((u,1])=p(u).
\]
Thus \(p\) is right--continuous on \([0,1)\). Since \(p\) is monotone, it belongs to \(BV([0,1])\).

Let
\[
\alpha_\nu(u):=\nu([0,u])\qquad(u\in[0,1]),
\]
and define the auxiliary function
\[
\widehat p(u):=\alpha_\nu(1)-\alpha_\nu(u)\qquad(u\in[0,1]).
\]
Then for every \(u\in[0,1)\),
\[
\widehat p(u)=\nu([0,1])-\nu([0,u])=\nu((u,1])=p(u),
\]
while
\[
\widehat p(1)=\alpha_\nu(1)-\alpha_\nu(1)=0.
\]
Hence \(p-\widehat p\) vanishes on \([0,1)\).

We claim that
\[
\int_0^1 p\,dL=\int_0^1 \widehat p\,dL
\qquad(\forall L\in\mathcal L).
\]
Indeed, let \(s:=p-\widehat p\). Then \(s(u)=0\) for \(u\in[0,1)\), and
\[
s(1)=p(1)-\widehat p(1)=\nu(\{1\}).
\]
Since \(s\) differs from the zero function at only one point, \(s\in BV([0,1])\). Fix \(L\in\mathcal L\) and \(\varepsilon>0\). By continuity of \(L\) at \(1\), choose \(\delta\in(0,1)\) such that
\[
0\le L(1)-L(1-\delta)<\varepsilon.
\]
Let \(P:0=t_0<\cdots<t_m=1\) be any partition containing \(1-\delta\), and let \(\xi_k\in[t_{k-1},t_k]\).
Then the Riemann--Stieltjes sum
\[
S(P,\xi;s,L):=\sum_{k=1}^m s(\xi_k)\bigl(L(t_k)-L(t_{k-1})\bigr)
\]
has at most one nonzero term, and therefore
\[
|S(P,\xi;s,L)|
\le |s(1)|\bigl(L(1)-L(1-\delta)\bigr)
<
|s(1)|\,\varepsilon.
\]
Letting the mesh tend to \(0\) yields
\[
\int_0^1 s\,dL=0,
\]
and hence
\[
\int_0^1 p\,dL=\int_0^1 \widehat p\,dL.
\]

Now fix \(L\in\mathcal L\). Since \(L\) is continuous and of bounded variation, and \(\alpha_\nu\) is increasing,
the Riemann--Stieltjes integration--by--parts formula gives
\[
\int_0^1 L\,d\alpha_\nu+\int_0^1 \alpha_\nu\,dL
=
L(1)\alpha_\nu(1)-L(0)\alpha_\nu(0).
\]
Using \(L(0)=0\) and \(L(1)=1\), we obtain
\[
\int_{[0,1]}L(u)\,\nu(du)
=
\int_0^1 L\,d\alpha_\nu
=
\alpha_\nu(1)-\int_0^1 \alpha_\nu\,dL.
\]
Also,
\[
\int_0^1 dL=L(1)-L(0)=1,
\]
so
\[
\int_0^1 \widehat p(u)\,dL(u)
=
\int_0^1\bigl(\alpha_\nu(1)-\alpha_\nu(u)\bigr)\,dL(u)
=
\alpha_\nu(1)\int_0^1 dL-\int_0^1 \alpha_\nu\,dL
=
\alpha_\nu(1)-\int_0^1 \alpha_\nu\,dL.
\]
Therefore
\[
\int_{[0,1]}L(u)\,\nu(du)=\int_0^1 \widehat p(u)\,dL(u)=\int_0^1 p(u)\,dL(u)
\qquad(\forall L\in\mathcal L).
\]
Combining this with \eqref{eq:U-minus-c-measure}, we obtain
\[
U(L)-c=\int_{[0,1]}L(u)\,\nu(du)=\int_0^1 p(u)\,dL(u)
\qquad(\forall L\in\mathcal L).
\]
Thus the functional
\[
V(L):=\int_{[0,1]}L(u)\,\nu(du)=\int_0^1 p(u)\,dL(u)
\qquad(L\in\mathcal L)
\]
represents \(\succsim\) on \(\mathcal L\).

\medskip
\noindent\textbf{Step 8: Uniqueness up to Positive Affine Transformation.}
Let \(p_1:=p\), and let \(p_2:[0,1]\to\mathbb R\) be another nonincreasing bounded--variation kernel,
right--continuous on \([0,1)\), satisfying
\[
p_2(1)=p_2(1^-),
\]
and such that
\[
L_1\succsim L_2
\Longleftrightarrow
\int_0^1 p_2\,dL_1\ge \int_0^1 p_2\,dL_2
\qquad(\forall L_1,L_2\in\mathcal L).
\]

For \(i=1,2\), define
\[
V_i(L):=\int_0^1 p_i(u)\,dL(u)\qquad(L\in\mathcal L).
\]
For \(L_1,L_2\in\mathcal L\) and \(\lambda\in[0,1]\), linearity of the Stieltjes integral gives
\[
V_i\bigl((1-\lambda)L_1+\lambda L_2\bigr)
=
(1-\lambda)V_i(L_1)+\lambda V_i(L_2).
\]
Thus \(V_1\) and \(V_2\) are affine and represent the same nontrivial preorder on \(\mathcal L\). Hence, by
Proposition~\ref{prop:HM-affine-Lorenz}\textup{(iv)}, there exist constants \(a>0\) and \(b\in\mathbb R\) such that
\[
V_2(L)=a\,V_1(L)+b
\qquad(\forall L\in\mathcal L).
\]
Equivalently,
\[
\int_0^1 p_2\,dL-a\int_0^1 p_1\,dL-b\int_0^1 dL=0
\qquad(\forall L\in\mathcal L).
\]
Since \(\int_0^1 dL=L(1)-L(0)=1\), this becomes
\begin{equation}\label{eq:uniq-r-annihilates}
\int_0^1 \bigl(p_2-a p_1-b\bigr)\,dL=0
\qquad(\forall L\in\mathcal L).
\end{equation}

Define
\[
r:=p_2-a p_1-b.
\]
Then \(r\in BV([0,1])\), \(r\) is right--continuous on \([0,1)\), and
\[
r(1)=r(1^-),
\]
because both \(p_1\) and \(p_2\) satisfy the endpoint condition.

Fix \(u\in[0,1)\) and define
\[
K_u(t):=\max\Bigl\{0,\frac{t-u}{1-u}\Bigr\}\qquad(t\in[0,1]).
\]
Then \(K_u\in\mathcal L\), \(K_u\) is constant on \([0,u]\), and affine on \([u,1]\) with slope \(1/(1-u)\).
Let \(P:0=t_0<\cdots<t_m=1\) be a partition such that \(u=t_j\) for some \(j\), and let \(\xi_i\in[t_{i-1},t_i]\).
The Riemann--Stieltjes sum for \(\int_0^1 r\,dK_u\) is
\[
S(P,\xi;r,K_u):=\sum_{i=1}^m r(\xi_i)\bigl(K_u(t_i)-K_u(t_{i-1})\bigr).
\]
Since \(K_u(t_i)-K_u(t_{i-1})=0\) for \(i\le j\), while
\[
K_u(t_i)-K_u(t_{i-1})=\frac{1}{1-u}(t_i-t_{i-1})\qquad(i\ge j+1),
\]
we have
\[
S(P,\xi;r,K_u)=\frac{1}{1-u}\sum_{i=j+1}^m r(\xi_i)\,(t_i-t_{i-1}).
\]
Because \(r\in BV([0,1])\), it has at most countably many discontinuities and is Riemann integrable on \([u,1]\).
Therefore, along partitions with mesh tending to \(0\) and containing \(u\) as a node,
\[
\int_0^1 r\,dK_u=\frac{1}{1-u}\int_u^1 r(t)\,dt.
\]
Applying \eqref{eq:uniq-r-annihilates} with \(L=K_u\) gives
\begin{equation}\label{eq:uniq-tail-int}
\int_u^1 r(t)\,dt=0\qquad(\forall u\in[0,1)).
\end{equation}

Now fix \(x\in[0,1)\) and choose any \(h\in(0,1-x)\). Then
\[
\int_x^{x+h}r(t)\,dt
=
\int_x^1 r(t)\,dt-\int_{x+h}^1 r(t)\,dt
=
0.
\]
Therefore
\begin{align*}
|r(x)|
&=
\left|r(x)-\frac1h\int_x^{x+h}r(t)\,dt\right|\\
&=
\left|\frac1h\int_x^{x+h}\bigl(r(x)-r(t)\bigr)\,dt\right|\\
&\le
\sup_{t\in[x,x+h]}|r(x)-r(t)|.
\end{align*}
Since \(r\) is right--continuous at \(x\), the right--hand side tends to \(0\) as \(h\downarrow0\). Hence
\[
r(x)=0\qquad(\forall x\in[0,1)).
\]
Since \(r(1)=r(1^-)\), we also obtain
\[
r(1)=\lim_{t\uparrow1}r(t)=0.
\]
Therefore
\[
p_2(u)=a\,p_1(u)+b\qquad(\forall u\in[0,1]).
\]
Since \(p_1=p\) and \(p_2=\widetilde p\), this concludes our proof of Primal Theorem.

\end{proof}

\begin{theorem}[Dual Lorenz--Stieltjes representation via dyadic discretization]
\label{thm:dual-A5}
Let $\succsim$ be a complete preorder on $\mathcal L^\uparrow$ satisfying:
\begin{enumerate}
\item[(D1)] \textbf{Dual independence:} Axiom~\ref{ax:HM-dual-independence}.
\item[(D2)] \textbf{Closed graph continuity:}  
The graph
\[
\{(L_1,L_2)\in\mathcal L^\uparrow\times\mathcal L^\uparrow:\ L_1\succsim L_2\}
\]
is closed in \( \mathcal L^\uparrow  \times \mathcal L^\uparrow  \), where $\mathcal L^\uparrow$  carries the topology induced by
\((C([0,1]),\|\cdot\|_\infty)\).
 
\item[(D3)] \textbf{Weak dominance:} $L_1\ge L_2$ pointwise $\Rightarrow L_1\succsim L_2$.
\item[(D4)] \textbf{Nontriviality:} there exist $L^+,L^-\in\mathcal L^\uparrow$ with $L^+\succ L^-$.
\end{enumerate}

Define the involution operator $\mathcal S:\mathcal L^\uparrow\to\mathcal L^\uparrow$ by
\[
(\mathcal SL)(u):=1-L^{-1}(1-u),\qquad u\in[0,1],
\]
and the transported preorder $\succsim^{\mathcal S}$ on $\mathcal L^\uparrow$ by
\[
M_1\succsim^{\mathcal S}M_2 \iff \mathcal SM_1\succsim \mathcal SM_2.
\]

Assume, in addition:
\begin{enumerate}
\item[(D5)] \textbf{Local Archimedean at equality after conjugacy:}
the transported preorder \(\succsim^{\mathcal S}\) satisfies
Axiom~\ref{ax:archimedean} on \(\mathcal L^\uparrow\) for some
anchor \(B^{\mathcal S}\in\mathcal L^\uparrow\) such that
\[
E\succ^{\mathcal S} B^{\mathcal S}.
\]
\end{enumerate}
Then there exist a finite nonnegative Borel measure \(\nu\) on \([0,1]\) and a constant \(c\in\mathbb R\) such that, defining
\[
q(u)=\nu([0,u])\qquad(0\le u<1),
\]
and fixing the endpoint value by
\[
q(1)=q(1^-),\qquad q(1^-):=\lim_{t\uparrow1}q(t),
\]
the function \(q\) is nondecreasing, of bounded variation, right--continuous on \([0,1)\), and the functional
\[
V(L):=c+\int_0^1 q(u)\,d\bigl(L^{-1}(u)\bigr)\qquad(L\in\mathcal L^\uparrow)
\]
represents \(\succsim\):
\[
L_1\succsim L_2\ \Longleftrightarrow\ V(L_1)\ge V(L_2).
\]

Equivalently, there exists
\[
c':=c+\nu([0,1])
\]
such that
\[
V(L)=c'-\int_{[0,1]}L^{-1}(u)\,\nu(du)\qquad(L\in\mathcal L^\uparrow).
\]

Moreover, if \(\widetilde c\in\mathbb R\) and \(\widetilde q:[0,1]\to\mathbb R\) is another nondecreasing
bounded--variation kernel, right--continuous on \([0,1)\), satisfying
\[
\widetilde q(1)=\widetilde q(1^-),\qquad
\widetilde q(1^-):=\lim_{t\uparrow1}\widetilde q(t),
\]
and such that
\[
\widetilde V(L):=\widetilde c+\int_0^1 \widetilde q(u)\,d\bigl(L^{-1}(u)\bigr)
\qquad(L\in\mathcal L^\uparrow)
\]
also represents \(\succsim\) on \(\mathcal L^\uparrow\), then there exist \(a>0\) and \(b\in\mathbb R\) such that
\[
\widetilde q(u)=a\,q(u)+b\qquad(u\in[0,1]).
\]
\end{theorem}

\begin{proof}
\medskip
\noindent\textbf{Step 1: The Involution Operator $\mathcal S$.}
For $L\in\mathcal L^\uparrow$ define
\begin{equation}\label{eq:S-def-dual}
(\mathcal SL)(u):=1-L^{-1}(1-u)\qquad(u\in[0,1]).
\end{equation}

\emph{(1.1) $\mathcal S(\mathcal L^\uparrow)\subset\mathcal L^\uparrow$.}
Fix $L\in\mathcal L^\uparrow$. Then $L^{-1}\in C([0,1])$ is strictly increasing with $L^{-1}(0)=0$, $L^{-1}(1)=1$.
For $a,b\in[0,1]$ and $\theta\in[0,1]$, set $x:=L^{-1}(a)$ and $y:=L^{-1}(b)$. Convexity of $L$ gives
\[
L\bigl(\theta x+(1-\theta)y\bigr)\le \theta L(x)+(1-\theta)L(y)=\theta a+(1-\theta)b.
\]
Since $L^{-1}$ is increasing,
\[
\theta L^{-1}(a)+(1-\theta)L^{-1}(b)=\theta x+(1-\theta)y \le L^{-1}\bigl(\theta a+(1-\theta)b\bigr),
\]
so $L^{-1}$ is concave. As $u\mapsto 1-u$ is affine, $u\mapsto L^{-1}(1-u)$ is concave, hence $u\mapsto -L^{-1}(1-u)$ is convex,
and therefore $\mathcal SL$ is convex. Also $L^{-1}(1-u)$ is strictly decreasing, hence $\mathcal SL$ is strictly increasing, and
\[
(\mathcal SL)(0)=1-L^{-1}(1)=0,\qquad (\mathcal SL)(1)=1-L^{-1}(0)=1.
\]
Thus $\mathcal SL\in\mathcal L^\uparrow$.

\emph{(1.2) Involution.}
Let $M:=\mathcal SL$. For $v\in[0,1]$, $v=M(u)$ iff $1-v=L^{-1}(1-u)$ iff $L(1-v)=1-u$ iff
\begin{equation}\label{eq:Sinv-explicit}
M^{-1}(v)=1-L(1-v).
\end{equation}
Hence, for $u\in[0,1]$,
\[
(\mathcal S(\mathcal SL))(u)=1-(\mathcal SL)^{-1}(1-u)
=1-\bigl(1-L(u)\bigr)=L(u),
\]
so $\mathcal S^2=\mathrm{id}$ on $\mathcal L^\uparrow$.

\emph{(1.3) Order Isomorphism.}
If $L_1,L_2\in\mathcal L^\uparrow$ satisfy $L_1\ge L_2$ pointwise, then for each $y\in[0,1]$ letting $x:=L_1^{-1}(y)$ gives
$L_2(x)\le L_1(x)=y$, hence (since $L_2$ is increasing) $L_2^{-1}(y)\ge x=L_1^{-1}(y)$. Thus $L_1^{-1}\le L_2^{-1}$ pointwise and
$\mathcal SL_1\ge \mathcal SL_2$ pointwise. If $\mathcal SL_1\ge \mathcal SL_2$ pointwise, then applying $\mathcal S$ again yields
$L_1=\mathcal S(\mathcal SL_1)\ge \mathcal S(\mathcal SL_2)=L_2$. Hence $\mathcal S$ is an order isomorphism
for the pointwise order on $\mathcal L^\uparrow$.

\emph{(1.4) Sup-norm identity.}
For $L_1,L_2\in\mathcal L^\uparrow$,
\begin{equation}\label{eq:S-isometry-dual}
\|\mathcal SL_1-\mathcal SL_2\|_\infty
=\sup_{u\in[0,1]}\bigl|L_1^{-1}(1-u)-L_2^{-1}(1-u)\bigr|
=\|L_1^{-1}-L_2^{-1}\|_\infty.
\end{equation}

\emph{(1.5) Continuity of inversion (hence of $\mathcal S$).}
Let $f_n,f\in C([0,1])$ be strictly increasing with $f_n(0)=f(0)=0$, $f_n(1)=f(1)=1$, and $\|f_n-f\|_\infty\to0$.
Fix $\varepsilon>0$ and set
\[
K_\varepsilon:=\{(x,y)\in[0,1]^2:\ |x-y|\ge\varepsilon\}.
\]
The map $(x,y)\mapsto |f(x)-f(y)|$ is continuous on the compact set $K_\varepsilon$, hence attains
\[
m_\varepsilon:=\min_{(x,y)\in K_\varepsilon}|f(x)-f(y)|.
\]
Strict monotonicity of $f$ implies $m_\varepsilon>0$. Let $\eta:=m_\varepsilon/3$ and take $n$ large so that $\|f_n-f\|_\infty\le\eta$.
Fix $u\in[0,1]$, set $x:=f^{-1}(u)$ and $x_n:=f_n^{-1}(u)$. If $|x_n-x|\ge\varepsilon$, then $(x_n,x)\in K_\varepsilon$, so
\[
|f(x_n)-f(x)|\ge m_\varepsilon.
\]
But $f(x)=u=f_n(x_n)$, hence
\[
|f(x_n)-f(x)|=|f(x_n)-u|
=|f(x_n)-f_n(x_n)|\le \|f-f_n\|_\infty\le\eta<m_\varepsilon,
\]
a contradiction. Therefore $|x_n-x|<\varepsilon$ for all $u$, i.e. $\|f_n^{-1}-f^{-1}\|_\infty\le\varepsilon$ for all large $n$.
Thus $\|f_n^{-1}-f^{-1}\|_\infty\to0$. Applying this to $f_n=L_n$ shows $L\mapsto L^{-1}$ is $\|\cdot\|_\infty$-continuous on $\mathcal L^\uparrow$,
hence $\mathcal S$ is $\|\cdot\|_\infty$-continuous on $\mathcal L^\uparrow$ by \eqref{eq:S-def-dual}. Since $L\mapsto L^{-1}$ is $\|\cdot\|_\infty$--continuous on $\mathcal L^\uparrow$ and $\mathcal S$ is built from inversion and affine maps,
$\mathcal S$ is $\|\cdot\|_\infty$--continuous. Because $\mathcal S^{-1}=\mathcal S$, it is a homeomorphism of $\mathcal L^\uparrow$.

\medskip
\noindent\textbf{Step 2: Mixture Conjugacy: Dual Mixtures $\leftrightarrow$ Ordinary Mixtures.}
For $L_1,L_3\in\mathcal L^\uparrow$ and $\lambda\in(0,1)$ define the dual mixture
\[
L_1\oplus_\lambda L_3:=\bigl(\lambda L_1^{-1}+(1-\lambda)L_3^{-1}\bigr)^{-1}.
\]
Since $L_1^{-1},L_3^{-1}$ are strictly increasing concave maps with endpoints $(0,0)$ and $(1,1)$,
their convex combination is strictly increasing concave, hence its inverse belongs to $\mathcal L^\uparrow$.

For $u\in[0,1]$ we compute
\begin{align}
\mathcal S(L_1\oplus_\lambda L_3)(u)
&=1-(L_1\oplus_\lambda L_3)^{-1}(1-u)\nonumber\\
&=1-\bigl(\lambda L_1^{-1}(1-u)+(1-\lambda)L_3^{-1}(1-u)\bigr)\nonumber\\
&=\lambda\bigl(1-L_1^{-1}(1-u)\bigr)+(1-\lambda)\bigl(1-L_3^{-1}(1-u)\bigr)\nonumber\\
&=\lambda\,\mathcal SL_1(u)+(1-\lambda)\,\mathcal SL_3(u).\label{eq:S-conjugacy-dual}
\end{align}
Applying $\mathcal S$ to \eqref{eq:S-conjugacy-dual} and using $\mathcal S^2=\mathrm{id}$ yields the \emph{reverse} conjugacy:
for all $M_1,M_3\in\mathcal L^\uparrow$ and $\lambda\in(0,1)$,
\begin{equation}\label{eq:reverse-conjugacy}
\mathcal S\bigl(\lambda M_1+(1-\lambda)M_3\bigr)
=(\mathcal SM_1)\oplus_\lambda(\mathcal SM_3).
\end{equation}

\medskip
\noindent\textbf{Step 3: Transported Preorder and Affine Representation.}
Define the transported preorder $\succsim^{\mathcal S}$ on $\mathcal L^\uparrow$ by
\[
M_1\succsim^{\mathcal S}M_2 \quad\Longleftrightarrow\quad \mathcal SM_1\succsim \mathcal SM_2.
\]
Since $\mathcal S$ is bijective (Step~1.2), $\succsim^{\mathcal S}$ is a complete preorder.
Nontriviality transfers from (D4): if $L^+\succ L^-$ then $\mathcal SL^+\succ^{\mathcal S}\mathcal SL^-$.

\emph{(3.1) Mixture independence.}
Fix $M_1,M_2,M_3\in\mathcal L^\uparrow$ and $\lambda\in(0,1)$, and set $L_i:=\mathcal SM_i$.
Then
\begin{align*}
M_1\succsim^{\mathcal S}M_2
&\Longleftrightarrow L_1\succsim L_2\\
&\overset{\textup{(D1)}}{\Longleftrightarrow} L_1\oplus_\lambda L_3\succsim L_2\oplus_\lambda L_3\\
&\Longleftrightarrow \mathcal S(L_1\oplus_\lambda L_3)\succsim^{\mathcal S}\mathcal S(L_2\oplus_\lambda L_3)\\
&\overset{\eqref{eq:S-conjugacy-dual}}{\Longleftrightarrow}
\bigl(\lambda M_1+(1-\lambda)M_3\bigr)\succsim^{\mathcal S}\bigl(\lambda M_2+(1-\lambda)M_3\bigr).
\end{align*}
i.e. $\succsim^{\mathcal S}$ satisfies ordinary mixture independence on $\mathcal L^\uparrow$.

\emph{(3.2) Weak dominance.}
If $M_1\ge M_2$ pointwise, then $\mathcal SM_1\ge \mathcal SM_2$ (Step~1.3),
hence $\mathcal SM_1\succsim \mathcal SM_2$ by (D3), so $M_1\succsim^{\mathcal S}M_2$.

\emph{(3.3) Closed graph.}
If $(M_n,N_n)\to(M,N)$ in $\|\cdot\|_\infty$ and $M_n\succsim^{\mathcal S}N_n$ for all $n$,
then $\mathcal SM_n\succsim \mathcal SN_n$ for all $n$.
Since $\mathcal S$ is a homeomorphism , $(\mathcal SM_n,\mathcal SN_n)\to(\mathcal SM,\mathcal SN)$.
Closed graph of $\succsim$ (D2) gives $\mathcal SM\succsim \mathcal SN$, hence $M\succsim^{\mathcal S}N$.

\emph{(3.4) Affine continuous representation.}
By Proposition~\ref{prop:HM-affine-Lorenz} applied to $(\mathcal L^\uparrow,\succsim^{\mathcal S})$,
there exists a nonconstant affine $\|\cdot\|_\infty$--continuous $W:\mathcal L^\uparrow\to\mathbb R$
such that
\[
M_1\succsim^{\mathcal S}M_2\Longleftrightarrow W(M_1)\ge W(M_2),
\qquad
W((1-\lambda)M_1+\lambda M_2)=(1-\lambda)W(M_1)+\lambda W(M_2).
\]
\medskip
\noindent\textbf{Step 4: Dyadic Discrete Representation for $W$ on $\mathcal L^\uparrow$.}
For each $m\in\mathbb N$ set
\[
n:=2^m,\qquad t_k:=\frac{k}{n}\qquad(k=0,1,\dots,n).
\]
Let
\[
X_m:=\{g\in C([0,1]): g \text{ is affine on each }[t_{k-1},t_k]\},
\qquad
\mathcal L_m^\uparrow:=\mathcal L^\uparrow\cap X_m.
\]
Let $P_m:C([0,1])\to X_m$ be polygonal interpolation: $(P_mf)(t_k)=f(t_k)$ and $P_mf$ affine on each interval.
Then $P_mf\to f$ uniformly for all $f\in C([0,1])$, and $P_mM\in\mathcal L_m^\uparrow$ for every $M\in\mathcal L^\uparrow$.

Apply Theorem~\ref{thm:dyadic-discrete-Lorenz-Stieltjes} with
\[
\mathcal K=\mathcal L^\uparrow,
\]
the preorder $\succsim^{\mathcal S}$, and the affine continuous representation $W$.
Then there exists $m_0$ such that for every $m\ge m_0$ there exists a finite nonnegative Borel measure
\(
\nu_m
\)
supported on $\{t_1,\dots,t_{n-1}\}$ such that
\begin{equation}\label{eq:disc-W-difference}
W(M)-W(N)=\int_{[0,1]}\bigl(M(u)-N(u)\bigr)\,\nu_m(du)
\qquad(\forall M,N\in\mathcal L_m^\uparrow).
\end{equation}

Define
\[
p_m(u):=\nu_m((u,1])\qquad(0\le u<1),\qquad p_m(1):=0.
\]
Then $p_m$ is a nonincreasing right--continuous step function on the dyadic grid.
Writing
\[
p_m|_{[t_{k-1},t_k)}\equiv p(m,k)\qquad(k=1,\dots,n),
\]
we have
\[
p(m,n)=0.
\]
Because $\nu_m$ is supported on $\{t_1,\dots,t_{n-1}\}$,
\[
p(m,k)=\sum_{j=k}^{n-1}\nu_m(\{t_j\}),
\qquad
\nu_m(\{t_k\})=p(m,k)-p(m,k+1)
\qquad(k=1,\dots,n-1).
\]
In particular,
\[
\nu_m([0,1])=p(m,1)\ge0,
\qquad
0\le p_m(u)\le p(m,1)\qquad(u\in[0,1]).
\]

If $M\in\mathcal L_m^\uparrow$, then discrete integration by parts gives
\begin{align}
\int_0^1 p_m\,dM
&=\sum_{k=1}^{n}p(m,k)\bigl(M(t_k)-M(t_{k-1})\bigr)\nonumber\\
&=\sum_{k=1}^{n-1}M(t_k)\bigl(p(m,k)-p(m,k+1)\bigr)+M(1)p(m,n)-M(0)p(m,1)\nonumber\\
&=\sum_{k=1}^{n-1}M(t_k)\,\nu_m(\{t_k\})\nonumber\\
&=\int_{[0,1]}M(u)\,\nu_m(du),\label{eq:disc-W-parts-dual}
\end{align}
using $M(0)=0$, $M(1)=1$, and $p(m,n)=0$.

Let
\[
E(u):=u.
\]
Since $E\in\mathcal L_m^\uparrow$, define
\[
c_m:=W(E)-\int_{[0,1]}E(u)\,\nu_m(du).
\]
Taking $N=E$ in \eqref{eq:disc-W-difference} yields
\[
W(M)-W(E)=\int_{[0,1]}\bigl(M(u)-E(u)\bigr)\,\nu_m(du),
\]
hence
\[
W(M)-c_m=\int_{[0,1]}M(u)\,\nu_m(du).
\]
Combining this with \eqref{eq:disc-W-parts-dual}, we obtain
\begin{equation}\label{eq:disc-W}
W(M)-c_m
=
\int_0^1 p_m(u)\,dM(u)
=
\int_{[0,1]}M(u)\,\nu_m(du)
\qquad(\forall M\in\mathcal L_m^\uparrow).
\end{equation}

\medskip
\noindent\textbf{Step 5: Uniform Mass bound from Archimedean continuity for \(\succsim^{\mathcal S}\).}
Assume \textup{(D5)}. Let \(B^{\mathcal S}\in\mathcal L^\uparrow\) and \(C,\delta>0\) be given by
Axiom~\ref{ax:archimedean} for the transported preorder \(\succsim^{\mathcal S}\), and set
\[
\Delta_{B^{\mathcal S}}:=W(E)-W(B^{\mathcal S})>0.
\]

\emph{(5.1) Global Bound.}
We claim that
\[
W(E)-W(M)\le C\Delta_{B^{\mathcal S}}\,\|E-M\|_\infty
\qquad(M\in\mathcal L^\uparrow).
\]
Indeed, if \(0<\|E-M\|_\infty<\delta\), then by Axiom~\ref{ax:archimedean} there exists
\(s_M\in[0,1]\) such that
\[
s_M\le C\|E-M\|_\infty
\qquad\text{and}\qquad
M\succsim^{\mathcal S} (1-s_M)E+s_M B^{\mathcal S}.
\]
Applying the affine representation \(W\) gives
\[
W(M)\ge (1-s_M)W(E)+s_MW(B^{\mathcal S})
= W(E)-s_M\Delta_{B^{\mathcal S}},
\]
hence
\[
W(E)-W(M)\le s_M\Delta_{B^{\mathcal S}}
\le C\Delta_{B^{\mathcal S}}\,\|E-M\|_\infty.
\]
Now let \(M\in\mathcal L^\uparrow\) be arbitrary. If \(M=E\), there is nothing to prove. Otherwise define
\[
\theta:=\min\Bigl\{1,\frac{\delta}{2\|E-M\|_\infty}\Bigr\}\in(0,1],
\qquad
M^\theta:=(1-\theta)E+\theta M.
\]
Then
\[
\|E-M^\theta\|_\infty=\theta\|E-M\|_\infty<\delta,
\]
so the previous estimate applied to \(M^\theta\) yields
\[
W(E)-W(M^\theta)\le C\Delta_{B^{\mathcal S}}\,\|E-M^\theta\|_\infty
= C\Delta_{B^{\mathcal S}}\,\theta\|E-M\|_\infty.
\]
Since
\[
W(M^\theta)=(1-\theta)W(E)+\theta W(M),
\]
we have
\[
W(E)-W(M^\theta)=\theta\bigl(W(E)-W(M)\bigr).
\]
Cancelling \(\theta>0\) yields
\[
W(E)-W(M)\le C\Delta_{B^{\mathcal S}}\,\|E-M\|_\infty
\qquad(M\in\mathcal L^\uparrow).
\tag{\(\star\)}
\]

\emph{(5.2) Strict corner curves (stay inside \(\mathcal L^\uparrow\)).}
Fix once and for all some \(\varepsilon\in(0,1)\). For \(j\in\{1,\dots,n-1\}\) define
\[
K_{m,j}(u):=\max\Bigl\{0,\frac{u-t_j}{1-t_j}\Bigr\}\in X_m,
\]
and define the strictified corners
\[
K_{m,1}^\varepsilon:=(1-\varepsilon)K_{m,1}+\varepsilon E,
\qquad
K_{m,n-1}^\varepsilon:=(1-\varepsilon)K_{m,n-1}+\varepsilon E.
\]
Then \(K_{m,1}^\varepsilon,K_{m,n-1}^\varepsilon\in\mathcal L_m^\uparrow\).

\emph{(5.3) A clean mass identity.}
Let \(S_m:=\sum_{k=1}^{n}p(m,k)\), so \(\int_0^1 p_m\,dE=\frac1n S_m\).
Since \(p(m,n)=0\), one checks from dyadic increments that
\[
\int_0^1 p_m\,dK_{m,n-1}^\varepsilon=\frac{\varepsilon}{n}S_m,
\qquad
\int_0^1 p_m\,dK_{m,1}^\varepsilon
=\frac{\varepsilon}{n}p(m,1)+\Bigl(\frac{1-\varepsilon}{n-1}+\frac{\varepsilon}{n}\Bigr)\sum_{k=2}^{n}p(m,k).
\]
Using \eqref{eq:disc-W} for \(E,K_{m,1}^\varepsilon,K_{m,n-1}^\varepsilon\) and eliminating \(c_m\) gives
\begin{equation}\label{eq:mass-identity}
\nu_m([0,1])=p(m,1)
=\frac{\bigl(W(E)-W(K_{m,n-1}^\varepsilon)\bigr)+(n-1)\bigl(W(E)-W(K_{m,1}^\varepsilon)\bigr)}{1-\varepsilon}.
\end{equation}

\emph{(5.4) Uniform bound.}
Since
\[
\|E-K_{m,1}^\varepsilon\|_\infty=(1-\varepsilon)\|E-K_{m,1}\|_\infty=\frac{1-\varepsilon}{n},
\]
and
\[
\|E-K_{m,n-1}^\varepsilon\|_\infty
=(1-\varepsilon)\|E-K_{m,n-1}\|_\infty
=(1-\varepsilon)\frac{n-1}{n}\le 1-\varepsilon,
\]
the global estimate \((\star)\) yields
\[
W(E)-W(K_{m,1}^\varepsilon)\le C\Delta_{B^{\mathcal S}}\,\frac{1-\varepsilon}{n},
\]
and
\[
W(E)-W(K_{m,n-1}^\varepsilon)\le C\Delta_{B^{\mathcal S}}(1-\varepsilon)\frac{n-1}{n}.
\]
Substituting these bounds into \eqref{eq:mass-identity} gives
\[
\nu_m([0,1])
\le
\frac{
C\Delta_{B^{\mathcal S}}(1-\varepsilon)\frac{n-1}{n}
+
(n-1)C\Delta_{B^{\mathcal S}}(1-\varepsilon)\frac{1}{n}
}{1-\varepsilon}
\le 2C\Delta_{B^{\mathcal S}}.
\]
Thus
\begin{equation}\label{eq:uniform-mass-dual}
0\le \nu_m([0,1])\le 2C\Delta_{B^{\mathcal S}}
\qquad(\forall m\ge m_0).
\end{equation}
Since
\[
c_m=W(E)-\int_{[0,1]}E(u)\,\nu_m(du),
\]
the same bound implies
\[
W(E)-2C\Delta_{B^{\mathcal S}}\le c_m\le W(E),
\]
so \((c_m)\) is bounded.

\medskip
\noindent\textbf{Step 6: Limit Measure for $W$ on $\mathcal L^\uparrow$.}
For $m\ge m_0$ define $T_m:C([0,1])\to\mathbb R$ by $T_m(f):=\int_{[0,1]} f\,d\nu_m$.
Then $T_m$ is positive and
\[
\|T_m\|=T_m(\mathbf 1)=\nu_m([0,1])\le M_0
\]
by \eqref{eq:uniform-mass-dual}.

Let $\mathscr D\subset C([0,1])$ be the countable $\|\cdot\|_\infty$--dense dyadic--rational polygonal class
(from the primal theorem’s density step), enumerate $\mathscr D=\{d_1,d_2,\dots\}$.
By diagonal selection choose a strictly increasing subsequence $(m_j)$ such that
$T_{m_j}(d_r)$ converges for every $r$.
Since $(c_m)$ is bounded, pass to a further subsequence (not relabeled) such that $c_{m_j}\to c\in\mathbb R$.

Define $T(f):=\lim_{j\to\infty}T_{m_j}(f)$ for $f\in C([0,1])$.
The uniform bound $\|T_{m_j}\|\le M_0$ and density of $\mathscr D$ imply this limit exists for all $f$,
and $T$ is a positive continuous linear functional with $\|T\|\le M_0$.
By Riesz--Markov, there exists a finite nonnegative Borel measure $\nu$ on $[0,1]$ such that
\[
T(f)=\int_{[0,1]} f\,d\nu\qquad(\forall f\in C([0,1])).
\]

Finally fix $M\in\mathcal L^\uparrow$ and set $M^{(j)}:=P_{m_j}M\in\mathcal L_{m_j}^\uparrow$.
Then $\|M^{(j)}-M\|_\infty\to0$ and continuity of $W$ gives $W(M^{(j)})\to W(M)$.
Using \eqref{eq:disc-W} at $m=m_j$,
\[
W(M^{(j)})-c_{m_j}=T_{m_j}(M^{(j)}).
\]
Moreover,
\[
T_{m_j}(M^{(j)})-T(M)=T_{m_j}(M^{(j)}-M)+\bigl(T_{m_j}(M)-T(M)\bigr),
\]
and $|T_{m_j}(M^{(j)}-M)|\le \|T_{m_j}\|\,\|M^{(j)}-M\|_\infty\le M_0\|M^{(j)}-M\|_\infty\to0$,
while $T_{m_j}(M)\to T(M)$ by definition of $T$.
Hence $T_{m_j}(M^{(j)})\to T(M)$ and letting $j\to\infty$ yields
\begin{equation}\label{eq:W-limit-measure}
W(M)-c=\int_{[0,1]} M\,d\nu\qquad(\forall M\in\mathcal L^\uparrow).
\end{equation}

\medskip
\noindent\textbf{Step 7: Pullback, Cumulative kernel, and Measure form.}
Define
\[
U(L):=W(\mathcal SL)\qquad(L\in\mathcal L^\uparrow).
\]
Then \(U\) represents \(\succsim\) on \(\mathcal L^\uparrow\).

Let \(\mu_0\) denote the finite nonnegative Borel measure obtained in Step~6, so that
\[
W(M)=c+\int_{[0,1]}M(u)\,\mu_0(du)\qquad(M\in\mathcal L^\uparrow).
\]
Hence, for \(L\in\mathcal L^\uparrow\),
\begin{align*}
U(L)
&=W(\mathcal SL)\\
&=c+\int_{[0,1]}\bigl(1-L^{-1}(1-u)\bigr)\,\mu_0(du)\\
&=c+\mu_0([0,1])-\int_{[0,1]}L^{-1}(1-u)\,\mu_0(du).
\end{align*}

Let \(\tau:[0,1]\to[0,1]\) be given by \(\tau(u):=1-u\), and define the pushforward measure
\[
\nu:=\mu_0\circ\tau^{-1}.
\]
Then
\[
\int_{[0,1]}L^{-1}(1-u)\,\mu_0(du)=\int_{[0,1]}L^{-1}(v)\,\nu(dv),
\]
so
\begin{equation}\label{eq:U-measure-inv}
U(L)=c+\nu([0,1])-\int_{[0,1]}L^{-1}(v)\,\nu(dv)
\qquad(L\in\mathcal L^\uparrow).
\end{equation}

Define \(q:[0,1]\to\mathbb R\) by
\[
q(u):=\nu([0,u])\qquad(0\le u<1),
\qquad
q(1):=q(1^-),
\]
where the left limit exists because the map \(u\mapsto \nu([0,u])\) is nondecreasing on \([0,1)\).

By continuity from below of the finite measure \(\nu\),
\[
q(1)=\lim_{t\uparrow1}\nu([0,t])=\nu\!\left(\bigcup_{t<1}[0,t]\right)=\nu([0,1)).
\]
For \(0\le u<1\),
\[
q(u)=\nu([0,u])\le \nu([0,1))=q(1).
\]
Hence \(q\) is nondecreasing on \([0,1]\).

Let \(u\in[0,1)\) and let \(u_n\downarrow u\). Then \([0,u_n]\downarrow [0,u]\), so continuity from above gives
\[
q(u_n)=\nu([0,u_n])\longrightarrow \nu([0,u])=q(u).
\]
Thus \(q\) is right--continuous on \([0,1)\). Since \(q\) is monotone, it belongs to \(BV([0,1])\).

Define the auxiliary function
\[
\widehat q(u):=\nu([0,u])\qquad(u\in[0,1]).
\]
Then \(\widehat q(u)=q(u)\) for \(u\in[0,1)\), while
\[
\widehat q(1)=\nu([0,1]).
\]
Hence \(q-\widehat q\) vanishes on \([0,1)\).

We claim that
\[
\int_0^1 q(u)\,d\bigl(L^{-1}(u)\bigr)
=
\int_0^1 \widehat q(u)\,d\bigl(L^{-1}(u)\bigr)
\qquad(\forall L\in\mathcal L^\uparrow).
\]
Indeed, let \(s:=q-\widehat q\). Then \(s(u)=0\) for \(u\in[0,1)\), and
\[
s(1)=q(1)-\widehat q(1)=\nu([0,1))-\nu([0,1])=-\nu(\{1\}).
\]
Since \(s\) differs from the zero function at only one point, \(s\in BV([0,1])\). Fix \(L\in\mathcal L^\uparrow\) and \(\varepsilon>0\).
By continuity of \(L^{-1}\) at \(1\), choose \(\delta\in(0,1)\) such that
\[
0\le L^{-1}(1)-L^{-1}(1-\delta)<\varepsilon.
\]
Let \(P:0=t_0<\cdots<t_m=1\) be any partition containing \(1-\delta\), and let \(\xi_k\in[t_{k-1},t_k]\).
Then the Riemann--Stieltjes sum
\[
S(P,\xi;s,L^{-1})
:=
\sum_{k=1}^m s(\xi_k)\bigl(L^{-1}(t_k)-L^{-1}(t_{k-1})\bigr)
\]
has at most one nonzero term, and therefore
\[
|S(P,\xi;s,L^{-1})|
\le |s(1)|\bigl(L^{-1}(1)-L^{-1}(1-\delta)\bigr)
<
|s(1)|\,\varepsilon.
\]
Letting the mesh tend to \(0\) yields
\[
\int_0^1 s\,d(L^{-1})=0,
\]
which proves the claim.

Since \(\nu\) is the Lebesgue--Stieltjes measure induced by \(\widehat q\), one has
\[
\int_0^1 L^{-1}(u)\,d\widehat q(u)=\int_{[0,1]}L^{-1}(u)\,\nu(du).
\]
Now fix \(L\in\mathcal L^\uparrow\). Since \(L^{-1}\) is continuous, increasing, and of bounded variation,
the integration--by--parts formula gives
\[
\int_0^1 \widehat q(u)\,d\bigl(L^{-1}(u)\bigr)+\int_0^1 L^{-1}(u)\,d\widehat q(u)
=
\widehat q(1)L^{-1}(1)-\widehat q(0)L^{-1}(0).
\]
Because \(\widehat q(1)=\nu([0,1])\), \(\widehat q(0)=\nu([0,0])\), and
\[
L^{-1}(0)=0,\qquad L^{-1}(1)=1,
\]
we obtain
\[
\int_0^1 \widehat q(u)\,d\bigl(L^{-1}(u)\bigr)
=
\nu([0,1])-\int_{[0,1]}L^{-1}(u)\,\nu(du).
\]
Therefore
\[
\int_0^1 q(u)\,d\bigl(L^{-1}(u)\bigr)
=
\nu([0,1])-\int_{[0,1]}L^{-1}(u)\,\nu(du).
\]
Combining this with \eqref{eq:U-measure-inv}, we obtain
\[
U(L)=c+\int_0^1 q(u)\,d\bigl(L^{-1}(u)\bigr)
\qquad(L\in\mathcal L^\uparrow).
\]
Equivalently, setting
\[
c':=c+\nu([0,1]),
\]
we have
\[
U(L)=c'-\int_{[0,1]}L^{-1}(u)\,\nu(du)
\qquad(L\in\mathcal L^\uparrow).
\]
This proves the existence part.

\medskip
\noindent\textbf{Step 8: Uniqueness up to Positive Affine Transformations.}
Let \(\widetilde c\in\mathbb R\) and let \(\widetilde q:[0,1]\to\mathbb R\) be another nondecreasing bounded--variation kernel,
right--continuous on \([0,1)\), satisfying
\(
\widetilde q(1)=\widetilde q(1^-)
\), and such that
\[
\widetilde V(L):=\widetilde c+\int_0^1 \widetilde q(u)\,d\bigl(L^{-1}(u)\bigr)
\qquad(L\in\mathcal L^\uparrow)
\]
also represents \(\succsim\) on \(\mathcal L^\uparrow\).

Set
\[
q_1:=q,\qquad q_2:=\widetilde q,\qquad c_1:=c,\qquad c_2:=\widetilde c.
\]
For \(i=1,2\), define
\[
V_i(L):=c_i+\int_0^1 q_i(u)\,d(L^{-1})(u)\qquad(L\in\mathcal L^\uparrow).
\]
For \(L_1,L_2\in\mathcal L^\uparrow\) and \(\lambda\in[0,1]\), the definition of the dual mixture gives
\[
(L_1\oplus_\lambda L_2)^{-1}=\lambda L_1^{-1}+(1-\lambda)L_2^{-1}.
\]
Hence
\begin{align*}
V_i(L_1\oplus_\lambda L_2)
&=
c_i+\int_0^1 q_i(u)\,d\bigl((L_1\oplus_\lambda L_2)^{-1}(u)\bigr)\\
&=
c_i+\int_0^1 q_i(u)\,d\bigl(\lambda L_1^{-1}(u)+(1-\lambda)L_2^{-1}(u)\bigr)\\
&=
c_i+\lambda\int_0^1 q_i(u)\,dL_1^{-1}(u)+(1-\lambda)\int_0^1 q_i(u)\,dL_2^{-1}(u)\\
&=
\lambda V_i(L_1)+(1-\lambda)V_i(L_2).
\end{align*}
Thus each \(V_i\) is affine with respect to dual mixtures.

Define
\[
\Phi_i(M):=V_i(\mathcal SM)\qquad(M\in\mathcal L^\uparrow,\ i=1,2).
\]
Then each \(\Phi_i\) represents \(\succsim^{\mathcal S}\). Moreover, for \(M_1,M_2\in\mathcal L^\uparrow\) and \(\lambda\in[0,1]\),
\[
\mathcal S\bigl(\lambda M_1+(1-\lambda)M_2\bigr)
=
(\mathcal SM_1)\oplus_\lambda(\mathcal SM_2)
\]
by \eqref{eq:reverse-conjugacy}. Therefore
\begin{align*}
\Phi_i\bigl(\lambda M_1+(1-\lambda)M_2\bigr)
&=
V_i\bigl(\mathcal S(\lambda M_1+(1-\lambda)M_2)\bigr)\\
&=
V_i\bigl((\mathcal SM_1)\oplus_\lambda(\mathcal SM_2)\bigr)\\
&=
\lambda V_i(\mathcal SM_1)+(1-\lambda)V_i(\mathcal SM_2)\\
&=
\lambda\Phi_i(M_1)+(1-\lambda)\Phi_i(M_2).
\end{align*}
Thus each \(\Phi_i\) is affine under ordinary mixtures on \(\mathcal L^\uparrow\).

Since \(\Phi_1\) and \(\Phi_2\) are affine and represent the same nontrivial preorder \(\succsim^{\mathcal S}\),
Proposition~\ref{prop:HM-affine-Lorenz}\textup{(iv)} yields constants \(a>0\) and \(d\in\mathbb R\) such that
\[
\Phi_2=a\,\Phi_1+d\qquad\text{on }\mathcal L^\uparrow.
\]
Since \(\mathcal S\) is bijective, this is equivalent to
\[
V_2=a\,V_1+d\qquad\text{on }\mathcal L^\uparrow.
\]
Define
\[
b:=a c_1+d-c_2.
\]
Then, for every \(L\in\mathcal L^\uparrow\),
\[
\int_0^1 q_2(u)\,d(L^{-1})(u)-a\int_0^1 q_1(u)\,d(L^{-1})(u)-b\int_0^1 d(L^{-1})(u)=0.
\]
Since
\[
\int_0^1 d(L^{-1})(u)=L^{-1}(1)-L^{-1}(0)=1,
\]
this becomes
\[
\int_0^1\bigl(q_2-a q_1-b\bigr)(u)\,d(L^{-1})(u)=0
\qquad(\forall L\in\mathcal L^\uparrow).
\]
Define
\[
r:=q_2-a q_1-b.
\]
Then \(r\in BV([0,1])\) and \(r\) is right--continuous on \([0,1)\).

Fix \(u\in(0,1)\) and \(\varepsilon\in(0,1)\). Set
\[
J_u(t):=\min\Bigl\{1,\frac{t}{u}\Bigr\},\qquad E(t):=t,\qquad
J_{u,\varepsilon}:=(1-\varepsilon)J_u+\varepsilon E
\qquad(t\in[0,1]).
\]
The map \(J_u\) is continuous, nondecreasing, concave, satisfies \(J_u(0)=0\) and \(J_u(1)=1\), and is piecewise affine
with slope \(1/u\) on \([0,u]\) and slope \(0\) on \([u,1]\). Hence \(J_{u,\varepsilon}\) is continuous and concave on \([0,1]\).
Moreover, for every \(0\le s<t\le1\),
\[
J_{u,\varepsilon}(t)-J_{u,\varepsilon}(s)
=
(1-\varepsilon)\bigl(J_u(t)-J_u(s)\bigr)+\varepsilon(t-s)
\ge\varepsilon(t-s)>0.
\]
Thus \(J_{u,\varepsilon}\) is strictly increasing. Since also
\[
J_{u,\varepsilon}(0)=0,\qquad J_{u,\varepsilon}(1)=1,
\]
it is a homeomorphism of \([0,1]\) onto \([0,1]\). Therefore
\[
L_{u,\varepsilon}:=J_{u,\varepsilon}^{-1}
\]
is continuous, strictly increasing, convex, satisfies \(L_{u,\varepsilon}(0)=0\), \(L_{u,\varepsilon}(1)=1\), and hence
\[
L_{u,\varepsilon}\in\mathcal L^\uparrow.
\]

Applying the preceding annihilation identity to \(L=L_{u,\varepsilon}\) gives
\[
0
=
\int_0^1 r(t)\,d\bigl(L_{u,\varepsilon}^{-1}\bigr)(t)
=
\int_0^1 r(t)\,dJ_{u,\varepsilon}(t).
\]
The function \(J_{u,\varepsilon}\) is given by
\[
J_{u,\varepsilon}(t)=
\begin{cases}
\Bigl(\dfrac{1-\varepsilon}{u}+\varepsilon\Bigr)t,
&0\le t\le u,\\[6pt]
1-\varepsilon+\varepsilon t,
&u\le t\le1.
\end{cases}
\]
Define
\[
m_{u,\varepsilon}(t):=
\Bigl(\dfrac{1-\varepsilon}{u}+\varepsilon\Bigr)\mathbf1_{[0,u)}(t)
+
\varepsilon\,\mathbf1_{[u,1]}(t).
\]
Then for every \(0\le a<b\le1\),
\[
J_{u,\varepsilon}(b)-J_{u,\varepsilon}(a)
=
\int_{(a,b]}m_{u,\varepsilon}(t)\,dt.
\]
Hence the Lebesgue--Stieltjes measure induced by \(J_{u,\varepsilon}\) is exactly \(m_{u,\varepsilon}(t)\,dt\). Since \(r\in BV([0,1])\), the function \(r\) is bounded and Borel measurable, so
\begin{align*}
0
&=
\int_0^1 r(t)\,dJ_{u,\varepsilon}(t)\\
&=
\int_{[0,1]}r(t)\,m_{u,\varepsilon}(t)\,dt\\
&=
\Bigl(\frac{1-\varepsilon}{u}+\varepsilon\Bigr)\int_0^u r(t)\,dt
+
\varepsilon\int_u^1 r(t)\,dt\\
&=
\frac{1-\varepsilon}{u}\int_0^u r(t)\,dt+\varepsilon\int_0^1 r(t)\,dt.
\end{align*}
Therefore
\[
\int_0^u r(t)\,dt
=
-\frac{u\varepsilon}{1-\varepsilon}\int_0^1 r(t)\,dt.
\]
Because \(r\in BV([0,1])\), the integral \(\int_0^1 r(t)\,dt\) is finite. Letting \(\varepsilon\downarrow0\) yields
\[
\int_0^u r(t)\,dt=0\qquad(\forall u\in(0,1)).
\]

Now fix \(x\in[0,1)\) and choose any \(h\in(0,1-x)\). Then
\[
\int_x^{x+h}r(t)\,dt
=
\int_0^{x+h}r(t)\,dt-\int_0^x r(t)\,dt
=
0.
\]
Therefore
\begin{align*}
|r(x)|
&=
\left|r(x)-\frac1h\int_x^{x+h}r(t)\,dt\right|\\
&=
\left|\frac1h\int_x^{x+h}\bigl(r(x)-r(t)\bigr)\,dt\right|\\
&\le
\sup_{t\in[x,x+h]}|r(x)-r(t)|.
\end{align*}
Since \(r\) is right--continuous at \(x\), the last line would tend to \(0\) as \(h\downarrow0\). Hence
\[
r(x)=0\qquad(\forall x\in[0,1)).
\]
Thus
\[
q_2(u)=a\,q_1(u)+b\qquad(\forall u\in[0,1)).
\]

Taking \(u\uparrow1\) in this identity and using the assumptions:
\(
q_1(1)=q_1(1^-),\quad q_2(1)=q_2(1^-),
\)
we obtain
\(
q_2(1)=a\,q_1(1)+b.
\)
Together with the identity on \([0,1)\), it yields
\[
q_2(u)=a\,q_1(u)+b\qquad(\forall u\in[0,1]).
\]
Since \(q_1=q\) and \(q_2=\widetilde q\), this finishes our proof of Dual Theorem.
\end{proof}

\paragraph{Domain bookkeeping for the dual functional.}
Recall the basic definitions of three Lorenz function domains:
\[
\begin{aligned}
\mathcal L &:= \Bigl\{L\in C([0,1]) : L(0)=0,\ L(1)=1,\ L \text{ is nondecreasing and convex}\Bigr\},\\
\mathcal L^\uparrow &:= \{L\in\mathcal L : L \text{ is strictly increasing}\},\\
\mathcal L^\circ &:= \bigl\{L\in C^2([0,1]) : L(0)=0,\ L(1)=1,\ L'(u)>0,\ L''(u)>0\ \forall u\in(0,1)\bigr\}.
\end{aligned}
\]
Then \(\mathcal L^\circ\subset\mathcal L^\uparrow\subset\mathcal L\).
For $L\in\mathcal L^\uparrow$, $L$ is a homeomorphism of $[0,1]$ onto $[0,1]$; hence $L^{-1}$ is continuous and increasing.

\begin{remark}[Conceptual Overview of Proof Strategy.]
The previous theorems have given us two integral representations with monotone kernels $p,q\in BV([0,1])$. 

At first glance it might be tempting to directly use analytical techniques on the kernels $p$ and $q$ of two integrals. However this is the first point where a genuine technical difficulty arises. The kernels are only assumed to be elements of $BV$, so there is no reliable way to manipulate $p'$ and $q'$ analytically. This prompts the need for extra measure--theoretic arguments.

The overarching theme of this theorem is to  establish the desired global links between the two functionals As a result, the proof could be roughly categorized into two parts.

We would first restrict attention to a smooth interior class of Lorenz curves $\mathcal L^\circ$. This restricted domain allows us to control how the functionals behave under small perturbations of Lorenz curves. After some preparatory arguments one can establish the existence of directional derivatives of the primal functional and its dual.
where $\sigma_L$ is the $L$--dependent measure obtained from $\nu$ through the inverse map $L^{-1}$.

Since both primal functional and dual functional represent the same preorder on $\mathcal L^\circ$, their directional derivatives must be sign aligned. A standard result from functional analysis forces proportionality: for each $L\in\mathcal L^\circ$ there exists a scalar $\lambda(L)>0$ such that
\[
\mu=\lambda(L)\sigma_L
\quad\text{on }(0,1).
\]
This identity is reminiscent of the typical Euler--Lagrange type equations in Functional Analysis. However it still depends on the chosen Lorenz curve $L$, so it does not yet provide a global relationship between the measures $\mu$ and $\nu$.

The general idea in the second half is to carve and vary Lorenz curves in controlled manner and then stitch them together, so it become possible to identify the global relationship between the measure distribution functions $\mu$ and $\nu$.  

To put it into simpler terms, we would construct parametrized Lorenz connector curves that allow different endpoint data values of the function domain to be adjusted independently. By stitching together different Lorenz curves one can compare the proportionality identities coming from different choices of $L$. This eventually forces both $\mu$ and $\nu$ to be atomless on $(0,1)$. The atomless property is crucial: if atoms were present, the measures $\sigma_L$ would have point masses that depend discontinuously on $L$ which in turn sabotages the establishment of universal ratio relationship between $\mu$ and $\nu$.

Once atoms are ruled out, it becomes possible to use analytical techniques to establish a universal tail ratio functional equation linking the distribution functions:
\[
\frac{\mu((0,u])}{\mu((u,1))}
=
\frac{A_1}{A_0}\cdot \frac{\nu((0,z])}{\nu((z,1))}
=
\frac{u(1-z)}{z(1-u)}\cdot \frac{\nu((0,z])}{\nu((z,1))}
\qquad\forall\,0<z<u<1.
\]
Because this ratio identity holds simultaneously for arbitrarily chosed combinations of points, it forms a functional equation on the distribution. Solving this functional equation yields  the unique admissible solution that both $\mu$ and $\nu$ are multiples of Lebesgue measure on $(0,1)$.

Substituting these measures back into the primal and dual representations forces the kernels $p$ and $q$ to be affine. Consequently both functionals collapse to positive affine transforms of the Lorenz area
\[
A(L)=\int_0^1 L(u)du
\]
and therefore the preorder coincides with the generalized Lorenz order.
\end{remark}

\begin{theorem}[Rigidity of compatible primal and dual Lorenz--Stieltjes representations]
\label{thm:main}
Let \(\succsim\) be a complete preorder on \(\mathcal L\).
Assume that \(\succsim\) is nontrivial on \(\mathcal L\), and that its restriction
to \(\mathcal L^\uparrow\) is nontrivial.

Suppose there exist \(c_1,c_2\in\mathbb R\) and functions
\(p,q:[0,1]\to\mathbb R\) such that:

\begin{enumerate}
\item \(p\) is of bounded variation, nonincreasing and right--continuous on \([0,1)\), and
\[
V(L):=c_1+\int_0^1 p(u)\,dL(u)
\qquad(L\in\mathcal L)
\]
represents \(\succsim\) on \(\mathcal L\);

\item \(q\) is of bounded variation, nondecreasing and right--continuous on \([0,1)\), and
\[
V^*(L):=c_2+\int_0^1 q(t)\,d\bigl(L^{-1}\bigr)(t)
\qquad(L\in\mathcal L^\uparrow)
\]
represents the restriction of \(\succsim\) to \(\mathcal L^\uparrow\).
\end{enumerate}

Fix the endpoint-normalized kernels
\[
\widehat p(u):=
\begin{cases}
p(u), & 0\le u<1,\\
\lim_{t\uparrow1}p(t), & u=1,
\end{cases}
\qquad
\widehat q(t):=
\begin{cases}
q(t), & 0\le t<1,\\
\lim_{s\uparrow1}q(s), & t=1.
\end{cases}
\]
Then there exist constants \(\beta_0,\gamma_0\in\mathbb R\) and
\(\beta_1,\gamma_1>0\) such that
\[
\widehat p(u)=\beta_0+\beta_1(1-u),
\qquad
\widehat q(t)=\gamma_0+\gamma_1 t
\qquad(u,t\in[0,1]).
\]

Consequently, there exist constants \(\beta_2,\gamma_2\in\mathbb R\) such that
\[
V(L)=\beta_2+\beta_1\int_0^1L(u)\,du
\qquad(L\in\mathcal L),
\]
and
\[
V^*(L)=\gamma_2+\gamma_1\int_0^1L(u)\,du
\qquad(L\in\mathcal L^\uparrow).
\]
In particular,
\[
L_1\succsim L_2
\iff
\int_0^1L_1(u)\,du\ge \int_0^1L_2(u)\,du,
\]
so \(\succsim\) coincides with the Lorenz-area order, equivalently the Gini order,
on \(\mathcal L\).
\end{theorem}

\medskip
\noindent\textbf{Step 1: BV kernels, local perturbations, and the measure Euler identity.}

\smallskip
\noindent\emph{(1.1) Associated Lebesgue--Stieltjes measures and renormalized integral forms.}
Define
\[
P(u):=p(0)-p(u),\qquad Q(t):=q(t)-q(0)\qquad(u,t\in[0,1]).
\]
Since \(p\) is nonincreasing and \(q\) is nondecreasing, while the endpoint conditions
\[
p(1)=p(1^-),\qquad q(1)=q(1^-)
\]
hold by assumptions \textup{(P)} and \textup{(D)}, the functions \(P\) and \(Q\) are increasing on \([0,1]\),
right--continuous on \([0,1)\), and satisfy
\[
P(0)=Q(0)=0,\qquad P(1)=P(1^-),\qquad Q(1)=Q(1^-).
\]

Let \(\mu\) and \(\nu\) be the finite nonnegative Lebesgue--Stieltjes measures induced by \(P\) and \(Q\). Equivalently,
\[
\mu([0,u])=P(u),\qquad \nu([0,t])=Q(t)\qquad(0\le u,t\le1).
\]
Hence, for every \(0\le a<b\le1\),
\[
\mu((a,b])=\mu([0,b])-\mu([0,a])=P(b)-P(a)=p(a)-p(b),
\]
\[
\nu((a,b])=\nu([0,b])-\nu([0,a])=Q(b)-Q(a)=q(b)-q(a).
\]

The endpoint atoms are zero. Indeed,
\[
\mu(\{0\})=\mu([0,0])=P(0)=0,\qquad
\nu(\{0\})=\nu([0,0])=Q(0)=0.
\]
Also,
\[
\mu([0,1))=\lim_{u\uparrow1}\mu([0,u])=\lim_{u\uparrow1}P(u)=P(1^-)=P(1)=\mu([0,1]),
\]
so
\[
\mu(\{1\})=\mu([0,1])-\mu([0,1))=0.
\]
Similarly,
\[
\nu([0,1))=\lim_{t\uparrow1}\nu([0,t])=\lim_{t\uparrow1}Q(t)=Q(1^-)=Q(1)=\nu([0,1]),
\]
hence
\[
\nu(\{1\})=\nu([0,1])-\nu([0,1))=0.
\]

Two identities will be used repeatedly later:
\[
\mu((u,1])=\mu([0,1])-\mu([0,u])=P(1)-P(u)=p(u)-p(1)
\qquad(0\le u\le1),
\]
and
\[
\nu([0,t])=Q(t)=q(t)-q(0)
\qquad(0\le t\le1).
\]

Now let \(L\in\mathcal L\). Since \(\mu\) is the Lebesgue--Stieltjes measure induced by \(P\),
\[
\int_0^1 L(u)\,dP(u)=\int_{[0,1]}L(u)\,\mu(du).
\]
Applying integration by parts to \((P,L)\), and using \(L(0)=0\), \(L(1)=1\), gives
\[
\int_0^1 P(u)\,dL(u)+\int_{[0,1]}L(u)\,\mu(du)=P(1).
\]
Because \(P=p(0)-p\) and \(\int_0^1 dL=1\),
\[
\int_0^1 P(u)\,dL(u)
=
p(0)\int_0^1 dL(u)-\int_0^1 p(u)\,dL(u)
=
p(0)-\int_0^1 p(u)\,dL(u).
\]
Substituting this into the previous identity yields
\[
p(0)-\int_0^1 p(u)\,dL(u)+\int_{[0,1]}L(u)\,\mu(du)=p(0)-p(1),
\]
hence
\[
\int_0^1 p(u)\,dL(u)=p(1)+\int_{[0,1]}L(u)\,\mu(du).
\]
Therefore
\[
V(L)=\bigl(c_1+p(1)\bigr)+\int_{[0,1]}L(u)\,\mu(du)
\qquad(L\in\mathcal L).
\]

Now let \(L\in\mathcal L^\uparrow\). Since \(\nu\) is the Lebesgue--Stieltjes measure induced by \(Q\),
\[
\int_0^1 L^{-1}(t)\,dQ(t)=\int_{[0,1]}L^{-1}(t)\,\nu(dt).
\]
Applying integration by parts to \((Q,L^{-1})\), and using \(L^{-1}(0)=0\), \(L^{-1}(1)=1\), gives
\[
\int_0^1 Q(t)\,d\bigl(L^{-1}\bigr)(t)+\int_{[0,1]}L^{-1}(t)\,\nu(dt)=Q(1).
\]
Because \(Q=q-q(0)\) and \(\int_0^1 d(L^{-1})=1\),
\[
\int_0^1 Q(t)\,d\bigl(L^{-1}\bigr)(t)
=
\int_0^1 q(t)\,d\bigl(L^{-1}\bigr)(t)-q(0).
\]
Substituting this into the previous identity yields
\[
\int_0^1 q(t)\,d\bigl(L^{-1}\bigr)(t)-q(0)+\int_{[0,1]}L^{-1}(t)\,\nu(dt)=q(1)-q(0),
\]
hence
\[
\int_0^1 q(t)\,d\bigl(L^{-1}\bigr)(t)=q(1)-\int_{[0,1]}L^{-1}(t)\,\nu(dt).
\]
Therefore
\[
V^*(L)=\bigl(c_2+q(1)\bigr)-\int_{[0,1]}L^{-1}(t)\,\nu(dt)
\qquad(L\in\mathcal L^\uparrow).
\]

Define
\[
\widetilde V(L):=\int_{[0,1]}L(u)\,\mu(du)\qquad(L\in\mathcal L),
\qquad
\widetilde V^*(L):=-\int_{[0,1]}L^{-1}(t)\,\nu(dt)\qquad(L\in\mathcal L^\uparrow).
\]
Since \(\mu(\{0\})=\mu(\{1\})=\nu(\{0\})=\nu(\{1\})=0\), these may equally be written as
\[
\widetilde V(L)=\int_{(0,1)}L(u)\,\mu(du),
\qquad
\widetilde V^*(L)=-\int_{(0,1)}L^{-1}(t)\,\nu(dt).
\]
Moreover,
\[
V=c_1+p(1)+\widetilde V\quad\text{on }\mathcal L,
\qquad
V^*=c_2+q(1)+\widetilde V^*\quad\text{on }\mathcal L^\uparrow.
\]
Thus \(V\) and \(\widetilde V\) represent the same preorder on \(\mathcal L\), and \(V^*\) and \(\widetilde V^*\)
represent the same preorder on \(\mathcal L^\uparrow\).

If \(\mu([0,1])=0\), then \(\widetilde V(L)=0\) for all \(L\in\mathcal L\), so \(V\) is constant on \(\mathcal L\),
contradicting Axiom~\ref{ax:HM-nontrivial}. Hence
\[
\mu([0,1])>0.
\]
Similarly, if \(\nu([0,1])=0\), then \(\widetilde V^*(L)=0\) for all \(L\in\mathcal L^\uparrow\), so \(V^*\) is constant
on \(\mathcal L^\uparrow\), contradicting the assumed existence of \(L^+,L^-\in\mathcal L^\uparrow\) with \(L^+\succ L^-\).
Hence
\[
\nu([0,1])>0.
\]

\smallskip
\noindent\emph{(1.2) Admissible local perturbations and the two G\^ateaux derivatives.}
Fix \(L\in\mathcal L^\circ\) and \(\eta\in C_c^\infty(0,1)\). If \(\eta\equiv0\), then all derivative statements below are trivial,
so assume \(\eta\not\equiv0\).

Let
\[
S:=\supp\eta=\overline{\{u\in(0,1):\eta(u)\neq0\}}.
\]
Since \(\eta\in C_c^\infty(0,1)\), the set \(S\) is a nonempty compact subset of \((0,1)\). Define
\[
s_-:=\min S,\qquad s_+:=\max S.
\]
Since \(
0<s_-\le s_+<1
\),we set
\[
\delta:=\frac12\min\{s_-,\,1-s_+\}\in(0,1/2).
\]
Therefore \(
S\subset[s_-,s_+]\subset(\delta,1-\delta) 
\). It's equivalent to:
\[
[0,\delta]\cup[1-\delta,1]\subset[0,1]\setminus S,
\]
and therefore
\[
\eta(u)=0\qquad\forall\,u\in[0,\delta]\cup[1-\delta,1].
\]

For \(\varepsilon\in\mathbb R\), define \(
L_\varepsilon:=L+\varepsilon\eta
\). Since \(\eta=0\) on \([0,\delta]\cup[1-\delta,1]\), we have
\[
L_\varepsilon=L\qquad\text{on }[0,\delta]\cup[1-\delta,1]
\]
for every \(\varepsilon\in\mathbb R\), and in particular
\[
L_\varepsilon(0)=L(0)=0,\qquad L_\varepsilon(1)=L(1)=1.
\]

Because \(L',L''\) are continuous and strictly positive on the compact interval \([\delta,1-\delta]\), the minima
\[
m_1:=\min_{u\in[\delta,1-\delta]}L'(u)>0,\qquad
m_2:=\min_{u\in[\delta,1-\delta]}L''(u)>0
\]
exist. Also set
\[
M_1:=\|\eta'\|_\infty,\qquad M_2:=\|\eta''\|_\infty.
\]
Choose
\[
\varepsilon_0:=\min\left\{1,\frac{m_1}{2(M_1+1)},\frac{m_2}{2(M_2+1)}\right\}>0.
\]
Then for every \(|\varepsilon|\le\varepsilon_0\) and every \(u\in[\delta,1-\delta]\),
\[
L_\varepsilon'(u)=L'(u)+\varepsilon\eta'(u)\ge m_1-|\varepsilon|M_1\ge \frac{m_1}{2}>0,
\]
and similarly
\[
L_\varepsilon''(u)=L''(u)+\varepsilon\eta''(u)\ge m_2-|\varepsilon|M_2\ge \frac{m_2}{2}>0.
\]
On \([0,\delta]\cup[1-\delta,1]\), one has \(L_\varepsilon=L\), hence
\[
L_\varepsilon'(u)=L'(u)>0,\qquad L_\varepsilon''(u)=L''(u)>0
\qquad(u\in(0,\delta)\cup(1-\delta,1)).
\]
Therefore, for every \(|\varepsilon|\le\varepsilon_0\),
\[
L_\varepsilon\in C^2([0,1]),\qquad L_\varepsilon'(u)>0,\qquad L_\varepsilon''(u)>0
\qquad(u\in(0,1)).
\]
Hence
\[
L_\varepsilon\in\mathcal L^\circ
\qquad\forall\,|\varepsilon|\le\varepsilon_0.
\]

We first compute the derivative of the primal renormalized functional. For \(|\varepsilon|\le\varepsilon_0\), linearity of the \(\mu\)-integral gives
\[
\widetilde V(L_\varepsilon)-\widetilde V(L)
=
\int_{(0,1)}(L+\varepsilon\eta)\,d\mu-\int_{(0,1)}L\,d\mu
=
\varepsilon\int_{(0,1)}\eta(u)\,\mu(du).
\]
Hence
\[
D\widetilde V(L)[\eta]=\int_{(0,1)}\eta(u)\,\mu(du).
\]

We now compute the derivative of the dual functional. For every \(|\varepsilon|\le\varepsilon_0\), the curve \(L_\varepsilon\in\mathcal L^\circ\) is strictly increasing,
so \(L_\varepsilon^{-1}:[0,1]\to[0,1]\) is well-defined and continuous.

Because \(\eta=0\) on \([0,\delta]\cup[1-\delta,1]\), we have
\[
L_\varepsilon(\delta)=L(\delta),\qquad L_\varepsilon(1-\delta)=L(1-\delta).
\]

We first claim that
\[
L_\varepsilon^{-1}(t)=L^{-1}(t)
\qquad
\forall\,t\in[0,L(\delta)]\cup[L(1-\delta),1].
\]
Fix \(t\in[0,L(\delta)]\), and set \(x:=L^{-1}(t)\). Then \(x\in[0,\delta]\), since \(t\le L(\delta)\) and \(L\) is strictly increasing.
Let \(y:=L_\varepsilon^{-1}(t)\), so \(L_\varepsilon(y)=t\). If \(y>\delta\), then strict monotonicity of \(L_\varepsilon\) gives
\[
t=L_\varepsilon(y)>L_\varepsilon(\delta)=L(\delta),
\]
contradicting \(t\le L(\delta)\). Hence \(y\in[0,\delta]\). On this interval, \(L_\varepsilon=L\), so
\[
t=L_\varepsilon(y)=L(y).
\]
By uniqueness of the inverse of the strictly increasing map \(L\), we conclude that
\[
y=L^{-1}(t)=x.
\]
Thus
\[
L_\varepsilon^{-1}(t)=L^{-1}(t)\qquad(t\in[0,L(\delta)]).
\]

The proof on \([L(1-\delta),1]\) is identical. Indeed, if \(t\in[L(1-\delta),1]\) and \(L_\varepsilon(y)=t\), then \(y\ge1-\delta\);
otherwise strict monotonicity would imply
\[
t=L_\varepsilon(y)<L_\varepsilon(1-\delta)=L(1-\delta),
\]
a contradiction. On \([1-\delta,1]\), again \(L_\varepsilon=L\), so \(y=L^{-1}(t)\).

Hence
\[
L_\varepsilon^{-1}(t)=L^{-1}(t)
\qquad
\forall\,t\in[0,L(\delta)]\cup[L(1-\delta),1].
\]
Therefore the inverse difference quotient is identically zero on these two outer intervals.

Now fix
\[
t\in[L(\delta),L(1-\delta)].
\]
Set
\[
x:=L^{-1}(t),\qquad x_\varepsilon:=L_\varepsilon^{-1}(t).
\]
Then \(x\in[\delta,1-\delta]\). We claim that \(x_\varepsilon\in[\delta,1-\delta]\) as well.
Indeed, if \(x_\varepsilon<\delta\), then
\[
t=L_\varepsilon(x_\varepsilon)<L_\varepsilon(\delta)=L(\delta),
\]
contradiction. If \(x_\varepsilon>1-\delta\), then
\[
t=L_\varepsilon(x_\varepsilon)>L_\varepsilon(1-\delta)=L(1-\delta),
\]
contradiction. Thus
\[
x_\varepsilon\in[\delta,1-\delta].
\]

Since
\[
L(x)=t,\qquad L_\varepsilon(x_\varepsilon)=t,
\]
we have
\[
0=L_\varepsilon(x_\varepsilon)-L(x)=L(x_\varepsilon)-L(x)+\varepsilon\eta(x_\varepsilon).
\]
By the mean value theorem, there exists \(\xi_\varepsilon\) between \(x\) and \(x_\varepsilon\) such that
\[
L(x_\varepsilon)-L(x)=L'(\xi_\varepsilon)(x_\varepsilon-x).
\]
Hence
\[
\frac{x_\varepsilon-x}{\varepsilon}
=
-\frac{\eta(x_\varepsilon)}{L'(\xi_\varepsilon)}.
\]

Let
\[
m:=\min_{u\in[\delta,1-\delta]}L'(u)>0.
\]
Then, for every \(t\in[L(\delta),L(1-\delta)]\),
\[
\left|
\frac{L_\varepsilon^{-1}(t)-L^{-1}(t)}{\varepsilon}
\right|
=
\left|
\frac{x_\varepsilon-x}{\varepsilon}
\right|
\le
\frac{\|\eta\|_\infty}{m}.
\]
Moreover,
\[
|x_\varepsilon-x|
\le
\frac{\|\eta\|_\infty}{m}\,|\varepsilon|
\longrightarrow0
\qquad(\varepsilon\to0).
\]
Thus \(x_\varepsilon\to x\), hence \(\eta(x_\varepsilon)\to\eta(x)\) and \(\xi_\varepsilon\to x\). Therefore,
for every fixed \(t\in[L(\delta),L(1-\delta)]\),
\[
\frac{L_\varepsilon^{-1}(t)-L^{-1}(t)}{\varepsilon}
\longrightarrow
-\frac{\eta(L^{-1}(t))}{L'(L^{-1}(t))}
\qquad(\varepsilon\to0).
\]

Combining this with the already-proved equality \(L_\varepsilon^{-1}=L^{-1}\) on
\([0,L(\delta)]\cup[L(1-\delta),1]\), we obtain pointwise convergence on all of \((0,1)\):
\[
\frac{L_\varepsilon^{-1}(t)-L^{-1}(t)}{\varepsilon}
\longrightarrow
-\frac{\eta(L^{-1}(t))}{L'(L^{-1}(t))}
\qquad(t\in(0,1)).
\]
Also, for all sufficiently small \(|\varepsilon|\),
\[
\left|
\frac{L_\varepsilon^{-1}(t)-L^{-1}(t)}{\varepsilon}
\right|
\le
\frac{\|\eta\|_\infty}{m}
\qquad(t\in(0,1)).
\]
Since \(\nu\) is finite, the dominating constant is \(\nu\)-integrable. Therefore, by dominated convergence,
\begin{align*}
D\widetilde V^*(L)[\eta]
&=
\lim_{\varepsilon\to0}
\frac{\widetilde V^*(L_\varepsilon)-\widetilde V^*(L)}{\varepsilon}\\
&=
\lim_{\varepsilon\to0}
\left(
-\int_{(0,1)}
\frac{L_\varepsilon^{-1}(t)-L^{-1}(t)}{\varepsilon}\,\nu(dt)
\right)\\
&=
\int_{(0,1)}
\frac{\eta(L^{-1}(t))}{L'(L^{-1}(t))}\,\nu(dt).
\end{align*}
Because \(L\in\mathcal L^\circ\), the inverse function theorem gives
\[
(L^{-1})'(t)=\frac{1}{L'(L^{-1}(t))}
\qquad(t\in(0,1)),
\]
and hence
\[
D\widetilde V^*(L)[\eta]
=
\int_{(0,1)}\eta(L^{-1}(t))(L^{-1})'(t)\,\nu(dt).
\]

\smallskip
\noindent\emph{(1.3) Euler-Lagrange Equation under Measure Theoretic setting.}
For this fixed \(L\in\mathcal L^\circ\), define
\[
\sigma_L(A):=
\int_{(0,1)}\mathbf 1_A(L^{-1}(t))(L^{-1})'(t)\,\nu(dt)
\qquad(A\in\mathcal B((0,1))).
\]
We claim that \(\sigma_L\) is a nonnegative Borel measure on \((0,1)\), finite on compact subsets.

Let \(A\in\mathcal B((0,1))\). Since \(L^{-1}\) is continuous on \((0,1)\), the map
\[
t\longmapsto \mathbf 1_A(L^{-1}(t))(L^{-1})'(t)
\]
is nonnegative and Borel measurable, so \(\sigma_L(A)\) is well-defined. Also,
\(
\sigma_L(\varnothing)=0
\). Now let \((A_k)_{k\ge1}\subset\mathcal B((0,1))\) be pairwise disjoint, and set
\[
A:=\bigcup_{k=1}^\infty A_k.
\]
For each \(n\), define
\[
f_n(t):=\sum_{k=1}^n \mathbf 1_{A_k}(L^{-1}(t))(L^{-1})'(t).
\]
Then
\[
f_n(t)\uparrow \mathbf 1_A(L^{-1}(t))(L^{-1})'(t)
\qquad(t\in(0,1)).
\]
By the monotone convergence theorem,
\begin{align*}
\sigma_L(A)
&=
\int_{(0,1)}\mathbf 1_A(L^{-1}(t))(L^{-1})'(t)\,\nu(dt)\\
&=
\lim_{n\to\infty}\int_{(0,1)}f_n(t)\,\nu(dt)\\
&=
\lim_{n\to\infty}\sum_{k=1}^n
\int_{(0,1)}\mathbf 1_{A_k}(L^{-1}(t))(L^{-1})'(t)\,\nu(dt)\\
&=
\sum_{k=1}^\infty \sigma_L(A_k).
\end{align*}
Hence \(\sigma_L\) is a nonnegative Borel measure on \((0,1)\).

Now let \(K\Subset(0,1)\) be compact. Since \(L'\) is continuous and strictly positive on \(K\), the minimum
\(
m_K:=\min_{u\in K}L'(u)>0
\)
exists. If \(L^{-1}(t)\in K\), then
\[
(L^{-1})'(t)=\frac{1}{L'(L^{-1}(t))}\le \frac{1}{m_K}.
\]
Therefore
\begin{align*}
\sigma_L(K)
&=
\int_{(0,1)}\mathbf 1_K(L^{-1}(t))(L^{-1})'(t)\,\nu(dt)\\
&\le
\frac{1}{m_K}\int_{(0,1)}\mathbf 1_K(L^{-1}(t))\,\nu(dt)\\
&\le
\frac{1}{m_K}\,\nu((0,1))<\infty.
\end{align*}
Thus \(\sigma_L\) is finite on compact subsets of \((0,1)\).

We next record the corresponding integration formula. Let
\[
s(u)=\sum_{j=1}^N a_j\,\mathbf 1_{A_j}(u)
\qquad(a_j\ge0,\ A_j\in\mathcal B((0,1)))
\]
be a nonnegative simple Borel function on \((0,1)\). Then
\begin{align*}
\int_{(0,1)}s(u)\,\sigma_L(du)
&=
\sum_{j=1}^N a_j\,\sigma_L(A_j)\\
&=
\sum_{j=1}^N a_j
\int_{(0,1)}\mathbf 1_{A_j}(L^{-1}(t))(L^{-1})'(t)\,\nu(dt)\\
&=
\int_{(0,1)}s(L^{-1}(t))(L^{-1})'(t)\,\nu(dt).
\end{align*}
Now let \(h:(0,1)\to[0,\infty]\) be Borel measurable, and choose a sequence of nonnegative simple functions \(s_n\uparrow h\).
Then
\[
s_n(L^{-1}(t))(L^{-1})'(t)\uparrow h(L^{-1}(t))(L^{-1})'(t),
\]
so another application of monotone convergence yields
\begin{equation}\label{eq:sigmaL-integration-formula}
\int_{(0,1)}h(u)\,\sigma_L(du)
=
\int_{(0,1)}h(L^{-1}(t))(L^{-1})'(t)\,\nu(dt)
\end{equation}
for every nonnegative Borel measurable \(h\).

Since \(\supp\eta\Subset(0,1)\), the measure \(\sigma_L\) is finite on \(\supp\eta\), and \(\eta\) is bounded.
Hence \(\eta^+,\eta^-\in L^1(\sigma_L)\). Applying \eqref{eq:sigmaL-integration-formula} to \(\eta^+\) and \(\eta^-\),
and subtracting, gives
\[
D\widetilde V^*(L)[\eta]
=
\int_{(0,1)}\eta(u)\,\sigma_L(du)
\qquad(\eta\in C_c^\infty(0,1)).
\]

Set
\[
X:=C_c^\infty(0,1),\qquad
F(\eta):=\int_{(0,1)}\eta(u)\,\mu(du),\qquad
G_L(\eta):=\int_{(0,1)}\eta(u)\,\sigma_L(du).
\]
We show
\[
F\not\equiv0,\qquad G_L\not\equiv0.
\]

Let
\[
\mathcal I:=\{(r,s): r,s\in\mathbb Q,\ 0<r<s<1\}.
\]
Then \(\mathcal I\) is countable, and \(
(0,1)=\bigcup_{(r,s)\in\mathcal I}(r,s)
\). Since \(\mu((0,1))>0\), there exists \((a,b)\in\mathcal I\) such that
\(
\mu((a,b))>0
\). Set
\[
\delta:=\min\{a,\ 1-b,\ (b-a)/2\}>0,
\qquad
\varepsilon_n:=\delta/n
\qquad(n\ge2),
\qquad
A_n:=(a+\varepsilon_n,\ b-\varepsilon_n).
\]
Then \(A_n\uparrow(a,b)\), so by continuity from below,
\[
\mu(A_n)\uparrow \mu((a,b))>0.
\]
Choose \(n_0\ge2\) such that \(\mu(A_{n_0})>0\), and set \(
\varepsilon:=\varepsilon_{n_0}.
\)
By \eqref{eq:interval-cutoff}--\eqref{eq:cutoff-plateau} Lemma~\ref{lem:mollifier-cutoffs}, there exists cutoff:
\(
\eta_\mu:=\eta_{a,b,\varepsilon}
\)
belongs to \(X=C_c^\infty(0,1)\), satisfies
\[
0\le \eta_\mu\le1,
\]
and
\[
\eta_\mu\equiv1
\qquad\text{on }[a+\varepsilon,\ b-\varepsilon].
\]
Hence
\[
F(\eta_\mu)
=
\int_{(0,1)}\eta_\mu(u)\,\mu(du)
\ge
\mu((a+\varepsilon,\ b-\varepsilon))
=
\mu(A_{n_0})>0.
\]
Thus \(F\not\equiv0\).

Applying the same countable-basis argument to \(\nu\), choose \(0<c<d<1\) it gives:
\(
\nu((c,d))>0
\). Define
\[
\delta':=\min\{c,\ 1-d,\ (d-c)/2\}>0,
\qquad
\varepsilon_n':=\delta'/n
\qquad(n\ge2),
\qquad
K_n:=(c+\varepsilon_n',\ d-\varepsilon_n').
\]
Then \(K_n\uparrow(c,d)\), so
\[
\nu(K_n)\uparrow \nu((c,d))>0.
\]
Choose \(n_1\ge2\) such that \(\nu(K_{n_1})>0\), set
\(
\varepsilon':=\varepsilon_{n_1}',
\)
and define
\[
K:=[c+\varepsilon',\ d-\varepsilon']\Subset(0,1).
\]
Then \(\nu(K)\ge \nu(K_{n_1})>0\).

Because \((L^{-1})'\) is continuous and strictly positive on \(K\), the minimum
\[
m_L:=\min_{t\in K}(L^{-1})'(t)>0
\]
exists. Set
\[
B_L:=L^{-1}(K)\Subset(0,1).
\]
Since \(L^{-1}\) is strictly increasing,
\[
\mathbf 1_{B_L}(L^{-1}(t))=\mathbf 1_K(t),
\]
and therefore
\[
\sigma_L(B_L)
=
\int_{(0,1)}\mathbf 1_{B_L}(L^{-1}(t))(L^{-1})'(t)\,\nu(dt)
=
\int_K (L^{-1})'(t)\,\nu(dt)
\ge
m_L\,\nu(K)>0.
\]

Let
\[
u_-:=\min B_L,\qquad u_+:=\max B_L.
\]
Define
\[
\alpha:=\frac{u_-}{2},\qquad
\beta:=\frac{1+u_+}{2},\qquad
\varepsilon:=\frac14\min\{u_-,\,1-u_+\}.
\]
Then
\[
0<\varepsilon<\min\{\alpha,1-\beta\},
\]
and
\[
B_L\subset[\alpha+\varepsilon,\ \beta-\varepsilon].
\]
Again by applying  \eqref{eq:interval-cutoff}--\eqref{eq:cutoff-plateau} of Lemma~\ref{lem:mollifier-cutoffs}, there exists cutoff
\(
\eta_L:=\eta_{\alpha,\beta,\varepsilon}
\)
belongs to \(X\), satisfies
\(
0\le \eta_L\le1,
\)
and
\(
\eta_L\equiv1~
 \text{on }[\alpha+\varepsilon,\ \beta-\varepsilon]\supset B_L.
\)
Hence
\[
G_L(\eta_L)
=
\int_{(0,1)}\eta_L(u)\,\sigma_L(du)
\ge
\sigma_L(B_L)>0.
\]
Thus \(G_L\not\equiv0\).

Because \(\widetilde V\) and \(\widetilde V^*\) represent the same preorder on \(\mathcal L^\circ\), for every \(\eta\in C_c^\infty(0,1)\)
and every sufficiently small \(\varepsilon\),
\[
\bigl(\widetilde V(L+\varepsilon\eta)-\widetilde V(L)\bigr)
\cdot
\bigl(\widetilde V^*(L+\varepsilon\eta)-\widetilde V^*(L)\bigr)\ge0.
\]
Dividing by \(\varepsilon^2\) and letting \(\varepsilon\to0\), we obtain
\[
D\widetilde V(L)[\eta]\cdot D\widetilde V^*(L)[\eta]\ge0
\qquad(\forall \eta\in C_c^\infty(0,1)).
\]
Using the derivative formulas proved above, this becomes
\[
F(\eta)\,G_L(\eta)\ge0
\qquad(\forall \eta\in X).
\]

\begin{lemma}[Co-Monotonicity $\Rightarrow$ Proportionality]\label{lem:alignment}
Let $X$ be a real vector space and $F,G:X\to\mathbb R$ be linear maps with $G\not\equiv 0$.
Assume
\[
F(x)\,G(x)\ \ge\ 0\qquad\forall x\in X.
\]
Then there exists $\lambda\ge 0$ such that $F=\lambda G$.
\end{lemma}

\begin{proof}
If $F\equiv 0$, the conclusion holds with $\lambda=0$, so suppose $F\not\equiv 0$.
Since $F$ and $G$ are nonzero linear functionals, there exist
$x_F,x_G\in X$ with $F(x_F)\neq 0$ and $G(x_G)\neq 0$.

We first show that there exists $x_0\in X$ such that
\[
F(x_0)\neq 0,\qquad G(x_0)\neq 0.
\]
Assume by contradiction that no such $x_0$ exists. Then for every $x\in X$,
\[
F(x)\neq 0\ \Longrightarrow\ G(x)=0,\qquad
G(x)\neq 0\ \Longrightarrow\ F(x)=0.
\]
In particular, $F(x_F)\neq 0$ implies $G(x_F)=0$, while $G(x_G)\neq 0$ implies $F(x_G)=0$.
Consider $x_F+t x_G$ for $t\in\mathbb R$. By linearity,
\[
F(x_F+t x_G)=F(x_F)+tF(x_G)=F(x_F),
\]
\[
G(x_F+t x_G)=G(x_F)+tG(x_G)=t\,G(x_G).
\]
Hence
\[
F(x_F+t x_G)\,G(x_F+t x_G)=F(x_F)\,t\,G(x_G).
\]
Since $F(x_F)\neq 0$ and $G(x_G)\neq 0$, choosing $t$ with sign opposite to $F(x_F)G(x_G)$ gives
\[
F(x_F+t x_G)\,G(x_F+t x_G)<0,
\]
contradicting the assumption that $F(x)G(x)\ge 0$ for all $x\in X$. Thus there exists $x_0\in X$ with
\[
F(x_0)\neq 0,\qquad G(x_0)\neq 0.
\]

Define $a:=F(x_0)$ and $b:=G(x_0)$. Let $y\in X$ be arbitrary and define
\[
P_y(t):=F(y+t x_0)\,G(y+t x_0),\qquad t\in\mathbb R.
\]
By assumption of Co-monotonicity, $P_y(t)\ge 0$ for all $t\in\mathbb R$. Using linearity,
\[
F(y+t x_0)=F(y)+t a,\qquad G(y+t x_0)=G(y)+t b,
\]
so
\[
P_y(t)=(F(y)+t a)(G(y)+t b)
= a b\,t^2 + (a G(y)+b F(y))\,t + F(y)G(y).
\]
Thus $P_y$ is a real quadratic polynomial which is nonnegative on $\mathbb R$; therefore its
discriminant is non-positive:
\[
\Delta_y
= (a G(y)+b F(y))^2 - 4 a b F(y)G(y)
= (a G(y)-b F(y))^2\ \le\ 0.
\]
Hence $a G(y)-b F(y)=0$ and thus:

\[
F(y)=\frac{a}{b}\,G(y)\qquad\forall y\in X,
\]
so $F=\lambda G$ with $\lambda:=a/b$.

Finally, the sign condition implies $\lambda\ge 0$. Indeed, pick any $z\in X$ with $G(z)\neq 0$.
Then
\[
0\ \le\ F(z)\,G(z)=\lambda\,G(z)^2,
\]
and since $G(z)^2>0$, we get $\lambda\ge 0$.

\end{proof}

Since \(F\not\equiv0\) and \(G_L\not\equiv0\), Lemma~\ref{lem:alignment} applies. Therefore there exists
\(\lambda(L)\ge0\) such that
\begin{equation}\label{eq:test-equality-expanded}
\int_{(0,1)}\eta(u)\,\mu(du)
=
\lambda(L)\int_{(0,1)}\eta(u)\,\sigma_L(du)
\qquad(\forall \eta\in C_c^\infty(0,1)).
\end{equation}
Because \(F\not\equiv0\), we must in fact have
\[
\lambda(L)>0.
\]

The measures \(\mu\) and \(\sigma_L\) are nonnegative Borel measures on \((0,1)\), finite on compact subsets.
Applying Lemma~\ref{lem:test-functions-determine-measures} to \eqref{eq:test-equality-expanded}, we conclude that
\begin{equation}\label{eq:measure-euler}
\mu=\lambda(L)\,\sigma_L
\qquad\text{as Borel measures on }(0,1).
\end{equation}

\medskip
\noindent\textbf{Step 2: Atomless of  Measure Distribution $\nu$ and $\mu$.}

\begin{proposition}[Parameterized Lorenz Connector Function]\label{convex-connector}
Let \(0\le \alpha<\beta\le 1\), and set \(h:=\beta-\alpha\). Let
\[
(y_\alpha,v_\alpha,s_\alpha)\in\mathbb R\times(0,\infty)\times(0,\infty),
\qquad
(y_\beta,v_\beta,s_\beta)\in\mathbb R\times(0,\infty)\times(0,\infty).
\]
Define
\[
V:=v_\beta-v_\alpha,
\qquad
Y:=y_\beta-y_\alpha-hv_\alpha.
\]
Assume
\[
V>0,
\qquad
0<Y<hV.
\]
Then there exists \(P\in C^2([\alpha,\beta])\) such that
\[
P(\alpha)=y_\alpha,\quad P'(\alpha)=v_\alpha,\quad P''(\alpha)=s_\alpha,
\]
\[
P(\beta)=y_\beta,\quad P'(\beta)=v_\beta,\quad P''(\beta)=s_\beta,
\]
and
\[
P'(u)>0,\qquad P''(u)>0\qquad (u\in(\alpha,\beta)).
\]
Moreover,
\[
P(u)=y_\alpha+v_\alpha(u-\alpha)+\int_\alpha^u (u-s)P''(s)\,ds
\qquad (u\in[\alpha,\beta]).
\]

Conversely, if \(P\in C^2([\alpha,\beta])\) satisfies the above endpoint conditions and
\(P''(u)>0\) for all \(u\in(\alpha,\beta)\), then necessarily
\[
v_\beta>v_\alpha,
\qquad
0<y_\beta-y_\alpha-hv_\alpha<h(v_\beta-v_\alpha).
\]
\end{proposition}

\begin{proof}
Set
\[
S:=s_\alpha+s_\beta>0,
\qquad
\tau:=\frac{Y}{V}\in(0,h).
\]

\smallskip
\noindent\emph{Necessity.}
Suppose that \(P\in C^2([\alpha,\beta])\) satisfies the stated endpoint conditions and that
\(P''(u)>0\) for all \(u\in(\alpha,\beta)\).
By the Fundamental Theorem of Calculus,
\[
\int_\alpha^\beta P''(s)\,ds
=
P'(\beta)-P'(\alpha)
=
v_\beta-v_\alpha
=
V.
\]
Hence \(V>0\). Next, define
\[
U(s):=P'(s)-P'(\alpha)=P'(s)-v_\alpha,
\qquad
W(s):=\beta-s.
\]
Then \(U'(s)=P''(s)\) and \(W'(s)=-1\). By integration by parts,
\[
\int_\alpha^\beta (\beta-s)P''(s)\,ds
=
\bigl[U(s)W(s)\bigr]_\alpha^\beta
-
\int_\alpha^\beta U(s)W'(s)\,ds
=
\int_\alpha^\beta \bigl(P'(s)-v_\alpha\bigr)\,ds.
\]
Using the Fundamental Theorem of Calculus once more,
\[
\int_\alpha^\beta (\beta-s)P''(s)\,ds
=
P(\beta)-P(\alpha)-h v_\alpha
=
y_\beta-y_\alpha-hv_\alpha
=
Y.
\]
Since \(P''(s)>0\) for \(s\in(\alpha,\beta)\) and \(0<\beta-s<h\) on \((\alpha,\beta)\),
\[
0
<
\int_\alpha^\beta (\beta-s)P''(s)\,ds
<
h\int_\alpha^\beta P''(s)\,ds.
\]
Therefore
\[
0<Y<hV.
\]

\smallskip
\noindent\emph{Constructive Proof for its existence.}
Set
\[
d:=\frac12\min\{\tau,h-\tau\}>0.
\]
Choose
\[
\varepsilon_*:=\min\left\{d,\ \frac{V}{4S},\ \frac{Y}{4hS},\ \frac{dV}{8hS}\right\}>0,
\qquad
m_*:=\min\left\{s_\alpha,\ s_\beta,\ \frac{V}{4h},\ \frac{Y}{2h^2},\ \frac{dV}{6h^2}\right\}>0.
\]
From now on fix the quantification domains:
\[
0<\varepsilon<\varepsilon_*,
\qquad
0<m<m_*.
\]

Recall that by Lemma~\ref{lem:mollifier-cutoffs} we have:
\[
\psi\in C_c^\infty(\mathbb R),
\qquad
\psi\ge0,
\qquad
\operatorname{supp}(\psi)=[-1,1],
\qquad
\int_{\mathbb R}\psi(t)\,dt=1.
\]
Since \(\psi(t)=A^{-1}\rho(1-t^2)\), the function \(\psi\) is even. Also, the boundary value is zero:\(
\psi(\pm1)=\frac{\rho(0)}{A}=0,
\quad
\psi(t)>0\quad (|t|<1).
\)
Define
\[
\psi_\varepsilon(t):=\frac1\varepsilon\,\psi\!\left(\frac{t}{\varepsilon}\right)
\qquad (t\in\mathbb R).
\]
Again by Lemma~\ref{lem:mollifier-cutoffs},
\[
\psi_\varepsilon\in C_c^\infty(\mathbb R),
\qquad
\psi_\varepsilon\ge0,
\qquad
\operatorname{supp}(\psi_\varepsilon)\subset[-\varepsilon,\varepsilon],
\qquad
\int_{\mathbb R}\psi_\varepsilon(t)\,dt=1.
\]

Since \(\min\{\tau,h-\tau\}\le h/2\), one has \(d\le h/4\). So it yields:
\[
0<\varepsilon<d\le \frac h4<\frac h2.
\]

Define
\[
\omega_a^\varepsilon(s):=\psi_\varepsilon\bigl(s-(\alpha+\varepsilon)\bigr),
\qquad
\omega_b^\varepsilon(s):=\psi_\varepsilon\bigl(s-(\beta-\varepsilon)\bigr),
\qquad s\in[\alpha,\beta].
\]
Then \(\omega_a^\varepsilon,\omega_b^\varepsilon\in C^\infty([\alpha,\beta])\), and \(
\omega_a^\varepsilon\ge0, \omega_b^\varepsilon\ge0
\). By definition:
\[
\operatorname{supp}(\omega_a^\varepsilon)\subset[\alpha,\alpha+2\varepsilon],
\qquad
\operatorname{supp}(\omega_b^\varepsilon)\subset[\beta-2\varepsilon,\beta].
\]
Moreover,
\[
\omega_a^\varepsilon(\alpha)=\psi_\varepsilon(-\varepsilon)=\frac1\varepsilon\psi(-1)=0,
\qquad
\omega_b^\varepsilon(\beta)=\psi_\varepsilon(\varepsilon)=\frac1\varepsilon\psi(1)=0.
\]
Since \(\alpha+2\varepsilon<\beta\) and \(\alpha<\beta-2\varepsilon\), one also has
\[
\omega_a^\varepsilon(\beta)=0,
\qquad
\omega_b^\varepsilon(\alpha)=0.
\]
Because \(\psi(t)>0\) for \(|t|<1\),
\[
\omega_a^\varepsilon(s)>0\quad (s\in(\alpha,\alpha+2\varepsilon)),
\qquad
\omega_b^\varepsilon(s)>0\quad (s\in(\beta-2\varepsilon,\beta)).
\]

Since \(\omega_a^\varepsilon\) vanishes outside \([\alpha,\alpha+2\varepsilon]\),
\[
\int_\alpha^\beta \omega_a^\varepsilon(s)\,ds
=
\int_\alpha^{\alpha+2\varepsilon} \omega_a^\varepsilon(s)\,ds.
\]
With the change of variables
\[
r=\frac{s-(\alpha+\varepsilon)}{\varepsilon},
\qquad
s=\alpha+\varepsilon+\varepsilon r,
\qquad
ds=\varepsilon\,dr,
\]
we get
\[
\int_\alpha^\beta \omega_a^\varepsilon(s)\,ds
=
\int_{-1}^{1} \frac1\varepsilon\psi(r)\,\varepsilon\,dr
=
\int_{-1}^{1}\psi(r)\,dr
=
1.
\]
Similarly,
\[
\int_\alpha^\beta \omega_b^\varepsilon(s)\,ds=1.
\]

Define
\[
\Gamma_a:=\int_\alpha^\beta (\beta-s)\omega_a^\varepsilon(s)\,ds,
\qquad
\Gamma_b:=\int_\alpha^\beta (\beta-s)\omega_b^\varepsilon(s)\,ds.
\]
Since \(\omega_a^\varepsilon\) vanishes outside \([\alpha,\alpha+2\varepsilon]\), the same change of
variables gives
\[
\Gamma_a
=
\int_{-1}^{1} \bigl(\beta-(\alpha+\varepsilon+\varepsilon r)\bigr)\psi(r)\,dr
=
\int_{-1}^{1} \bigl(h-\varepsilon-\varepsilon r\bigr)\psi(r)\,dr.
\]
Because \(\psi\) is even, the function \(r\mapsto r\psi(r)\) is odd, hence
\[
\int_{-1}^{1} r\psi(r)\,dr=0.
\]
Therefore
\[
\Gamma_a
=
(h-\varepsilon)\int_{-1}^{1}\psi(r)\,dr
-
\varepsilon\int_{-1}^{1} r\psi(r)\,dr
=
h-\varepsilon.
\]
Likewise,
\[
\Gamma_b
=
\int_{-1}^{1} \bigl(\beta-(\beta-\varepsilon+\varepsilon r)\bigr)\psi(r)\,dr
=
\int_{-1}^{1} \bigl(\varepsilon-\varepsilon r\bigr)\psi(r)\,dr
=
\varepsilon.
\]

Define
\[
\chi_a^\varepsilon(s):=\int_s^\beta \omega_a^\varepsilon(t)\,dt,
\qquad
\chi_b^\varepsilon(s):=\int_\alpha^s \omega_b^\varepsilon(t)\,dt,
\qquad s\in[\alpha,\beta].
\]
Then \(\chi_a^\varepsilon,\chi_b^\varepsilon\in C^\infty([\alpha,\beta])\), and we have \(
0\le \chi_a^\varepsilon\le1,
\quad
0\le \chi_b^\varepsilon\le1.
\)
Also,
\[
\chi_a^\varepsilon(\alpha)=\int_\alpha^\beta \omega_a^\varepsilon(t)\,dt=1,
\qquad
\chi_a^\varepsilon(\beta)=0,
\]
\[
\chi_b^\varepsilon(\alpha)=0,
\qquad
\chi_b^\varepsilon(\beta)=\int_\alpha^\beta \omega_b^\varepsilon(t)\,dt=1.
\]
Moreover,
\[
\operatorname{supp}(\chi_a^\varepsilon)\subset[\alpha,\alpha+2\varepsilon],
\qquad
\operatorname{supp}(\chi_b^\varepsilon)\subset[\beta-2\varepsilon,\beta].
\]
Indeed, if \(s>\alpha+2\varepsilon\), then \([s,\beta]\cap\operatorname{supp}(\omega_a^\varepsilon)=\varnothing\),
so \(\chi_a^\varepsilon(s)=0\). The proof for \(\chi_b^\varepsilon\) is analogous.

By Fubini's theorem on the triangle
\[
T_a:=\{(s,t)\in[\alpha,\beta]^2:\ \alpha\le s\le t\le \beta\},
\]
one has
\[
\int_\alpha^\beta \chi_a^\varepsilon(s)\,ds
=
\int_\alpha^\beta \int_s^\beta \omega_a^\varepsilon(t)\,dt\,ds
=
\int_\alpha^\beta \int_\alpha^t \omega_a^\varepsilon(t)\,ds\,dt
=
\int_\alpha^\beta (t-\alpha)\omega_a^\varepsilon(t)\,dt.
\]
Since \(\omega_a^\varepsilon\) vanishes outside \([\alpha,\alpha+2\varepsilon]\),
\[
\int_\alpha^\beta (t-\alpha)\omega_a^\varepsilon(t)\,dt
=
\int_\alpha^{\alpha+2\varepsilon} (t-\alpha)\omega_a^\varepsilon(t)\,dt.
\]
Using the change of variables
\[
r=\frac{t-(\alpha+\varepsilon)}{\varepsilon},
\qquad
t=\alpha+\varepsilon+\varepsilon r,
\qquad
dt=\varepsilon\,dr,
\]
we obtain
\[
\int_\alpha^\beta \chi_a^\varepsilon(s)\,ds
=
\int_{-1}^{1} \varepsilon(1+r)\psi(r)\,dr
=
\varepsilon\int_{-1}^{1}\psi(r)\,dr
+
\varepsilon\int_{-1}^{1}r\psi(r)\,dr
=
\varepsilon.
\]

Similarly, on the triangle
\[
T_b:=\{(t,s)\in[\alpha,\beta]^2:\ \alpha\le t\le s\le \beta\},
\]
Fubini's theorem gives
\[
\int_\alpha^\beta \chi_b^\varepsilon(s)\,ds
=
\int_\alpha^\beta \int_\alpha^s \omega_b^\varepsilon(t)\,dt\,ds
=
\int_\alpha^\beta \int_t^\beta \omega_b^\varepsilon(t)\,ds\,dt
=
\int_\alpha^\beta (\beta-t)\omega_b^\varepsilon(t)\,dt.
\]
Since \(\omega_b^\varepsilon\) vanishes outside \([\beta-2\varepsilon,\beta]\),
\[
\int_\alpha^\beta (\beta-t)\omega_b^\varepsilon(t)\,dt
=
\int_{\beta-2\varepsilon}^{\beta} (\beta-t)\omega_b^\varepsilon(t)\,dt.
\]
With the change of variables
\[
r=\frac{t-(\beta-\varepsilon)}{\varepsilon},
\qquad
t=\beta-\varepsilon+\varepsilon r,
\qquad
dt=\varepsilon\,dr,
\]
we get
\[
\int_\alpha^\beta \chi_b^\varepsilon(s)\,ds
=
\int_{-1}^{1} \varepsilon(1-r)\psi(r)\,dr
=
\varepsilon\int_{-1}^{1}\psi(r)\,dr
-
\varepsilon\int_{-1}^{1}r\psi(r)\,dr
=
\varepsilon.
\]

Define the baseline
\[
B_{\varepsilon,m}(s)
:=
m+(s_\alpha-m)\chi_a^\varepsilon(s)+(s_\beta-m)\chi_b^\varepsilon(s),
\qquad s\in[\alpha,\beta].
\]
Then \(B_{\varepsilon,m}\in C^\infty([\alpha,\beta])\), and since \(m\le s_\alpha\) and
\(m\le s_\beta\),
\[
B_{\varepsilon,m}(s)\ge m
\qquad (s\in[\alpha,\beta]).
\]
Also,
\[
B_{\varepsilon,m}(\alpha)=m+(s_\alpha-m)\cdot 1+(s_\beta-m)\cdot 0=s_\alpha,
\]
\[
B_{\varepsilon,m}(\beta)=m+(s_\alpha-m)\cdot 0+(s_\beta-m)\cdot 1=s_\beta.
\]

For \(f\in C([\alpha,\beta])\), write
\[
\mathcal I_0(f):=\int_\alpha^\beta f(s)\,ds,
\qquad
\mathcal I_1(f):=\int_\alpha^\beta (\beta-s)f(s)\,ds.
\]
Using the identities just proved,
\[
\mathcal I_0(B_{\varepsilon,m})
=
mh+(s_\alpha-m)\int_\alpha^\beta \chi_a^\varepsilon(s)\,ds
+(s_\beta-m)\int_\alpha^\beta \chi_b^\varepsilon(s)\,ds
\]
\[
=
mh+(s_\alpha-m)\varepsilon+(s_\beta-m)\varepsilon
=
m(h-2\varepsilon)+\varepsilon S
<
mh+\varepsilon S
<
\frac V4+\frac V4
=
\frac V2.
\]

For the weighted moment,
\[
\mathcal I_1(B_{\varepsilon,m})
=
\int_\alpha^\beta (\beta-s)m\,ds
+
(s_\alpha-m)\mathcal I_1(\chi_a^\varepsilon)
+
(s_\beta-m)\mathcal I_1(\chi_b^\varepsilon).
\]
Since \(0\le \beta-s\le h\) on \([\alpha,\beta]\),
\[
\mathcal I_1(\chi_a^\varepsilon)
\le
h\int_\alpha^\beta \chi_a^\varepsilon(s)\,ds
=
h\varepsilon,
\qquad
\mathcal I_1(\chi_b^\varepsilon)
\le
h\int_\alpha^\beta \chi_b^\varepsilon(s)\,ds
=
h\varepsilon.
\]
Therefore
\[
\mathcal I_1(B_{\varepsilon,m})
\le
\frac{mh^2}{2}+h\varepsilon(s_\alpha-m)+h\varepsilon(s_\beta-m)
\le
\frac{mh^2}{2}+h\varepsilon S
<
\frac Y4+\frac Y4
=
\frac Y2.
\]

Define
\[
K_1:=V-\mathcal I_0(B_{\varepsilon,m})>0,
\qquad
K_2:=Y-\mathcal I_1(B_{\varepsilon,m})>0,
\qquad
r:=\frac{K_2}{K_1}.
\]
Since \(Y=\tau V\),
\[
r-\tau
=
\frac{Y-\mathcal I_1(B_{\varepsilon,m})}{V-\mathcal I_0(B_{\varepsilon,m})}-\tau
=
\frac{\tau\,\mathcal I_0(B_{\varepsilon,m})-\mathcal I_1(B_{\varepsilon,m})}
{V-\mathcal I_0(B_{\varepsilon,m})}.
\]
Hence
\[
|r-\tau|
\le
\frac{\tau\,\mathcal I_0(B_{\varepsilon,m})+\mathcal I_1(B_{\varepsilon,m})}
{V-\mathcal I_0(B_{\varepsilon,m})}.
\]
Because \(\mathcal I_0(B_{\varepsilon,m})<V/2\), one has
\[
V-\mathcal I_0(B_{\varepsilon,m})>\frac V2.
\]
Also, since \(\tau<h\),
\[
\tau\,\mathcal I_0(B_{\varepsilon,m})+\mathcal I_1(B_{\varepsilon,m})
<
h(mh+\varepsilon S)+\frac{mh^2}{2}+h\varepsilon S
=
\frac32 mh^2+2h\varepsilon S.
\]
Therefore
\[
|r-\tau|
<
\frac{2}{V}\left(\frac32 mh^2+2h\varepsilon S\right)
=
\frac{3mh^2}{V}+\frac{4h\varepsilon S}{V}
<
\frac d2+\frac d2
=
d.
\]
Thus
\[
\tau-d<r<\tau+d.
\]

Since
\[
2d=\min\{\tau,h-\tau\}\le \tau,
\qquad
2d=\min\{\tau,h-\tau\}\le h-\tau,
\]
one has
\[
d\le \tau-d,
\qquad
d\le h-\tau-d.
\]
Because \(\varepsilon<d\), it follows that
\[
\varepsilon<\tau-d,
\qquad
\tau+d<h-\varepsilon.
\]
Hence
\[
\Gamma_b=\varepsilon<\tau-d<r<\tau+d<h-\varepsilon=\Gamma_a.
\]

Now solve the linear system
\[
c_a+c_b=K_1,
\qquad
\Gamma_a c_a+\Gamma_b c_b=K_2.
\]
Since \(\Gamma_a-\Gamma_b=h-2\varepsilon>0\), the unique solution is
\[
c_a
=
K_1\,\frac{r-\Gamma_b}{\Gamma_a-\Gamma_b},
\qquad
c_b
=
K_1\,\frac{\Gamma_a-r}{\Gamma_a-\Gamma_b}.
\]
Because \(K_1>0\) and \(\Gamma_b<r<\Gamma_a\), one has
\[
c_a\ge0,
\qquad
c_b\ge0.
\]

Define
\[
X (s):=B_{\varepsilon,m}(s)+c_a\omega_a^\varepsilon(s)+c_b\omega_b^\varepsilon(s),
\qquad s\in[\alpha,\beta].
\]
Then \(X \in C^\infty([\alpha,\beta])\), and
\[
X (s)\ge B_{\varepsilon,m}(s)\ge m>0
\qquad (s\in[\alpha,\beta]).
\]
Because
\[
\omega_a^\varepsilon(\alpha)=\omega_b^\varepsilon(\alpha)=0,
\qquad
\omega_a^\varepsilon(\beta)=\omega_b^\varepsilon(\beta)=0,
\]
one has
\[
X (\alpha)=B_{\varepsilon,m}(\alpha)=s_\alpha,
\qquad
X (\beta)=B_{\varepsilon,m}(\beta)=s_\beta.
\]
Moreover,
\[
\mathcal I_0(X)
=
\mathcal I_0(B_{\varepsilon,m})+c_a\mathcal I_0(\omega_a^\varepsilon)+c_b\mathcal I_0(\omega_b^\varepsilon)
=
\mathcal I_0(B_{\varepsilon,m})+c_a+c_b
=
V,
\]
\[
\mathcal I_1(X )
=
\mathcal I_1(B_{\varepsilon,m})+c_a\Gamma_a+c_b\Gamma_b
=
\mathcal I_1(B_{\varepsilon,m})+K_2
=
Y.
\]

Define
\[
P'(u):=v_\alpha+\int_\alpha^u X (s)\,ds,
\qquad
P(u):=y_\alpha+\int_\alpha^u P'(t)\,dt,
\qquad u\in[\alpha,\beta].
\]
Since \(X \in C([\alpha,\beta])\), one has \(P\in C^2([\alpha,\beta])\) and
\[
P''(u)=X (u)
\qquad (u\in[\alpha,\beta]).
\]
In particular,
\[
P''(u)=X (u)>0
\qquad (u\in(\alpha,\beta)).
\]
Therefore \(P'\) is strictly increasing on \([\alpha,\beta]\). Since
\[
P'(\alpha)=v_\alpha>0,
\]
it follows that
\[
P'(u)\ge v_\alpha>0
\quad (u\in[\alpha,\beta]).
\]

The endpoint conditions at \(u=\alpha\) are immediate:
\[
P(\alpha)=y_\alpha,
\qquad
P'(\alpha)=v_\alpha,
\qquad
P''(\alpha)=X(\alpha)=s_\alpha.
\]
At \(u=\beta\),
\[
P'(\beta)
=
v_\alpha+\int_\alpha^\beta X(s)\,ds
=
v_\alpha+\mathcal I_0(X)
=
v_\alpha+V
=
v_\beta,
\]
and
\[
P(\beta)
=
y_\alpha+\int_\alpha^\beta P'(t)\,dt
=
y_\alpha+\int_\alpha^\beta \left(v_\alpha+\int_\alpha^t X (s)\,ds\right)dt
\]
\[
=
y_\alpha+h v_\alpha+\int_\alpha^\beta (\beta-s)X (s)\,ds
=
y_\alpha+h v_\alpha+\mathcal I_1(X )
=
y_\beta.
\]
Also,
\[
P''(\beta)=X (\beta)=s_\beta.
\]

Finally, for \(u\in[\alpha,\beta]\),
\[
P(u)
=
y_\alpha+\int_\alpha^u P'(t)\,dt
=
y_\alpha+\int_\alpha^u \left(v_\alpha+\int_\alpha^t X (s)\,ds\right)dt
\]
\[
=
y_\alpha+v_\alpha(u-\alpha)+\int_\alpha^u\int_\alpha^t X (s)\,ds\,dt.
\]
By Fubini's theorem on the triangle
\[
\{(s,t):\ \alpha\le s\le t\le u\},
\]
we obtain
\[
\int_\alpha^u\int_\alpha^t X (s)\,ds\,dt
=
\int_\alpha^u\int_s^u X (s)\,dt\,ds
=
\int_\alpha^u (u-s)X(s)\,ds
=
\int_\alpha^u (u-s)P''(s)\,ds.
\]
Hence
\[
P(u)=y_\alpha+v_\alpha(u-\alpha)+\int_\alpha^u (u-s)P''(s)\,ds.
\]
This completes the proof.
\end{proof}

Assume for the sake of argument that there exists $t_0\in(0,1)$ such that $\nu(\{t_0\})>0$.
Fix an arbitrary $u_0\in(t_0,1)$ and set $z:=t_0$. Define the secant slopes
\[
A_0:=\frac{z}{u_0},\qquad A_1:=\frac{1-z}{1-u_0}.
\]
Since $0<z<u_0<1$, we have $A_0>0$, $A_1>0$, and $A_0<A_1$.

Choose endpoint slopes
\[
v_0:=\frac12A_0,\qquad v_1:=\frac12(A_0+A_1),\qquad v_2:=A_1+1,
\]
and choose any endpoint curvatures $s_0,s_1,s_2>0$.

Apply Proposition~\ref{convex-connector} on $[0,u_0]$ with $\alpha=0$, $\beta=u_0$, and endpoint triples
\[
(y_\alpha,v_\alpha,s_\alpha)=(0,v_0,s_0),\qquad (y_\beta,v_\beta,s_\beta)=(z,v_1,s_1).
\]
Here $h=u_0$ and the secant slope is $(y_\beta-y_\alpha)/h=z/u_0=A_0$. Since $v_0=\tfrac12A_0<A_0$ and
$v_1=\tfrac12(A_0+A_1)>A_0$, the compatibility inequalities hold and the proposition yields a function
$P_1\in C^2([0,u_0])$ satisfying
\[
P_1(0)=0,\quad P_1'(0)=v_0,\quad P_1''(0)=s_0,\qquad
P_1(u_0)=z,\quad P_1'(u_0)=v_1,\quad P_1''(u_0)=s_1,
\]
with $P_1'(u)>0$ and $P_1''(u)>0$ for all $u\in(0,u_0)$.

Apply Proposition~\ref{convex-connector} on $[u_0,1]$ with $\alpha=u_0$, $\beta=1$, and endpoint triples
\[
(y_\alpha,v_\alpha,s_\alpha)=(z,v_1,s_1),\qquad (y_\beta,v_\beta,s_\beta)=(1,v_2,s_2).
\]
Here $h=1-u_0$ and the secant slope is $(y_\beta-y_\alpha)/h=(1-z)/(1-u_0)=A_1$. Since
$v_1=\tfrac12(A_0+A_1)<A_1$ and $v_2=A_1+1>A_1$, the compatibility inequalities hold and the proposition yields
$P_2\in C^2([u_0,1])$ satisfying
\[
P_2(u_0)=z,\quad P_2'(u_0)=v_1,\quad P_2''(u_0)=s_1,\qquad
P_2(1)=1,\quad P_2'(1)=v_2,\quad P_2''(1)=s_2,
\]
with $P_2'(u)>0$ and $P_2''(u)>0$ for all $u\in(u_0,1)$.

Define $L:[0,1]\to\mathbb R$ by
\[
L(u):=
\begin{cases}
P_1(u),&u\in[0,u_0],\\
P_2(u),&u\in[u_0,1].
\end{cases}
\]
Because $P_1$ and $P_2$ match in value, first derivative, and second derivative at $u_0$ by construction, we have
$L\in C^2([0,1])$, and
\[
L(0)=0,\qquad L(1)=1,\qquad L'(u)>0,\qquad L''(u)>0\quad(u\in(0,1)).
\]
Hence $L\in\mathcal L^\circ$ and in particular $L$ is strictly increasing. Since $L(u_0)=z=t_0$, we have
$L^{-1}(t_0)=u_0$, and by the inverse function theorem $L^{-1}\in C^2((0,1))$ with
\[
(L^{-1})'(t)=\frac{1}{L'(L^{-1}(t))}\qquad(t\in(0,1)),
\]
so $(L^{-1})'(t_0)=1/L'(u_0)\in(0,\infty)$.

Now evaluate $\sigma_L$ at the singleton $\{u_0\}$:
\[
\sigma_L(\{u_0\})
=\int_{(0,1)} \mathbf 1_{\{u_0\}}(L^{-1}(t))\,(L^{-1})'(t)\,\nu(dt).
\]
Since $L$ is strictly increasing, $L^{-1}(t)=u_0$ holds if and only if $t=L(u_0)=t_0$, hence
$\mathbf 1_{\{u_0\}}(L^{-1}(t))=\mathbf 1_{\{t_0\}}(t)$ and therefore
\[
\sigma_L(\{u_0\})=(L^{-1})'(t_0)\,\nu(\{t_0\})>0.
\]
By the measure Euler identity \eqref{eq:measure-euler} (applied to this $L\in\mathcal L^\circ$),
there exists $\lambda(L)>0$ such that $\mu=\lambda(L)\sigma_L$ on $(0,1)$, hence
\[
\mu(\{u_0\})=\lambda(L)\,\sigma_L(\{u_0\})>0.
\]
Because $u_0\in(t_0,1)$ was arbitrary, this would imply that $\mu$ has an atom at every point of the uncountable set $(t_0,1)$.
This contradicts the finiteness of $\mu$: indeed, for each $n\in\mathbb N$ define
\[
A_n:=\bigl\{x\in(0,1):\ \mu(\{x\})\ge 1/n\bigr\}.
\]
If $A_n$ were infinite, then for every $k\in\mathbb N$ one could choose $k$ distinct points $x_1,\dots,x_k\in A_n$, and by finite additivity,
\[
\mu((0,1))\ \ge\ \mu\!\left(\bigcup_{i=1}^k\{x_i\}\right)\ =\ \sum_{i=1}^k\mu(\{x_i\})\ \ge\ \frac{k}{n},
\]
which is impossible as $k\to\infty$. Hence each $A_n$ is finite. Therefore the set of all atoms
$\{x\in(0,1):\mu(\{x\})>0\}=\bigcup_{n\ge 1}A_n$ is countable, so it cannot contain $(t_0,1)$.
This contradiction shows that $\nu(\{t\})=0$ for every $t\in(0,1)$.

Finally, fix any $u\in(0,1)$ and any $L\in\mathcal L^\circ$. Since $\nu$ has no atoms on $(0,1)$,
\[
\sigma_L(\{u\})=\int_{(0,1)}\mathbf 1_{\{u\}}(L^{-1}(t))\,(L^{-1})'(t)\,\nu(dt)
=(L^{-1})'(L(u))\,\nu(\{L(u)\})=0,
\]
and applying $\mu=\lambda(L)\sigma_L$ yields $\mu(\{u\})=0$. Thus $\mu$ has no atoms on $(0,1)$ as well.

\medskip
\noindent\textbf{Step 3: Piecewise $C^2$ Lorenz connectors and the Universal ratio identity.}

Fix arbitrary numbers $0<z<u<1$ and set
\[
A_0:=\frac{z}{u},\qquad A_1:=\frac{1-z}{1-u},\qquad M:=\mu((0,1)),\qquad N:=\nu((0,1)).
\]
Then $0<A_0<A_1$ and $M,N<\infty$. Here $\delta$ controls the width of the transition zone around $u$, $\eta$ controls the slope perturbation in the construction, $\varepsilon$ is reserved for the final accuracy in the limiting argument, and $\gamma$ denotes the size of a small neighborhood of $z$.

Set
\[
\delta_*:=\min\Bigl\{\frac{u}{2},\frac{1-u}{2}\Bigr\},
\qquad
\eta_*:=\min\Bigl\{\frac{A_0}{2},\frac{A_1-A_0}{2}\Bigr\}.
\]
A pair $(\delta,\eta)$ will be called admissible if
\begin{equation}\label{eq:delta-eta-bounds-step3}
0<\delta<\delta_*,
\qquad
0<\eta<\eta_*.
\end{equation}
All limits below are taken through admissible pairs.

Choose an admissible pair \((\delta,\eta)\) and set nodes:
\[
u_0:=0,\qquad u_1:=u-\delta,\qquad u_2:=u,\qquad u_3:=u+\delta,\qquad u_4:=1.
\]
Then
\[
0<u_1<u_2<u_3<1.
\]
Set
\[
y_0:=0,\qquad
y_1:=z-\delta(A_0+\eta),\qquad
y_2:=z,\qquad
y_3:=z+\delta(A_1-\eta),\qquad
y_4:=1.
\]
Then
\[
0=y_0<y_1<y_2<y_3<y_4=1.
\]
Indeed, \(y_1<z<y_3\) is immediate. Since \(\eta<\eta_*\le A_0/2\), we have
\(A_0+\eta<3A_0/2\), and since \(\delta<\delta_*\le u/2\), using \(z=A_0u\),
\[
\delta(A_0+\eta)<\frac{u}{2}\cdot\frac{3A_0}{2}=\frac34A_0u<z,
\]
hence \(y_1=z-\delta(A_0+\eta)>0\). Also, since \(\eta>0\) and
\(\delta<\delta_*\le (1-u)/2<1-u\),
\[
\delta(A_1-\eta)<\delta A_1<(1-u)A_1=1-z,
\]
hence \(y_3=z+\delta(A_1-\eta)<1\).

Define
\[
v_0:=A_0-\eta,\qquad
v_1:=A_0,\qquad
v_2:=\frac{A_0+A_1}{2},\qquad
v_3:=A_1,\qquad
v_4:=A_1+\eta.
\]
Because
\[
0<\eta<\eta_*\le \min\left\{\frac{A_0}{2},\frac{A_1-A_0}{2}\right\},
\]
we have
\[
0<v_0<v_1<v_2<v_3<v_4.
\]
 
Use the endpoint triples
\[
(y_0,v_0,a),\qquad
(y_1,v_1,b),\qquad
(y_2,v_2,c),\qquad
(y_3,v_3,d),\qquad
(y_4,v_4,e).
\]

\medskip
\noindent\textbf{Assembling Lorenz Curves with Analytical Estimates}
We verify that Proposition~\ref{convex-connector} applies on the four intervals
\[
[u_0,u_1],\qquad [u_1,u_2],\qquad [u_2,u_3],\qquad [u_3,u_4],
\]
with endpoint triples
\[
(y_0,v_0,a),\qquad
(y_1,v_1,b),\qquad
(y_2,v_2,c),\qquad
(y_3,v_3,d),\qquad
(y_4,v_4,e).
\]

On \([u_0,u_1]\), using \(z=A_0u\),
\[
\frac{y_1-y_0}{u_1-u_0}
=
\frac{z-\delta(A_0+\eta)}{u-\delta}
=
\frac{A_0(u-\delta)-\delta\eta}{u-\delta}
=
A_0-\frac{\delta\eta}{u-\delta}
<
A_0
=
v_1,
\]
and
\[
\frac{y_1-y_0}{u_1-u_0}>v_0
\iff
A_0-\frac{\delta\eta}{u-\delta}>A_0-\eta
\iff
\frac{\delta}{u-\delta}<1
\iff
\delta<\frac{u}{2},
\]
which holds because \(\delta<\delta_*\le u/2\).

On \([u_1,u_2]\), we have \(u_2-u_1=\delta\) and
\[
\frac{y_2-y_1}{u_2-u_1}
=
\frac{z-(z-\delta(A_0+\eta))}{\delta}
=
A_0+\eta,
\]
so
\[
v_1=A_0<A_0+\eta<\frac{A_0+A_1}{2}=v_2
\]
because \(\eta<\eta_*\le (A_1-A_0)/2\).

On \([u_2,u_3]\), we have \(u_3-u_2=\delta\) and
\[
\frac{y_3-y_2}{u_3-u_2}
=
\frac{z+\delta(A_1-\eta)-z}{\delta}
=
A_1-\eta,
\]
so
\[
v_2=\frac{A_0+A_1}{2}<A_1-\eta<v_3=A_1
\]
because \(\eta<\eta_*\le (A_1-A_0)/2\).

On \([u_3,u_4]\), using \(1-z=A_1(1-u)\),
\[
\frac{y_4-y_3}{u_4-u_3}
=
\frac{1-(z+\delta(A_1-\eta))}{1-u-\delta}
=
\frac{(1-z)-\delta A_1+\delta\eta}{1-u-\delta}
=
A_1+\frac{\delta\eta}{1-u-\delta}
>
A_1
=
v_3,
\]
and
\[
\frac{y_4-y_3}{u_4-u_3}<v_4=A_1+\eta
\iff
\frac{\delta}{1-u-\delta}<1
\iff
\delta<\frac{1-u}{2},
\]
which holds because \(\delta<\delta_*\le (1-u)/2\).

Now let's apply Proposition~\ref{convex-connector} on \([u_0,u_1]\) with endpoint triples
\[
(y_0,v_0,a),\qquad (y_1,v_1,b).
\]
There exists \(P_0\in C^2([u_0,u_1])\) such that
\[
P_0(u_0)=y_0,\qquad P_0'(u_0)=v_0,\qquad P_0''(u_0)=a,
\]
\[
P_0(u_1)=y_1,\qquad P_0'(u_1)=v_1,\qquad P_0''(u_1)=b,
\]
and
\[
P_0'(t)>0,\qquad P_0''(t)>0\qquad (t\in(u_0,u_1)).
\]

Apply Proposition~\ref{convex-connector} on \([u_1,u_2]\) with endpoint triples
\[
(y_1,v_1,b),\qquad (y_2,v_2,c).
\]
There exists \(P_1\in C^2([u_1,u_2])\) such that
\[
P_1(u_1)=y_1,\qquad P_1'(u_1)=v_1,\qquad P_1''(u_1)=b,
\]
\[
P_1(u_2)=y_2,\qquad P_1'(u_2)=v_2,\qquad P_1''(u_2)=c,
\]
and
\[
P_1'(t)>0,\qquad P_1''(t)>0\qquad (t\in(u_1,u_2)).
\]

Apply Proposition~\ref{convex-connector} on \([u_2,u_3]\) with endpoint triples
\[
(y_2,v_2,c),\qquad (y_3,v_3,d).
\]
There exists \(P_2\in C^2([u_2,u_3])\) such that
\[
P_2(u_2)=y_2,\qquad P_2'(u_2)=v_2,\qquad P_2''(u_2)=c,
\]
\[
P_2(u_3)=y_3,\qquad P_2'(u_3)=v_3,\qquad P_2''(u_3)=d,
\]
and
\[
P_2'(t)>0,\qquad P_2''(t)>0\qquad (t\in(u_2,u_3)).
\]

Apply Proposition~\ref{convex-connector} on \([u_3,u_4]\) with endpoint triples
\[
(y_3,v_3,d),\qquad (y_4,v_4,e).
\]
There exists \(P_3\in C^2([u_3,u_4])\) such that
\[
P_3(u_3)=y_3,\qquad P_3'(u_3)=v_3,\qquad P_3''(u_3)=d,
\]
\[
P_3(u_4)=y_4,\qquad P_3'(u_4)=v_4,\qquad P_3''(u_4)=e,
\]
and
\[
P_3'(t)>0,\qquad P_3''(t)>0\qquad (t\in(u_3,u_4)).
\]

Define \(L_{\delta,\eta}:[0,1]\to\mathbb R\) by
\[
L_{\delta,\eta}(t):=
\begin{cases}
P_0(t),& t\in[u_0,u_1],\\
P_1(t),& t\in[u_1,u_2],\\
P_2(t),& t\in[u_2,u_3],\\
P_3(t),& t\in[u_3,u_4].
\end{cases}
\]
At the gluing points,
\[
P_0(u_1)=y_1=P_1(u_1),\qquad
P_0'(u_1)=v_1=P_1'(u_1),\qquad
P_0''(u_1)=b=P_1''(u_1),
\]
\[
P_1(u_2)=y_2=P_2(u_2),\qquad
P_1'(u_2)=v_2=P_2'(u_2),\qquad
P_1''(u_2)=c=P_2''(u_2),
\]
\[
P_2(u_3)=y_3=P_3(u_3),\qquad
P_2'(u_3)=v_3=P_3'(u_3),\qquad
P_2''(u_3)=d=P_3''(u_3).
\]
Hence \(L_{\delta,\eta}\in C^2([0,1])\). Moreover,
\[
L_{\delta,\eta}(0)=0,\qquad L_{\delta,\eta}(1)=1,
\]
and
\[
L_{\delta,\eta}'(t)>0,\qquad L_{\delta,\eta}''(t)>0\qquad (t\in(0,1)).
\]
Therefore
\(
L_{\delta,\eta}\) belongs to \(\mathcal L^\circ
\). Since \(u_2=u\) and \(y_2=z\), the stitched Lorenz curve gives:
\begin{equation}\label{eq:Lu=z-step3} L_{\delta,\eta}(u)=L_{\delta,\eta}(u_2)=y_2=z. \end{equation}

Since $L_{\delta,\eta}''>0$ on $(0,1)$, the derivative $L_{\delta,\eta}'$ is strictly increasing on $[0,1]$, so
\[
\min_{s\in[0,1]}L_{\delta,\eta}'(s)=L_{\delta,\eta}'(0)=v_0=A_0-\eta>\frac{A_0}{2}.
\]
For convenience, denote
\[
f_{\delta,\eta}(t):=(L_{\delta,\eta}^{-1})'(t),\qquad t\in(0,1).
\]
Then, for every $t\in(0,1)$,
\begin{equation}\label{eq:uniform-bound-step3}
0<f_{\delta,\eta}(t)
=\frac{1}{L_{\delta,\eta}'(L_{\delta,\eta}^{-1}(t))}
\le \frac{1}{A_0-\eta}
<\frac{2}{A_0}.
\end{equation}

Fix $t\in(0,z)$ and set
\[
\delta_t^-:=\min\Bigl\{\delta_*,\frac{z-t}{A_1}\Bigr\}>0.
\]
If $(\delta,\eta)$ is admissible and $0<\delta<\delta_t^-$, then $\eta<\eta_*\le (A_1-A_0)/2$ implies $A_0+\eta<A_1$, and therefore
\[
y_1=z-\delta(A_0+\eta)>z-\delta A_1>t.
\]
Because $L_{\delta,\eta}$ is strictly increasing, $t<y_1=L_{\delta,\eta}(u_1)$ implies
\[
L_{\delta,\eta}^{-1}(t)<u_1.
\]
On $[0,u_1]$, the derivative $L_{\delta,\eta}'$ increases from $v_0=A_0-\eta$ to $v_1=A_0$, hence
\[
A_0-\eta<L_{\delta,\eta}'(L_{\delta,\eta}^{-1}(t))<A_0,
\]
and therefore
\[
\frac{1}{A_0}<f_{\delta,\eta}(t)<\frac{1}{A_0-\eta}.
\]
Consequently,
\[
0<f_{\delta,\eta}(t)-\frac{1}{A_0}
<\frac{1}{A_0-\eta}-\frac{1}{A_0}
=\frac{\eta}{A_0(A_0-\eta)}
\le \frac{2\eta}{A_0^2}.
\]
Since $\eta\to 0$ as $(\delta,\eta)\to(0,0)$ through admissible pairs, it follows that
\[
f_{\delta,\eta}(t)\longrightarrow \frac{1}{A_0}
\qquad\text{for every }t\in(0,z).
\]

Fix $t\in(z,1)$ and set
\[
\delta_t^+:=\min\Bigl\{\delta_*,\frac{t-z}{A_1}\Bigr\}>0.
\]
If $(\delta,\eta)$ is admissible and $0<\delta<\delta_t^+$, then
\[
y_3=z+\delta(A_1-\eta)\le z+\delta A_1<t,
\]
so strict monotonicity gives
\[
L_{\delta,\eta}^{-1}(t)>u_3.
\]
On $[u_3,1]$, the derivative $L_{\delta,\eta}'$ increases from $v_3=A_1$ to $v_4=A_1+\eta$, hence
\[
A_1<L_{\delta,\eta}'(L_{\delta,\eta}^{-1}(t))<A_1+\eta,
\]
and therefore
\[
\frac{1}{A_1+\eta}<f_{\delta,\eta}(t)<\frac{1}{A_1}.
\]
Consequently,
\[
0<\frac{1}{A_1}-f_{\delta,\eta}(t)
<\frac{1}{A_1}-\frac{1}{A_1+\eta}
=\frac{\eta}{A_1(A_1+\eta)}
\le \frac{\eta}{A_1^2}.
\]
Since $\eta\to 0$ as $(\delta,\eta)\to(0,0)$ through admissible pairs, it follows that
\[
f_{\delta,\eta}(t)\longrightarrow \frac{1}{A_1}
\qquad\text{for every }t\in(z,1).
\]

Apply \eqref{eq:measure-euler} to $L_{\delta,\eta}$. There exists $\lambda_{\delta,\eta}>0$ such that
\[
\mu=\lambda_{\delta,\eta}\,\sigma_{L_{\delta,\eta}}
\qquad\text{on }(0,1).
\]
Evaluating on $(0,u]$ and $(u,1)$ and using \eqref{eq:Lu=z-step3} together with strict monotonicity, we obtain
\[
\mathbf 1_{(0,u]}(L_{\delta,\eta}^{-1}(t))=\mathbf 1_{(0,z]}(t),
\qquad
\mathbf 1_{(u,1)}(L_{\delta,\eta}^{-1}(t))=\mathbf 1_{(z,1)}(t),
\]
hence
\[
\mu((0,u])=\lambda_{\delta,\eta}\int_{(0,z]} f_{\delta,\eta}(t)\,\nu(dt),
\qquad
\mu((u,1))=\lambda_{\delta,\eta}\int_{(z,1)} f_{\delta,\eta}(t)\,\nu(dt).
\]

At the distinguished point $t=z$, we have $L_{\delta,\eta}^{-1}(z)=u$ and
\[
L_{\delta,\eta}'(u)=v_2=\frac{A_0+A_1}{2},
\]
so
\[
f_{\delta,\eta}(z)=\frac{1}{L_{\delta,\eta}'(u)}=\frac{2}{A_0+A_1}.
\]
Therefore
\[
\int_{(0,z]} f_{\delta,\eta}(t)\,\nu(dt)
=
\int_{(0,z)} f_{\delta,\eta}(t)\,\nu(dt)
+\frac{2}{A_0+A_1}\,\nu(\{z\}).
\]
By the atomless property established in step 2, $\nu(\{z\})=0$, therefore:
\[
\int_{(0,z]} f_{\delta,\eta}(t)\,\nu(dt)
=
\int_{(0,z)} f_{\delta,\eta}(t)\,\nu(dt).
\]
Denote
\[
I^-_{\delta,\eta}:=\int_{(0,z)} f_{\delta,\eta}(t)\,\nu(dt),
\qquad
I^+_{\delta,\eta}:=\int_{(z,1)} f_{\delta,\eta}(t)\,\nu(dt).
\]
Then
\[
\mu((0,u])=\lambda_{\delta,\eta} I^-_{\delta,\eta},
\qquad
\mu((u,1))=\lambda_{\delta,\eta} I^+_{\delta,\eta},
\]
Solving equation by eliminating $\lambda_{\delta,\eta}$ gives:
\begin{equation}\label{eq:ratio-delta-eps}
\mu((0,u])\,I^+_{\delta,\eta}
=
\mu((u,1))\,I^-_{\delta,\eta}.
\end{equation}

Let $\varepsilon>0$ be arbitrary. Since
\[
(z-\gamma,z)\downarrow\varnothing,
\qquad
(z,z+\gamma)\downarrow\varnothing
\qquad(\gamma\downarrow 0),
\]
and $N=\nu((0,1))<\infty$, choose
\[
0<\gamma<\min\{z,1-z\}
\]
such that
\[
\frac{3}{A_0}\,\nu((z-\gamma,z))<\frac{\varepsilon}{2},
\qquad
\frac{3}{A_0}\,\nu((z,z+\gamma))<\frac{\varepsilon}{2}.
\]
Set
\[
\delta_0:=\min\Bigl\{\delta_*,\frac{\gamma}{A_1}\Bigr\}.
\]
Choose $\eta_0\in(0,\eta_*)$ so small that
\[
\frac{2\eta_0 N}{A_0^2}<\frac{\varepsilon}{2},
\qquad
\frac{\eta_0 N}{A_1^2}<\frac{\varepsilon}{2}.
\]
If
\[
0<\delta<\delta_0,
\qquad
0<\eta<\eta_0,
\]
then $(\delta,\eta)$ is admissible, and for every $t\in(0,z-\gamma]$ we have
\[
\delta<\frac{\gamma}{A_1}\le \frac{z-t}{A_1},
\]
so the preceding left-hand estimate yields
\[
\left|f_{\delta,\eta}(t)-\frac{1}{A_0}\right|
\le \frac{2\eta}{A_0^2}
\qquad (t\in(0,z-\gamma]).
\]
Likewise, for every $t\in[z+\gamma,1)$ we have
\[
\delta<\frac{\gamma}{A_1}\le \frac{t-z}{A_1},
\]
so the preceding right-hand estimate yields
\[
\left|f_{\delta,\eta}(t)-\frac{1}{A_1}\right|
\le \frac{\eta}{A_1^2}
\qquad (t\in[z+\gamma,1)).
\]

Using these bounds together with \eqref{eq:uniform-bound-step3}, we obtain
\[
\begin{aligned}
\left|I^-_{\delta,\eta}-\frac{1}{A_0}\nu((0,z))\right|
&\le
\int_{(0,z-\gamma]}
\left|f_{\delta,\eta}(t)-\frac{1}{A_0}\right|\nu(dt)
+
\int_{(z-\gamma,z)}
\left|f_{\delta,\eta}(t)-\frac{1}{A_0}\right|\nu(dt)
\\
&\le
\frac{2\eta}{A_0^2}\,\nu((0,z-\gamma])
+
\left(\frac{2}{A_0}+\frac{1}{A_0}\right)\nu((z-\gamma,z))
\\
&\le
\frac{2\eta_0 N}{A_0^2}
+
\frac{3}{A_0}\nu((z-\gamma,z))
<\varepsilon.
\end{aligned}
\]
Similarly,
\[
\begin{aligned}
\left|I^+_{\delta,\eta}-\frac{1}{A_1}\nu((z,1))\right|
&\le
\int_{(z,z+\gamma)}
\left|f_{\delta,\eta}(t)-\frac{1}{A_1}\right|\nu(dt)
+
\int_{[z+\gamma,1)}
\left|f_{\delta,\eta}(t)-\frac{1}{A_1}\right|\nu(dt)
\\
&\le
\left(\frac{2}{A_0}+\frac{1}{A_1}\right)\nu((z,z+\gamma))
+
\frac{\eta}{A_1^2}\,\nu([z+\gamma,1))
\\
&\le
\frac{3}{A_0}\nu((z,z+\gamma))
+
\frac{\eta_0 N}{A_1^2}
<\varepsilon.
\end{aligned}
\]
Hence, given $(\delta,\eta)$ are governed by the admissible quantification, when they tend to 0, we have: 
\[
\lim_{\substack{(\delta,\eta)\to(0,0)}}
I^-_{\delta,\eta}
=
\frac{1}{A_0}\nu((0,z)),
\quad
\lim_{\substack{(\delta,\eta)\to(0,0)}}
I^+_{\delta,\eta}
=
\frac{1}{A_1}\nu((z,1)).
\]
 
Recall that $\nu(\{z\})=0$, passing to the limit in \eqref{eq:ratio-delta-eps} gives
\[
\mu((0,u])\cdot \frac{1}{A_1}\,\nu((z,1))
=
\mu((u,1))\cdot \frac{1}{A_0}\,\nu((0,z))+\mu((u,1))\cdot \frac{1}{A_0}\,\nu(\{z\}).
\]
Therefore:
\begin{equation}\label{eq:tail-cross}
\mu((0,u])\cdot \frac{1}{A_1}\,\nu((z,1))
=
\mu((u,1))\cdot \frac{1}{A_0}\,\nu((0,z]).
\end{equation}
Since $0\le \mu((0,u])\le M$, $0\le \mu((u,1))\le M$, $0\le \nu((0,z])\le N$, and $0\le \nu((z,1))\le N$, all four tail terms in \eqref{eq:tail-cross} are finite. 

In next step, we would justify the claims that
\[
\mu((0,u])>0,\qquad \mu((u,1))>0,\qquad \nu((0,z])>0,\qquad \nu((z,1))>0,
\]
So that division operation against \eqref{eq:tail-cross} is legitimate. Once this is done, \eqref{eq:tail-cross} becomes desired universal ratio identity:
\begin{equation}\label{eq:ratio-final}
\frac{\mu((0,u])}{\mu((u,1))}
=
\frac{A_1}{A_0}\cdot \frac{\nu((0,z])}{\nu((z,1))}
=
\frac{u(1-z)}{z(1-u)}\cdot \frac{\nu((0,z])}{\nu((z,1))}
\qquad\forall\,0<z<u<1.
\end{equation}

\medskip
\noindent\textbf{Step 4: Forms of Measure Distribution Function $\mu$ and $\nu$.}

We now analyze \eqref{eq:ratio-delta-eps}--\eqref{eq:ratio-final}.

\smallskip
\emph{Step 4.1. Strict positivity of all interior tails.}

\smallskip
\emph{Step 4.1. Strict positivity of all interior tails.}

Set
\[
M:=\mu([0,1]),\qquad N:=\nu([0,1]).
\]
By Step~1, the measures $\mu$ and $\nu$ are finite and nonzero on $[0,1]$, and moreover
\[
\mu(\{0\})=\mu(\{1\})=\nu(\{0\})=\nu(\{1\})=0.
\]
Hence
\[
0<M=\mu((0,1))=\mu([0,1])<\infty,
\qquad
0<N=\nu((0,1))=\nu([0,1])<\infty.
\]

By Step~2, both $\mu$ and $\nu$ are atomless on $(0,1)$. Therefore, for every $u,z\in(0,1)$,
\[
\mu(\{u\})=0,\qquad \nu(\{z\})=0.
\]
Consequently, since
\[
(0,1)=(0,u]\sqcup (u,1),
\qquad
(0,1)=(0,z]\sqcup (z,1),
\]
we obtain
\begin{equation}\label{eq:split-total-mass-step41}
M=\mu((0,u])+\mu((u,1)),
\qquad
N=\nu((0,z])+\nu((z,1))
\qquad (u,z\in(0,1)).
\end{equation}

Recall from Step~3 that for every pair $(z,u)$ with $0<z<u<1$ one has
\begin{equation}\label{eq:tail-cross-explicit-step41}
\mu((0,u])\frac{1-u}{1-z}\,\nu((z,1))
=
\mu((u,1))\frac{u}{z}\,\nu((0,z]).
\end{equation}

We shall prove:
\[
\mu((u,1))>0,\qquad \mu((0,u])>0 \qquad \forall\,u\in(0,1),
\]
\[
\nu((0,z])>0,\qquad \nu((z,1))>0 \qquad \forall\,z\in(0,1).
\]

\noindent\textbf{Claim 1.} For every \(u\in(0,1)\), \(\mu((u,1))>0\).

\smallskip
\noindent\emph{Proof.}
Fix \(u\in(0,1)\) and suppose, towards a contradiction, that \(\mu((u,1))=0\). Then \eqref{eq:split-total-mass-step41} gives \(\mu((0,u])=M>0\).

For each \(n\in\mathbb N\), define \(z_n:=\frac{u}{n+1}\). Then \(0<z_n<u\) and \(z_n\downarrow0\). Applying \eqref{eq:tail-cross-explicit-step41} with \(z=z_n\) yields
\[
\mu((0,u])\frac{1-u}{1-z_n}\,\nu((z_n,1))
=
\mu((u,1))\frac{u}{z_n}\,\nu((0,z_n])=0.
\]
Since \(\mu((0,u])=M>0\) and \((1-u)/(1-z_n)>0\), it follows that
\[
\nu((z_n,1))=0\qquad\forall n\in\mathbb N.
\]

We now verify carefully that \((z_n,1)\uparrow(0,1)\). Indeed, because \(z_n\downarrow0\), the intervals are increasing:
\[
(z_1,1)\subset (z_2,1)\subset\cdots.
\]
Moreover, if \(t\in(0,1)\), then \(z_n\to0\), so there exists \(n\) large enough that \(z_n<t\); hence \(t\in(z_n,1)\). Thus
\[
\bigcup_{n=1}^\infty (z_n,1)=(0,1).
\]

By continuity from below of the finite measure \(\nu\),
\[
N
=\nu((0,1))
=\nu\!\left(\bigcup_{n=1}^\infty (z_n,1)\right)
=\lim_{n\to\infty}\nu((z_n,1))
=0,
\]
contradicting \(N>0\).

Therefore \(\mu((u,1))>0\) for every \(u\in(0,1)\).
\hfill$\square$

\medskip
\noindent\textbf{Claim 2.} For every \(u\in(0,1)\), \(\mu((0,u])>0\).

\smallskip
\noindent\emph{Proof.}
Fix \(u\in(0,1)\) and suppose, towards a contradiction, that \(\mu((0,u])=0\). Then \eqref{eq:split-total-mass-step41} gives \(\mu((u,1))=M>0\).

For each \(n\in\mathbb N\), define
\[
z_n:=u\left(1-\frac{1}{n+1}\right)=\frac{nu}{n+1}.
\]
Then \(0<z_n<u\) and \(z_n\uparrow u\). Applying \eqref{eq:tail-cross-explicit-step41} with \(z=z_n\) yields
\[
0
=
\mu((0,u])\frac{1-u}{1-z_n}\,\nu((z_n,1))
=
\mu((u,1))\frac{u}{z_n}\,\nu((0,z_n]).
\]
Since \(\mu((u,1))=M>0\) and \(u/z_n>0\), it follows that
\[
\nu((0,z_n])=0\qquad\forall n\in\mathbb N.
\]

We now verify  that \((0,z_n]\uparrow(0,u)\). Indeed, because \(z_n\uparrow u\), the sets are increasing:
\[
(0,z_1]\subset (0,z_2]\subset\cdots.
\]
If \(t\in(0,u)\), then \(z_n\to u\), so there exists \(n\) large enough that \(t\le z_n\); hence \(t\in(0,z_n]\). Thus
\[
\bigcup_{n=1}^\infty (0,z_n]=(0,u).
\]

By continuity from below of \(\nu\),
\[
\nu((0,u))
=
\nu\!\left(\bigcup_{n=1}^\infty (0,z_n]\right)
=
\lim_{n\to\infty}\nu((0,z_n])
=
0.
\]
Since \(\nu(\{u\})=0\) by Step~2, we obtain
\[
\nu((0,u])=\nu((0,u))+\nu(\{u\})=0.
\]
Hence, by \eqref{eq:split-total-mass-step41}, \(\nu((u,1))=N>0\).

Now let \(u'\in(u,1)\) be arbitrary. Since \(0<u<u'<1\), we may apply \eqref{eq:tail-cross-explicit-step41} with the pair \((z,u)=(u,u')\), obtaining
\[
\mu((0,u'])\frac{1-u'}{1-u}\,\nu((u,1))
=
\mu((u',1))\frac{u'}{u}\,\nu((0,u]).
\]
But we have already shown that \(\nu((0,u])=0\), so the right-hand side is \(0\). Because \((1-u')/(1-u)>0\) and \(\nu((u,1))>0\), it follows that
\[
\mu((0,u'])=0\qquad\forall\,u'\in(u,1).
\]

Now choose the sequence
\(
u'_n:=1-\frac{1-u}{n+1}\qquad(n\in\mathbb N)
\). Then \((u'_n)_{n\ge1}\subset(u,1)\) and \(u'_n\uparrow1\). Hence the sets \((0,u'_n]\) are increasing and
\[
\bigcup_{n=1}^\infty (0,u'_n]=(0,1).
\]
Since \(\mu((0,u'_n])=0\) for every \(n\), continuity from below gives
\[
M
=
\mu((0,1))
=
\mu\!\left(\bigcup_{n=1}^\infty (0,u'_n]\right)
=
\lim_{n\to\infty}\mu((0,u'_n])
=
0,
\]
contradicting \(M>0\).

Therefore \(\mu((0,u])>0\) for every \(u\in(0,1)\).
\hfill$\square$

\medskip
\noindent\textbf{Claim 3.} For every $z\in(0,1)$, $\nu((0,z])>0$.

\smallskip
\noindent\emph{Proof.}
Fix $z\in(0,1)$, and choose any $u\in(z,1)$. By Claims~1 and~2, $\mu((0,u])>0$ and $\mu((u,1))>0$.

Suppose, towards a contradiction, that $\nu((0,z])=0$. Then \eqref{eq:tail-cross-explicit-step41} gives
\[
\mu((0,u])\frac{1-u}{1-z}\,\nu((z,1))
=
\mu((u,1))\frac{u}{z}\,\nu((0,z])=0.
\]
Since $\mu((0,u])>0$ and $(1-u)/(1-z)>0$, it follows that $\nu((z,1))=0$. Therefore, using \eqref{eq:split-total-mass-step41}, we get $N=\nu((0,z])+\nu((z,1))=0+0=0$, contradicting $N>0$.

Hence $\nu((0,z])>0$ for every $z\in(0,1)$.
\hfill$\square$

\medskip
\noindent\textbf{Claim 4.} For every $z\in(0,1)$, $\nu((z,1))>0$.

\smallskip
\noindent\emph{Proof.}
Fix $z\in(0,1)$, and choose any $u\in(z,1)$. By Claim~1, $\mu((u,1))>0$, and by Claim~3, $\nu((0,z])>0$.

Suppose, towards a contradiction, that $\nu((z,1))=0$. Then \eqref{eq:tail-cross-explicit-step41} yields
\[
0
=
\mu((0,u])\frac{1-u}{1-z}\,\nu((z,1))
=
\mu((u,1))\frac{u}{z}\,\nu((0,z]).
\]
Since $\mu((u,1))>0$, $u/z>0$, and $\nu((0,z])>0$, the right-hand side is strictly positive, a contradiction.

Therefore $\nu((z,1))>0$ for every $z\in(0,1)$.
\hfill$\square$

\medskip
Combining Claims~1--4, we conclude that for every $u\in(0,1)$ and every $z\in(0,1)$,
\[
0<\mu((0,u])<M,\qquad 0<\mu((u,1))<M,
\]
\[
0<\nu((0,z])<N,\qquad 0<\nu((z,1))<N.
\]
So, all ratios appearing in \eqref{eq:ratio-final} are well-defined and strictly positive.

\smallskip
\emph{(4.2) Constancy of \(K\) and identification of \(\mu\) and \(\nu\) on \((0,1)\).}
Define, for \(u\in(0,1)\),
\[
r(u):=\frac{1-u}{u}\cdot \frac{\mu((0,u])}{\mu((u,1))},
\qquad
s(u):=\frac{1-u}{u}\cdot \frac{\nu((0,u])}{\nu((u,1))}.
\]
By Step~4.1, all four tails are strictly positive, so \(r(u)\) and \(s(u)\) are well-defined and strictly positive on \((0,1)\).
Equation \eqref{eq:ratio-final} is exactly the statement that
\[
r(u)=s(z)\qquad\forall\,0<z<u<1.
\]

Fix \(0<z_1<z_2<1\). For any \(u\in(z_2,1)\), we have \(z_1<u\) and \(z_2<u\), hence
\[
s(z_1)=r(u)=s(z_2).
\]
Therefore \(s\) is constant on \((0,1)\). Denote this constant by \(K>0\).

Now fix \(u\in(0,1)\) and choose any \(z\in(0,u)\). Then
\[
r(u)=s(z)=K.
\]
Thus \(r\) is also constant on \((0,1)\), with the same constant \(K\). Equivalently,
\begin{equation}\label{eq:tail-ratio-K}
\frac{\mu((0,u])}{\mu((u,1))}=K\,\frac{u}{1-u},
\qquad
\frac{\nu((0,u])}{\nu((u,1))}=K\,\frac{u}{1-u}
\qquad(u\in(0,1)).
\end{equation}

Set
\[
M:=\mu((0,1))\in(0,\infty),
\qquad
N:=\nu((0,1))\in(0,\infty).
\]
Since \(\mu\) and \(\nu\) are atomless on \((0,1)\), for every \(u\in(0,1)\),
\[
M=\mu((0,u])+\mu((u,1)),
\qquad
N=\nu((0,u])+\nu((u,1)).
\]
Solving \eqref{eq:tail-ratio-K} for the left tails gives
\begin{equation}\label{eq:muCDF-step43}
\mu((0,u])=M\,\frac{Ku}{1+(K-1)u},
\qquad
\nu((0,u])=N\,\frac{Ku}{1+(K-1)u}
\qquad(u\in(0,1)).
\end{equation}

Then \(m\) and \(n\) are continuous and strictly positive on \((0,1)\). Moreover, for every \(0\le a<b<1\),
\[
\mu((a,b])=\mu((0,b])-\mu((0,a]) =\int_a^b m(u)\,du, \quad 
\nu((a,b])=\nu((0,b])-\nu((0,a]) =\int_a^b n(t)\,dt.
\]
Let
\(
\mathcal P:=\{\varnothing\}\cup\{(a,b]:\,0\le a<b<1\}.
\)
Then \(\mathcal P\) is a \(\pi\)-system generating \(\mathcal B((0,1))\), and the two finite Borel measures
\[
A\longmapsto \mu(A),
\qquad
A\longmapsto \int_A m(u)\,du
\]
agree on \(\mathcal P\). By uniqueness of finite measures,
\begin{align*}
\mu(A)&=\int_A m(u)\,du
\qquad(A\in\mathcal B((0,1))).\\
\nu(B)&=\int_B n(t)\,dt
\qquad(B\in\mathcal B((0,1))).
\end{align*}
Equivalently,
\[
\mu(du)=m(u)\,du,
\qquad
\nu(dt)=n(t)\,dt
\qquad\text{on }(0,1).
\]

Now fix \(L\in\mathcal L^\circ\). From Step~1, there exists \(\lambda(L)>0\) such that
\[
\mu=\lambda(L)\,\sigma_L
\qquad\text{on }(0,1),
\]
where
\[
\sigma_L(A):=
\int_{(0,1)}\mathbf 1_A(L^{-1}(t))(L^{-1})'(t)\,\nu(dt)
\qquad(A\in\mathcal B((0,1))).
\]
Using \(\nu(dt)=n(t)\,dt\), the change of variables \(t=L(u)\), and
\(
(L^{-1})'(L(u))\,L'(u)=1
\). Therefore, for every Borel set \(A\subset(0,1)\):
\begin{align*}
\sigma_L(A)
&=
\int_{(0,1)}\mathbf 1_A(L^{-1}(t))(L^{-1})'(t)\,n(t)\,dt\\
&=
\int_{(0,1)}\mathbf 1_A(u)\,(L^{-1})'(L(u))\,n(L(u))\,L'(u)\,du\\
&=
\int_A n(L(u))\,du.
\end{align*}
Thus for every \(0<a<b<1\),
\[
\int_a^b m(u)\,du
=
\mu((a,b])
=
\lambda(L)\,\sigma_L((a,b])
=
\lambda(L)\int_a^b n(L(u))\,du.
\]
Define
\(
f_L(u):=m(u)-\lambda(L)\,n(L(u))
\qquad(u\in(0,1)).
\)
Since \(m\) and \(n\) are continuous on \((0,1)\) and \(L\in\mathcal L^\circ\), the function
\(f_L\) is continuous on \((0,1)\). Moreover, for every \(0<a<b<1\),
\(
\int_a^b f_L(u)\,du=0
\).
We claim that \(f_L\equiv 0\) on \((0,1)\). Suppose not. Then there exists \(u_0\in(0,1)\)
with \(f_L(u_0)\neq 0\).

If \(f_L(u_0)>0\), continuity yields \(\delta>0\) such that
\[
(u_0-\delta,u_0+\delta)\subset(0,1)
\quad\text{and}\quad
f_L(u)\ge \frac{f_L(u_0)}{2}>0
\qquad\forall u\in(u_0-\delta,u_0+\delta).
\]
Hence
\[
\int_{u_0-\delta}^{u_0+\delta} f_L(u)\,du
\ge
\int_{u_0-\delta}^{u_0+\delta} \frac{f_L(u_0)}{2}\,du
=
\delta f_L(u_0)>0,
\]
contradicting \(\int_a^b f_L(u)\,du=0\) for all \(0<a<b<1\).

If \(f_L(u_0)<0\), the same argument applied to \(-f_L\) gives a contradiction.

Therefore \(f_L(u)=0\) for all \(u\in(0,1)\), it gives:
\begin{equation}\label{eq:density-EL-step43}
m(u)=\lambda(L)\,n(L(u))
\qquad(u\in(0,1)).
\end{equation}

Choose any \(0<y_1<y_2<1\), and define
\begin{equation}\label{eq:R-def-step43}
R:=\frac{n(y_1)}{n(y_2)}>0.
\end{equation}

Fix \(u_2\in(y_2,1)\), and define
\[
\theta(u_2):=\frac{y_1}{y_2}u_2,
\qquad
\phi(u_2):=u_2-\frac{y_2-y_1}{1-y_2}(1-u_2),
\qquad
I(u_2):=(\theta(u_2),\phi(u_2)).
\]
Let
\[
A:=\frac{y_2-y_1}{1-y_2}>0.
\]
Then
\[
\phi(u_2)=(1+A)u_2-A,
\qquad
\theta(u_2)=\frac{y_1}{y_2}u_2.
\]
Thus \(\theta\) and \(\phi\) are continuous and strictly increasing, and
\[
\theta(y_2)=\phi(y_2)=y_1.
\]
Also,
\[
\phi'(u_2)-\theta'(u_2)
=
(1+A)-\frac{y_1}{y_2}
=
\frac{y_2-y_1}{y_2(1-y_2)}>0.
\]
Hence
\[
\phi(u_2)>\theta(u_2)
\qquad\forall\,u_2>y_2,
\]
so \(I(u_2)\) is a nonempty open interval.

We claim that, for any \(u_1\in I(u_2)\),
\begin{equation}\label{eq:secant-chain-step43}
\frac{y_1}{u_1}
<
\frac{y_2-y_1}{u_2-u_1}
<
\frac{1-y_2}{1-u_2}.
\end{equation}
Indeed,
\[
\frac{y_1}{u_1}<\frac{y_2-y_1}{u_2-u_1}
\iff
y_1(u_2-u_1)<u_1(y_2-y_1)
\iff
y_1u_2<y_2u_1
\iff
\theta(u_2)<u_1,
\]
and
\[
\frac{y_2-y_1}{u_2-u_1}<\frac{1-y_2}{1-u_2}
\iff
(y_2-y_1)(1-u_2)<(1-y_2)(u_2-u_1)
\iff
u_1<\phi(u_2).
\]
Thus \eqref{eq:secant-chain-step43} holds exactly because \(u_1\in I(u_2)\).

Now fix \(u_2\in(y_2,1)\) and \(u_1\in I(u_2)\), and set
\[
A_0:=\frac{y_1}{u_1},
\qquad
A_1:=\frac{y_2-y_1}{u_2-u_1},
\qquad
A_2:=\frac{1-y_2}{1-u_2}.
\]
Then \eqref{eq:secant-chain-step43} gives
\[
0<A_0<A_1<A_2.
\]
Choose slopes
\[
v_0:=\frac12A_0,
\qquad
v_1:=\frac12(A_0+A_1),
\qquad
v_2:=\frac12(A_1+A_2),
\qquad
v_3:=A_2+1,
\]
and choose arbitrary
\[
s_0,s_1,s_2,s_3>0.
\]
Then
\[
v_0<A_0<v_1,
\qquad
v_1<A_1<v_2,
\qquad
v_2<A_2<v_3.
\]
Hence Proposition~\ref{convex-connector} applies successively on the intervals
\[
[0,u_1],\quad [u_1,u_2],\quad [u_2,1],
\]
with endpoint values
\[
(0,v_0,s_0),\qquad
(y_1,v_1,s_1),\qquad
(y_2,v_2,s_2),\qquad
(1,v_3,s_3).
\]
After gluing the three pieces, we obtain a curve
\(
L_{u_1,u_2}
\) that satisfy the conditions of $\mathcal L^\circ$
such that:
\[
L_{u_1,u_2}(u_1)=y_1,
\qquad
L_{u_1,u_2}(u_2)=y_2.
\]

Applying \eqref{eq:density-EL-step43} to \(L=L_{u_1,u_2}\) at the points \(u_1\) and \(u_2\), we obtain
\[
m(u_1)=\lambda(L_{u_1,u_2})\,n(y_1),
\qquad
m(u_2)=\lambda(L_{u_1,u_2})\,n(y_2).
\]
Eliminating \(\lambda(L_{u_1,u_2})\) gives
\begin{equation}\label{eq:m-relation-step43}
m(u_1)=R\,m(u_2)
\qquad
\forall\,u_2\in(y_2,1),\ \forall\,u_1\in I(u_2).
\end{equation}

Fix \(u_2\in(y_2,1)\), and define
\[
u_M:=\frac{\theta(u_2)+\phi(u_2)}{2}\in I(u_2).
\]
Since \(u_M-\theta(u_2)>0\) and \(\phi(u_2)-u_M>0\), continuity of \(\theta\) and \(\phi\) yields \(\delta>0\) such that
\[
|u_2'-u_2|<\delta
\quad\Longrightarrow\quad
\theta(u_2')<u_M<\phi(u_2'),
\]
that is,
\[
u_M\in I(u_2')
\qquad\text{whenever }|u_2'-u_2|<\delta.
\]
Applying \eqref{eq:m-relation-step43} with \(u_1=u_M\), first for \(u_2\), and then for such \(u_2'\), gives
\[
m(u_M)=R\,m(u_2),
\qquad
m(u_M)=R\,m(u_2').
\]
Since \(R>0\), it gives:
\[
m(u_2)=m(u_2')
\qquad\text{whenever }|u_2'-u_2|<\delta.
\]
Thus \(m\) is locally constant at every point of \((y_2,1)\). Since \(m\) is continuous and \((y_2,1)\) is connected,
\(m\) is constant on \((y_2,1)\). Therefore there exists \(C>0\) such that
\begin{equation}\label{eq:m-constant-tail-step43}
m(u)=C
\qquad(u\in(y_2,1)).
\end{equation}

Now fix \(u\in(y_2,1)\). Since \(\mu(du)=m(u)\,du\) on \((0,1)\), \eqref{eq:m-constant-tail-step43} gives
\[
\mu((u,1))
=
\int_u^1 m(s)\,ds
=
C(1-u).
\]
Because
\[
M=\mu((0,u])+\mu((u,1)),
\]
we obtain
\[
\mu((0,u])=M-C(1-u).
\]
Substituting these expressions into the first identity of \eqref{eq:tail-ratio-K}, we find
\[
\frac{M-C(1-u)}{C(1-u)}=K\frac{u}{1-u}
\qquad(u\in(y_2,1)).
\]
Multiplying by \(1-u\) yields
\[
\frac{M}{C}-1=(K-1)u
\qquad(u\in(y_2,1)).
\]
The left-hand side is independent of \(u\), whereas the right-hand side is affine in \(u\). Since \(y_2<1\), the interval \((y_2,1)\) contains more than one point, so necessarily
\[
K=1.
\]

Returning to \eqref{eq:muCDF-step43}, we obtain
\[
\mu((0,u])=Mu,
\qquad
\nu((0,u])=Nu
\qquad(u\in(0,1)).
\]
Hence, for every \(0\le a<b<1\),
\[
\mu((a,b])=M(b-a),
\qquad
\nu((a,b])=N(b-a).
\]
By the same uniqueness argument on the \(\pi\)-system \(\mathcal P\), it follows that
\[
\mu(du)=M\,du,
\qquad
\nu(dt)=N\,dt
\qquad\text{on }(0,1).
\]

Finally, substituting these densities into \eqref{eq:density-EL-step43} gives
\[
M=\lambda(L)\,N
\qquad\forall\,L\in\mathcal L^\circ,
\]
and therefore
\[
\lambda(L)=\frac{M}{N}
\qquad\forall\,L\in\mathcal L^\circ.
\]
Thus the proportionality factor $\lambda$ in \eqref{eq:measure-euler} is independent of \(L\).

\medskip
\noindent\textbf{Step 5: Recovery of the kernels and completion of the proof.}

By Step~4, there exist constants \(M,N>0\) such that
\[
\mu(du)=M\,du,
\qquad
\nu(dt)=N\,dt
\qquad\text{on }(0,1).
\]
Moreover, Step~1 established
\[
\mu(\{0\})=\mu(\{1\})=\nu(\{0\})=\nu(\{1\})=0.
\]
Hence, for every \(u,t\in[0,1]\),
\[
\mu([0,u])=Mu,
\qquad
\nu([0,t])=Nt.
\]
Indeed, for \(0<u<1\),
\[
\mu([0,u])=\mu(\{0\})+\mu((0,u])=0+Mu=Mu,
\]
and similarly \(\nu([0,t])=Nt\) for \(0<t<1\); the cases \(u=0\) and \(t=0\) are immediate, while
\[
\mu([0,1])=\lim_{u\uparrow1}\mu([0,u])=M,
\qquad
\nu([0,1])=\lim_{t\uparrow1}\nu([0,t])=N
\]
by continuity from below.

From Step~1, we already know that
\[
\mu((u,1])=p(u)-p(1)
\qquad(0\le u\le1),
\]
and
\[
\nu([0,t])=q(t)-q(0)
\qquad(0\le t\le1).
\]
Therefore, for every \(u,t\in[0,1]\),
\[
p(u)=p(1)+\mu((u,1])=p(1)+M(1-u),
\]
\[
q(t)=q(0)+\nu([0,t])=q(0)+Nt.
\]
Define
\[
\beta_0:=p(1),\qquad \beta_1:=M,
\qquad
\gamma_0:=q(0),\qquad \gamma_1:=N.
\]
Since \(M,N>0\), we have
\(
\beta_1,\gamma_1>0
\). Thus
\[
p(u)=\beta_0+\beta_1(1-u)
\qquad(u\in[0,1]),
\quad
q(t)=\gamma_0+\gamma_1 t
\qquad(t\in[0,1]).
\]
This proves the affine forms of both kernels functions.

\smallskip
\noindent\emph{Primal functional.}
Step~1 showed that, for every \(L\in\mathcal L\),
\[
V(L)=\bigl(c_1+p(1)\bigr)+\int_{[0,1]}L(u)\,\mu(du).
\]
Since \(\mu(du)=M\,du\) on \([0,1]\), this becomes
\[
V(L)=\bigl(c_1+\beta_0\bigr)+\beta_1\int_0^1 L(u)\,du
\qquad(L\in\mathcal L).
\]
Define
\[
\beta_2:=c_1+\beta_0.
\]
Recalling the notation
\[
A(L):=\int_0^1 L(u)\,du,
\]
we obtain
\[
V(L)=\beta_2+\beta_1A(L)
\qquad(L\in\mathcal L).
\]

\smallskip
\noindent\emph{Dual functional.}
Step~1 also showed that, for every \(L\in\mathcal L^\uparrow\),
\[
V^*(L)=\bigl(c_2+q(1)\bigr)-\int_{[0,1]}L^{-1}(t)\,\nu(dt).
\]
Since \(\nu(dt)=N\,dt\) on \([0,1]\), this gives
\[
V^*(L)=c_2+q(1)-N\int_0^1 L^{-1}(t)\,dt
\qquad(L\in\mathcal L^\uparrow).
\]

We next prove the standard area identity (which is obviously graphically):
\begin{equation}\label{eq:area-inverse-identity}
\int_0^1 L^{-1}(t)\,dt
=
1-\int_0^1 L(u)\,du
\qquad(L\in\mathcal L^\uparrow).
\end{equation}
Fix \(L\in\mathcal L^\uparrow\). Since \(L\) is a strictly increasing homeomorphism of \([0,1]\),
\[
u\le L^{-1}(t)\iff L(u)\le t
\qquad(u,t\in[0,1]).
\]
Therefore, by Tonelli's theorem,
\begin{align*}
\int_0^1 L^{-1}(t)\,dt
&=
\int_0^1\int_0^{L^{-1}(t)}1\,du\,dt\\
&=
\int_0^1\int_{L(u)}^1 1\,dt\,du\\
&=
\int_0^1 \bigl(1-L(u)\bigr)\,du\\
&=
1-\int_0^1 L(u)\,du.
\end{align*}
This proves \eqref{eq:area-inverse-identity}.

Substituting \eqref{eq:area-inverse-identity} into the dual representation yields
\[
V^*(L)
=
c_2+q(1)-N\bigl(1-A(L)\bigr)
=
\bigl(c_2+q(1)-N\bigr)+N A(L).
\]
Since \(q(t)=q(0)+Nt\), we have
\[
q(1)=q(0)+N=\gamma_0+\gamma_1,
\]
and hence
\[
q(1)-N=q(0)=\gamma_0.
\]
Therefore
\[
V^*(L)=\bigl(c_2+\gamma_0\bigr)+\gamma_1A(L)
\qquad(L\in\mathcal L^\uparrow).
\]
Define
\[
\gamma_2:=c_2+\gamma_0.
\]
Then
\[
V^*(L)=\gamma_2+\gamma_1A(L)
\qquad(L\in\mathcal L^\uparrow).
\]

Combining the primal and dual formulas, we have proved
\[
V(L)=\beta_2+\beta_1A(L)
\qquad(L\in\mathcal L),
\]
\[
V^*(L)=\gamma_2+\gamma_1A(L)
\qquad(L\in\mathcal L^\uparrow),
\]
with \(\beta_1,\gamma_1>0\).

Finally, since \(V\) represents \(\succsim\) on \(\mathcal L\) and \(\beta_1>0\), for all \(L_1,L_2\in\mathcal L\),
\[
L_1\succsim L_2
\iff
V(L_1)\ge V(L_2)
\iff
A(L_1)\ge A(L_2).
\]
Thus \(\succsim\) coincides with the Lorenz-area order on \(\mathcal L\). Equivalently, if
\[
J(L):=1-2A(L)
\]
denotes the Gini coefficient, then
\[
L_1\succsim L_2
\iff
J(L_1)\le J(L_2).
\]
This completes the proof.

\subsubsection{No Leveling-Down}
Having just established an axiomatic characterization of the Gini index as the inequality measure
$F_{Egal}$, we are one step away froma coherent egalitarian SWF! A natural requirement that many
might find appealing is $F_{aggregate}(X) = [1 -\lambda G(X)]$ . $F_{util}(X)$ should always be increasing. To be
more specific, the objective is to ensure added Penalization 'score' due to inequality should never outweigh the gain in utility, making the contribution of increasing individual utility xi negative to $F_{aggregate}(X)$.

\begin{theorem}[No Leveling Down Theorem]
Fix an integer $n\ge 2$. Let $X=(x_1,\dots,x_n)$ satisfy $0\le x_1\le \cdots \le x_n$ and set $S:=\sum_{i=1}^n x_i>0$. Define
\[
F_{\mathrm{util}}(X):=\sum_{i=1}^n x_i=S,\qquad 
G(X):=\frac{1}{2nS}\sum_{i=1}^n\sum_{j=1}^n |x_i-x_j|,
\qquad
F_{\mathrm{aggregate}}(X):=\bigl(1-\lambda G(X)\bigr)S,
\]
where $\lambda\in \mathbb{R}_+:=[0,\infty)$. Motivated by the requirement that the inequality penalty never outweigh the utility gain, we seek to find the exact $\lambda$ for which $F_{\mathrm{aggregate}}$ in terms of utility profile:
\[
\frac{\partial F_{\mathrm{aggregate}}}{\partial x_k}(X)\ge 0\quad\text{for all }k\in\{1,\dots,n\}.
\]
Then $F_{\mathrm{aggregate}}$ is coordinatewise nondecreasing if and only if
\[
\lambda\le \frac{n}{n-1}.
\]
\end{theorem}

\begin{proof}

Since $x_1\le \cdots \le x_n$, for $i<j$ we have $|x_i-x_j|=x_j-x_i$. Hence
\begin{align*}
\sum_{i=1}^n\sum_{j=1}^n |x_i-x_j|
&=2\sum_{j=1}^n\sum_{i=1}^{j-1}(x_j-x_i)
=2\sum_{j=1}^n\Bigl((j-1)x_j-\sum_{i=1}^{j-1}x_i\Bigr).
\end{align*}
Interchanging the order in the double sum,
\[
\sum_{j=1}^n\sum_{i=1}^{j-1}x_i=\sum_{i=1}^n (n-i)x_i,
\]
so
\begin{align*}
\sum_{i=1}^n\sum_{j=1}^n |x_i-x_j|
&=2\sum_{i=1}^n\bigl((i-1)-(n-i)\bigr)x_i
=2\sum_{i=1}^n (2i-n-1)x_i.
\end{align*}
Dividing by $2nS$:
\begin{equation}\label{eq:G-rank}
G(X)=\frac{\sum_{i=1}^n (2i-n-1)x_i}{nS}
=\frac{2\sum_{i=1}^n i x_i}{nS}-\frac{n+1}{n}.
\end{equation}
Introduce
\[
A:=\sum_{j=1}^n (n+1-j)x_j.
\]
Since $\sum_{i=1}^n i x_i+\sum_{i=1}^n (n+1-i)x_i=(n+1)S$, we have $\sum_{i=1}^n i x_i=(n+1)S-A$. Substituting this into \eqref{eq:G-rank} yields the weight-quotient form
\begin{equation}\label{eq:G-A-over-S}
G(X)=\frac{1}{n}\Bigl(n+1-\frac{2A}{S}\Bigr).
\end{equation}
Multiplying \eqref{eq:G-A-over-S} by $S$ gives the identity
\begin{equation}\label{eq:SG}
S\,G(X)=\frac{(n+1)S-2A}{n}.
\end{equation}

Given $\dfrac{\partial A}{\partial x_k}=n+1-k$ and $\dfrac{\partial S}{\partial x_k}=1$, so by the quotient rule
\begin{align}
\frac{\partial}{\partial x_k}\Bigl(\frac{A}{S}\Bigr)
&=\frac{(n+1-k)S-A}{S^2},\label{eq:partial-A-over-S}\\[2mm]
\frac{\partial G}{\partial x_k}
&=-\frac{2}{n}\cdot\frac{(n+1-k)S-A}{S^2}
=\frac{2}{n}\cdot\frac{A-S(n+1-k)}{S^2}.\label{eq:partial-G}
\end{align}
It is convenient to multiply \eqref{eq:partial-G} by $S$ and eliminate $A/S$ via \eqref{eq:G-A-over-S}. Since
\[
\frac{A}{S}=\frac{n+1-nG(X)}{2},
\]
we obtain the closed form
\begin{equation}\label{eq:S-partial-G}
S\,\frac{\partial G}{\partial x_k}
=\frac{2}{n}\Bigl(\frac{A}{S}-(n+1-k)\Bigr)
=\frac{1}{n}\Bigl(2k-(n+1)-nG(X)\Bigr).
\end{equation}
In particular, for $k=n$,
\begin{equation}\label{eq:S-partial-G-top}
S\,\frac{\partial G}{\partial x_n}
=\frac{1}{n}\bigl(n-1-nG(X)\bigr)
\end{equation}

Now we need to ensure  $F_{\mathrm{aggregate}}$ as increasing function of Utility. By the product rule,
\begin{align}
\frac{\partial F_{\mathrm{aggregate}}}{\partial x_k}
&=\frac{\partial}{\partial x_k}\Bigl((1-\lambda G)S\Bigr)
=(1-\lambda G)\cdot \frac{\partial S}{\partial x_k}
+S\cdot\frac{\partial(1-\lambda G)}{\partial x_k}\nonumber\\
&=(1-\lambda G)\cdot 1-\lambda S\frac{\partial G}{\partial x_k}.\label{eq:partial-F-start}
\end{align}
Insert \eqref{eq:S-partial-G} into \eqref{eq:partial-F-start}:
\begin{align*}
\frac{\partial F_{\mathrm{aggregate}}}{\partial x_k}
&=1-\lambda G-\lambda\cdot\frac{1}{n}\Bigl(2k-(n+1)-nG\Bigr)\\
&=1-\frac{\lambda}{n}\bigl(2k-(n+1)\bigr)
\end{align*}
Thus, for every $k\in\{1,\dots,n\}$ and every $X$ with fixed ranking,
\begin{equation}\label{eq:partial-F-closed}
{\;\displaystyle \frac{\partial F_{\mathrm{aggregate}}}{\partial x_k}
=1-\frac{\lambda}{n}\bigl(2k-(n+1)\bigr).\;}
\end{equation}
Note that \eqref{eq:partial-F-closed} is independent of $X$ and $G$.

This corresponds to the intuitive understanding that within fixed profile of numbers, increasing the number assigned to the individual who already has the highest number would worsen inequality.Therefore
\[
\frac{\partial F_{\mathrm{aggregate}}}{\partial x_k}\ge 0\ \ \forall k
\quad\Longleftrightarrow\quad
\frac{\partial F_{\mathrm{aggregate}}}{\partial x_n}\ge 0
\quad\Longleftrightarrow\quad
1-\frac{\lambda}{n}(n-1)\ge 0
\quad\Longleftrightarrow\quad
\lambda\le \frac{n}{n-1}.
\]
\end{proof}

\section{Generalized Atkinson-index}

While Temkin’s principles offer a promising vision of how to approach inequality, issues arise when they are tied to statistical formulas that don’t match the theoretical intentions. These statistical measures, despite their apparent utility, fail to accurately represent the egalitarian judgments that Temkin advocates. Our final result, Generalized Atkinson Index, provides a more robust and flexible foundations given it is built from the ground up to address inequality in a manner that avoids any inconsistency. The following theorem intended to establish the non-obvious truth that the underlying motivation of measuring inequality by comparing with mean value is actually by a function that will be called generalized-Atkinson index.

The following characterization theorem had been attempted in \cite{Lan2010AnAT}. Again, we note that the package of axioms in its original form, could not achieve the result as it intended for two independent flaws.

We work throughout with a single positive fairness measure function on $\mathbb R_+:=[0,\infty)$:
\[
f:\bigsqcup_{n\ge 1}(\mathbb R_+^n\setminus\{0\})\to(0,\infty)  
\]

For a nonzero vector $x=(x_1,\dots,x_n)\in\mathbb R_+^n\setminus\{0\}$, write
\[
\wt(x):=\sum_{i=1}^n x_i,
\qquad
p_i(x):=\frac{x_i}{\wt(x)}.
\]
If $u\in\mathbb R_+^m$ and $v\in\mathbb R_+^n$, write $[u,v]\in\mathbb R_+^{m+n}$ for their concatenation.

\begin{axiom}[Continuity]\label{ax:continuity}
For every $n\ge 1$, the restriction of $f$ to $\mathbb R_+^n\setminus\{0\}$ is continuous.
\end{axiom}

\begin{axiom}[Bounded equal-profile growth]\label{ax:boundedness}
\[
\lim_{n\to\infty}\frac{f(\ones{n+1})}{f(\ones{n})}=1.
\]
\end{axiom}

\begin{axiom}[Homogeneity]\label{ax:homogeneity}
For every $n\ge 1$, every $x\in\mathbb R_+^n\setminus\{0\}$, and every $t>0$,
\[
f(tx)=f(x).
\]
\end{axiom}

\begin{axiom}[Minimal Egalitarianism]\label{ax:strict-equalization}
\[
f(1,0)<f\!\left(\frac12,\frac12\right).
\]
\end{axiom}

\begin{incorrectaxiom}[Partition] For a partitioned system, let \( x = [x_1, x_2] \) and \( y = [y_1, y_2] \) be resource allocation vectors satisfying \( \sum_j x^i_j = \sum_j y^i_j \), \( i = 1, 2 \). The fairness ratio satisfies: 
\[ \frac{f(x)}{f(y)} = g^{-1} \left( s_1 \cdot g \left( \frac{f(x_1)}{f(y_1)} \right) + s_2 \cdot g \left( \frac{f(x_2)}{f(y_2)} \right) \right)
\] 
where \( s_i > 0 \), \( \sum_i s_i = 1 \), and \( g \) is continuous, strictly monotonic function. \end{incorrectaxiom}

The earlier axioms, such as continuity and boundedness, are ethically neutral; they are introduced solely as structural constraints on the functional form. By contrast, the third and fourth axioms can be interpreted as embodying egalitarian principles: the homogeneity axiom indicates that a uniform proportional increase in utility leaves the evaluation of the function unchanged, while the fourth reflects a minimal preference for equality over inequality. The partition axiom is derived from the classical Kolmogorov approach to the characterization of mean functions\cite{kolmogorov1991notion}. However, as  will be shown, the partition axiom in its current form does not work as intended.

Consider that in the original proof of \cite{Lan2010AnAT}, starting from the chain of algebraic derivations in line \textup{(114)} , the substitution:
\[
K=\left(\frac{u}{v}\right)^{1/\rho},
\qquad
X_i=\left(\frac{a_i}{b_i}\right)^{\rho}
\]
is followed by the identity
\[
\frac{a_i^r v^r}{b_i^r u^r}
=
\left(\frac{X_i}{K}\right)^{r/\rho},
\]
which is false. In fact, it's easy to see:
\[
\frac{a_i^r v^r}{b_i^r u^r}
=
X_i^{r/\rho}\left(\frac{v}{u}\right)^r
=
X_i^{r/\rho}K^{-\rho r},
\]
whereas
\[
\left(\frac{X_i}{K}\right)^{r/\rho}
=
X_i^{r/\rho}K^{-r/\rho}.
\]
Thus the exponents of $K$ do not match. This algebraic mistake propagates to the later derivations and culminates in the wrong equation~(121): the same expression appears on both sides, multiplied only by a power of $K$. Consequently the intended results of main theorem about the unique admissible functional representations does not follow.

A more serious problem is the original proof does not derive the power-type weight from the stated partition axiom. The original partition axiom only asserts that there exists a comparison-mean representation for each admissible partitioned comparison, it does not determine the corresponding weights as functions of the block totals. In particular, the original partition axiom only asserted that, for each admissible
comparison
\[
x=[x^1,x^2],\qquad y=[y^1,y^2],
\qquad
w(x^i)=w(y^i)\ \ (i=1,2),
\]
there exist some weights \(s_1,s_2>0\), \(s_1+s_2=1\), such that
\[
\frac{f(x)}{f(y)}
=
g^{-1}\!\left(
s_1\,g\!\left(\frac{f(x^1)}{f(y^1)}\right)
+
s_2\,g\!\left(\frac{f(x^2)}{f(y^2)}\right)
\right).
\]
That statement never identified the form of the weight functions. So it doesn't imply the formula
\[
s_i=\frac{w_i^\rho}{w_1^\rho+w_2^\rho},
\qquad
w_i:=w(x^i)=w(y^i),
\]
Which had been thoroughly used in the rest of original proof. Thus the crucial power-type weight was not proved from the axioms. It was inserted into the argument as an additional assumption. 

The seemingly obvious repair would be to strengthen the partition axiom by assuming that for some fixed exponent $\rho$, the total-based weight satisfies:
\[
\frac{f(x)}{f(y)}
=
g^{-1}\!\left(
\frac{w_1^\rho}{w_1^\rho+w_2^\rho}\,
g\!\left(\frac{f(x^1)}{f(y^1)}\right)
+
\frac{w_2^\rho}{w_1^\rho+w_2^\rho}\,
g\!\left(\frac{f(x^2)}{f(y^2)}\right)
\right)
\]
whenever
\[
x=[x^1,x^2],\qquad y=[y^1,y^2],\qquad \wt(x^i)=\wt(y^i)=:w_i\ \ (i=1,2).
\]

However, as an assumption this condition is indeed wrong. To see this, consider the positive power family that we intend to derive as representations:
\[
f(z)=\left(\sum_{i=1}^n p_i(z)^\rho\right)^{1/\beta},
\qquad
p_i(z):=\frac{z_i}{\sum_j z_j},
\qquad
\beta\neq 0,
\qquad
\rho\neq 1.
\]
For each \(m\in\mathbb N\), let
\[
u_m:=\left(\frac1m,\dots,\frac1m\right)\in\mathbb R_+^m.
\]
Then
\[
f(u_m)^\beta
=
m\left(\frac1m\right)^\rho
=
m^{\,1-\rho}.
\]

Now consider the two admissible comparisons
\[
x_A^1=u_1,\quad y_A^1=u_2,
\qquad
x_A^2=u_1,\quad y_A^2=u_1,
\]
and
\[
x_B^1=u_2,\quad y_B^1=u_4,
\qquad
x_B^2=u_1,\quad y_B^2=u_1.
\]
In both cases the two block totals are equal to \((1,1)\):
\[
w(x_A^1)=w(y_A^1)=w(x_A^2)=w(y_A^2)=1,
\]
\[
w(x_B^1)=w(y_B^1)=w(x_B^2)=w(y_B^2)=1.
\]
Moreover, the blockwise fairness ratios are identical:
\[
\left(\frac{f(x_A^1)}{f(y_A^1)}\right)^\beta
=
\frac{f(u_1)^\beta}{f(u_2)^\beta}
=
\frac{1}{2^{\,1-\rho}}
=
2^{\,\rho-1},
\]
\[
\left(\frac{f(x_B^1)}{f(y_B^1)}\right)^\beta
=
\frac{f(u_2)^\beta}{f(u_4)^\beta}
=
\frac{2^{\,1-\rho}}{4^{\,1-\rho}}
=
2^{\,\rho-1},
\]
and
\[
\left(\frac{f(x_A^2)}{f(y_A^2)}\right)^\beta
=
\left(\frac{f(x_B^2)}{f(y_B^2)}\right)^\beta
=
1.
\]
Hence any full-ratio law whose selector depends only on the two block totals \((w_1,w_2)\) would force the same overall ratio in the two comparisons. In particular, this would be true for the canonical choice
\[
\lambda(w_1,w_2)=\frac{w_1^\theta}{w_1^\theta+w_2^\theta}.
\]

However, the two whole-profile ratios are different. Indeed:
\[
[x_A^1,x_A^2]=(1,1),
\qquad
p([x_A^1,x_A^2])=\left(\frac12,\frac12\right),
\]
so
\[
f([x_A^1,x_A^2])^\beta
=
2\left(\frac12\right)^\rho
=
2^{\,1-\rho}.
\]
Also,
\[
[y_A^1,y_A^2]=\left(\frac12,\frac12,1\right),
\qquad
p([y_A^1,y_A^2])=\left(\frac14,\frac14,\frac12\right),
\]
hence
\[
f([y_A^1,y_A^2])^\beta
=
2\left(\frac14\right)^\rho+\left(\frac12\right)^\rho
=
2^{-\rho}\bigl(1+2^{\,1-\rho}\bigr).
\]
Therefore
\[
\left(\frac{f([x_A^1,x_A^2])}{f([y_A^1,y_A^2])}\right)^\beta
=
\frac{2^{\,1-\rho}}{2^{-\rho}(1+2^{\,1-\rho})}
=
\frac{2}{1+2^{\,1-\rho}}.
\]

Similarly,
\[
[x_B^1,x_B^2]=\left(\frac12,\frac12,1\right),
\qquad
p([x_B^1,x_B^2])=\left(\frac14,\frac14,\frac12\right),
\]
so
\[
f([x_B^1,x_B^2])^\beta
=
2\left(\frac14\right)^\rho+\left(\frac12\right)^\rho
=
2^{-\rho}\bigl(1+2^{\,1-\rho}\bigr).
\]
On the other hand,
\[
[y_B^1,y_B^2]=\left(\frac14,\frac14,\frac14,\frac14,1\right),
\qquad
p([y_B^1,y_B^2])=\left(\frac18,\frac18,\frac18,\frac18,\frac12\right),
\]
so
\[
f([y_B^1,y_B^2])^\beta
=
4\left(\frac18\right)^\rho+\left(\frac12\right)^\rho
=
2^{-\rho}\bigl(1+4^{\,1-\rho}\bigr).
\]
Thus
\[
\left(\frac{f([x_B^1,x_B^2])}{f([y_B^1,y_B^2])}\right)^\beta
=
\frac{1+2^{\,1-\rho}}{1+4^{\,1-\rho}}.
\]
Denote
\(
a:=2^{\,1-\rho}>0
\).
Then the two overall ratios become
\[
\frac{2}{1+a}
\qquad\text{and}\qquad
\frac{1+a}{1+a^2}.
\]
They are equal iff:
\[
\frac{2}{1+a}=\frac{1+a}{1+a^2}
\iff
2(1+a^2)=(1+a)^2
\iff
(a-1)^2=0
\iff
a=1
\iff
\rho=1.
\]
Hence the two overall ratios are different for every nontrivial power branch \(\rho\neq 1\).

This counterexample intends to highlight conceptual source of the failure point of original axiomatic structure. The problem is that two vector blocks can have the same total value while carrying different internal distribution. A function that depends only on totals cannot reacts to that distinction, while the nontrivial power family that the theorem intends to establish must be sensitive to the internal distribution of partitioned blocks.

Here in this paper, we adopt a different repair strategy. We add a separate permutation axiom, expressing the standard requirement that the value of the fairness measure should not depend on the ordering of coordinates. We then replace the old existential selector statement with the following \emph{reference-sensitive partition} principle
\[
\frac{f([x^1,x^2])}{f([y^1,y^2])}
=
g^{-1}\!\left(
\sigma(y^1,y^2)\,
g\!\left(\frac{f(x^1)}{f(y^1)}\right)
+
\bigl(1-\sigma(y^1,y^2)\bigr)\,
g\!\left(\frac{f(x^2)}{f(y^2)}\right)
\right),
\]
under the same block-total matching condition. This new axiom is stronger than the original partition axiom: it requires, for each pair of block lengths, a single continuous function depending on the entire reference partition \((y^1,y^2)\), rather than merely asserting the existence of some weights for each individual comparison. However, it does not impose any specific power-law type formula on the weight. Instead, regrouping arguments force the specific form of power law selector function to arise from an additive block content. Thus the desired formulaic form of weights could therefore be derived internally as part of the proof rather than imposed from the outside as an axiom. The proof will show that there exists a positive additive block functional \(\mu\) such that
\[
\sigma(y^1,y^2)
=
\frac{\mu(y^1)}{\mu(y^1)+\mu(y^2)},
\]
and then that necessarily
\[
\mu(z)=C\sum_j z_j^\rho
\]
for some \(\rho\in\mathbb R\) and \(C>0\). Consequently,
\[
\sigma(y^1,y^2)
=
\frac{\sum_j (y_j^1)^\rho}
{\sum_j (y_j^1)^\rho+\sum_j (y_j^2)^\rho}.
\]
In the special case \(y^i=[w_i]\), this reduces to
\[
\frac{w_1^\rho}{w_1^\rho+w_2^\rho}.
\]
This explains why the old argument looked plausible at first glance: it implicitly treated this one-dimensional special case as if it were the general law.

Here in this paper, we propose a new strategy by adding a separate permutation axiom, expressing the mundane structural requirement that the value of function should not depend on how the coordinate of elements. This axiom is very common among similar settings. and we replace the old existential selector by the almost identical \emph{reference-local partition} principle
\[
\frac{f([x^1,x^2])}{f([y^1,y^2])}
=
g^{-1}\!\left(
\sigma(y^1,y^2)\,
g\!\left(\frac{f(x^1)}{f(y^1)}\right)
+
\bigl(1-\sigma(y^1,y^2)\bigr)\,
g\!\left(\frac{f(x^2)}{f(y^2)}\right)
\right),
\]
Under the same block-total matching condition. This axiom does not assume the weights need to be power-sum type functions. It is slightly stronger in it encodes the condition so that the function depends on all the information available from \emph{reference partition} denominator rather than their aggregate weight. Permutation then forces invariance under internal relabeling of each vector block, while regrouping forces the selector to come from an additive block content. Therefore after some  operations from functional equations the weights
\[
\sigma(y^1,y^2)
=
\frac{\sum_j (y_j^1)^\rho}
{\sum_j (y_j^1)^\rho+\sum_j (y_j^2)^\rho}
\]
could therefore be derived internally as part of the proof rather than imposed from the outside as an axiom. In the special case \(y^i=[w_i]\), we would recover the:
\[
\frac{w_1^\rho}{w_1^\rho+w_2^\rho},
\]
Which explains why the old argument looked plausible at first glance: it had confused a special case with the general selector law.

We formally introduce the new axioms:
\begin{axiom}[Anonymity/Permutation-Invariance]\label{ax:anonymity}
For every $n\ge 1$, every permutation $\pi$ of $\{1,\dots,n\}$, and every $x\in\mathbb R_+^n\setminus\{0\}$,
\[
f(x_{\pi(1)},\dots,x_{\pi(n)})=f(x_1,\dots,x_n).
\]
\end{axiom}

\begin{axiom}[Reference-Sensitive Partition ]\label{ax:partition}
There exist a continuous strictly monotone function $g:(0,\infty)\to\mathbb R$ and, for every pair of block lengths $(m,n)$, a continuous selector
\[
\sigma_{m,n}:(0,\infty)^m\times(0,\infty)^n\to(0,1)
\]
such that the following holds. Whenever
\[
x=[x^1,x^2],\qquad y=[y^1,y^2]
\]
are \emph{positive} partitioned profiles with matched block totals
\[
\wt(x^i)=\wt(y^i)\qquad(i=1,2),
\]
one has
\[
\frac{f(x)}{f(y)}
=
g^{-1}\!\left(
\sigma(y^1,y^2)\,g\!\left(\frac{f(x^1)}{f(y^1)}\right)
+
\bigl(1-\sigma(y^1,y^2)\bigr)\,g\!\left(\frac{f(x^2)}{f(y^2)}\right)
\right),
\]
where, to lighten notation, $\sigma(y^1,y^2)$ denotes the value of the appropriate selector $\sigma_{m,n}$.
\end{axiom}

The next lemma is a positive-real number specialization of Theorem~2 from the classical paper 
\cite {WirsingZagier2001} of Wirsing and Zagier; we include a self-contained (in the sense that it does not require special techniques and results from analytical number theory) proof
tailored to our setting. For background on multiplicative methods in number theory, see
\cite{Tenenbaum2015,MontgomeryVaughan2007}.

\begin{lemma}[Erd\H{o}s--Wirsing rigidity for a completely multiplicative scale]\label{lem:erdos_wirsing_power}
Let $U:\mathbb N\to(0,\infty)$ satisfy:
\begin{enumerate}
\item[\textnormal{(i)}] \textbf{Complete multiplicativity:}\quad $U(mn)=U(m)\,U(n)$ for all $m,n\in\mathbb N$.
\item[\textnormal{(ii)}] \textbf{Asymptotically unit relative increment:}\quad $\displaystyle \lim_{n\to\infty}\frac{U(n+1)}{U(n)}=1$.
\end{enumerate}
Then there exists a constant $r\in\mathbb R$ such that
\[
U(n)=n^{\,r}\qquad\forall n\in\mathbb N.
\]
In particular, if $U(n):=\frac{f(\mathbf 1_n)}{f(\mathbf 1_1)}$ and \textnormal{(i)--(ii)} hold for this sequence, then
\[
f(\mathbf 1_n)=n^{\,r}\,f(\mathbf 1_1)\qquad\forall n\in\mathbb N.
\]

\smallskip
\emph{Interpretation.}  This lemma would help us establish a very useful property for multiplicative function: a completely multiplicative scale whose consecutive ratios asymptotically stabilize must be a pure power law.
\end{lemma}

\begin{deferredproof}{lem:erdos_wirsing_power}
Define
\[
F(n):=\log U(n)\qquad(n\in\mathbb N).
\]
Then $F:\mathbb N\to\mathbb R$ satisfies:

\smallskip
\noindent {Step 1: $F$ is completely additive.}
For all $m,n\in\mathbb N$,
\[
F(mn)=\log U(mn)=\log\bigl(U(m)U(n)\bigr)=\log U(m)+\log U(n)=F(m)+F(n).
\]
Also $F(1)=\log U(1)=\log(U(1\cdot 1))=\log(U(1)^2)$ forces $U(1)=1$, hence $F(1)=0$.

\smallskip
\noindent {Step 2: $\Delta F(n)\to 0$.}
Let $\Delta F(n):=F(n+1)-F(n)$. Then
\[
\Delta F(n)=\log U(n+1)-\log U(n)=\log\!\left(\frac{U(n+1)}{U(n)}\right).
\]
Since $\frac{U(n+1)}{U(n)}\to 1$ by (ii) and $\log$ is continuous at $1$, we have
\[
\Delta F(n)\xrightarrow[n\to\infty]{}0.
\]

\smallskip
\noindent {Step 3: g-adic contraction.}
Fix an integer $g\ge 2$. For each $n\in\mathbb N$, define a finite chain
\[
n_0:=n,\qquad n_{i}:=\left\lfloor\frac{n_{i-1}}{g}\right\rfloor \quad (i\ge 1),
\]
and remainders
\[
r_i:=n_{i-1}-g n_i\in\{0,1,\dots,g-1\}.
\]
Let $k=k_g(n)$ be the first index such that $n_k<g$ (so $n_k\in\{1,2,\dots,g-1\}$).

\smallskip
\noindent {Step 3a: $g^k\le n<g^{k+1}$.}
From $n_{i-1}=g n_i+r_i\ge g n_i$ we get $n_i\le n_{i-1}/g$, hence $n_k\le n/g^k$. Since $n_k\ge 1$,
\[
n\ge g^k.
\]
Also $n_{i-1}=g n_i+r_i\le g n_i+(g-1)<g(n_i+1)$. Iterating,
\[
n=n_0<g^k(n_k+1)\le g^k\cdot g=g^{k+1}.
\]
Thus $g^k\le n<g^{k+1}$.

\smallskip
\noindent {Step 3b: one-step approximation.}
Let $\varepsilon>0$. Since $\Delta F(n)\to 0$, there exists $N=N(\varepsilon,g)\in\mathbb N$ such that
\[
m\ge N\quad\Longrightarrow\quad |\Delta F(m)|=|F(m+1)-F(m)|\le \frac{\varepsilon}{g}.
\tag{$\ast$}
\]
Fix an index $i\in\{1,\dots,k\}$ with $g n_i\ge N$. Then
\[
n_{i-1}=g n_i+r_i.
\]
Using complete additivity,
\[
F(g n_i)=F(g)+F(n_i).
\]
Hence
\[
F(n_{i-1})-F(n_i)-F(g)=F(g n_i+r_i)-F(g n_i).
\]
If $r_i=0$ this is $0$. If $r_i\ge 1$, then
\[
F(g n_i+r_i)-F(g n_i)=\sum_{j=0}^{r_i-1}\Bigl(F(g n_i+j+1)-F(g n_i+j)\Bigr).
\]
For each $0\le j\le r_i-1$ we have $g n_i+j\ge g n_i\ge N$, so by ($\ast$),
\[
\bigl|F(g n_i+j+1)-F(g n_i+j)\bigr|\le \frac{\varepsilon}{g}.
\]
Therefore
\[
\bigl|F(n_{i-1})-F(n_i)-F(g)\bigr|
\le r_i\cdot\frac{\varepsilon}{g}\le (g-1)\cdot\frac{\varepsilon}{g}<\varepsilon.
\tag{$\dagger$}
\]

\smallskip
\noindent{Step 4: summing along the chain.}
Write the telescoping identity
\[
F(n)=F(n_0)=F(n_k)+\sum_{i=1}^{k}\bigl(F(n_{i-1})-F(n_i)\bigr)
      =F(n_k)+kF(g)+\sum_{i=1}^{k}e_i,
\]
where
\[
e_i:=F(n_{i-1})-F(n_i)-F(g).
\]
Split $\{1,\dots,k\}=I_{\mathrm{big}}\cup I_{\mathrm{small}}$ where
\[
I_{\mathrm{big}}:=\{\,i:\ g n_i\ge N\,\},\qquad I_{\mathrm{small}}:=\{1,\dots,k\}\setminus I_{\mathrm{big}}.
\]
For $i\in I_{\mathrm{big}}$, $(\dagger)$ gives $|e_i|<\varepsilon$, hence
\[
\sum_{i\in I_{\mathrm{big}}}|e_i|\le \varepsilon\,|I_{\mathrm{big}}|\le \varepsilon k.
\tag{1}
\]
For $i\in I_{\mathrm{small}}$ we have $g n_i<N$, hence $n_i\le N-1$. Consequently,
\[
n_{i-1}=g n_i+r_i\le g(N-1)+(g-1)=gN-1.
\]
Thus, for $i\in I_{\mathrm{small}}$, both $(n_{i-1},n_i)$ lie in the finite set
\[
\{1,2,\dots,gN-1\}\times\{1,2,\dots,N-1\}.
\]
Define the finite constant
\[
B:=\max_{\substack{1\le a\le gN-1\\1\le b\le N-1}}
\bigl|F(a)-F(b)-F(g)\bigr|<\infty.
\]
Then $|e_i|\le B$ for all $i\in I_{\mathrm{small}}$. Also, since $n_i$ decreases by a factor $\ge g$ each step until it falls below $N$, the number $|I_{\mathrm{small}}|$ is bounded by a constant depending only on $(N,g)$; one crude bound is
\[
|I_{\mathrm{small}}|\le 1+\left\lceil \log_g N\right\rceil=:M.
\]
Hence
\[
\sum_{i\in I_{\mathrm{small}}}|e_i|\le BM.
\tag{2}
\]
Moreover, $n_k\in\{1,\dots,g-1\}$, so
\[
|F(n_k)|\le \max_{1\le t\le g-1}|F(t)|=:C_0<\infty.
\tag{3}
\]
Combining (1)--(3),
\[
|F(n)-kF(g)|
\le |F(n_k)|+\sum_{i=1}^{k}|e_i|
\le C_0+\varepsilon k+BM.
\]
Thus, for each fixed $g\ge 2$ and each $\varepsilon>0$, there exists a constant $C=C(\varepsilon,g)$ such that
\[
|F(n)-k_g(n)F(g)|\le \varepsilon\,k_g(n)+C
\qquad\forall n\in\mathbb N.
\tag{4}
\]

\smallskip
\noindent {Step 5: identify the logarithmic slope.}
From $g^{k}\le n<g^{k+1}$ we have
\[
k\log g\le \log n<(k+1)\log g
\quad\Longrightarrow\quad
0\le \frac{\log n}{\log g}-k<1.
\tag{5}
\]
Divide (4) by $\log n$ and use $k\le \frac{\log n}{\log g}$ from (5):
\[
\left|\frac{F(n)}{\log n}-\frac{kF(g)}{\log n}\right|
\le \varepsilon\,\frac{k}{\log n}+\frac{C}{\log n}
\le \varepsilon\cdot\frac{1}{\log g}+\frac{C}{\log n}.
\tag{6}
\]
Also, from (5),
\[
\left|\frac{k}{\log n}-\frac{1}{\log g}\right|
=\frac{1}{\log n}\left|\frac{\log n}{\log g}-k\right|
\le \frac{1}{\log n}.
\]
Therefore
\[
\left|\frac{kF(g)}{\log n}-\frac{F(g)}{\log g}\right|
\le |F(g)|\cdot\left|\frac{k}{\log n}-\frac{1}{\log g}\right|
\le \frac{|F(g)|}{\log n}.
\tag{7}
\]
Combine (6) and (7):
\[
\left|\frac{F(n)}{\log n}-\frac{F(g)}{\log g}\right|
\le \frac{\varepsilon}{\log g}+\frac{C+|F(g)|}{\log n}.
\]
Let $n\to\infty$ (with $g,\varepsilon$ fixed):
\[
\limsup_{n\to\infty}\left|\frac{F(n)}{\log n}-\frac{F(g)}{\log g}\right|
\le \frac{\varepsilon}{\log g}.
\]
Since $\varepsilon>0$ is arbitrary,
\[
\lim_{n\to\infty}\frac{F(n)}{\log n}=\frac{F(g)}{\log g}.
\tag{8}
\]

\smallskip
\noindent {Step 6: independence of $g$ and exact identification of $F$.}
Equation (8) holds for every integer $g\ge 2$. In particular, for $g=2$ and any $g\ge 2$,
\[
\frac{F(g)}{\log g}
=\lim_{n\to\infty}\frac{F(n)}{\log n}
=\frac{F(2)}{\log 2}.
\]
Define
\[
r:=\frac{F(2)}{\log 2}.
\]
Then for every $g\ge 2$,
\[
F(g)=r\log g.
\]
Since $g$ here is an arbitrary integer $\ge 2$, this gives
\[
F(n)=r\log n\qquad\forall n\ge 2,
\]
and also $F(1)=0=r\log 1$.

\smallskip
\noindent {Step 7: return to $U$.}
For all $n\in\mathbb N$,
\[
U(n)=\exp(F(n))=\exp(r\log n)=n^{\,r}.
\]
This proves the lemma.
\end{deferredproof}

The following proposition is also a repackaging and combination of established results and techniques from literature on functional equation and theory of (quasi)mean functions; see, e.g.,\cite{HardyLittlewoodPolya1952}, \cite{aczel1975measures}.

\begin{proposition}[Homogeneous weighted quasi-arithmetic means]\label{prop:homogeneous-bajraktarevic}
Fix \(\alpha\in\mathbb R\). Let \(g:(0,\infty)\to\mathbb R\) be continuous and strictly monotone, and define
\[
M_g^{(\alpha)}(u,v)
:=
g^{-1}\!\left(
\frac{u^\alpha g(u)+v^\alpha g(v)}{u^\alpha+v^\alpha}
\right)
\qquad(u,v>0).
\]
If
\[
M_g^{(\alpha)}(cu,cv)=c\,M_g^{(\alpha)}(u,v)
\qquad\forall\,c,u,v>0,
\]
then, up to an affine change of generator, either
\[
g(x)=\log x
\qquad\text{or}\qquad
g(x)=x^\beta\quad(\beta\neq 0).
\]
\end{proposition}

\begin{proof}
We first prove that affine changes of the generator do not alter the mean. Indeed, if
\[
h(x)=A\,g(x)+B
\qquad(A\neq 0,\ B\in\mathbb R),
\]
then
\begin{align*}
M_h^{(\alpha)}(u,v)
&=
h^{-1}\!\left(
\frac{u^\alpha h(u)+v^\alpha h(v)}{u^\alpha+v^\alpha}
\right)\\
&=
g^{-1}\!\left(
\frac{
\displaystyle
\frac{u^\alpha(A g(u)+B)+v^\alpha(A g(v)+B)}{u^\alpha+v^\alpha}
-B
}{A}
\right)\\
&=
g^{-1}\!\left(
\frac{u^\alpha g(u)+v^\alpha g(v)}{u^\alpha+v^\alpha}
\right)
=
M_g^{(\alpha)}(u,v).
\end{align*}
Thus it is enough to show that \(g\) itself must be of the form
\[
g(x)=A\log x+B
\qquad\text{or}\qquad
g(x)=A x^\beta+B
\quad(\beta\neq 0),
\]
for some constants \(A\neq 0\), \(B\in\mathbb R\).

Set
\[
J:=g((0,\infty)).
\]
Since \(g\) is continuous and strictly monotone, \(J\) is an interval, and
\[
g^{-1}:J\to(0,\infty)
\]
is continuous and strictly monotone.

Fix \(c>0\). Define
\[
B_c:J\to\mathbb R,
\qquad
B_c(s):=g\!\bigl(c\,g^{-1}(s)\bigr),
\]
and also
\[
\omega:J\to(0,\infty),
\qquad
\omega(s):=\bigl(g^{-1}(s)\bigr)^\alpha.
\]
Because \(g^{-1}\) is continuous and strictly positive, \(\omega\) is continuous and strictly positive on \(J\).

We claim that
\begin{equation}\label{eq:weighted-jensen-inside-proof}
B_c\!\left(
\frac{\omega(s)s+\omega(t)t}{\omega(s)+\omega(t)}
\right)
=
\frac{\omega(s)B_c(s)+\omega(t)B_c(t)}{\omega(s)+\omega(t)}
\qquad\forall\,s,t\in J.
\end{equation}
To see this, let
\[
u:=g^{-1}(s),\qquad v:=g^{-1}(t).
\]
Then
\[
\omega(s)=u^\alpha,\qquad \omega(t)=v^\alpha,
\]
and
\[
\frac{\omega(s)s+\omega(t)t}{\omega(s)+\omega(t)}
=
\frac{u^\alpha g(u)+v^\alpha g(v)}{u^\alpha+v^\alpha}
=
g\!\bigl(M_g^{(\alpha)}(u,v)\bigr).
\]
Hence
\begin{align*}
B_c\!\left(
\frac{\omega(s)s+\omega(t)t}{\omega(s)+\omega(t)}
\right)
&=
g\!\left(
c\,g^{-1}\!\left(
\frac{u^\alpha g(u)+v^\alpha g(v)}{u^\alpha+v^\alpha}
\right)
\right)\\
&=
g\!\bigl(c\,M_g^{(\alpha)}(u,v)\bigr).
\end{align*}
By the assumed homogeneity of \(M_g^{(\alpha)}\),
\[
c\,M_g^{(\alpha)}(u,v)=M_g^{(\alpha)}(cu,cv),
\]
so
\begin{align*}
g\!\bigl(c\,M_g^{(\alpha)}(u,v)\bigr)
&=
g\!\bigl(M_g^{(\alpha)}(cu,cv)\bigr)\\
&=
\frac{(cu)^\alpha g(cu)+(cv)^\alpha g(cv)}{(cu)^\alpha+(cv)^\alpha}\\
&=
\frac{u^\alpha g(cu)+v^\alpha g(cv)}{u^\alpha+v^\alpha}\\
&=
\frac{\omega(s)B_c(s)+\omega(t)B_c(t)}{\omega(s)+\omega(t)}.
\end{align*}
This proves \eqref{eq:weighted-jensen-inside-proof}.

We now show that every continuous \(B:J\to\mathbb R\) satisfying
\begin{equation}\label{eq:weighted-jensen-general}
B\!\left(
\frac{\omega(x)x+\omega(y)y}{\omega(x)+\omega(y)}
\right)
=
\frac{\omega(x)B(x)+\omega(y)B(y)}{\omega(x)+\omega(y)}
\qquad\forall\,x,y\in J
\end{equation}
must be affine on \(J\).

Fix a compact interval \([a,b]\subset J\) with \(a<b\). For \(x,y\in[a,b]\), define
\[
T(x,y):=\frac{\omega(x)x+\omega(y)y}{\omega(x)+\omega(y)}.
\]
If \(x<y\), then \(T(x,y)\in(x,y)\), because \(\omega(x),\omega(y)>0\). Indeed,
\[
T(x,y)-x=\frac{\omega(y)}{\omega(x)+\omega(y)}(y-x)>0,
\]
\[
y-T(x,y)=\frac{\omega(x)}{\omega(x)+\omega(y)}(y-x)>0.
\]
Now apply \eqref{eq:weighted-jensen-general} to \(x<y\). Since
\[
B(T(x,y))
=
\frac{\omega(x)B(x)+\omega(y)B(y)}{\omega(x)+\omega(y)},
\]
we get
\[
B(T(x,y))-B(x)
=
\frac{\omega(y)}{\omega(x)+\omega(y)}\bigl(B(y)-B(x)\bigr),
\]
\[
B(y)-B(T(x,y))
=
\frac{\omega(x)}{\omega(x)+\omega(y)}\bigl(B(y)-B(x)\bigr).
\]
Together with
\[
T(x,y)-x=\frac{\omega(y)}{\omega(x)+\omega(y)}(y-x),
\qquad
y-T(x,y)=\frac{\omega(x)}{\omega(x)+\omega(y)}(y-x),
\]
this yields
\begin{equation}\label{eq:equal-slopes}
\frac{B(T(x,y))-B(x)}{T(x,y)-x}
=
\frac{B(y)-B(x)}{y-x}
=
\frac{B(y)-B(T(x,y))}{y-T(x,y)}.
\end{equation}

Because \(\omega\) is continuous and strictly positive on the compact interval \([a,b]\), there exist constants
\[
m:=\min_{u\in[a,b]}\omega(u)>0,
\qquad
M:=\max_{u\in[a,b]}\omega(u)<\infty.
\]
Set
\[
\theta:=\frac{m}{m+M}\in\left(0,\frac12\right].
\]
Then for every \(x<y\) in \([a,b]\),
\[
\theta\le \frac{\omega(y)}{\omega(x)+\omega(y)}\le 1-\theta,
\qquad
\theta\le \frac{\omega(x)}{\omega(x)+\omega(y)}\le 1-\theta,
\]
and therefore
\begin{equation}\label{eq:uniform-split-inside}
\theta(y-x)\le T(x,y)-x\le (1-\theta)(y-x),
\qquad
\theta(y-x)\le y-T(x,y)\le (1-\theta)(y-x).
\end{equation}

We now construct a nested sequence of partitions of \([a,b]\). Let
\[
P_0:=\{a,b\}.
\]
Given a partition
\[
P_n=\{x_0^{(n)}<x_1^{(n)}<\cdots<x_{N_n}^{(n)}\},
\]
define \(P_{n+1}\) by inserting, for each adjacent pair \((x_{i-1}^{(n)},x_i^{(n)})\), the point
\[
T\!\left(x_{i-1}^{(n)},x_i^{(n)}\right).
\]
Thus \(P_n\subset P_{n+1}\) for every \(n\).

Let
\[
\operatorname{mesh}(P_n):=\max_{1\le i\le N_n}\bigl(x_i^{(n)}-x_{i-1}^{(n)}\bigr).
\]
By \eqref{eq:uniform-split-inside}, every child interval created from a parent interval of length \(\ell\) has length at most \((1-\theta)\ell\). Hence, inductively,
\[
\operatorname{mesh}(P_n)\le (1-\theta)^n(b-a)\xrightarrow[n\to\infty]{}0.
\]
Since the partitions are nested and their meshes tend to \(0\), the set of partition nodes
\[
D:=\bigcup_{n\ge 0}P_n
\]
is dense in \([a,b]\).

Set
\[
k:=\frac{B(b)-B(a)}{b-a}.
\]
We claim that every adjacent interval in every partition \(P_n\) has secant slope \(k\). This is proved by induction on \(n\). It is true for \(P_0\) by definition of \(k\). Assume it holds for every adjacent interval in \(P_n\). Let \([p,q]\) be an adjacent interval in \(P_n\), and set \(r:=T(p,q)\). Then \eqref{eq:equal-slopes} gives
\[
\frac{B(r)-B(p)}{r-p}
=
\frac{B(q)-B(p)}{q-p}
=
\frac{B(q)-B(r)}{q-r}.
\]
By the induction hypothesis, the middle quotient equals \(k\), so both child intervals \([p,r]\) and \([r,q]\) also have secant slope \(k\). Thus the claim holds for \(P_{n+1}\), and the induction is complete.

Now let \(x\in D\). Then \(x\) belongs to some partition \(P_n\). Writing the partition points of \(P_n\) as
\[
a=t_0<t_1<\cdots<t_j=x<\cdots<t_{N_n}=b,
\]
and summing the increments along adjacent intervals, we obtain
\begin{align*}
B(x)-B(a)
&=
\sum_{i=1}^{j}\bigl(B(t_i)-B(t_{i-1})\bigr)
=
\sum_{i=1}^{j}k\,(t_i-t_{i-1})\\
&=
k(x-a).
\end{align*}
Hence
\[
B(x)=B(a)+k(x-a)
\qquad\forall\,x\in D.
\]
Since \(D\) is dense in \([a,b]\) and \(B\) is continuous, it follows that
\[
B(x)=B(a)+k(x-a)
\qquad\forall\,x\in[a,b].
\]
Thus \(B\) is affine on every compact subinterval of \(J\).

We now pass from local affinity to global affinity. Choose two distinct points \(s_0,t_0\in J\), and define
\[
A:=\frac{B(t_0)-B(s_0)}{t_0-s_0}.
\]
Let \(x\in J\) with \(x\neq s_0\). The compact interval
\[
I_x:=[\min\{x,s_0,t_0\},\,\max\{x,s_0,t_0\}]
\]
is contained in \(J\), and \(B\) is affine on \(I_x\). Therefore every secant slope on \(I_x\) is equal, so in particular
\[
\frac{B(x)-B(s_0)}{x-s_0}
=
\frac{B(t_0)-B(s_0)}{t_0-s_0}
=
A.
\]
Thus
\[
B(x)=A x+\bigl(B(s_0)-A s_0\bigr)
\qquad\forall\,x\in J,\ x\neq s_0.
\]
This identity is also true at \(x=s_0\), so \(B\) is globally affine on \(J\).

We may now apply this conclusion to \(B=B_c\). There exist constants \(a(c),b(c)\in\mathbb R\) such that
\[
B_c(s)=a(c)s+b(c)
\qquad\forall\,s\in J.
\]
Returning to the variable \(x=g^{-1}(s)\), we obtain
\begin{equation}\label{eq:affine-scaling-generator-final}
g(cx)=a(c)\,g(x)+b(c)
\qquad\forall\,c,x>0.
\end{equation}

Since the functions \(x\mapsto g(cx)\) and \(x\mapsto g(x)\) have the same strict monotonicity, the affine coefficient must be positive:
\[
a(c)>0
\qquad\forall\,c>0.
\]
Indeed, if \(a(c)<0\), then the right-hand side of \eqref{eq:affine-scaling-generator-final} would have the opposite monotonicity from \(g(x)\), which is impossible.

Next we show that \(a\) and \(b\) are continuous. Choose \(x_0>0\) with \(x_0\neq 1\). Since \(g\) is strictly monotone,
\[
g(x_0)\neq g(1).
\]
Evaluating \eqref{eq:affine-scaling-generator-final} at \(x=x_0\) and at \(x=1\), we get
\[
g(cx_0)=a(c)g(x_0)+b(c),
\qquad
g(c)=a(c)g(1)+b(c).
\]
Subtracting,
\[
a(c)=\frac{g(cx_0)-g(c)}{g(x_0)-g(1)}.
\]
Hence \(a\) is continuous. Then
\[
b(c)=g(c)-a(c)g(1),
\]
so \(b\) is continuous as well.

Now compute \(g(cdx)\) in two ways. First, by \eqref{eq:affine-scaling-generator-final},
\[
g(cdx)=a(cd)\,g(x)+b(cd).
\]
Second,
\[
g(cdx)=a(c)\,g(dx)+b(c)=a(c)\bigl(a(d)g(x)+b(d)\bigr)+b(c).
\]
Since \(g\) is nonconstant, comparison of coefficients yields
\begin{equation}\label{eq:a-multiplicative-final}
a(cd)=a(c)a(d),
\qquad
b(cd)=a(c)b(d)+b(c).
\end{equation}

Because \(a>0\) and continuous, the first identity implies
\[
a(c)=c^\lambda
\qquad\forall\,c>0
\]
for some \(\lambda\in\mathbb R\). To see this, define
\[
\Phi(t):=\log a(e^t)\qquad(t\in\mathbb R).
\]
Then \(\Phi\) is continuous and additive:
\[
\Phi(s+t)=\Phi(s)+\Phi(t).
\]
Hence \(\Phi(q)=q\Phi(1)\) for every rational \(q\), and continuity extends this to all \(t\in\mathbb R\). Thus
\[
\Phi(t)=\lambda t,
\]
and therefore \(a(c)=c^\lambda\).

We now distinguish two cases.

If \(\lambda=0\), then \(a(c)=1\) for all \(c>0\), and \eqref{eq:a-multiplicative-final} reduces to
\[
b(cd)=b(c)+b(d).
\]
Define
\[
\Psi(t):=b(e^t)\qquad(t\in\mathbb R).
\]
Then \(\Psi\) is continuous and additive, so by the same argument as above there exists \(q\in\mathbb R\) such that
\[
\Psi(t)=qt,
\qquad\text{i.e.}\qquad
b(c)=q\log c.
\]
Substituting into \eqref{eq:affine-scaling-generator-final},
\[
g(cx)=g(x)+q\log c
\qquad\forall\,c,x>0.
\]
Taking \(x=1\), we obtain
\[
g(c)=g(1)+q\log c.
\]
If \(q=0\), then \(g(c)=g(1)\) for every \(c>0\), contradicting strict monotonicity. Hence \(q\neq 0\), and \(g\) is affine-equivalent to \(\log x\).

If \(\lambda\neq 0\), then \(a(c)=c^\lambda\), and the second identity in \eqref{eq:a-multiplicative-final} becomes
\[
b(cd)=c^\lambda b(d)+b(c).
\]
Interchanging \(c\) and \(d\), we also have
\[
b(cd)=d^\lambda b(c)+b(d).
\]
Subtracting,
\[
(c^\lambda-1)b(d)=(d^\lambda-1)b(c)
\qquad\forall\,c,d>0.
\]
Choose \(d_0>0\) with \(d_0\neq 1\). Since \(\lambda\neq 0\), we have \(d_0^\lambda\neq 1\). Define
\[
\gamma:=\frac{b(d_0)}{d_0^\lambda-1}.
\]
Then
\[
b(c)=\gamma(c^\lambda-1)
\qquad\forall\,c>0.
\]
Substituting into \eqref{eq:affine-scaling-generator-final},
\[
g(cx)=c^\lambda g(x)+\gamma(c^\lambda-1).
\]
Define
\[
\widetilde g(x):=g(x)+\gamma.
\]
Then
\[
\widetilde g(cx)=c^\lambda \widetilde g(x)
\qquad\forall\,c,x>0.
\]
Taking \(x=1\),
\[
\widetilde g(c)=c^\lambda \widetilde g(1).
\]
If \(\widetilde g(1)=0\), then \(\widetilde g\equiv 0\), so \(g\) would be constant, impossible. Hence \(\widetilde g(1)\neq 0\), and therefore
\[
g(x)=A x^\lambda+B
\qquad\forall\,x>0
\]
for suitable constants \(A\neq 0\), \(B\in\mathbb R\). Thus \(g\) is affine-equivalent to \(x^\lambda\). Renaming \(\lambda\) as \(\beta\), we obtain \(g(x)=x^\beta\) with \(\beta\neq 0\), up to affine change of generator.

This completes the proof.
\end{proof}

\begin{theorem}[Generalized Atkinson-Index characterization]\label{thm:single-function-atkinson}
Assume Axioms~\ref{ax:continuity}, \ref{ax:homogeneity}, \ref{ax:boundedness}, \ref{ax:strict-equalization}, \ref{ax:anonymity}, and \ref{ax:partition}. Then, there exists $r>0$ and exactly one of the following two possibilities.

\smallskip
\noindent\textup{(i) Logarithmic branch.}
\[
f(x)=\exp\!\left(-r\sum_{i=1}^n p_i(x)\log p_i(x)\right)=\prod_{i=1}^n p_i(x)^{-r\,p_i(x)}
\qquad\bigl(x\in\mathbb R_+^n\setminus\{0\}\bigr),
\]
with the usual convention $0\log 0:=0$ (equivalently, by continuous extension from the positive orthant), and the selector in Axiom~\ref{ax:partition} is
\[
\sigma(y^1,y^2)=\frac{\wt(y^1)}{\wt(y^1)+\wt(y^2)}.
\]

\smallskip
\noindent\textup{(ii) Power branch.}
There exist $\beta\neq 0$ and $\rho>0$ such that
\[
\rho+\beta r=1,
\]
\[
f(x)=\left(\sum_{i=1}^n p_i(x)^\rho\right)^{1/\beta}
\qquad\bigl(x\in\mathbb R_+^n\setminus\{0\}\bigr),
\]
and the selector in Axiom~\ref{ax:partition} is
\[
\sigma(y^1,y^2)
=
\frac{\sum_j (y_j^1)^\rho}{\sum_j (y_j^1)^\rho+\sum_j (y_j^2)^\rho}.
\]
Conversely, every function in one of these two families satisfies Axioms~\ref{ax:continuity}, \ref{ax:boundedness}, \ref{ax:homogeneity}, \ref{ax:strict-equalization}, \ref{ax:anonymity}, and \ref{ax:partition}.
 
\end{theorem}
\begin{proof}
Multiplying $f$ by a positive constant leaves every axiom unchanged, because Axiom~\ref{ax:partition} uses only fairness ratios, Axiom~\ref{ax:boundedness} uses a quotient of two equal-profile values, and Axiom~\ref{ax:strict-equalization} is order-theoretic. We therefore normalize
\[
f(1)=1.
\]
By Axiom~\ref{ax:homogeneity}, this implies
\[
f(t)=1\qquad\forall t>0.
\]

\smallskip
\noindent\emph{Step 1: binary direct-product identity.}
Let $y\in(0,\infty)^m$ and $z=(z_1,z_2)\in(0,\infty)^2$. Set
\[
x:=[z_1y,z_2y],\qquad c:=[z_1\wt(y),z_2\wt(y)].
\]
Then
\[
\wt(x^1)=z_1\wt(y)=\wt(c^1),
\qquad
\wt(x^2)=z_2\wt(y)=\wt(c^2),
\]
so Axiom~\ref{ax:partition} applies to the pair $(x,c)$ and yields
\[
\frac{f(x)}{f(c)}
=
g^{-1}\!\left(
\sigma(c^1,c^2)\,g\!\left(\frac{f(z_1y)}{f(z_1\wt(y))}\right)
+
\bigl(1-\sigma(c^1,c^2)\bigr)\,g\!\left(\frac{f(z_2y)}{f(z_2\wt(y))}\right)
\right).
\]
By Axiom~\ref{ax:homogeneity} and scalar normalization,
\[
f(z_i y)=f(y),\qquad f(z_i\wt(y))=1\qquad(i=1,2).
\]
Hence the right-hand side is $g^{-1}(g(f(y)))=f(y)$. Since $f(c)=f(z_1,z_2)$ by homogeneity, we get
\[
f([z_1y,z_2y])=f(y)\,f(z_1,z_2).
\tag{1}\label{eq:binary-direct-product}
\]

\smallskip
\noindent\emph{Step 2: equal allocations are multiplicative.}
Define
\[
U(n):=f(\ones{n})\qquad(n\in\mathbb N).
\]
Taking $y=\ones{n}$ and $z=(1,1)$ in \eqref{eq:binary-direct-product}, we obtain
\[
U(2n)=U(2)U(n).
\tag{2}\label{eq:base-multiplicativity}
\]
Now suppose that for some fixed $k\ge2$,
\[
U(kn)=U(k)U(n)\qquad\forall n\in\mathbb N.
\tag{3}\label{eq:induction-hypothesis}
\]
Apply Axiom~\ref{ax:partition} to
\[
x=[\ones{kn},\ones{n}],
\qquad
y=[n\ones{k},n].
\]
Then
\[
\wt(x^1)=kn=\wt(y^1),
\qquad
\wt(x^2)=n=\wt(y^2),
\]
so
\begin{align*}
\frac{U((k+1)n)}{U(k+1)}
&=
g^{-1}\!\left(
\sigma(y^1,y^2)\,g\!\left(\frac{U(kn)}{U(k)}\right)
+
\bigl(1-\sigma(y^1,y^2)\bigr)\,g\!\left(\frac{U(n)}{1}\right)
\right)\\
&=
g^{-1}\!\left(
\sigma(y^1,y^2)\,g(U(n))
+
\bigl(1-\sigma(y^1,y^2)\bigr)\,g(U(n))
\right)\\
&=U(n),
\end{align*}
where \eqref{eq:induction-hypothesis} was used in the second line. Thus
\[
U((k+1)n)=U(k+1)U(n).
\]
By induction on $k$, $U$ is completely multiplicative:
\[
U(mn)=U(m)U(n)\qquad\forall m,n\in\mathbb N.
\]
Together with Axiom~\ref{ax:boundedness}, the previously established Erd\H{o}s--Wirsing rigidity lemma gives a constant $r\in\mathbb R$ such that
\[
U(n)=n^r,
\qquad	 
f(\ones{n})=n^r\qquad\forall n\in\mathbb N.
\tag{4}\label{eq:equal-profile-power-law}
\]

\smallskip
\noindent\emph{Step 3: attainable block ratios fill an interval.}
Consider the continuous function
\[
\varphi:(0,1)\to(0,\infty),
\qquad
\varphi(t):=f(t,1-t).
\]
By Axiom~\ref{ax:strict-equalization} and continuity at the boundary point $(1,0)$,
\[
\varphi\!\left(\frac12\right)=f\!\left(\frac12,\frac12\right)>f(1,0)=\lim_{t\uparrow1}\varphi(t).
\]
Hence $\varphi$ is nonconstant, so its image contains a nondegenerate interval $I\subset(0,\infty)$. Fix such an interval.

Now let $y\in(0,\infty)^m$ and let $u\in I$. Choose $t\in(0,1)$ with $\varphi(t)=u$, and set
\[
\widetilde y:=[ty,(1-t)y].
\]
Then $\wt(\widetilde y)=\wt(y)$, and by \eqref{eq:binary-direct-product},
\[
f(\widetilde y)=f(y)\,f(t,1-t)=u\,f(y).
\]
Therefore every $u\in I$ occurs as a refinement ratio
\[
\frac{f(\widetilde y)}{f(y)}=u
\]
of every positive block $y$.

\smallskip
\noindent\emph{Step 4: invariances of the selector $\sigma$.}
Let $u$ and $v$ be positive blocks. Choose distinct $a,b\in I$, and pick
$s,t\in(0,1)$ such that
\[
\varphi(s)=a,
\qquad
\varphi(t)=b.
\]
Set
\[
x^1:=[su,(1-s)u],
\qquad
x^2:=[tv,(1-t)v].
\]
By \eqref{eq:binary-direct-product},
\[
\frac{f(x^1)}{f(u)}=f(s,1-s)=\varphi(s)=a,
\qquad
\frac{f(x^2)}{f(v)}=f(t,1-t)=\varphi(t)=b.
\]
If $\pi$ and $\tau$ are internal permutations of the coordinates of $u$ and $v$, define
\[
\pi_*x^1:=[s\,\pi u,(1-s)\,\pi u],
\qquad
\tau_*x^2:=[t\,\tau v,(1-t)\,\tau v].
\]
By Axiom~\ref{ax:anonymity},
\[
f([\pi_*x^1,\tau_*x^2])=f([x^1,x^2]),
\qquad
f([\pi u,\tau v])=f([u,v]).
\]
Hence
\[
\frac{f([x^1,x^2])}{f([u,v])}
=
\frac{f([\pi_*x^1,\tau_*x^2])}{f([\pi u,\tau v])}.
\]
Applying Axiom~\ref{ax:partition} to both sides gives
\[
\sigma(u,v)g(a)+(1-\sigma(u,v))g(b)
=
\sigma(\pi u,\tau v)g(a)+(1-\sigma(\pi u,\tau v))g(b).
\]
Because $g(a)\neq g(b)$, one gets
\[
\sigma(\pi u,\tau v)=\sigma(u,v).
\tag{5}\label{eq:internal-permutation-invariance}
\]
Similarly, swapping the two blocks and using Axiom~\ref{ax:anonymity},
\[
\sigma(u,v)g(a)+(1-\sigma(u,v))g(b)
=
\sigma(v,u)g(b)+(1-\sigma(v,u))g(a),
\]
whence
\[
\sigma(v,u)=1-\sigma(u,v).
\tag{6}\label{eq:block-swap}
\]
Finally, scale both the compared and reference partitions by the same $\lambda>0$. Homogeneity leaves every fairness ratio unchanged, so
\[
\sigma(\lambda u,\lambda v)g(a)+(1-\sigma(\lambda u,\lambda v))g(b)
=
\sigma(u,v)g(a)+(1-\sigma(u,v))g(b),
\]
and therefore
\[
\sigma(\lambda u,\lambda v)=\sigma(u,v)
\qquad\forall \lambda>0.
\tag{7}\label{eq:common-scale-invariance}
\]

\smallskip
\noindent\emph{Step 5: regrouping identities.}
Let $u,v,w$ be positive blocks. Choose arbitrary $a,b,c\in I$, and refinements $x^1,x^2,x^3$ of $u,v,w$ such that
\[
\frac{f(x^1)}{f(u)}=a,
\qquad
\frac{f(x^2)}{f(v)}=b,
\qquad
\frac{f(x^3)}{f(w)}=c.
\]
Compute
\[
\frac{f([x^1,x^2,x^3])}{f([u,v,w])}
\]
by the two bracketings $([u,v],w)$ and $(u,[v,w])$. Writing
\[
A:=\sigma([u,v],w),\quad B:=\sigma(u,v),\quad C:=\sigma(v,w),\quad D:=\sigma(u,[v,w]),
\]
Axiom~\ref{ax:partition} gives
\[
A\bigl(Bg(a)+(1-B)g(b)\bigr)+(1-A)g(c)
=
Dg(a)+(1-D)\bigl(Cg(b)+(1-C)g(c)\bigr).
\]
Now fix $b$ and $c$ and vary $a$ through $I$. Since $g(I)$ is a nondegenerate interval, the coefficient of $g(a)$ must be the same on both sides. Doing the same for $b$ and $c$, we obtain
\[
\sigma([u,v],w)\,\sigma(u,v)=\sigma(u,[v,w]),
\tag{8}\label{eq:regroup-1}
\]
\[
\sigma([u,v],w)\,\bigl(1-\sigma(u,v)\bigr)
=
\bigl(1-\sigma(u,[v,w])\bigr)\sigma(v,w),
\tag{9}\label{eq:regroup-2}
\]
\[
1-\sigma([u,v],w)
=
\bigl(1-\sigma(u,[v,w])\bigr)\bigl(1-\sigma(v,w)\bigr).
\tag{10}\label{eq:regroup-3}
\]

\smallskip
\noindent\emph{Step 6: $\sigma$ comes from an additive positive block content.}
Let $e:=(1)$, and define
\[
\mu(z):=\frac{\sigma(z,e)}{1-\sigma(z,e)}\qquad(z\text{ a positive block}).
\tag{11}\label{eq:mu-definition}
\]
Apply \eqref{eq:regroup-2} and \eqref{eq:regroup-3} to the triple $(u,v,e)$. With the abbreviations
\[
p:=\sigma(u,v),\quad q:=\sigma([u,v],e),\quad r:=\sigma(v,e),\quad d:=\sigma(u,[v,e]),
\]
these identities become
\[
q(1-p)=(1-d)r,
\qquad
1-q=(1-d)(1-r).
\]
Dividing the first by the second yields
\[
\frac{q}{1-q}(1-p)=\frac{r}{1-r},
\]
i.e.
\[
\mu([u,v])\bigl(1-\sigma(u,v)\bigr)=\mu(v).
\tag{12}\label{eq:mu-first-relation}
\]
Now swap $u$ and $v$. Because $[u,v]$ and $[v,u]$ differ only by a permutation, \eqref{eq:internal-permutation-invariance} gives
\[
\mu([u,v])=\mu([v,u]).
\]
Using also \eqref{eq:block-swap}, the swapped version of \eqref{eq:mu-first-relation} becomes
\[
\mu([u,v])\sigma(u,v)=\mu(u).
\tag{13}\label{eq:mu-second-relation}
\]
Adding \eqref{eq:mu-first-relation} and \eqref{eq:mu-second-relation}, we get
\[
\mu([u,v])=\mu(u)+\mu(v),
\tag{14}\label{eq:mu-additive}
\]
and then \eqref{eq:mu-second-relation} yields
\[
\sigma(u,v)=\frac{\mu(u)}{\mu(u)+\mu(v)}.
\tag{15}\label{eq:sigma-via-mu}
\]

\smallskip
\noindent\emph{Step 7: $\mu$ is a power sum.}
From \eqref{eq:sigma-via-mu} and the common-scale invariance \eqref{eq:common-scale-invariance},
\[
\frac{\mu(\lambda u)}{\mu(\lambda u)+\mu(\lambda v)}
=
\frac{\mu(u)}{\mu(u)+\mu(v)}
\qquad(\lambda>0),
\]
which implies
\[
\frac{\mu(\lambda u)}{\mu(\lambda v)}=\frac{\mu(u)}{\mu(v)}.
\]
In particular, with $e=(1)$ as above,
\[
\mu(\lambda u)=c(\lambda)\,\mu(u),
\qquad
c(\lambda):=\frac{\mu(\lambda e)}{\mu(e)}.
\tag{16}\label{eq:c-lambda}
\]
Applying \eqref{eq:c-lambda} twice gives
\[
\mu(\lambda\tau u)=c(\lambda)c(\tau)\mu(u),
\]
while applying it once with $\lambda\tau$ gives
\[
\mu(\lambda\tau u)=c(\lambda\tau)\mu(u).
\]
Thus
\[
c(\lambda\tau)=c(\lambda)c(\tau).
\]
Since $c$ is continuous, there exists $\rho\in\mathbb R$ such that
\[
c(\lambda)=\lambda^\rho.
\tag{17}\label{eq:c-is-power}
\]
Now define the scalar function
\[
m(t):=\mu((t))
\qquad(t>0).
\]
By \eqref{eq:mu-additive},
\[
\mu(z_1,\dots,z_m)=\sum_{j=1}^m m(z_j).
\]
By \eqref{eq:c-is-power},
\[
m(\lambda t)=\lambda^\rho m(t).
\]
Taking $t=1$ shows that there is a constant $C>0$ such that
\[
m(t)=Ct^\rho.
\]
Hence
\[
\mu(z)=C\sum_j z_j^\rho.
\]
The constant $C$ cancels from \eqref{eq:sigma-via-mu}, so the selector is now explicit:
\[
\sigma(y^1,y^2)
=
\frac{\sum_j (y_j^1)^\rho}{\sum_j (y_j^1)^\rho+\sum_j (y_j^2)^\rho}.
\tag{18}\label{eq:explicit-sigma}
\]

\smallskip
\noindent\emph{Step 8: the recursive identity and the exact two-user reduction.}
Let $x=[x^1,x^2]$ be a positive partition, and write
\[
w_i:=\wt(x^i)\qquad(i=1,2).
\]
Applying Axiom~\ref{ax:partition} to the pair
\[
x=[x^1,x^2],\qquad y=[w_1,w_2],
\]
and using scalar normalization together with \eqref{eq:explicit-sigma}, we obtain
\[
\frac{f(x)}{f(w_1,w_2)}
=
g^{-1}\!\left(
\frac{w_1^\rho}{w_1^\rho+w_2^\rho}\,g\bigl(f(x^1)\bigr)
+
\frac{w_2^\rho}{w_1^\rho+w_2^\rho}\,g\bigl(f(x^2)\bigr)
\right).
\tag{19}\label{eq:recursive-identity}
\]

Now let $x_1,x_2\in\mathbb Q_{>0}$, and choose $m\in\mathbb N$ so that
\[
N_1:=mx_1\in\mathbb N,
\qquad
N_2:=mx_2\in\mathbb N.
\]
Apply Axiom~\ref{ax:partition} to
\[
X=[\ones{N_1},\ones{N_2}],
\qquad
Y=[x_1\ones{m},x_2\ones{m}].
\]
Then
\[
\wt(X^1)=N_1=\wt(Y^1),
\qquad
\wt(X^2)=N_2=\wt(Y^2),
\]
so the axiom applies. By \eqref{eq:binary-direct-product} and \eqref{eq:equal-profile-power-law},
\[
f(X)=f(\ones{N_1+N_2})=m^r(x_1+x_2)^r,
\]
\[
f(Y)=f(\ones{m})\,f(x_1,x_2)=m^r f(x_1,x_2).
\]
Also,
\[
\frac{f(X^1)}{f(Y^1)}
=
\frac{f(\ones{N_1})}{f(x_1\ones{m})}
=
\frac{(mx_1)^r}{m^r}
=
x_1^r,
\]
\[
\frac{f(X^2)}{f(Y^2)}=x_2^r,
\]
and by \eqref{eq:explicit-sigma},
\[
\sigma(Y^1,Y^2)
=
\frac{m x_1^\rho}{m x_1^\rho+m x_2^\rho}
=
\frac{x_1^\rho}{x_1^\rho+x_2^\rho}.
\]
Therefore
\[
\frac{m^r(x_1+x_2)^r}{m^r f(x_1,x_2)}
=
g^{-1}\!\left(
\frac{x_1^\rho}{x_1^\rho+x_2^\rho}\,g(x_1^r)
+
\frac{x_2^\rho}{x_1^\rho+x_2^\rho}\,g(x_2^r)
\right).
\]
Cancelling $m^r$ and extending by continuity from rational to arbitrary positive $x_1,x_2$, we obtain the exact two-user reduction
\[
\frac{(x_1+x_2)^r}{f(x_1,x_2)}
=
g^{-1}\!\left(
\frac{x_1^\rho}{x_1^\rho+x_2^\rho}\,g(x_1^r)
+
\frac{x_2^\rho}{x_1^\rho+x_2^\rho}\,g(x_2^r)
\right)
\qquad(x_1,x_2>0).
\tag{20}\label{eq:two-user-reduction}
\]

\smallskip
\noindent\emph{Step 9: the branch $r=0$ is impossible under Axiom~\ref{ax:strict-equalization}.}
If $r=0$, then \eqref{eq:two-user-reduction} becomes
\[
\frac{1}{f(x_1,x_2)}
=
g^{-1}\!\left(
\frac{x_1^\rho}{x_1^\rho+x_2^\rho}\,g(1)
+
\frac{x_2^\rho}{x_1^\rho+x_2^\rho}\,g(1)
\right)
=
1,
\]
so
\[
f(x_1,x_2)=1\qquad\forall x_1,x_2>0.
\]
Then the recursive identity \eqref{eq:recursive-identity} shows by induction on the dimension that $f\equiv1$ on every positive orthant, and by continuity on every $\mathbb R_+^n\setminus\{0\}$. This contradicts Axiom~\ref{ax:strict-equalization}. Hence
\[
r\neq0.
\tag{21}\label{eq:r-nonzero}
\]

\smallskip
\noindent\emph{Step 10: the induced weighted quasi-arithmetic mean is homogeneous.}
Set
\[
\alpha:=\frac{\rho}{r},
\]
and define
\[
M_g^{(\alpha)}(u,v)
:=
g^{-1}\!\left(
\frac{u^\alpha g(u)+v^\alpha g(v)}{u^\alpha+v^\alpha}
\right)
\qquad(u,v>0).
\]
Then \eqref{eq:two-user-reduction} says exactly
\[
M_g^{(\alpha)}(x_1^r,x_2^r)
=
\frac{(x_1+x_2)^r}{f(x_1,x_2)}.
\tag{22}\label{eq:master-mean-identity}
\]
Using Axiom~\ref{ax:homogeneity},
\[
M_g^{(\alpha)}((tx_1)^r,(tx_2)^r)
=
\frac{(tx_1+tx_2)^r}{f(tx_1,tx_2)}
=
t^r\frac{(x_1+x_2)^r}{f(x_1,x_2)}
=
t^r M_g^{(\alpha)}(x_1^r,x_2^r).
\]
Because $r\neq0$, the map $x\mapsto x^r$ is a homeomorphism of $(0,\infty)$ onto itself. Hence
\[
M_g^{(\alpha)}(cu,cv)=c\,M_g^{(\alpha)}(u,v)
\qquad\forall c,u,v>0.
\tag{23}\label{eq:homogeneous-induced-mean}
\]
By Proposition~\ref{prop:homogeneous-bajraktarevic}, exactly one of the following cases holds, up to affine transformations
\[
g(x)=\log x,
\quad\text{or}\quad
g(x)=x^\beta\quad(\beta\neq0).
\]

Indeed, if
\[
h(x)=A\,g(x)+B
\qquad(A\neq 0,\ B\in\mathbb R),
\]
then for every \(\lambda\in(0,1)\) and every \(u,v>0\),
\[
h^{-1}\!\bigl(\lambda h(u)+(1-\lambda)h(v)\bigr)
=
g^{-1}\!\bigl(\lambda g(u)+(1-\lambda)g(v)\bigr).
\]
Thus the partition axiom itself also holds are unchanged when \(g\) is replaced by an affine transform.

\smallskip
\noindent\emph{Step 11: the logarithmic branch.}
Assume first that $g(x)=\log x$. Then \eqref{eq:two-user-reduction} becomes
\[
\log\frac{(x_1+x_2)^r}{f(x_1,x_2)}
=
r\,\frac{x_1^\rho\log x_1+x_2^\rho\log x_2}{x_1^\rho+x_2^\rho},
\]
whence
\[
f(x_1,x_2)
=
\exp\!\left(
-r\,
\frac{x_1^\rho\log\frac{x_1}{x_1+x_2}+x_2^\rho\log\frac{x_2}{x_1+x_2}}{x_1^\rho+x_2^\rho}
\right).
\tag{24}\label{eq:two-user-log-formula}
\]
We now compare the two recursive decompositions of the three-point profile $(1,1,t)$.

For the split $[1,1]\mid[t]$, \eqref{eq:recursive-identity} gives
\[
\log f(1,1,t)
=
\log f(2,t)+\frac{2^\rho}{2^\rho+t^\rho}\log f(1,1).
\]
Using \eqref{eq:two-user-log-formula},
\[
\log f(2,t)
=
r\log(2+t)-r\frac{2^\rho\log 2+t^\rho\log t}{2^\rho+t^\rho},
\qquad
\log f(1,1)=r\log 2,
\]
so
\[
\log f(1,1,t)
=
r\log(2+t)-r\frac{t^\rho\log t}{2^\rho+t^\rho}.
\tag{25}\label{eq:left-log-branch}
\]
For the split $[1]\mid[1,t]$, \eqref{eq:recursive-identity} gives
\[
\log f(1,1,t)
=
\log f(1,1+t)+\frac{(1+t)^\rho}{1+(1+t)^\rho}\log f(1,t).
\]
Again by \eqref{eq:two-user-log-formula},
\[
\log f(1,1+t)
=
r\log(2+t)-r\frac{(1+t)^\rho\log(1+t)}{1+(1+t)^\rho},
\]
\[
\log f(1,t)
=
r\log(1+t)-r\frac{t^\rho\log t}{1+t^\rho},
\]
and therefore
\[
\log f(1,1,t)
=
r\log(2+t)-r\frac{(1+t)^\rho t^\rho\log t}{(1+t^\rho)(1+(1+t)^\rho)}.
\tag{26}\label{eq:right-log-branch}
\]
Comparing \eqref{eq:left-log-branch} and \eqref{eq:right-log-branch}, and cancelling $r\log t$ for $t\neq1$, we obtain
\[
\frac{1}{2^\rho+t^\rho}
=
\frac{(1+t)^\rho}{(1+t^\rho)(1+(1+t)^\rho)}.
\]
By continuity this also holds at $t=1$, where it becomes
\[
\frac{1}{2^\rho+1}=\frac{2^\rho}{2(1+2^\rho)}.
\]
Hence $2=2^\rho$, so
\[
\rho=1.
\tag{27}\label{eq:rho-equals-one}
\]
Substituting \eqref{eq:rho-equals-one} into \eqref{eq:two-user-log-formula} gives
\[
f(x_1,x_2)
=
\exp\!\left(-r\sum_{i=1}^2 p_i(x)\log p_i(x)\right).
\tag{28}\label{eq:two-user-log-final}
\]
Now define the Shannon entropy of the share vector by
\[
H(x):=-\sum_{i=1}^n p_i(x)\log p_i(x).
\]
Its chain rule is
\[
H([x^1,x^2])
=
H(w_1,w_2)+\frac{w_1}{w_1+w_2}H(x^1)+\frac{w_2}{w_1+w_2}H(x^2).
\]
Because \eqref{eq:recursive-identity} with $\rho=1$ has exactly the same form for $\log f$, and because the base case is \eqref{eq:two-user-log-final}, an induction on the dimension yields
\[
f(x)=e^{rH(x)}=\prod_{i=1}^n p_i(x)^{-r\,p_i(x)}.
\tag{29}\label{eq:logarithmic-family}
\]
Since boundary continuity gives $f(1,0)=1$, Axiom~\ref{ax:strict-equalization} now reads
\[
1<f\!\left(\frac12,\frac12\right)=2^r,
\]
so $r>0$.

\smallskip
\noindent\emph{Step 12: the power branch.}
Assume now that $g(x)=x^\beta$ with $\beta\neq0$. Then \eqref{eq:two-user-reduction} becomes
\[
\left(\frac{(x_1+x_2)^r}{f(x_1,x_2)}\right)^\beta
=
\frac{x_1^{\rho+\beta r}+x_2^{\rho+\beta r}}{x_1^\rho+x_2^\rho}.
\]
Set
\(
\lambda:=\rho+\beta r 
\). Then
\[
f(x_1,x_2)
=
(x_1+x_2)^r\left(\frac{x_1^\rho+x_2^\rho}{x_1^\lambda+x_2^\lambda}\right)^{1/\beta}.
\tag{30}\label{eq:two-user-power-general}
\]
We again compare the two recursive decompositions of $(1,1,t)$. Since $g(x)=x^\beta$, \eqref{eq:recursive-identity} is equivalent to
\[
f([x^1,x^2])^\beta
=
f(w_1,w_2)^\beta\,
\frac{w_1^\rho f(x^1)^\beta+w_2^\rho f(x^2)^\beta}{w_1^\rho+w_2^\rho}.
\tag{31}\label{eq:recursive-power-form}
\]
For the split $[1,1]\mid[t]$, this gives
\[
f(1,1,t)^\beta
=
f(2,t)^\beta\,\frac{2^\rho f(1,1)^\beta+t^\rho}{2^\rho+t^\rho}.
\]
Using $f(1,1)=2^r$ and \eqref{eq:two-user-power-general}, we get
\[
f(2,t)^\beta=(2+t)^{\beta r}\frac{2^\rho+t^\rho}{2^\lambda+t^\lambda},
\qquad
2^\rho f(1,1)^\beta=2^{\rho+\beta r}=2^\lambda,
\]
therefore
\[
f(1,1,t)^\beta
=
(2+t)^{\beta r}\frac{2^\lambda+t^\rho}{2^\lambda+t^\lambda}.
\tag{32}\label{eq:left-power-branch}
\]
For the split $[1]\mid[1,t]$, \eqref{eq:recursive-power-form} gives
\[
f(1,1,t)^\beta
=
f(1,1+t)^\beta\,\frac{1+(1+t)^\rho f(1,t)^\beta}{1+(1+t)^\rho}.
\]
Using \eqref{eq:two-user-power-general} again,
\[
f(1,1+t)^\beta=(2+t)^{\beta r}\frac{1+(1+t)^\rho}{1+(1+t)^\lambda},
\]
\[
f(1,t)^\beta=(1+t)^{\beta r}\frac{1+t^\rho}{1+t^\lambda}=(1+t)^{\lambda-\rho}\frac{1+t^\rho}{1+t^\lambda}.
\]
Hence
\[
1+(1+t)^\rho f(1,t)^\beta
=
\frac{1+t^\lambda+(1+t)^\lambda(1+t^\rho)}{1+t^\lambda},
\]
and therefore
\[
f(1,1,t)^\beta
=
(2+t)^{\beta r}
\frac{1+t^\lambda+(1+t)^\lambda(1+t^\rho)}{(1+t^\lambda)(1+(1+t)^\lambda)}.
\tag{33}\label{eq:right-power-branch}
\]
Comparing \eqref{eq:left-power-branch} and \eqref{eq:right-power-branch} yields
\[
\frac{2^\lambda+t^\rho}{2^\lambda+t^\lambda}
=
\frac{1+t^\lambda+(1+t)^\lambda(1+t^\rho)}{(1+t^\lambda)(1+(1+t)^\lambda)}.
\]
After cross-multiplication and simplification,
\[
(t^\lambda-t^\rho)\bigl(2^\lambda(1+t)^\lambda-t^\lambda-(1+t)^\lambda-1\bigr)=0.
\]
Now, by \eqref{eq:r-nonzero},
\[
\lambda-\rho=\beta r\neq0,
\]
so for $t\neq1$, the first factor is nonzero. Hence
\[
2^\lambda(1+t)^\lambda=t^\lambda+(1+t)^\lambda+1
\qquad(t\neq1).
\]
By continuity this also holds at $t=1$, where it becomes
\[
4^\lambda=2^\lambda+2.
\]
Setting $u:=2^\lambda>0$, we get
\[
u^2-u-2=0,
\]
so $u=2$, i.e.
\[
2^\lambda=2,
\qquad
\lambda=1.
\]
Therefore
\[
\rho+\beta r=1.
\tag{34}\label{eq:parameter-relation}
\]
Substituting \eqref{eq:parameter-relation} into \eqref{eq:two-user-power-general}, we obtain
\[
f(x_1,x_2)
=
(x_1+x_2)^{r-1/\beta}(x_1^\rho+x_2^\rho)^{1/\beta},
\qquad
\rho=1-\beta r.
\tag{35}\label{eq:two-user-power-final}
\]
We now prove by induction on $n$ that
\[
f(x_1,\dots,x_n)
=
\left(\sum_{i=1}^n x_i^\rho\right)^{1/\beta}
\left(\sum_{i=1}^n x_i\right)^{r-1/\beta}.
\tag{36}\label{eq:n-user-power-form}
\]
The case $n=2$ is exactly \eqref{eq:two-user-power-final}. Assume \eqref{eq:n-user-power-form} holds for $n=k$, and write
\[
S_k:=\sum_{i=1}^k x_i,
\qquad
A_k:=\sum_{i=1}^k x_i^\rho.
\]
Applying \eqref{eq:recursive-power-form} to the split
\[
[x_1,\dots,x_k]\mid[x_{k+1}],
\]
we get
\[
f(x_1,\dots,x_{k+1})
=
f(S_k,x_{k+1})
\left(
\frac{S_k^\rho f(x_1,\dots,x_k)^\beta+x_{k+1}^\rho}{S_k^\rho+x_{k+1}^\rho}
\right)^{1/\beta}.
\]
By the induction hypothesis,
\[
f(x_1,\dots,x_k)^\beta=A_k S_k^{\beta r-1}.
\]
Since \eqref{eq:parameter-relation} implies $\beta r-1=-\rho$, we have
\[
S_k^\rho f(x_1,\dots,x_k)^\beta=A_k.
\]
Substituting this identity and the two-user formula \eqref{eq:two-user-power-final}, we obtain
\begin{align*}
f(x_1,\dots,x_{k+1})
&=
(S_k+x_{k+1})^{r-1/\beta}(S_k^\rho+x_{k+1}^\rho)^{1/\beta}
\left(
\frac{A_k+x_{k+1}^\rho}{S_k^\rho+x_{k+1}^\rho}
\right)^{1/\beta}\\
&=
(S_k+x_{k+1})^{r-1/\beta}(A_k+x_{k+1}^\rho)^{1/\beta},
\end{align*}
which is exactly \eqref{eq:n-user-power-form} for $k+1$. Thus \eqref{eq:n-user-power-form} holds for all $n\ge2$.

Dividing by the total sum, \eqref{eq:n-user-power-form} becomes
\[
f(x)=\left(\sum_{i=1}^n p_i(x)^\rho\right)^{1/\beta}.
\tag{37}\label{eq:power-family}
\]
Taking $x=(t,1-t)$ with $0<t<1$ in \eqref{eq:power-family}, we obtain
\[
f(t,1-t)=\bigl(t^\rho+(1-t)^\rho\bigr)^{1/\beta}.
\]
If $\rho<0$, then
\[
t^\rho+(1-t)^\rho \xrightarrow[t\downarrow0]{} +\infty.
\]
Hence
\[
f(t,1-t)\xrightarrow[t\downarrow0]{}
\begin{cases}
+\infty,& \beta>0,\\
0,& \beta<0,
\end{cases}
\]
contradicting Axiom~\ref{ax:continuity} at the boundary point $(0,1)$, because the limit must equal the finite positive number $f(0,1)$.
If $\rho=0$, then
\[
f(t,1-t)=2^{1/\beta}\qquad(0<t<1).
\]
Letting $t\downarrow0$ and using continuity at $(0,1)$, we obtain
\[
f(0,1)=2^{1/\beta}.
\]
By Axiom~\ref{ax:anonymity},
\[
f(1,0)=f(0,1)=2^{1/\beta}=f\!\left(\frac12,\frac12\right),
\]
contradicting Axiom~\ref{ax:strict-equalization}. Therefore
\[
\rho>0.
\tag{38}\label{eq:rho-positive}
\]
Now the same two-point identity and $\rho>0$ imply
\[
\lim_{t\downarrow0}f(t,1-t)=1.
\]
By continuity at $(0,1)$ and Axiom~\ref{ax:anonymity},
\[
f(1,0)=f(0,1)=1.
\]
Also,
\[
f\!\left(\frac12,\frac12\right)
=
\left(2\left(\frac12\right)^\rho\right)^{1/\beta}
=
2^{(1-\rho)/\beta}
=
2^r,
\]
where \eqref{eq:parameter-relation} was used in the last step. Hence Axiom~\ref{ax:strict-equalization} implies $r>0$.

\smallskip
\noindent\emph{Step 13: converse verification.}
For the logarithmic family \eqref{eq:logarithmic-family}, Axioms~\ref{ax:continuity}--\ref{ax:anonymity} are immediate. Also
\[
f(\ones{n})=n^r,
\]
so Axiom~\ref{ax:boundedness} holds. Let
\[
x=[x^1,x^2],\qquad y=[y^1,y^2],\qquad w_i:=\wt(x^i)=\wt(y^i),\qquad W:=w_1+w_2.
\]
The Shannon chain rule gives
\[
\log f(x)
=
\log f(w_1,w_2)+\frac{w_1}{W}\log f(x^1)+\frac{w_2}{W}\log f(x^2),
\]
and the same identity for $y$. Subtracting,
\[
\log\frac{f(x)}{f(y)}
=
\frac{w_1}{W}\log\frac{f(x^1)}{f(y^1)}
+
\frac{w_2}{W}\log\frac{f(x^2)}{f(y^2)}.
\]
Thus Axiom~\ref{ax:partition} holds with $g(t)=\log t$ and
\[
\sigma(y^1,y^2)=\frac{w(y^1)}{w(y^1)+w(y^2)}.
\]
Since $r>0$,
\[
f(1,0)=1<2^r=f\!\left(\frac12,\frac12\right),
\]
so Axiom~\ref{ax:strict-equalization} holds.

For the power family \eqref{eq:power-family} with \eqref{eq:parameter-relation} and \eqref{eq:rho-positive}, Axioms~\ref{ax:continuity}--\ref{ax:anonymity} are again immediate, and
\[
f(\ones{n})
=
\left(n\cdot n^{-\rho}\right)^{1/\beta}
=
n^{(1-\rho)/\beta}
=
n^r,
\]
so Axiom~\ref{ax:boundedness} holds. Define
\[
\mu(z):=\sum_j z_j^\rho.
\]
Then
\[
f(z)^\beta=\frac{\mu(z)}{\wt(z)^\rho}.
\]
For matched positive partitions as above,
\[
f(x)^\beta=\frac{\mu(x^1)+\mu(x^2)}{W^\rho},
\qquad
f(y)^\beta=\frac{\mu(y^1)+\mu(y^2)}{W^\rho},
\]
so
\[
\left(\frac{f(x)}{f(y)}\right)^\beta
=
\frac{\mu(x^1)+\mu(x^2)}{\mu(y^1)+\mu(y^2)}.
\]
Because $\mu(x^i)=w_i^\rho f(x^i)^\beta$ and $\mu(y^i)=w_i^\rho f(y^i)^\beta$, this becomes
\[
\left(\frac{f(x)}{f(y)}\right)^\beta
=
\frac{\mu(y^1)}{\mu(y^1)+\mu(y^2)}
\left(\frac{f(x^1)}{f(y^1)}\right)^\beta
+
\frac{\mu(y^2)}{\mu(y^1)+\mu(y^2)}
\left(\frac{f(x^2)}{f(y^2)}\right)^\beta.
\]
Thus Axiom~\ref{ax:partition} holds with $g(t)=t^\beta$ and
\[
\sigma(y^1,y^2)=\frac{\mu(y^1)}{\mu(y^1)+\mu(y^2)}
=
\frac{\sum_j (y_j^1)^\rho}{\sum_j (y_j^1)^\rho+\sum_j (y_j^2)^\rho}.
\]
Finally,
\[
f(1,0)=1<2^r=f\!\left(\frac12,\frac12\right),
\]
so Axiom~\ref{ax:strict-equalization} holds. This completes the proof.
\end{proof}

\subsection{Generalized Atkinson-index weighted $F_{\text{Aggregate}}$}
Just like gini index, we also want to understand the relationship between fairness measure \(f(x)\) and relevant weighted SWF.
So we now relate the power branch of Theorem~\ref{thm:single-function-atkinson} to the Atkinson index and the corresponding aggregate-welfare functional. Let
\(
\mu:=\frac1n\sum_{i=1}^n x_i
\)
be the arithmetic mean. For
\[
\epsilon\in(-\infty,1)\setminus\{0\},
\]
define the equally distributed equivalent level by
\[
y_e:=\left(\frac1n\sum_{i=1}^n x_i^{1-\epsilon}\right)^{\frac{1}{1-\epsilon}}
\]
and the Atkinson index by
\[
A(\epsilon):=1-\frac{y_e}{\mu}.
\]

Assume \(f\) belongs to the power branch of Theorem~\ref{thm:single-function-atkinson},
\[
f(x)=\left(\sum_{i=1}^n p_i(x)^\rho\right)^{1/\beta},
\qquad
p_i(x):=\frac{x_i}{\sum_{j=1}^n x_j},
\qquad
\rho>0,
\qquad
\beta\neq 0,
\]
and set
\(
\epsilon:=1-\rho.
\)
Then
\begin{align*}
\sum_{i=1}^n p_i(x)^\rho& =
\frac{\sum_{i=1}^n x_i^\rho}{\left(\sum_{i=1}^n x_i\right)^\rho}=
\frac{n\,y_e^{\,\rho}}{(n\mu)^\rho}\\
&=
n^{1-\rho}\left(\frac{y_e}{\mu}\right)^\rho =
n^{1-\rho}(1-A(\epsilon))^\rho.
\end{align*}
Therefore
\[
f(x)=n^{(1-\rho)/\beta}(1-A(\epsilon))^{\rho/\beta}.
\]
Using the parameter relation \(r=(1-\rho)/\beta\) from \eqref{eq:parameter-relation}, this becomes
\[
f(x)=n^r(1-A(\epsilon))^{\rho/\beta},
\]
and hence
\[
1-A(\epsilon)
=
\left(\frac{f(x)}{n^r}\right)^{\beta/\rho}.
\]
Multiplying by \(\sum_{i=1}^n x_i=n\mu\), we obtain the associated aggregate-welfare representation
\[
(1-A(\epsilon))\sum_{i=1}^n x_i
=
\left(\frac{f(x)}{n^r}\right)^{\beta/\rho}\sum_{i=1}^n x_i
=
n y_e.
\]

In the classical Atkinson slice one has
\[
0<\epsilon<1,
\qquad
\beta=\epsilon,
\qquad
\rho=1-\epsilon,
\qquad
r=1,
\]
so
\[
f_\epsilon(x)
=
\left(\sum_{i=1}^n p_i(x)^{1-\epsilon}\right)^{1/\epsilon}
=
n(1-A(\epsilon))^{\frac{1-\epsilon}{\epsilon}},
\]
and therefore
\[
(1-A(\epsilon))\sum_{i=1}^n x_i
=
\left(\frac{f_\epsilon(x)}{n}\right)^{\frac{\epsilon}{1-\epsilon}}
\sum_{i=1}^n x_i
=
n y_e.
\]

Thus, on the power branch, the fairness measure \(f\) determines the Atkinson equality factor \(1-A(\epsilon)\) by a simple power transformation, and the induced aggregate welfare recovers the Atkinson-weighted equally distributed equivalent total \(n y_e\). In the classical case \(0<\epsilon<1\), this transformation is increasing.

\section{Conclusion}
When we first confront the problem of \textit{unequal} inequality, it appears daunting, seemingly pointing to a unstructured landscape with messy contradictions. This perception is further amplified by the multitude of seemingly plausible inequality measurable, each capturing some aspect of inequality. However, as we delved deeper into these measures, it became clear that most of them are untenable due to their structural properties. They either remain invariant under too many or too few transformations. Through this process of elimination, we narrowed down the viable candidates for formalizing egalitarianism. Gini index emerged as a particularly distinctive measure, reflecting deviation from counterfactual equality under ratio-invariance. Atkinson index, on the other hand, generalized a range of indices that are built on deviation from the mean through a parameterized approach. While this still doesn't guarantee that all the potential plurality of inequality-measure could been elegantly unified, it undeniably reveals a surprising picture that seems more coherent.

Throughout the paper, we have made many cases for the merits of the axiomatic methodology. It  transform pre-formal intuitions into precise formulations, thereby enabling the evaluation of theoretical consistency. However, it must be acknowledged that the representation theorem of one-person utility theory and utilitarianism , while valuable, appear tautological in nature—offering results that are predictable and plain. The real shift in perspective occurs when we turn our attention to egalitarianism, where the demand for consistency becomes more pronounced. We began by reviewing attempts to use statistical measures and rank-discounting schemes as practical substitutes. As demonstrated, these methods  result in inconsistencies precisely because they lack  solid axiomatic foundation. In the following section, we present a formalized approach to egalitarianism through the Gini index and Atkinson index, offering a coherent, axiomatic framework that addresses the deficiencies highlighted here. These representation theorems replace the flawed, ad-hoc methods with mathematically rigorous theory of egalitarianism.

\appendix

\section{Deferred Proofs}
\AllDeferredProofs

\nocite{*} 
\printbibliography

@inproceedings{Lan2010AnAT,
  title={An Axiomatic Theory of Fairness in Resource Allocation},
  author={Tian Lan and Mung Chiang}, 
  year={2010},
  url={https://www.princeton.edu/~chiangm/fairness.pdf}
}

@book{dudley2018real,
  title={Real analysis and probability},
  author={Dudley, Richard M},
  year={2018},
  publisher={Chapman and Hall/CRC}
}

@incollection{kolmogorov1991notion,
  author    = {Kolmogorov, A. N.},
  title     = {On the Notion of Mean},
  booktitle = {Mathematics and Mechanics: Selected Works of A. N. Kolmogorov},
  publisher = {Kluwer Academic Publishers},
  pages     = {144--146},
  year      = {1991},
  note      = {Translation of the 1930 paper}
}

@article{WirsingZagier2001,
  author  = {Wirsing, Eduard and Zagier, Don},
  title   = {Multiplicative Functions with Difference Tending to Zero},
  journal = {Acta Arithmetica},
  volume  = {100},
  number  = {1},
  pages   = {75--78},
  year    = {2001},
  doi     = {10.4064/aa100-1-6}
}

@book{MontgomeryVaughan2007,
  author    = {Montgomery, Hugh L. and Vaughan, Robert C.},
  title     = {Multiplicative Number Theory I: Classical Theory},
  series    = {Cambridge Studies in Advanced Mathematics},
  volume    = {97},
  publisher = {Cambridge University Press},
  address   = {Cambridge},
  year      = {2007}
}

@book{Tenenbaum2015,
  author    = {Tenenbaum, G{\'e}rald},
  title     = {Introduction to Analytic and Probabilistic Number Theory},
  edition   = {3},
  series    = {Graduate Studies in Mathematics},
  volume    = {163},
  publisher = {American Mathematical Society},
  address   = {Providence, RI},
  year      = {2015},
  doi       = {10.1090/GSM/163}
}

@book{AmbrosioFuscoPallara2000,
  author    = {Ambrosio, Luigi and Fusco, Nicola and Pallara, Diego},
  title     = {Functions of Bounded Variation and Free Discontinuity Problems},
  publisher = {Oxford University Press},
  address   = {Oxford},
  year      = {2000},
  doi       = {10.1093/oso/9780198502456.001.0001}
}

@book{BrennerScott2008,
  author    = {Brenner, Susanne C. and Scott, L. Ridgway},
  title     = {The Mathematical Theory of Finite Element Methods},
  edition   = {3},
  series    = {Texts in Applied Mathematics},
  publisher = {Springer},
  address   = {New York},
  year      = {2008},
  doi       = {10.1007/978-0-387-75934-0}
}

@book{SchaeferWolff1999,
  author    = {Schaefer, Helmut H. and Wolff, Manfred P.},
  title     = {Topological Vector Spaces},
  edition   = {2},
  series    = {Graduate Texts in Mathematics},
  volume    = {3},
  publisher = {Springer},
  year      = {1999}
}

@incollection{Weymark1991-WEYARO,
	author = {John A. Weymark},
	booktitle = {Interpersonal Comparisons of Well-Being},
	editor = {Jon Elster and John Roemer},
	pages = {255--320},
	publisher = {Cambridge University Press},
	title = {A Reconsideration of the Harsanyi?Sen Debate on Utilitarianism},
	year = {1991}
}

@book{hardy1952inequalities,
  title={Inequalities},
  author={Hardy, G.H.},
  year={1952},
  publisher={Cambridge University Press}
}

@incollection{DAspremont2008,
	author = {Claude D'Aspremont and Philippe Mongin},
	booktitle = {Justice, Political Liberalism, and Utilitarianism},
	editor = {M. Fleurbaey M. Salles and J. Weymark},
	pages = {Ch. 11},
	publisher = {Cambridge University Press},
	title = {A Welfarist Version of Harsanyi's Theorem},
	year = {2008}
}

@article{Thomson1976-THOKLD-2,
	author = {Judith Jarvis Thomson},
	doi = {10.5840/monist197659224},
	journal = {The Monist},
	number = {2},
	pages = {204--217},
	publisher = {The Hegeler Institute},
	title = {Killing, Letting Die, and the Trolley Problem},
	volume = {59},
	year = {1976}
}

@InCollection{sep-consequentialism,
	author       =	{Sinnott-Armstrong, Walter},
	title        =	{{Consequentialism}},
	booktitle    =	{The {Stanford} Encyclopedia of Philosophy},
	editor       =	{Edward N. Zalta and Uri Nodelman},
	howpublished =	{\url{https://plato.stanford.edu/archives/win2023/entries/consequentialism/}},
	year         =	{2023},
	edition      =	{{W}inter 2023},
	publisher    =	{Metaphysics Research Lab, Stanford University}
}

@book{RockafellarWets1998,
  author    = {Rockafellar, R. Tyrrell and Wets, Roger J.-B.},
  title     = {Variational Analysis},
  series    = {Grundlehren der mathematischen Wissenschaften},
  publisher = {Springer},
  address   = {Berlin and Heidelberg},
  year      = {1998},
  doi       = {10.1007/978-3-642-02431-3}
}

@book{Scanlon1998,
	address = {Cambridge},
	author = {Thomas Scanlon},
	publisher = {Harvard University Press},
	title = {What We Owe to Each Other},
	year = {1998}
}

@Inbook{Bossert2004,
author="Bossert, Walter
and Weymark, John A.",

title="Utility in Social Choice",
bookTitle="Handbook of Utility Theory: Volume 2 Extensions",
year="2004",
publisher="Springer US",
address="Boston, MA",
pages="1099--1177",

doi="10.1007/978-1-4020-7964-1_7",
url="https://doi.org/10.1007/978-1-4020-7964-1_7"
}

@article{Foot1967-FOOTPO-2,
	author = {Philippa Foot},
	journal = {Oxford Review},
	pages = {5--15},
	title = {The Problem of Abortion and the Doctrine of the Double Effect},
	volume = {5},
	year = {1967}
}

@article{thomson1984trolley,
	author = {Judith Thomson},
	doi = {10.2307/796133},
	journal = {Yale Law Journal},
	number = {6},
	pages = {1395--1415},
	title = {The Trolley Problem},
	volume = {94},
	year = {1985}
}

@incollection{McCarthy2016-MCCPIE-2,
	author = {David McCarthy},
	booktitle = {The Oxford Handbook of Probability and Philosophy},
	editor = {Alan H\'{a}jek and Christopher Hitchcock},
	pages = {705--737},
	publisher = {Oxford University Press},
	title = {Probability in Ethics},
	year = {2016}
}

@book{Broome1991-BROWGE,
	author = {John Broome},
	editor = {},
	publisher = {Wiley-Blackwell},
	title = {Weighing Goods: Equality, Uncertainty and Time},
	year = {1991}
}

@book{fishburn1979utility,
  title={Utility theory for decision making},
  author={ Peter C. Fishburn},
  year={1979},
  publisher={Krieger NY}
}

@article{Aaberge,
author = {Aaberge, Rolf},
year = {2001},
month = {02},
pages = {115-132},
title = {Axiomatic Characterization of the Gini Coefficient and Lorenz Curve Orderings},
volume = {101},
journal = {Journal of Economic Theory},
doi = {10.1006/jeth.2000.2749}
}

@article{McCarthy2015,
	author = {David McCarthy},
	doi = {10.1093/mind/fzv028},
	journal = {Mind},
	number = {496},
	pages = {1045--1109},
	title = {Distributive Equality},
	volume = {124},
	year = {2015}
}

@article{McCarthy2017,
	author = {David McCarthy},
	doi = {10.1017/s0266267116000225},
	journal = {Economics and Philosophy},
	number = {2},
	pages = {215--57},
	publisher = {Cambridge University Press (Cup)},
	title = {The Priority View},
	volume = {33},
	year = {2017}
}

@book{Temkin1993,
	author = {Larry S. Temkin},
	publisher = {Oxford University Press},
	title = {Inequality},
	year = {1993}
}

@book{Adler2011-ADLWAF,
	author = {Matthew Adler},
	editor = {},
	publisher = {Oxford University Press},
	title = {Well-Being and Fair Distribution: Beyond Cost-Benefit Analysis},
	year = {2011}
}

@book{Rockafellar1970,
  author    = {Rockafellar, R. Tyrrell},
  title     = {Convex Analysis},
  series    = {Princeton Mathematical Series},
  volume    = {28},
  publisher = {Princeton University Press},
  address   = {Princeton, NJ},
  year      = {1970}
}

@book{brezis2010functional,
  title={Functional Analysis, Sobolev Spaces and Partial Differential Equations},
  author={Brezis, H.},
  isbn={9780387709130},
  lccn={2010938382},
  series={Universitext},
  url={https://books.google.com.hk/books?id=GAA2XqOIIGoC},
  year={2010},
  publisher={Springer New York}
}

@article{Debreu1959,
  author = {Gerard Debreu},
  title = {Topological Methods in Cardinal Utility Theory},
  year = {1959},
  journal = {Cowles Foundation Discussion Papers},
  volume = {299},
  url = {https://elischolar.library.yale.edu/cowles-discussion-paper-series/299},
}

@book{aczel1975measures,
  title={On Measures of Information and Their Characterizations},
  author={Acz{\'e}l, J. and Dar{\'o}czy, Z.},
  series={Mathematics in Science and Engineering : a series of monographs and textbooks},
  year={1975},
  publisher={Academic Press}
}

@article{Dalton1920,
  author  = {Dalton, Hugh},
  title   = {The Measurement of the Inequality of Incomes},
  journal = {The Economic Journal},
  year    = {1920},
  volume  = {30},
  number  = {119},
  pages   = {348--361}
}

@article{Atkinson1970,
  author  = {Atkinson, Anthony B.},
  title   = {On the Measurement of Inequality},
  journal = {Journal of Economic Theory},
  year    = {1970},
  volume  = {2},
  number  = {3},
  pages   = {244--263}
}

@book{Sen1973,
  author    = {Sen, Amartya},
  title     = {On Economic Inequality},
  publisher = {Oxford University Press},
  year      = {1973}
}

@book{Cowell2011,
  author    = {Cowell, Frank A.},
  title     = {Measuring Inequality},
  edition   = {3},
  publisher = {Oxford University Press},
  year      = {2011}
}

@article{Gastwirth1972,
  author  = {Gastwirth, Joseph L.},
  title   = {The Estimation of the {L}orenz Curve and the {G}ini Index},
  journal = {The Review of Economics and Statistics},
  year    = {1972},
  volume  = {54},
  number  = {3},
  pages   = {306--316}
}

@article{Shorrocks1983,
  author  = {Shorrocks, Anthony F.},
  title   = {Ranking Income Distributions},
  journal = {Economica},
  year    = {1983},
  volume  = {50},
  number  = {197},
  pages   = {3--17}
}

@article{DonaldsonWeymark1980,
  author  = {Donaldson, David and Weymark, John A.},
  title   = {A Single-Parameter Generalization of the {G}ini Indices of Inequality},
  journal = {Journal of Economic Theory},
  year    = {1980},
  volume  = {22},
  number  = {1},
  pages   = {67--86}
}

@article{Weymark1981,
  author  = {Weymark, John A.},
  title   = {Generalized {G}ini Inequality Indices},
  journal = {Mathematical Social Sciences},
  year    = {1981},
  volume  = {1},
  number  = {4},
  pages   = {409--430}
}

@article{Yitzhaki1983,
  author  = {Yitzhaki, Shlomo},
  title   = {On an Extension of the {G}ini Inequality Index},
  journal = {International Economic Review},
  year    = {1983},
  volume  = {24},
  number  = {3},
  pages   = {617--628}
}

@book{HardyLittlewoodPolya1952,
  author    = {Hardy, G. H. and Littlewood, J. E. and P{\'o}lya, G.},
  title     = {Inequalities},
  edition   = {2},
  publisher = {Cambridge University Press},
  year      = {1952}
}

@book{MarshallOlkinArnold2011,
  author    = {Marshall, Albert W. and Olkin, Ingram and Arnold, Barry C.},
  title     = {Inequalities: Theory of Majorization and Its Applications},
  edition   = {2},
  publisher = {Springer},
  year      = {2011}
}

@article{HersteinMilnor1953,
  author  = {Herstein, I. N. and Milnor, John},
  title   = {An Axiomatic Approach to Measurable Utility},
  journal = {Econometrica},
  year    = {1953},
  volume  = {21},
  number  = {2},
  pages   = {291--297}
}

@book{Fishburn1970,
  author    = {Fishburn, Peter C.},
  title     = {Utility Theory for Decision Making},
  publisher = {Wiley},
  year      = {1970}
}

@book{Kreps1988,
  author    = {Kreps, David M.},
  title     = {Notes on the Theory of Choice},
  publisher = {Westview Press},
  year      = {1988}
}

@book{KrantzLuceSuppesTversky1971,
  author    = {Krantz, David H. and Luce, R. Duncan and Suppes, Patrick and Tversky, Amos},
  title     = {Foundations of Measurement, Volume I: Additive and Polynomial Representations},
  publisher = {Academic Press},
  year      = {1971}
}

@book{RudinFunctionalAnalysis1991,
  author    = {Rudin, Walter},
  title     = {Functional Analysis},
  edition   = {2},
  publisher = {McGraw--Hill},
  year      = {1991}
}

@book{ConwayFunctionalAnalysis1990,
  author    = {Conway, John B.},
  title     = {A Course in Functional Analysis},
  edition   = {2},
  publisher = {Springer},
  year      = {1990}
}

@book{RudinRealComplex1987,
  author    = {Rudin, Walter},
  title     = {Real and Complex Analysis},
  edition   = {3},
  publisher = {McGraw-Hill},
  address   = {New York},
  year      = {1987}
}

@book{Hormander2003,
  author    = {H{\"o}rmander, Lars},
  title     = {The Analysis of Linear Partial Differential Operators I: Distribution Theory and Fourier Analysis},
  series    = {Classics in Mathematics},
  publisher = {Springer},
  address   = {Berlin and Heidelberg},
  year      = {2003},
  doi       = {10.1007/978-3-642-61497-2}
}

@book{KreyszigFunctionalAnalysis1978,
  author    = {Kreyszig, Erwin},
  title     = {Introductory Functional Analysis with Applications},
  publisher = {Wiley},
  year      = {1978}
}

@book{AliprantisBorder2006,
  author    = {Aliprantis, Charalambos D. and Border, Kim C.},
  title     = {Infinite Dimensional Analysis: A Hitchhiker's Guide},
  edition   = {3},
  publisher = {Springer},
  year      = {2006}
}

@book{FollandRealAnalysis1999,
  author    = {Folland, Gerald B.},
  title     = {Real Analysis: Modern Techniques and Their Applications},
  edition   = {2},
  publisher = {Wiley},
  year      = {1999}
}

@book{RoydenFitzpatrick2010,
  author    = {Royden, H. L. and Fitzpatrick, P. M.},
  title     = {Real Analysis},
  edition   = {4},
  publisher = {Pearson},
  year      = {2010}
}

@book{RudinPMA1976,
  author    = {Rudin, Walter},
  title     = {Principles of Mathematical Analysis},
  edition   = {3},
  publisher = {McGraw--Hill},
  year      = {1976}
}

@book{Apostol1974,
  author    = {Apostol, Tom M.},
  title     = {Mathematical Analysis},
  edition   = {2},
  publisher = {Addison--Wesley},
  year      = {1974}
}

@book{Billingsley1995,
  author    = {Billingsley, Patrick},
  title     = {Probability and Measure},
  edition   = {3},
  publisher = {Wiley},
  year      = {1995}
}

@book{Bogachev2007,
  author    = {Bogachev, Vladimir I.},
  title     = {Measure Theory},
  publisher = {Springer},
  year      = {2007}
}

@book{EvansPDE2010,
  author    = {Evans, Lawrence C.},
  title     = {Partial Differential Equations},
  edition   = {2},
  publisher = {American Mathematical Society},
  year      = {2010}
}

@book{SteinShakarchi2005,
  author    = {Stein, Elias M. and Shakarchi, Rami},
  title     = {Real Analysis: Measure Theory, Integration, and Hilbert Spaces},
  publisher = {Princeton University Press},
  year      = {2005}
}

@article{Yaari1987,
  author  = {Yaari, Menahem E.},
  title   = {The Dual Theory of Choice under Risk},
  journal = {Econometrica},
  year    = {1987},
  volume  = {55},
  number  = {1},
  pages   = {95--115}
}

@article{Schmeidler1989,
  author  = {Schmeidler, David},
  title   = {Subjective Probability and Expected Utility without Additivity},
  journal = {Econometrica},
  year    = {1989},
  volume  = {57},
  number  = {3},
  pages   = {571--587}
}

@article{Acerbi2002,
  author  = {Acerbi, Carlo},
  title   = {Spectral Measures of Risk: A Coherent Representation of Subjective Risk Aversion},
  journal = {Journal of Banking \& Finance},
  year    = {2002},
  volume  = {26},
  number  = {7},
  pages   = {1505--1518}
}

@article{Kusuoka2001,
  author  = {Kusuoka, Shigeo},
  title   = {On Law Invariant Coherent Risk Measures},
  journal = {Advances in Mathematical Economics},
  year    = {2001},
  volume  = {3},
  pages   = {83--95}
}

@book{deBoor2001,
  author    = {de Boor, Carl},
  title     = {A Practical Guide to Splines},
  edition   = {Revised},
  publisher = {Springer},
  year      = {2001}
}
\end{document}